\documentclass[iop,twocolappendix,numberedappendix,appendixfloats]{emulateapj}

\usepackage{graphicx,threeparttablex,longtable,array} 
\usepackage{amssymb,amsmath,multirow,gensymb,mathrsfs}
\usepackage{enumitem,multirow,rotating}
\usepackage[caption=false]{subfig} 
\usepackage{natbib}
\usepackage{ragged2e,epstopdf}
\defcitealias{Sadavoy18b}{Paper I}
\defcitealias{Sadavoy18c}{Paper II}
\newcommand{\Jybeam}{\mbox{Jy beam$^{-1}$}}
\newcommand{\mJybeam}{\mbox{mJy beam$^{-1}$}}
\newcommand{\uJybeam}{\mbox{$\mu$Jy beam$^{-1}$}}
\newcommand{\um}{\mbox{$\mu$m}}

\newcommand{\Msun}{\mbox{M$_{\odot}$}}
\newcommand{\Mjupiter}{\mbox{M$_{\text{Jupiter}}$}}
\newcommand{\cmg}{\mbox{cm$^2$ g$^{-1}$}}

\newcommand{\Tbol}{\mbox{T$_\text{bol}$}}

\newcommand{\Lsun}{\mbox{L$_{\odot}$}}

\newcommand{\Pobs}{\mbox{$\mathcal{P}_{obs}$}}
\newcommand{\PI}{\mbox{$\mathcal{P}_I$}}
\newcommand{\sPI}{\mbox{$\sigma_{\mathcal{P}I}$}}
\newcommand{\sPIsq}{\mbox{$\sigma^2_{\mathcal{P}I}$}}
\newcommand{\PF}{\mbox{$\mathcal{P}_F$}}
\newcommand{\sPF}{\mbox{$\sigma_{\mathcal{P}F}$}}
\newcommand{\gdratio}{\mbox{$\mathcal{R}_{GD}$}}

\begin{document}

\title{Dust Polarization Toward Embedded Protostars in Ophiuchus with ALMA. III. Survey Overview}

\author{Sarah I. Sadavoy\altaffilmark{1$\dagger$,2},
Ian W. Stephens\altaffilmark{1},
Philip C. Myers\altaffilmark{1},
Leslie Looney\altaffilmark{3},
John Tobin\altaffilmark{4},
Woojin Kwon\altaffilmark{5,6},
Beno\^{i}t Commer\c{c}on\altaffilmark{7},
Dominique Segura-Cox\altaffilmark{8},
Thomas Henning\altaffilmark{9},
Patrick Hennebelle\altaffilmark{10,11},
	 }

\footnotetext[$\dagger$]{Hubble Fellow}
	 
\altaffiltext{1}{Harvard-Smithsonian Center for Astrophysics, 60 Garden Street, Cambridge, MA, 02138, USA}
\altaffiltext{2}{Department for Physics, Engineering Physics and Astrophysics, Queen's University, Kingston, ON, K7L 3N6, Canada}
\altaffiltext{3}{Department of Astronomy, University of Illinois, 1002 West Green Street, Urbana, IL, 61801, USA}
\altaffiltext{4}{National Radio Astronomy Observatory, Charlottesville, VA 22903, USA}
\altaffiltext{5}{Korea Astronomy and Space Science Institute (KASI), 776 Daedeokdae-ro, Yuseong-gu, Daejeon 34055, Republic of Korea}
\altaffiltext{6}{Korea University of Science and Technology (UST), 217 Gajang-ro, Yuseong-gu, Daejeon 34113, Republic of Korea}
\altaffiltext{7}{Universit\'{e} Lyon I, 46 All\'{e}e d'Italie, Ecole Normale Sup\'{e}rieure de Lyon, Lyon, Cedex 07, 69364, France}
\altaffiltext{8}{Centre for Astrochemical Studies, Max-Planck-Institute for Extraterrestrial Physics, Giessenbachstrasse 1, 85748, Garching, Germany}
\altaffiltext{9}{Max-Planck-Institut f\"{u}r Astronomie (MPIA), K\"{o}nigstuhl 17, D-69117 Heidelberg, Germany}
\altaffiltext{10}{Universit\'{e} Paris Diderot, AIM, Sorbonne Paris Cit\'{e}, CEA, CNRS, 91191, Gif-sur-Yvette, France}
\altaffiltext{11}{LERMA (UMR CNRS 8112), Ecole Normale Sup\'{e}rieure, 75231, Paris Cedex, France}


\date{Received ; accepted}

\begin{abstract}
We present 0.25\arcsec\ resolution (35 au) ALMA 1.3 mm dust polarization observations for 37 young stellar objects (YSOs) in the Ophiuchus molecular cloud.  These data encompass all the embedded protostars in the cloud and several Flat and Class II objects to produce the largest, homogeneous study of dust polarization on disk scales to date.  The goal of this study is to study dust polarization morphologies down to disk scales.  We find that 14/37 (38\%) of the observed YSOs are detected in polarization at our sensitivity.  Nine of these sources have uniform polarization angles and four sources have azimuthal polarization structure.  We find that the sources with uniform polarization tend to have steeper inclinations ($> 60$\degree) than those with azimuthal polarization ($< 60$\degree).  Overall, the majority (9/14) of the detected sources have polarization morphologies and disk properties consistent with dust self-scattering processes in optically thick disks.   The remaining sources may be instead tracing magnetic fields.  Their inferred field directions from rotating the polarization vectors by 90\degree\ are mainly poloidal or hourglass shaped.  We find no evidence of a strong toroidal field component toward any of our disks. For the 23 YSOs that are undetected in polarization, roughly half of them have 3-sigma upper limits of $< 2\%$.  These sources also tend to have inclinations $< 60$\degree\ and they are generally compact.  Since lower inclination sources tend to have azimuthal polarization, these YSOs may be undetected in polarization due to unresolved polarization structure within our beam.  We propose that disks with inclinations $> 60$\degree\ are the best candidates for future polarization studies of dust self-scattering as these systems will generally show uniform polarization vectors that do not require very high resolution to resolve. We release the continuum and polarization images for all the sources with this publication.  Data from the entire survey can be obtained from Dataverse.

\end{abstract} 


\section{Introduction\label{Intro}}

Interstellar magnetic fields in molecular clouds are most often characterized through sensitive observations of dust polarization.  The polarization signature is attributed to non-spherical dust grains that partially align with their short axes parallel to an external magnetic field due to radiative alignment torques (RATs) from an anisotropic radiation field \citep{Andersson15}.  Thus, dust polarization is expected to trace the morphology of the plane-of-sky magnetic field, with the polarization vectors parallel to the field direction from dust extinction and perpendicular to the field direction from thermal dust emission.  

One key goal of dust polarization studies is to trace magnetic field structure from the scales of molecular clouds ($\sim 10$ pc) to the scales of planet-forming disks ($\lesssim 100$ au).  In particular, dust polarization observations at early stages of the star formation process are necessary to understand the role of magnetic fields in both star and disk formation.  Numerous observations of embedded young stars, hereafter protostars, show polarization on the scales of their surrounding dense cores or dense envelopes \citep[e.g.,][]{Matthews09, Dotson10, Hull14, Galametz18}.  These detections suggest that the natal environment that produces the young stars and their disks are magnetized.  Nevertheless, there has been limited work tracing dust polarization down to the scales of the disks.  Polarization detections toward protostellar disks require high resolution observations at (sub)millimeter wavelengths to resolve the disk through the dense cloud and the surrounding dusty envelope.  Previous studies using the CARMA and SMA detected polarization only toward a few of the brightest protostellar disks \citep[e.g.,][]{Rao14, Stephens14, SeguraCox15, FernandezLopez16}.   The polarization detections from these studies had limited sensitivity and resolution, making their inferred field morphologies inconclusive.

The Atacama Large Millimeter/submillimeter Array (ALMA) has changed the landscape for observations of dust polarization on disk scales.  ALMA has the resolution and sensitivity to detect dust polarization toward a large number of disks for the first time.  Initial studies with ALMA have detected dust polarization toward a wide range of protostellar and protoplanetary disks \citep[e.g.,][]{Kataoka16hd, Stephens17, Alves18, Ohashi18}.  These polarization signatures, however, can arise from mechanisms other than grain alignment with a magnetic field.  In particular, large dust grains in disks can produce detectable polarization via self-scattering processes \citep[e.g.,][]{Kataoka15, Kataoka16, Pohl16, Yang16, Yang17} or dust grains can align themselves with the gradient of the radiation field \citep[hereafter, k-RAT alignment, e.g.,][]{LazarianHoang07,Tazaki17} or via collisions with gas flows \citep{Gold52, Yang19} and thereby produce a polarized signature.  A number of observations show polarization consistent with these other mechanisms \citep[e.g.,][]{Kataoka16hd, Kataoka17, Hull18, Harris18, Harrison19}.  Only a few studies have found polarization attributed to magnetic fields in disks \citep[e.g.,][]{Lee18, Sadavoy18b,Alves18, Ohashi18, Kwon19}. 

Most ALMA studies of dust polarization focused on one disk or a small sample of disks, and they also primarily selected disks that are among the biggest and brightest systems.   As a result, these studies are non-representative of typical disk properties.  To improve upon these initial studies, we conducted the first large, homogeneous dust polarization study of young protostellar disks with ALMA.  For this project, we observed all the embedded stars in the Ophiuchus molecular cloud in Band 6 (1.3 mm) dust polarization at a common resolution and sensitivity.  Since Ophiuchus is a nearby molecular clouds \citep[$d = 140$ pc;][]{OrtizLeon18}, it is an excellent target to obtain high resolution dust polarization observations of protostellar disks and their inner envelopes.   

We presented the first results in \citep[][hereafter, Paper I]{Sadavoy18b} and \citet[][hereafter, Paper II]{Sadavoy18c}.  Here, we present the observations from the entire study and release the full data products.   This paper is structured as follows; in Section \ref{data} we describe the source selection, observations, imaging techniques, and the polarization debias corrections.  In Section \ref{results}, we give an overview of the continuum and polarization detection statistics.  In Section \ref{products}, we describe the data products released with this paper and show the polarization maps for the detected sources.  In Section \ref{mech_discuss}, we employ a morphological analysis to determine the polarization mechanisms behind the polarization detections.  We focus primarily on polarization from magnetic fields and polarization from self-scattering.    In Section \ref{discussion}, we discuss poloidal and toroidal fields in disks and compare our small-scale observations to observations of magnetic fields on larger scales.   We also discuss the disk properties and the protostellar multiplicity in Ophiuchus.  Finally, we give our conclusions in Section \ref{summary}.


\section{Data}\label{data}

\subsection{Source Selection}

We selected 26 protostellar systems (Class 0 and Class I) from the surveys of \citet{Enoch09}, \citet{Evans09}, and \citet{ConnelleyGreene10}.   Table \ref{source_list} lists the sources in our sample.  Column 1 gives the field name based on their numerical identification in the cores to disks (c2d) survey \citep{Evans09} or in common literature.  Since VLA 1623 and IRAS 16293-2422 have bright companions at $\gtrsim$ 5\arcsec\ separation, we observed both sources with separated pointings so that we could detect their polarization within the inner third of the primary beam (as required by ALMA specifications for polarization data).  These are denoted with ``a'' and ``b'' in the field name.  Column 2 gives other common names of the sources from the literature.   Columns 3 and 4 give the phase center for each field, and column 5 gives the region of Ophiuchus in which the source is found following the boundaries in \citet{Young06} and \citet{Pattle15}.  Column 6 lists the classification of the source from the literature.  Finally, column 7 lists other known YSOs that are detected in each field.  

{\setlength{\extrarowheight}{0.8pt}%
\begin{table*}[h!]
\caption{Source List}\label{source_list}
\begin{tabular}{lllllll}
\hline\hline
Field & Central Source Name(s)\tablenotemark{a}	& \multicolumn{2}{c}{Phase Center (ICRS)}	& 	Region & Class\tablenotemark{b}	 & Other Known Sources\tablenotemark{c}\\
		  &							& RA (h,m,s)	& Dec ($\degree$,$\prime$,$\prime\prime$) &	           & 						& \\
\hline
c2d\_811	& GSS 30 IRS1, Oph-emb-8		& 16:26:21.35	&   -24:23:04.3			& L1688 Oph A	& I 			& GSS 30 IRS3\\
c2d\_822	& Oph-emb-9, GY 30			& 16:26:25.46	&  -24:23:01.3			& L1688 Oph A	& I 			& $\cdots$ \\
c2d\_831	& GY 91, Oph-emb-22			& 16:26:40.46	&  -24:27:14.3			& L1688 Oph A & I 			& $\cdots$\\ 
c2d\_857	& WL 16, GY 182,  Oph-emb-21 	& 16:27:02.32	&  -24:37:27.2			& L1688 Oph E	& I 			& $\cdots$\\
c2d\_862	& Oph-emb-6, GY 197, LFAM 26	& 16:27:05.24	&  -24:36:29.6			& L1688 Oph E	& I 			& $\cdots$\\
c2d\_867	& WL 17,	GY 205, Oph-emb-20  	& 16:27:06.75	&  -24:38:14.8			& L1688 Oph E	& I 			& $\cdots$\\ 
c2d\_871	& Elias 29, Oph-emb-16, WL 15,	& 16:27:09.40	&  -24:37:18.6			& L1688 Oph E	& I 			& $\cdots$\\
c2d\_885	& IRS 37, ISO Oph 124, Oph-emb-11 & 16:27:17.58	&  -24:28:56.2			& L1688 Oph B	& I 			& IRS 39 \\
c2d\_890	& IRS 42, GY 252, Oph-emb-28 	& 16:27:21.45	&  -24:41:43.0			& L1688 Oph F	& I 			& $\cdots$\\
c2d\_892	& Oph-emb-5					& 16:27:21.82	&  -24:27:27.6			& L1688 Oph B	& I 			& $\cdots$\\
c2d\_894	& Oph-emb-12, CRBR 2422.8		& 16:27:24.58	&  -24:41:03.1			& L1688 Oph F	& I 			& $\cdots$\\
c2d\_899	& IRS 43, GY 265, Oph-emb-14, YLW 15 	& 16:27:26.92	&  -24:40:50.58		& L1688 Oph F	& 0/I 		& GY 263 \\
c2d\_901	& IRS 44, GY 269, Oph-emb-13 	& 16:27:27.99	&  -24:39:33.4			& L1688 Oph F	& I 			& $\cdots$\\
c2d\_902	& IRS 45, Elias 32, Oph-emb-19 	& 16:27:28.44	&  -24:27:20.8			& L1688 Oph B	& I 			& VSSG 18 B \\
c2d\_904	& IRS 47, GY 279, Oph-emb-26 	& 16:27:30.17	&  -24:27:43.2			& L1688 Oph B	& I 			&  $\cdots$ \\
c2d\_954	& Oph-emb-1, Oph MMS 126		& 16:28:21.58	&  -24:36:23.6			& L1688		& 0 			&  $\cdots$\\
c2d\_963	& Oph-emb-18					& 16:28:57.85	&  -24:40:54.9			& L1688		& I 			&  $\cdots$\\
c2d\_989	& IRS 63, Oph-emb-17			& 16:31:35.65	&  -24:01:29.3			& L1709		& I 			&  $\cdots$\\
c2d\_990	& Oph-emb-4					& 16:31:36.77	&  -24:04:19.8			& L1709  		& I 			&  $\cdots$\\
c2d\_991	& Oph-emb-25, ISO Oph 200		& 16:31:43.75	&  -24:55:24.6			& L1689S		& I 			&  $\cdots$\\
c2d\_996	& Oph-emb-7					& 16:31:52.06	&  -24:57:26.0			& L1689S		& I 			&  $\cdots$\\
c2d\_998	& Oph-emb-15					& 16:31:52.45	&  -24:55:36.2			& L1689S		& I 			&  $\cdots$\\
c2d\_1003	& IRS 67, Oph-emb-10			& 16:32:00.99	&  -24:56:42.6			& L1689S		& I 			&  $\cdots$\\
c2d\_1008a & IRAS 16293-2422A, Oph-emb-2	& 16:32:22.87	&  -24:28:36.45	  		& L1689N 	  & 0 			& IRAS 16293-2422B\\
c2d\_1008b & IRAS 16293-2422B, Oph-emb-2	& 16:32:22.62	&  -24:28:32.5	  		& L1689N 	  & 0	 		& IRAS 16293-2422A\\
VLA1623a	& VLA 1623W, Oph-emb-3		& 16:26:25.64	&  -24:24:29.3	  		& L1688 Oph A  & 0			& VLA 1623A/B\\
VLA1623b	& VLA 1623A/B, Oph-emb-3		& 16:26:26.35	&  -24:24:30.55	  		& L1688 Oph A  & 0 			& VLA 1623W, VLA 1623NE\\
IRAS16288& ISO Oph 210, IRAS 16266-2450E & 16:32:02.22	&  -24:56:16.8			& L1689S		& I 			& $\cdots$\\
\hline
\end{tabular}
\begin{tablenotes}[normal,flushleft]
\item \tablenotemark{a} Common names for the sources at the phase center of the field.  Names taken from SIMBAD \citep{Wenger00} and are ordered first by the name we adopt in the paper and then other common names in alphabetical order. ``Oph-emb-'' is from \citet{Enoch09}, ``GY'' is from \citet{GreeneYoung92}, ``WL'' is from \citet{WilkingLada83}, ``LFAM'' is from \citet{Leous91}, ``Elias'' is from \citet{Elias78}, ``IRS'' is from \citet{Wilking89}, ``ISO Oph'' is from \citet{Bontemps01}, VSSG is from \citet{Shirono11}, and ``Oph MMS'' is from \citet{Stanke06}.
\item \tablenotemark{b} Original source classification as Class 0 or Class I based on \citet{Enoch09}, \citet{Evans09}, \citet{Connelley08}, and \citet{HsiehLai13}.  See text for details.
\item \tablenotemark{c} Additional known sources in each field that were detected.
\end{tablenotes}
\end{table*}
}

This sample includes all the known Class 0 objects and the embedded Class I sources in Ophiuchus.  Source classifications were determined using the standard definitions of the infrared spectral index, $\alpha_{IR}$, and the bolometric temperature as summarized in \citet{Evans09}.  For the Class I sources, we selected all the stars that had unambiguous envelope detections based on previous single-dish observations \citep[e.g.,][]{Enoch09} to ensure that the stars are young and still embedded.  Three YSOs (c2d\_839, c2d\_914, c2d\_922) have ``envelope'' designations \citep{Enoch09}, but appear more evolved with $\alpha_{IR} < 0.3$ and $\Tbol \gtrsim 600$ K \citep{Evans09, Dunham15}.  We excluded these sources from our sample.  We also added ISO Oph 210 (IRAS 16266-2450E\footnote{This field was mislabeled as IRAS16288 for the ALMA observations.  We continue to use the mislabeled name for consistency with the archive.}) to our source list.  This object was not featured in the ``c2d'' catalogue, but is listed as a YSO in \citet{HsiehLai13}.  

We note, however, that the original classifications for the YSOs in Ophiuchus have come under considerable question \citep[e.g.,][]{McClure10}.  In particular, \citet{McClure10} found that 16/26 ``embedded'' objects in Ophiuchus were at more evolved stages using infrared spectroscopy.  They attributed the difference to substantial foreground extinctions such that measurements of the infrared spectral index are unreliable for Ophiuchus.  In Appendix \ref{indiv}, we revisit the source classifications for the targets in our sample using archival data to help inform their evolutionary stages.

\begin{figure*}[h!]
\includegraphics{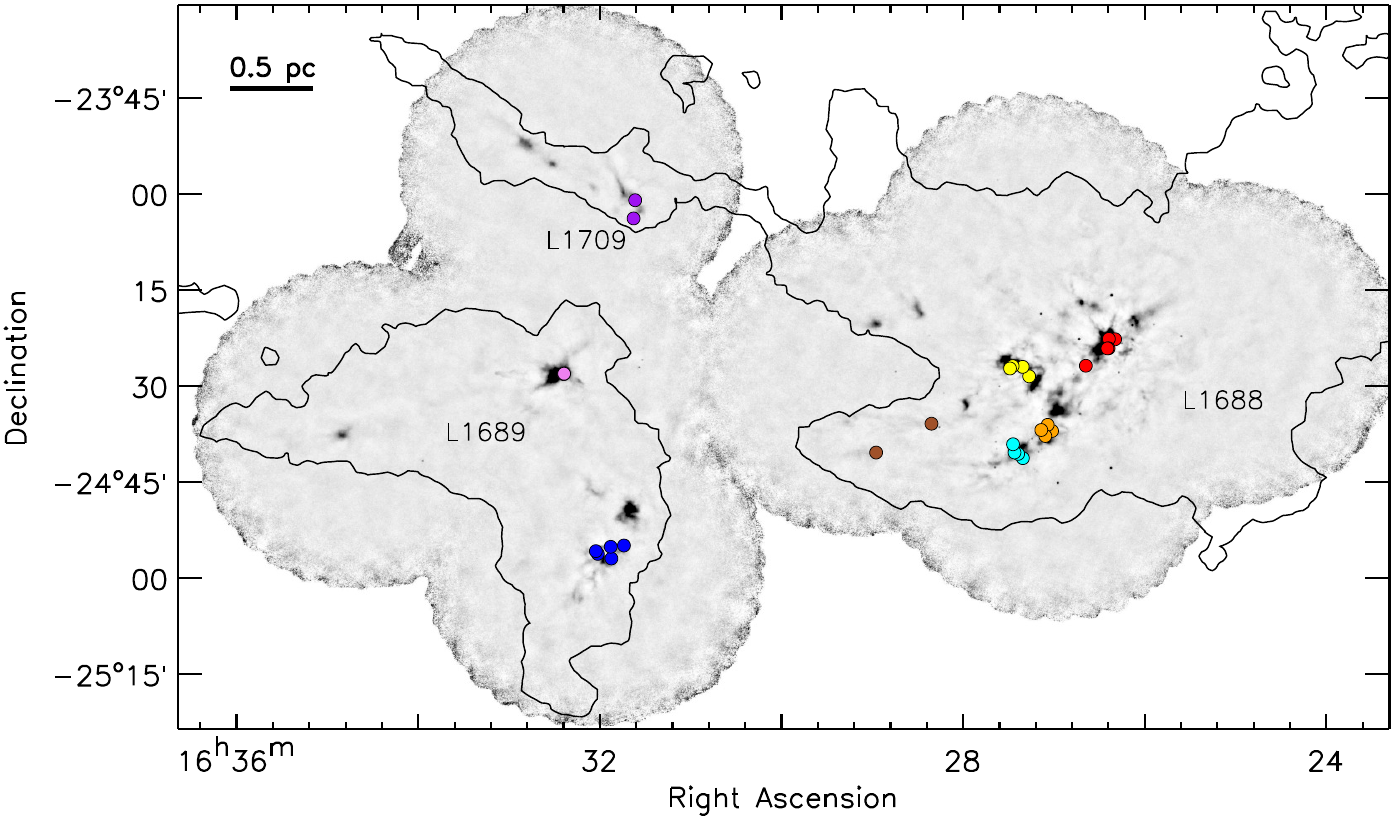}
\caption{SCUBA-2 850 \um\ map of the Ophiuchus molecular cloud \citep{Pattle15, Kirk18} with coloured points showing the 28 fields in our survey.  The points are coded for Oph A (red), Oph B (yellow), Oph E (orange), Oph F (cyan), L1688 (brown), L1689S (blue), L1689N (pink), and L1709 (purple).  The symbol size does not represent the primary beam.  Contours show $A_V = 5$ from the COMPLETE survey near-infrared (2MASS) extinction map \citep{Ridge06}.  The main star-forming complexes in Ophiuchus are also labeled.   \label{pointings}}
\end{figure*}

Figure \ref{pointings} shows the position of each of our 28 fields on a SCUBA-2 850 \um\ from \citet{Pattle15}.  The SCUBA-2 data are part of data release 3 (DR3), with details on the reduction and imaging given in \citet{Kirk18}.  The full Ophiuchus data set has been cropped to focus on L1688, L1689, and L1709.  The pointings are colour-coded by the regions given in Table \ref{source_list}.

\subsection{ALMA Observations}

The 28 pointings were observed at 1.3 mm in full polarization on 2017 May 20, July 11, and July 14 on shared tracks as part of the Cycle 3 program 2015.1.01112.S\footnote{This project was previously observed on 2016 July 22, 25, and 26, but these observations failed QA0 due to a faint polarization calibrator.  We do not include these observations in our analysis.}.   The baselines ranged from 15.1 m to 1124.3 m for the May observations and 16.7 m to 2647.3 m for the July observations.  The precipitable water vapour ranged from 0.38 mm on 11 July, 0.94 mm on 20 May, and 1.22 mm on 14 July.  There were 46 antenna on 20 May, 43 antenna on 11 July, and 42 antenna on 14 July.  The correlator was configured to the standard full polarization setting for Band 6, with each baseband set to 1.875 GHz bandwidth and 64 channels,  centered at 224 GHz, 226 GHz, 240 GHz, and 252 GHz. 

For all tracks, J1517-2422 was used for bandpass calibration, J1625-2527 was the calibrator for complex gain and phase, and J1549+0237 was the polarization leakage calibrator.  J1517-2422 was also used for absolute flux calibration on the 20 May and 14 July executions, with  J1733-1304 as the absolute flux calibration on 11 July.  To ensure sufficient parallactic angle coverage with J1549+0237, each track was observed over $3-4.5$ hours, using two or three consecutive sessions.  Typically, all the sessions contained observations of each calibrator.  The third session from 11 July did not include the bandpass and flux calibrators, and instead used the bandpass solutions from the second session and the complex gain calibrator for absolute flux calibration. The total time spent on each target field is $\approx 7$ minutes.  

The observations were manually calibrated using CASA 4.7.2 by the observatory.  The standard calibrations (bandpass, flux, gain) were applied first, followed by the final polarization calibration that was performed on each session separately.  The expected flux uncertainty for the final observations is $\sim 10$\%.  We examined the observed flux for the bandpass and leakage calibrators against the expected flux for the date of observations (using analysis utilities task \texttt{getALMAflux} in CASA) and found that they agreed within 10\%.  The complex gain calibrator had larger uncertainties of $\lesssim 20$\%, but this source is monitored less frequently with only one detection between our observation epochs and a single detection several months before our observations.   Hereafter, we quote statistical uncertainties unless stated otherwise.

\subsection{ALMA Imaging}

We imaged each field interactively using \texttt{clean}.  Many of the fields have peak Stokes I signal to noise (S/N) $>100$ and allowed self calibration.  We tested phase-only self calibration to the Stokes I data for all the observations with S/N $>20$ and only applied the ones with good gain solutions based on a visual inspection.  The Stokes Q, U, and V data did not benefit from self calibration.  We ran self-calibration iteratively with decreasing solution intervals until the Stokes I noise did not improve.  Table \ref{obs_summary} summarizes the self calibration iterations for each of the fields.  For most sources one or two rounds of phase-only self calibration were necessary to reach a consistent noise level.  The first round of self calibration used solution intervals equivalent to the entire scan length and the second round used solution intervals of 30.25 s, which corresponds to roughly half a scan.  For c2d\_989, VLA1623a, and VLA1623b, we applied a third round of phase-only self calibration with a solution interval of 15 s, or 5 integrations.  Additional iterations with shorter solution intervals did not improve the map noise.  For c2d1008a and c2d1008b, we applied for the third round an amplitude and phase self calibration with a solution interval equal to the scan length \citepalias[see also,][]{Sadavoy18b, Sadavoy18c}.  In all self calibration rounds, we visually inspected the data after applying the solutions to ensure the noise was dropping with each iteration.  

{\setlength{\extrarowheight}{0.8pt}%
\begin{table}[h!]
\caption{Imaging Summary}\label{obs_summary}
\begin{tabular}{lllllcl}
\hline\hline
Field & Self Cal\tablenotemark{a}	&	$\sigma_I$\tablenotemark{b}	&	$\sigma_Q$\tablenotemark{b}	&	$\sigma_U$\tablenotemark{b}	&   $\sPI$\tablenotemark{c}  & Beam\tablenotemark{d}	\\
		  &				&	\multicolumn{4}{c}{(\uJybeam)} 		& 	(arcsec)	\\
\hline
c2d\_811	& 	2p			& 	32			& 	25			& 	27			&  26 	& 0.27$\times$0.21 \\ 
c2d\_822	& 	1p			& 	34			&  	25			& 	25			&  25 	& 0.27$\times$0.21 \\
c2d\_831	& 	2p			& 	34			&  	28			&	27			& 27	         & 0.29$\times$0.24 \\
c2d\_857	& 	1p			& 	30			&  	26			& 	26			& 26		& 0.27$\times$0.20 \\
c2d\_862	& 	2p			& 	35			&  	28			& 	28			& 28		& 0.28$\times$0.24 \\
c2d\_867	& 	2p			& 	34			&  	27			& 	28			& 27		& 0.27$\times$0.21 \\
c2d\_871	& 	2p			& 	31			&  	27			& 	27			& 27		& 0.26$\times$0.20 \\
c2d\_885	& 	1p			& 	30			&  	28			& 	28			& 28		& 0.28$\times$0.24 \\
c2d\_890	& 	1p			& 	30			&  	28			& 	27 			& 28		& 0.27$\times$0.21 \\
c2d\_892	& 	none			& 	26			& 	26			& 	26			& 26		& 0.27$\times$0.21 \\
c2d\_894	& 	1p			& 	31			&  	27			& 	27 			& 27		& 0.27$\times$0.21 \\
c2d\_899	& 	1p			& 	30			&  	26			& 	26			& 26		& 0.27$\times$0.21 \\
c2d\_901	& 	1p			& 	29			&  	25			& 	25 			& 25		& 0.27$\times$0.21 \\
c2d\_902	& 	none			& 	29			&  	26			& 	26			& 26		& 0.27$\times$0.21 \\
c2d\_904	& 	1p			& 	29			&  	26			& 	26 			& 26		& 0.27$\times$0.21 \\
c2d\_954	& 	2p			& 	30			&  	26			& 	26			& 26		& 0.27$\times$0.21 \\
c2d\_963	& 	none			& 	29			& 	26			& 	26			& 26		& 0.27$\times$0.21 \\
c2d\_989	& 	3p			& 	71			& 	27			& 	27 			& 27		& 0.27$\times$0.20 \\
c2d\_990	& 	1p			& 	31			&  	27			& 	26			& 27		& 0.28$\times$0.23 \\
c2d\_991	& 	1p			& 	31			&  	26			&	27			& 27		& 0.27$\times$0.20 \\
c2d\_996	& 	none			& 	31			&  	27			& 	27 			& 27		& 0.27$\times$0.20 \\
c2d\_998	& 	none			& 	30			&  	26			& 	26			& 26		& 0.27$\times$0.20 \\
c2d\_1003	& 	2p			& 	48			& 	28			& 	28 			& 28		& 0.26$\times$0.20 \\
c2d\_1008a & 	2p,1ap		& 	280			& 	28			& 	30			& 29		& 0.28$\times$0.24 \\
c2d\_1008b & 	2p,1ap		& 	250			&  	29			& 	29			& 29		& 0.28$\times$0.23 \\
c2d\_1008\tablenotemark{e} & 	2p,1ap	& 	280	&  	25			& 	25			& 25		& 0.28$\times$0.23 \\
VLA1623a	& 	3p			& 	56			&  	27	  		& 	27			& 27		& 0.27$\times$0.21 \\
VLA1623b	& 	3p			& 	71			&  	27	  		& 	27 			& 27		& 0.27$\times$0.21 \\
IRAS16288& 	none			& 	32			&  	26			& 	26			& 26		& 0.27$\times$0.21\\
\hline
\end{tabular}
\begin{tablenotes}[normal,flushleft]
\item \tablenotemark{a} Successive self calibration iterations with phase (p) or amplitude and phase (ap).  Sources that were not self calibrated have ``none''.
\item \tablenotemark{b} Map sensitivity at the phase center.
\item \tablenotemark{c} The sensitivity in polarized intensity is estimated from the average of $\sigma_Q$ and $\sigma_U$.
\item \tablenotemark{d} Beam size of the field in the final images.
\item \tablenotemark{e} For the mosaic map of fields c2d\_1008a and c2d\_1008b.   
\end{tablenotes}
\end{table}
}

For each round of self calibration and for the final deep map, we applied \texttt{clean} with Briggs weighting and a robust parameter of 0.5 and a UVtaper of 0.1\arcsec.  The UV taper was applied to better recover extended emission.  We used interactive \texttt{clean} for all the Stokes I maps and non-interactive \texttt{clean} for Stokes Q and U.  Interactive \texttt{clean} was necessary for the Stokes I maps as some fields contained faint, diffuse extended emission.  We also used the multi-scale option during the Stokes I \texttt{clean} for those sources with substantial extended emission.  Table \ref{obs_summary} gives the map sensitivities for the Stokes I, Q, and U maps.  We exclude the Stokes V data, because they are cannot be calibrated.  The typical map sensitivity is 27 \uJybeam, although several fields are dynamic range limited and have substantially higher rms values.  The map resolution for each field is typically $0.27\arcsec \times 0.21\arcsec$ and the maximum recoverable scale is $\sim 2.6\arcsec$, which means we are sensitive to physical scales between $\sim 33-360$ au (assuming a distance of 140 pc).

For the c2d\_1008a and c2d\_1008b fields, we performed a final deep \texttt{clean} on the self-calibrated data with both fields to produce a mosaic.  The emission detected in both fields overlap within the inner half of their primary beams, and the individual fields gave consistent Stokes maps (see Appendix \ref{offaxis}).  For the mosaic field, we use \texttt{clean} with the same robust, multiscale, and uvtaper values as the individual fields.  The map sensitivities for the mosaicked Stokes Q and U maps improved by $\sim 14$\%, but we found a similar map sensitivity for mosaic Stokes I as the map is still limited by dynamic range.   Hereafter, we use the field name ``c2d\_1008'' for the mosaic maps.

\subsection{Debias Correction}\label{debias}

Polarization is measured from the quadrature sum of Stokes Q and U ($\Pobs = \sqrt{Q^2+U^2}$), which always yields positive values.  The quadrature sum subsequently biases the measured polarized intensity to higher values.  This effect is most pronounced for weak Stokes Q or U data (e.g., $\PI/\sPI < 4$), where noise will more significantly boost the inferred polarized signal \citep[e.g.,][]{Vaillancourt06}.

We follow \citet{HullPlambeck15} and calculate the debiased polarization intensities using a probability density function (PDF). Briefly, this method computes the probability that the observed, uncorrected polarization $\PI_{,obs}$ has a true, debiased polarization $\PI$ with,
\begin{multline}
\mbox{PDF}(\PI|\Pobs,\sPI) = \\
 \frac{\Pobs}{\sPIsq}I_0(x)\exp\left[-\frac{(\Pobs^2+\PI^2)}{2\sPIsq}\right], \label{eq_pdf}
 \end{multline}
where \sPI\ is the sensitivity of the polarization data and $I_0(x)$ is the Bessel function for the Bessel parameter, $x~=~\Pobs\PI/\sPIsq$.  The statistically most likely debiased polarization (\PI) will produce a peak in Equation \ref{eq_pdf}.  Thus, we measured \PI\ by taking the minimum of $-\mbox{PDF}(\PI|\Pobs,\sPI)$ using the routine \texttt{tnmin}\footnote{This routine uses a Truncated-Newton method to minimize a function \citep{Markwardt09}. } in IDL for each pixel with S/N $<9$; above this threshold, we used the standard maximum likelihood calculation \citep{Vaillancourt06},
\begin{equation}
\PI_{,bright} = \sqrt{Q^2+U^2 - \sPIsq},
\end{equation}
as the two debiasing methods are identical for such bright emission.  For $\sPI$, we use the mean rms from the Stokes Q and U observations, taking into account that the map sensitivities vary across the primary beam.

The polarization position angle, $\theta$, and its uncertainty $\sigma_{\theta}$ are given by \citep[e.g.,][]{Coude19},
\begin{eqnarray}
\theta &=& \frac{1}{2}\tan^{-1}\frac{U}{Q} \label{ang_eq} \\
\sigma_{\theta} &=& \frac{1}{2}\frac{\sqrt{(U\sigma_Q)^2 + (Q\sigma_U)^2}}{Q^2 + U^2}.
\end{eqnarray}
The polarization position angles are defined from $-90\degree$ to 90\degree, North to East.   The polarization fraction, \PF, and its uncertainty, \sPF, are defined by,
\begin{eqnarray} 
\PF &=& \frac{\PI}{I} \\
\sPF &=& \PF\sqrt{\left(\frac{\sPI}{\PI}\right)^2+\left(\frac{\sigma_I}{I_{ }}\right)^2}.
\end{eqnarray}
We calculate $\sigma_{\theta}$ and $\sPF$ for each pixel to account for the changing map sensitivity across the primary beam.   We note that the ALMA Technical Handbook for Cycle 6 gives a 1$\sigma$ instrumental polarization error of $0.03$\%\ for compact sources and a 1$\sigma$ error of 0.1\%\ for extended sources within the inner third of the primary beam  FWHM.  This instrument polarization error is not included in \sPF.

Hereafter, we use the term ``e-vector'' to correspond to the observed polarization position angles (e.g., no rotation has been applied) and the term ``b-vector'' when the position angles are rotated by 90\degree\ to show the inferred field direction.  We note, however, that the position angles are not full vectors because we cannot determine the true direction.


\section{Results}\label{results}

\subsection{Stokes I Overview}\label{cont_overview}

In Stokes I emission, we identify 41 distinct compact sources in the 27 fields.  Table \ref{cont_results} lists the coordinates and broad properties of each detected Stokes I object.  Column 1 gives the field name and Column 2 gives the source identification based on a common name in the literature (see Table \ref{source_list}).  For multiple sources in the field, the sources are ordered by their distance from the phase center.  Columns 3 and 4 give the coordinates of the peak emission.  Column 5 gives the source classification adopted for this study (see Appendix \ref{indiv}), where Class 0 sources and Class I sources are protostars accreting material from dense envelopes, Class II sources are pre-main sequence stars that are no longer accreting, and Flat spectrum sources are the transition stage between Class I and Class II when the envelope is being cleared out \citep[e.g., see][for more details]{Evans09}.  Columns 6 and 7 give the peak flux density and its corresponding error, $\sigma_{peak}$, which we take as the Stokes I uncertainty at the position of the source.  For sources near the phase center, this value is equivalent to $\sigma_I$ given in Table \ref{obs_summary}.  Columns 8-11 give the total flux, size, and position angle from simple Gaussian fits to the sources in the image plane using \texttt{imfit} in CASA. We use Gaussian fits to approximate the source properties, but note that some objects may be disks, envelopes, or a combination of both. 

{\setlength{\extrarowheight}{0.8pt}%
\begin{table*}
\caption{Continuum Results}\label{cont_results}
\begin{tabular}{lllllllllll}
\hline\hline
\multirow{2}{*}{Field} & \multirow{2}{*}{Source\tablenotemark{a}}	& \multicolumn{2}{c}{Peak Position (J2000)}	& 	\multirow{2}{*}{ID\tablenotemark{b}}  & Peak\tablenotemark{c}	& $\sigma_{peak}$\tablenotemark{c}	 & Flux\tablenotemark{d} & a\tablenotemark{d} & b\tablenotemark{d} & PA\tablenotemark{d} \\
		  &						& RA (h,m,s)	& Dec ($\degree$,$\prime$,$\prime\prime$) &	& \multicolumn{2}{c}{(\mJybeam)}   & {(mJy)} &  (mas) & (mas) & (deg) \\
\hline
c2d\_811	& GSS 30 IRS 1		& 16:26:21.357 	& -24:23:04.899 	& I 	& 12.8 	& 0.032	& 13.4$\pm$0.2 & 85.1$\pm$8.2 & 23.3$\pm$14 & 117$\pm$7 \\
		& GSS 30 IRS 3 		& 16:26:21.719 	& -24:22:50.967 	& I 	& 51.8 	& 0.076	& 158.5$\pm$2.0 & 585$\pm$8.6 & 192$\pm$4.0 & 110$\pm$0.4 \\
c2d\_822	& Oph-emb-9 			& 16:26:25.474		& -24:23:01.845	& I 	& 28.6 	& 0.034	& 45.0$\pm$0.2 & 229$\pm$1.4 & 89.6$\pm$3.0 & 28$\pm$0.9  \\
c2d\_831	& GY 91 				& 16:26:40.469		& -24:27:14.953 	& F 	& 14.8 	& 0.034	& 88.8$\pm$6.5 & 791$\pm$64 & 670$\pm$57 & 155$\pm$29 \\
c2d\_857	& WL 16 				& 16:27:02.327		& -24:37:27.709	& II 	& 4.4		& 0.030	& 4.35$\pm$0.07 & $\cdots$ & $\cdots$ & $\cdots$  \\
c2d\_862	& Oph-emb-6 			& 16:27:05.250		& -24:36:30.163	& I 	& 26.0	& 0.035	& 53.1$\pm$0.3 & 413$\pm$2.7 & 100$\pm$3.0 & 169$\pm$0.1 \\
		& ALMAJ162705.5 		& 16:27:05.509		& -24:36:32.269	& G	& 0.34	& 0.038	& 0.43$\pm$ 0.02 & $\cdots$ & $\cdots$ & $\cdots$  \\
c2d\_867	& WL 17 				& 16:27:06.764 	& -24:38:15.489	& F 	& 27.2	& 0.034	& 51.3$\pm$0.6 & 232$\pm$6.3 & 183$\pm$7.1 & 62$\pm$6  \\
c2d\_871	& Elias 29 			& 16:27:09.415		& -24:37:19.253	& I 	& 15.9 	& 0.031	& 17.2$\pm$0.2 & 74.9$\pm$8.4 & 63.0$\pm$7.0 & 105$\pm$55 \\
c2d\_885	& IRS 37-A 			& 16:27:17.581 	& -24:28:56.835 	&I 	& 9.8 	& 0.030	& 11.1$\pm$0.1	& 120$\pm$4.1 & 42.8$\pm$15 & 8$\pm$4 \\
		& IRS 37-B			& 16:27:17.442 	& -24:28:56.565 	&I	& 0.89 	& 0.030	& 1.02$\pm$0.06 & 100$\pm$36 & 90$\pm$54 & 171$\pm$82 \\
		& IRS 37-C			& 16:27:17.417 	& -24:28:55.053 	&I	& 0.81 	& 0.030	& 0.91$\pm$0.06 & $\cdots$ & $\cdots$ & $\cdots$  \\
		& ALMAJ162717.7		& 16:27:17.722 	& -24:28:52.839 	&G	& 0.27 	& 0.032	& 0.34$\pm$0.05 & $\cdots$ & $\cdots$ & $\cdots$  \\
		& IRS 39				& 16:27:18.472		& -24:29:06.393	& II	& 0.76	& 0.091	& 0.76$\pm$0.10 & $\cdots$ & $\cdots$ & $\cdots$  \\ 
c2d\_890	& IRS 42				& 16:27:21.456		& -24:41:43.545	& II	&11.5	& 0.030	& 12.2$\pm$0.06 & $\cdots$ & $\cdots$ & $\cdots$  \\
c2d\_892	& Oph-emb-5 			& $\cdots$ 		& $\cdots$ 		& S	& $\cdots$ & $\cdots$ & $\cdots$ & $\cdots$ & $\cdots$ & $\cdots$  \\
c2d\_894	& Oph-emb-12 			& 16:27:24.587		& -24:41:03.717 	& I	& 4.3		& 0.031 	& 4.43$\pm$0.07 & $\cdots$ & $\cdots$ & $\cdots$  \\
c2d\_899	& IRS 43-A 			& 16:27:26.906		& -24:40:50.729 	& I	& 13.3	& 0.030  	& 15.1$\pm$0.3 & 117$\pm$12 & 67$\pm$13 & 121$\pm$12 \\
		& IRS 43-B 			& 16:27:26.914		& -24:40:51.305	& I	& 1.6		& 0.030  	& 1.88$\pm$0.05 & 109$\pm$14 & 92$\pm$13 & 119$\pm$72 \\
		& GY 263 				& 16:27:26.605		& -24:40:45.617	& II	& 7.0		& 0.036  	& 16.2$\pm$0.4 & 384$\pm$11 & 130$\pm$7.1 & 127$\pm$1 \\
c2d\_901	& IRS 44 				& 16:27:27.987		& -24:39:33.945	& I	& 10.8	& 0.029 	& 12.0$\pm$0.3 & 113$\pm$12 & 65$\pm$12 & 119$\pm$13  \\
c2d\_902	& IRS 45	 			& 16:27:28.441		& -24:27:21.669	& I	& 2.0	 	& 0.029 	& 2.25$\pm$0.07 & 99$\pm$12 & 55$\pm$45 & 25$\pm$26 \\
		& VSSG 18 B 			& 16:27:29.208		& -24:27:17.475  	& II	& 1.0 	& 0.048 	& 1.23$\pm$0.06 & 121$\pm$27 & 99$\pm$33 & 128$\pm$81  \\ 
c2d\_904	& IRS 47 				& 16:27:30.173 	& -24:27:43.835	& I	& 6.8 	& 0.029 	& 9.7$\pm$0.3	& 232$\pm$17 & 13$\pm$53 & 61$\pm$5 \\
		& ALMAJ162729.7 		& 16:27:29.750 	& -24:27:35.825 	& G	& 0.52 	& 0.041	& 0.79$\pm$0.10 & 186$\pm$47 & 164$\pm$61 & 77$\pm$89 \\
c2d\_954	& Oph-emb-1 			& 16:28:21.620 	& -24:36:24.199 	& 0	& 11.0	& 0.030 	& 12.5$\pm$0.1 & 134$\pm$4.8 & 48.3$\pm$5.4 & 115$\pm$2 \\
c2d\_963	& Oph-emb-18 			& 16:28:57.868 	& -24:40:55.373 	& II	& 1.9 	& 0.029 	& 2.3$\pm$0.1 & 143$\pm$22 & 83$\pm$49 & 30$\pm$26 \\
c2d\_989	& IRS 63 				& 16:31:35.657 	& -24:01:29.935 	& I	& 90.8 	& 0.071 	& 312$\pm$4.9 & 485$\pm$9.2 & 330$\pm$7.1 & 148$\pm$2  \\
c2d\_990	& Oph-emb-4 			& 16:31:36.782 	& -24:04:20.363 	& II	& 9.2 	& 0.031 	& 13.1$\pm$0.1& 256$\pm$4.1 & 55.6$\pm$6.7 & 78$\pm$1  \\
c2d\_991	& Oph-emb-25 			& 16:31:43.755 	& -24:55:24.947 	& F	& 8.6 	& 0.031 	& 9.1$\pm$0.07 & 71.0$\pm$6.7 & 32.2$\pm$13.4 & 142$\pm$9 \\
c2d\_996	& Oph-emb-7 			& 16:31:52.045 	& -24:57:26.383	& II	& 0.38 	& 0.031 	& 0.34$\pm$0.03  & $\cdots$ & $\cdots$ & $\cdots$  \\
c2d\_998	& Oph-emb-15 			& 16:31:52.444 	& -24:55:36.511 	& I	& 3.3		& 0.030 	& 3.7$\pm$0.1 	& 112$\pm$18 & 46$\pm$37 & 179$\pm$16 \\
c2d\_1003	& IRS 67-A 			& 16:32:00.987 	& -24:56:42.767 	& I	& 8.1		 & 0.048	& 8.5$\pm$0.1 & 59$\pm$13 & 50$\pm$20 & 74$\pm$70 \\
		& IRS 67-B 			& 16:32:00.977		& -24:56:43.487 	& I	& 46.9	 & 0.048	& 53.6$\pm$0.5  & 113$\pm$5.5 & 66.2$\pm$5.8 & 91$\pm$6 \\
c2d\_1008\tablenotemark{e} & IRAS 16293B  & 16:32:22.612 & -24:28:32.610  & 0	& 482.2  	& 0.286    &1400$\pm$31 & 363$\pm$12 & 342$\pm$11 & 125$\pm$42\\
		   & IRAS 16293A 		& 16:32:22.874		& -24:28:36.714 		& 0	& 184.2	& 0.308	& 1200$\pm$54 & 802$\pm$38 & 434$\pm$24 & 53$\pm$3\\
VLA1623a	& VLA 16239W 		& 16:26:25.631 	& -24:24:29.611 	& 0	& 16.3 	& 0.056	& 65.5$\pm$1.3 & 705$\pm$15 & 105$\pm$11 & 10$\pm$0.5 \\
VLA1623b	& VLA 1623B 			& 16:26:26.307		& -24:24:30.699	& 0	& 66.9	& 0.071	& 128$\pm$2.7 & 314$\pm$8.4 & 103$\pm$14 & 43$\pm$2 \\
		& VLA 1623A 			& 16:26:26.393		& -24:24:30.843	& 0	& 62.0	& 0.071	& 136$\pm$5.5 & 361$\pm$19 & 158$\pm$17 & 76$\pm$4 \\
		& VLA 1623NE 			& 16:26:27.423		& -24:24:18.279	& II	& 11.5	& 0.394	& 49$\pm$2 & 579$\pm$27 & 267$\pm$14 & 129$\pm$2 \\
IRAS16288	& ISO Oph 210 	& 16:32:02.214 	& -24:56:17.309 	& F	& 3.9 	& 0.032	& 5.2$\pm$0.1 & 185$\pm$11 & 103$\pm$8 & 108$\pm$4\\
			& ALMAJ163203.3	& 16:32:03.310		& -24:56:14.429	& G	& 0.62	& 0.084	& 0.86$\pm$0.13 & 174$\pm$51 & 50$\pm$85 & 11$\pm$65 \\ 
\hline
\end{tabular}
\begin{tablenotes}[normal,flushleft]
\item \tablenotemark{a} Identifier for individual detections.  Sources are ordered by closest to the phase center.
\item \tablenotemark{b} Adopted source classification for this study, where ``0'' indicates Class 0 YSO, ``I'' indicates Class I YSO, ``F'' indicates Flat YSO, ``II'' indicates Class II or Class III YSO, ``G'' indicates a galaxy, and ``S'' indicates a background star.  See Appendix \ref{indiv} for details.
\item \tablenotemark{c} Peak 1.3 mm flux density in the primary beam corrected maps.  The error in the peak flux represents the Stokes I uncertainty reported in Table \ref{obs_summary} scaled by the primary beam correction at the position of the source.
\item \tablenotemark{d} Gaussian fit results to the Stoke I continuum emission.  Values correspond to the total flux density and the deconvolved Gaussian semi-major axis ($a$), semi-minor axis ($b$), and position angle (PA).  Position angle is measured north to east.  Unresolved sources use ellipses.
\item \tablenotemark{e} Values correspond to an approximate Gaussian fit to the emission $> 100$ m\Jybeam.  
\end{tablenotes}
\end{table*}
}

Table \ref{det_summary} summarizes the detection properties for each field.  Column 2 gives the number of sources detected in each field and Column 3 indicates whether or not extended emission is seen in the field.  We define extended emission as non-compact or non-Gaussian flux either connected to a compact source or is found between sources.   In most cases, the extended emission, when detected, is spatially filtered such that we do not recover all the flux.    Column 4 gives notes on previous detections from the literature, where ``identified disk'' indicates that the source emission was previously attributed to a disk and ``potential disk'' indicates the previous disk classification was ambiguous.  Note that not all ``identified disks'' have been confirmed with Keplerian rotation.  

{\setlength{\extrarowheight}{0.8pt}%
\begin{table*}
\caption{Stokes I Detection Summary}\label{det_summary}
\begin{tabular}{llll}
\hline\hline
Field & Sources			&	Extended Emission\tablenotemark{b}		& 	Notes\tablenotemark{c}	\\
\hline
c2d\_811	& 	2			& 	Y			& Identified disk for GSS 30 IRS3 (1)			\\ 
c2d\_822	& 	1			& 	N			& Identified disk (2) 			\\
c2d\_831	& 	1			& 	N			& Identified disk (2,3) 			\\ 
c2d\_857	& 	1			& 	N			& Identified disk (4) 			\\  
c2d\_862	& 	2			& 	N			& Identified disk (2), one new source	\\
c2d\_867	& 	1			& 	N			& Identified disk (2,5) 	\\
c2d\_871	& 	1			& 	Y			& Identified disk (6,7)		\\
c2d\_885	& 	5			& 	N			& Four sources known  (2), one new source \\
c2d\_890	& 	1			& 	N			& Identified disk (2,8)		\\
c2d\_892	& 	0			& 	N			& $\cdots$	\\ 
c2d\_894	& 	1			& 	N			& Identified disk (8)	\\
c2d\_899	& 	3			& 	Y			& Known multiple (9), known circumbinary structure (10) \\  
c2d\_901	& 	1			& 	N			& $\cdots$ \\
c2d\_902	& 	2			& 	N			& $\cdots$ \\ 
c2d\_904	& 	2			& 	Y			& One new source \\ 
c2d\_954	& 	1			& 	N			& Potential disk (11)\\
c2d\_963	& 	1			& 	N			& $\cdots$\\
c2d\_989	& 	1			& 	N			& Identified disk (6,12) \\
c2d\_990	& 	1			& 	N			& $\cdots$ \\
c2d\_991	& 	1			& 	N			& Identified disk (2) \\
c2d\_996	& 	1			& 	N			& $\cdots$ \\
c2d\_998	& 	1			& 	N			& $\cdots$  \\
c2d\_1003	& 	2			& 	Y			& Known multiple (13), known circumbinary structure (14) \\ 
c2d\_1008\tablenotemark{d} & 	2	&Y			& Known multiple with substantial extended emission (11,15,16)    \\
VLA1623a	& 	3			& 	Y			& Known multiple, three identified disks (11,15,17)\\
VLA1623b	& 	4			& 	Y			& Known multiple, three identified disks (11,15,17) \\
IRAS16288& 	2			& 	N			& One new source \\ 
\hline
\end{tabular}
\begin{tablenotes}[normal,flushleft]
\item \tablenotemark{a} Number of compact, Gaussian-like objects in the field.
\item \tablenotemark{b} Extended (non-Gaussian) emission extending from a source or between sources.
\item \tablenotemark{c} (1) \citealt{Jorgensen09}, (2) \citealt{Cieza19}, (3) \citealt{vanderMarel19}, (4) \citealt{ResslerBarsony03}, (5) \citealt{SheehanEisner17}, (6) \citealt{Lommen08}, (7) \citealt{Miotello14}, (8) \citealt{vanKempen09}, (9) \citealt{Girart00}, (10) \citealt{Brinch16}, (11) \citealt{Chen13}, (12) \citealt{Andrews07}, (13) \citealt{ConnelleyGreene10}, (14) \citealt{ArturV18}, (15) \citealt{Looney00}, (16) \citealt{Jorgensen16}, (17) \citealt{Murillo13}
\item \tablenotemark{d} Applicable to both the individual c2d\_1008a and c2d\_1008b fields.
\end{tablenotes}
\end{table*}
}

Figure \ref{multiples} shows a wide view of the twelve fields that either had multiple sources or extended dust emission.  Eleven fields have multiplicity and eight fields have extended emission.  Most of the fields with extended emission also had multiple sources detected.   Field c2d\_871 was the only field to have a single detection and also extended emission.   The multiplicity statistics will be discussed in Section \ref{mf} and the extended emission will be discussed with their corresponding source in Appendix~\ref{indiv}.     

\begin{figure*}[h!]
\centering
\includegraphics[scale=0.49]{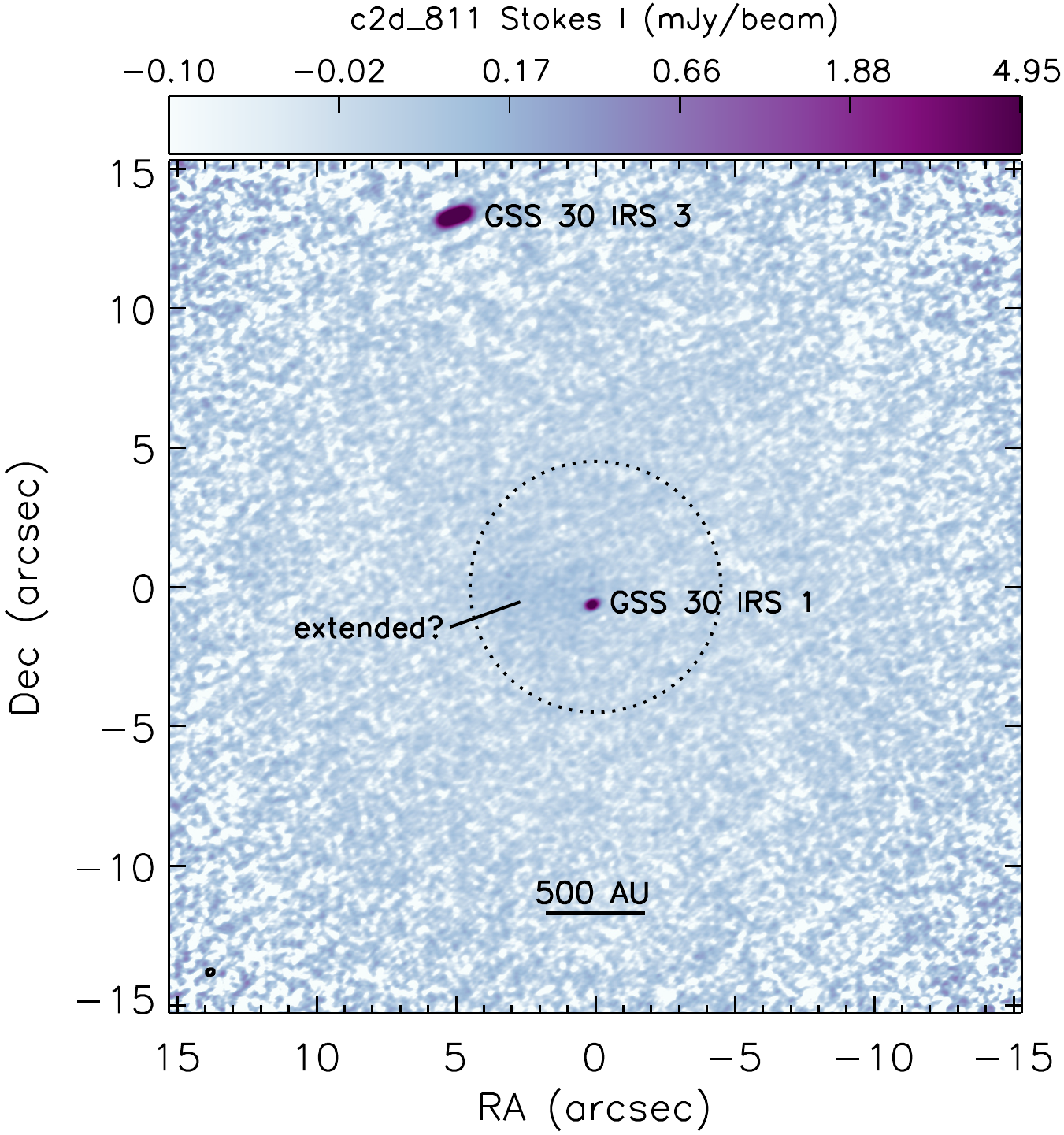}
\qquad
\includegraphics[scale=0.49]{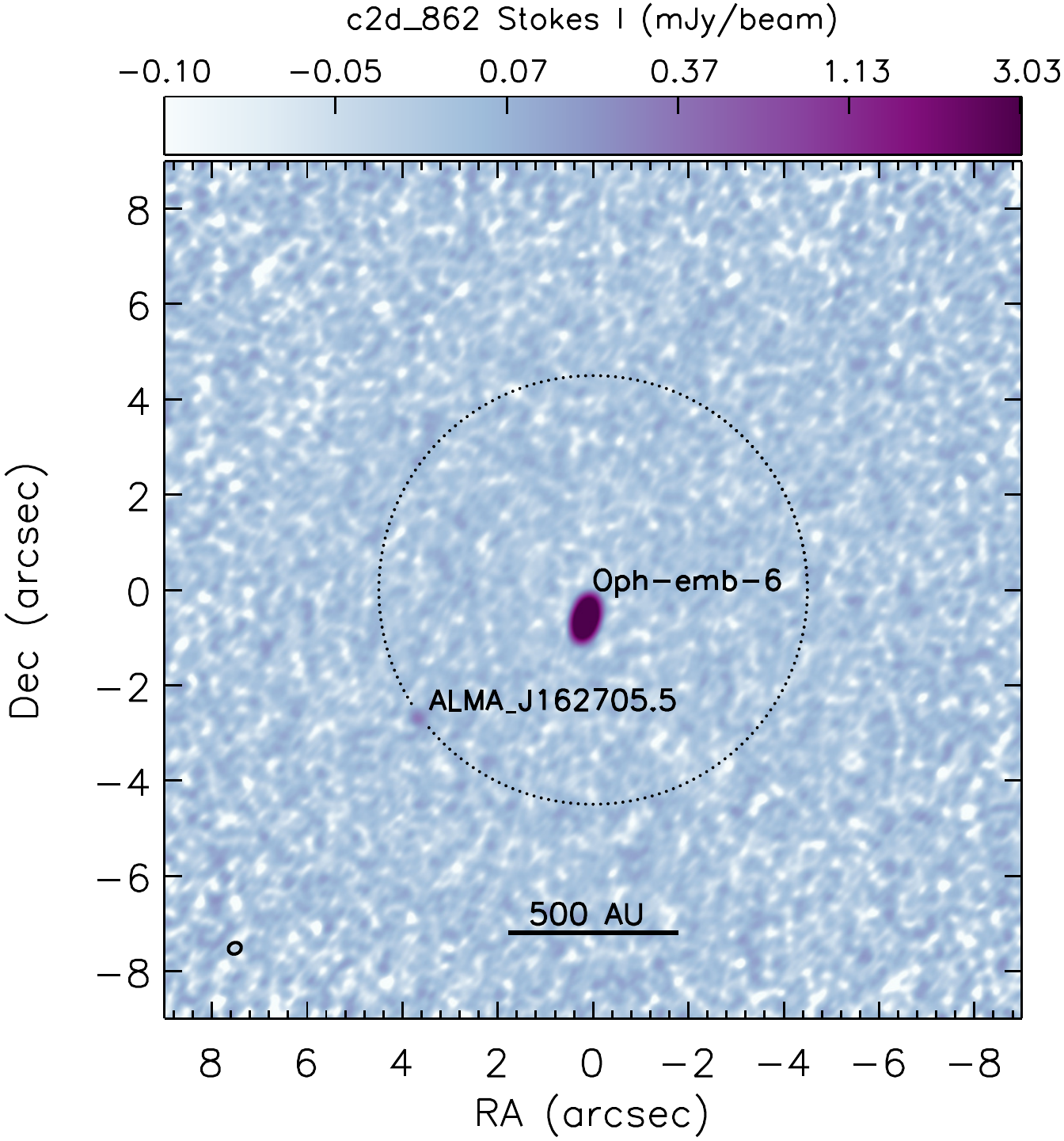}
\qquad
\includegraphics[scale=0.49]{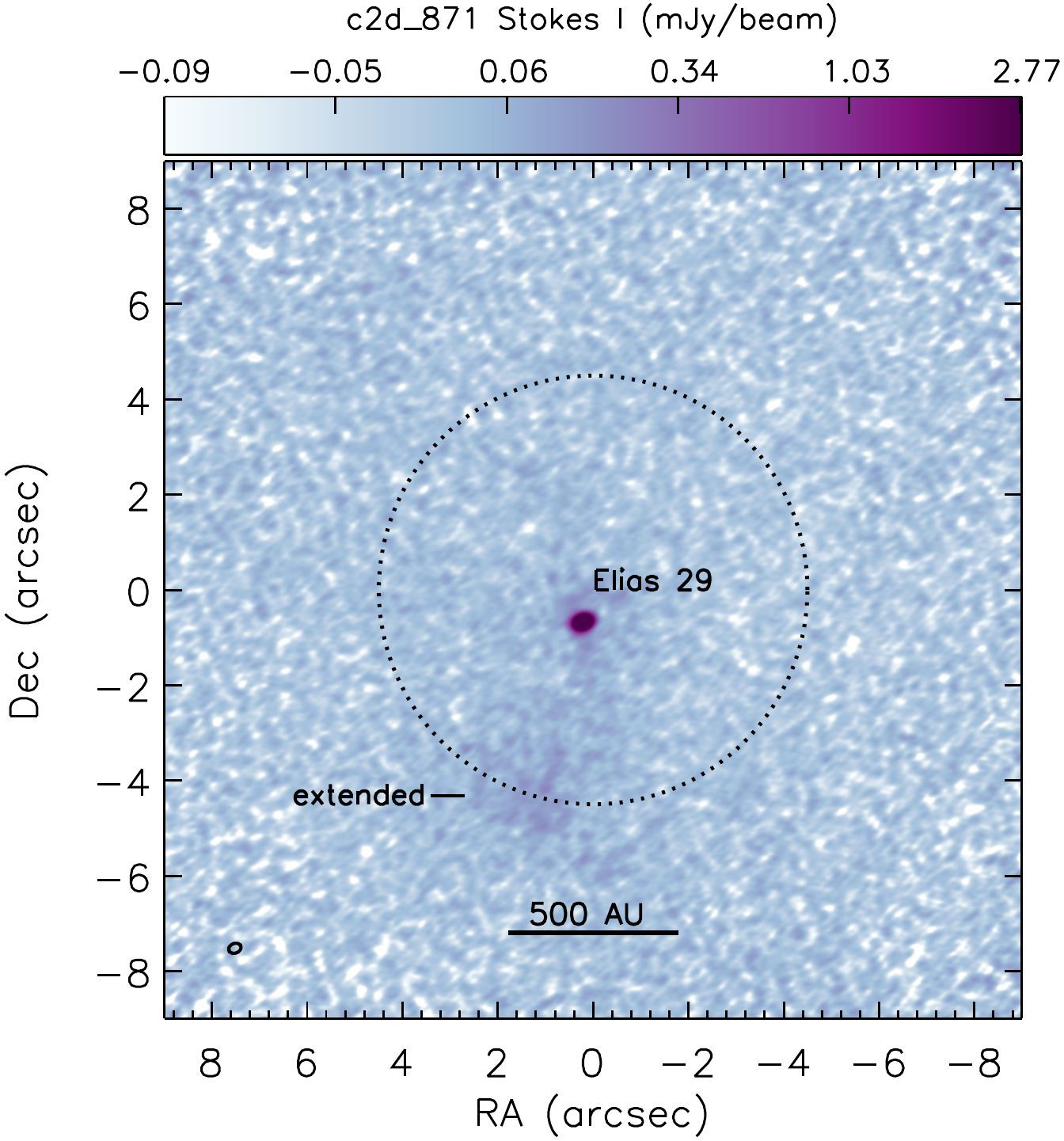}
\qquad
\includegraphics[scale=0.49]{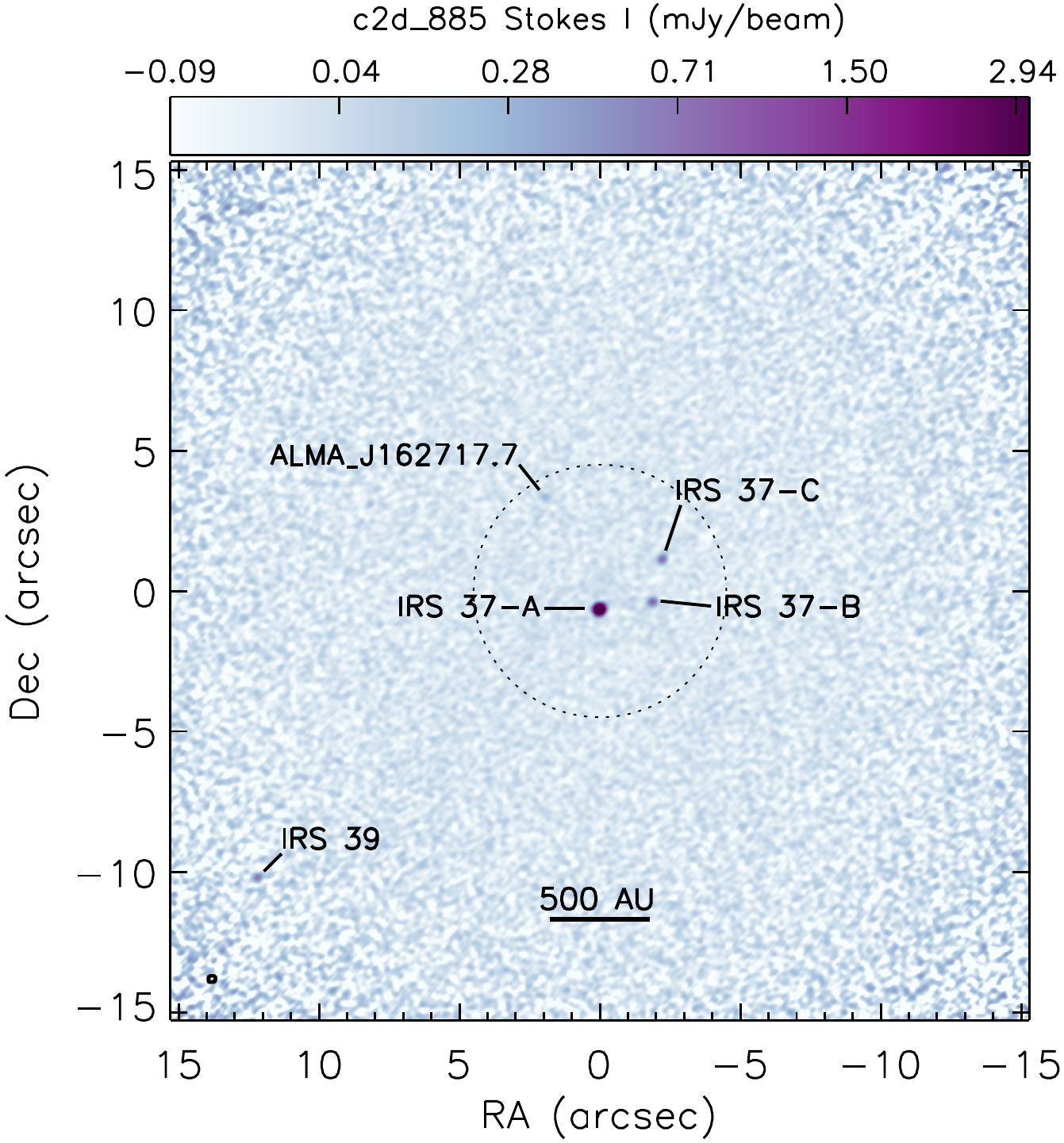}
\qquad
\includegraphics[scale=0.49]{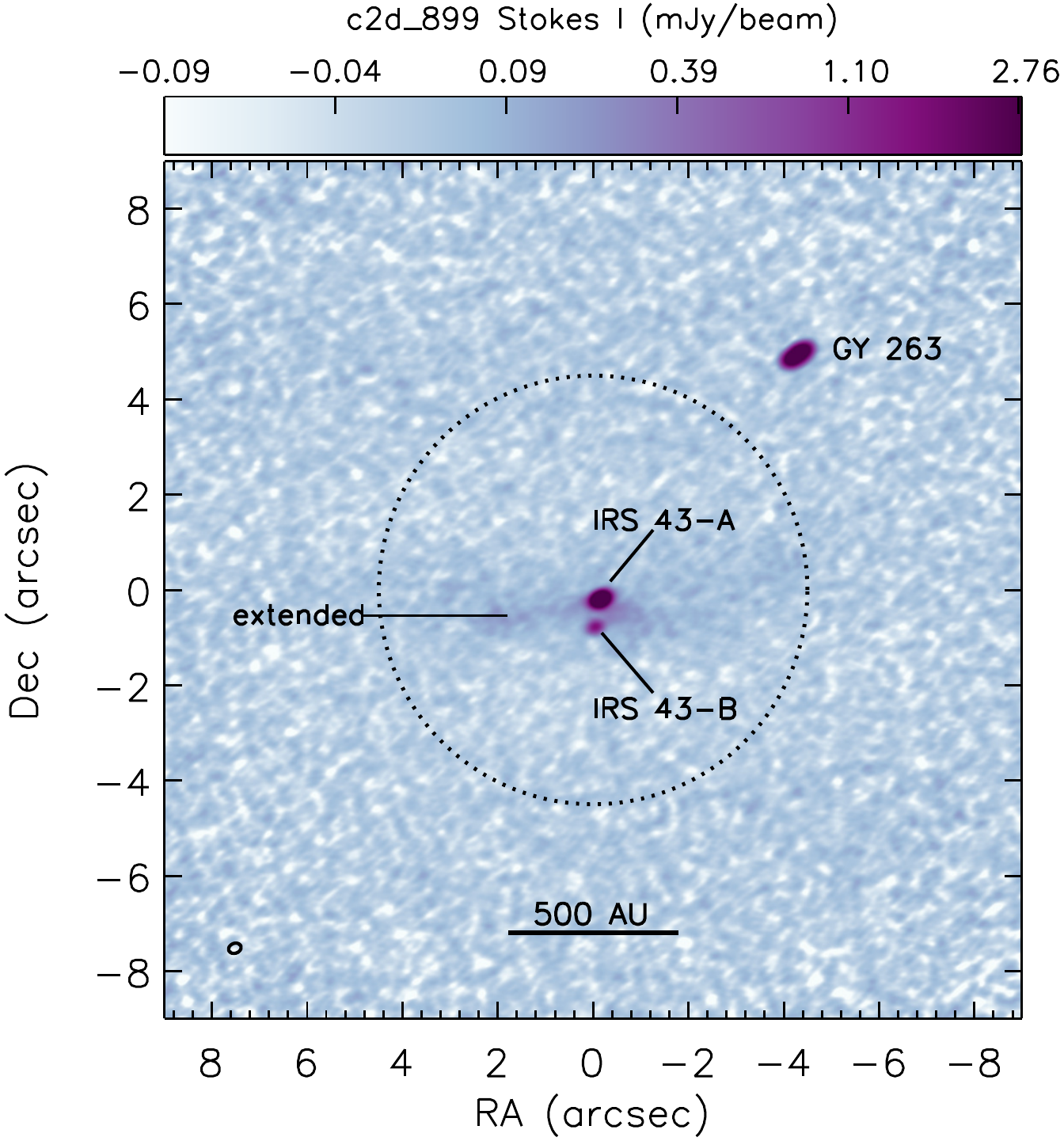}
\qquad
\includegraphics[scale=0.49]{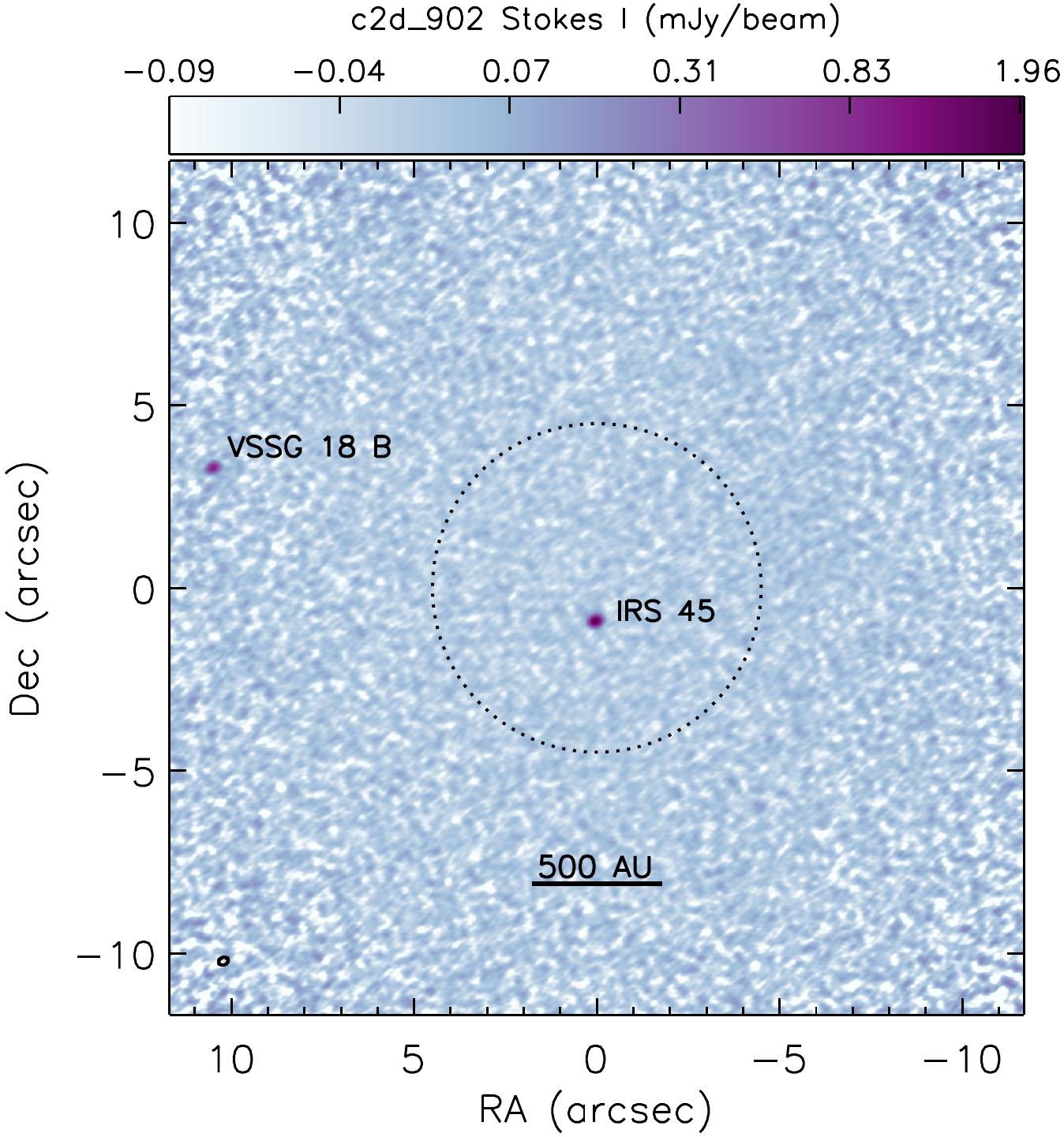}
\caption{Fields with multiple sources or extended emission.  The dotted circle shows the inner third of the primary beam FWHM (diameter $\approx 8.5$\arcsec) at the phase center of each field.  Small circle in the bottom-left corner shows the beam.  Source names are labeled following Table \ref{cont_results}.  Featured are fields c2d\_811, c2d\_862, c2d\_871, c2d\_885, c2d\_899, c2d\_902.  Note that the colour backgrounds apply different log scales to better show diffuse features.  Map centers correspond to the phase center in Table \ref{source_list}.
\label{multiples}}
\end{figure*}
\begin{figure*}
\ContinuedFloat
\centering
\includegraphics[scale=0.49]{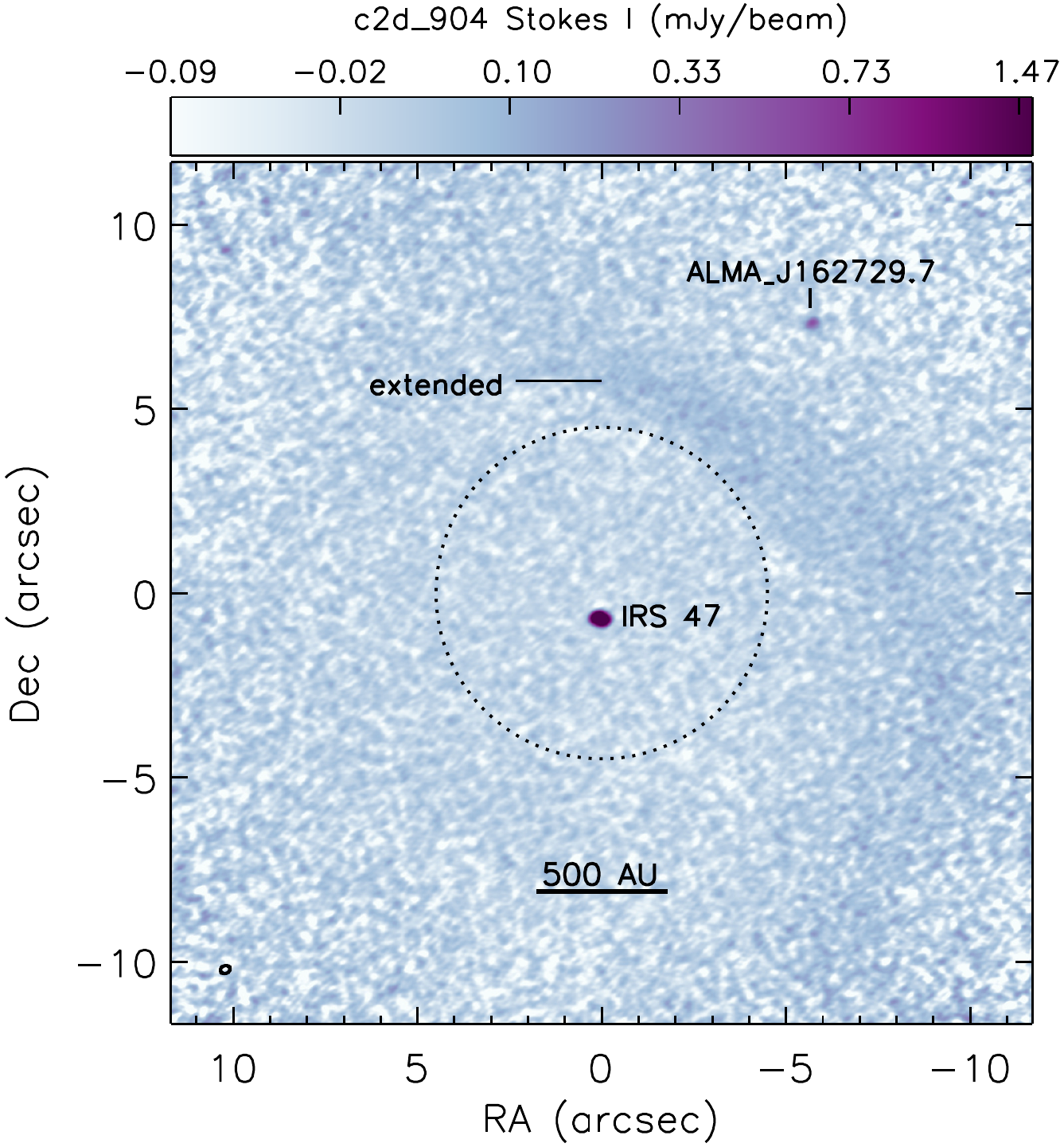}
\qquad
\includegraphics[scale=0.49]{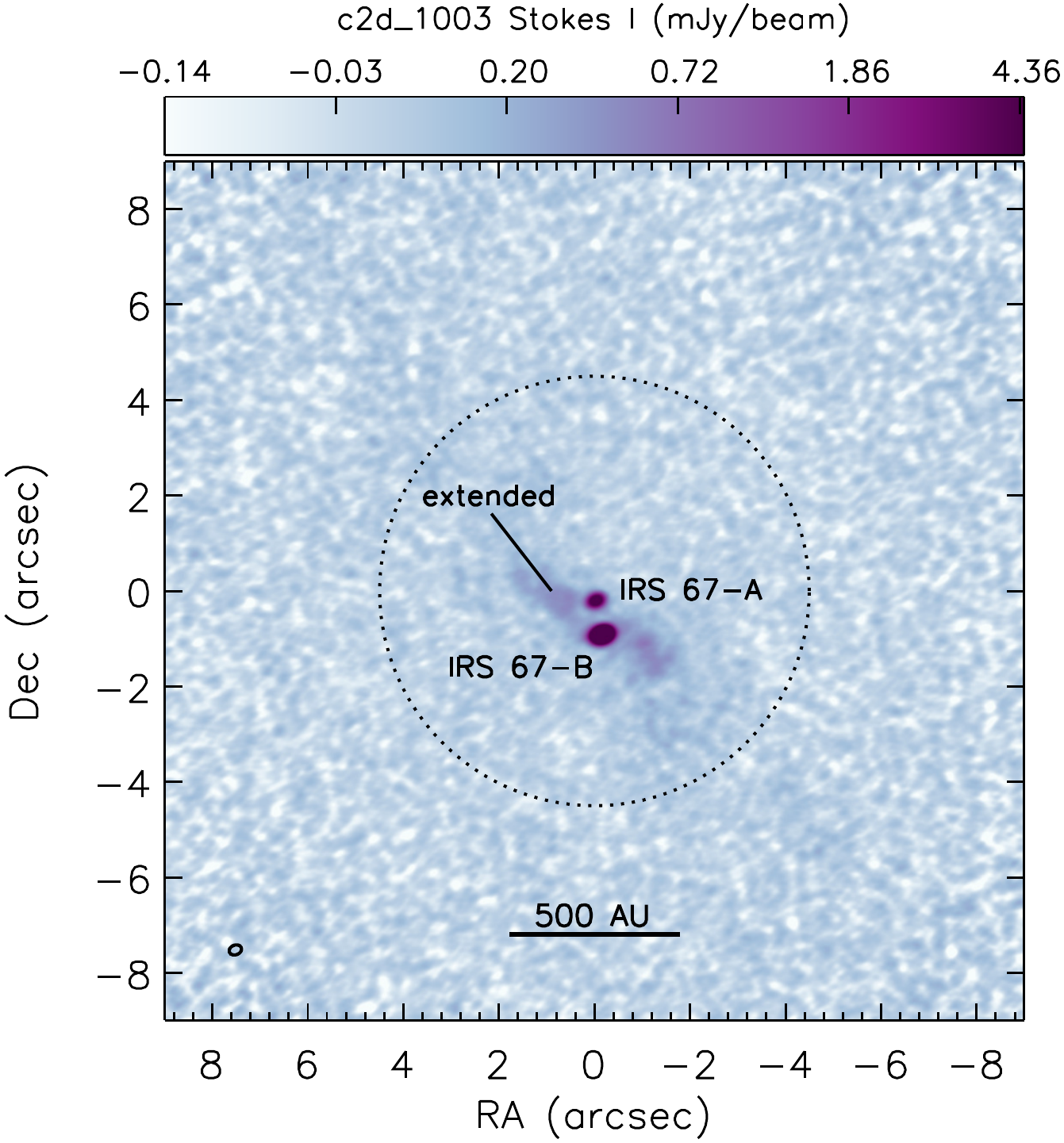}
\qquad
\includegraphics[scale=0.49]{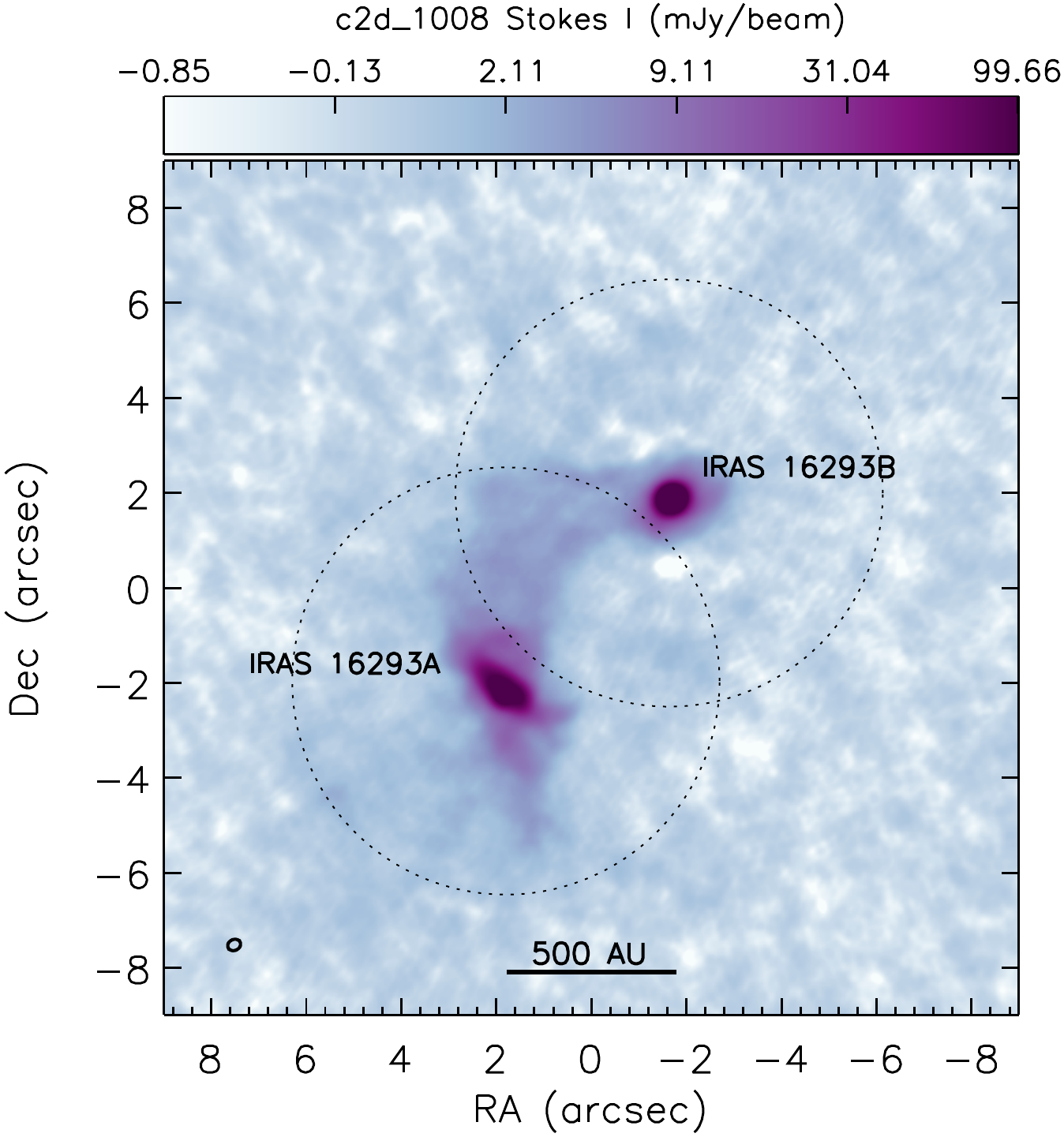}
\qquad
\includegraphics[scale=0.49]{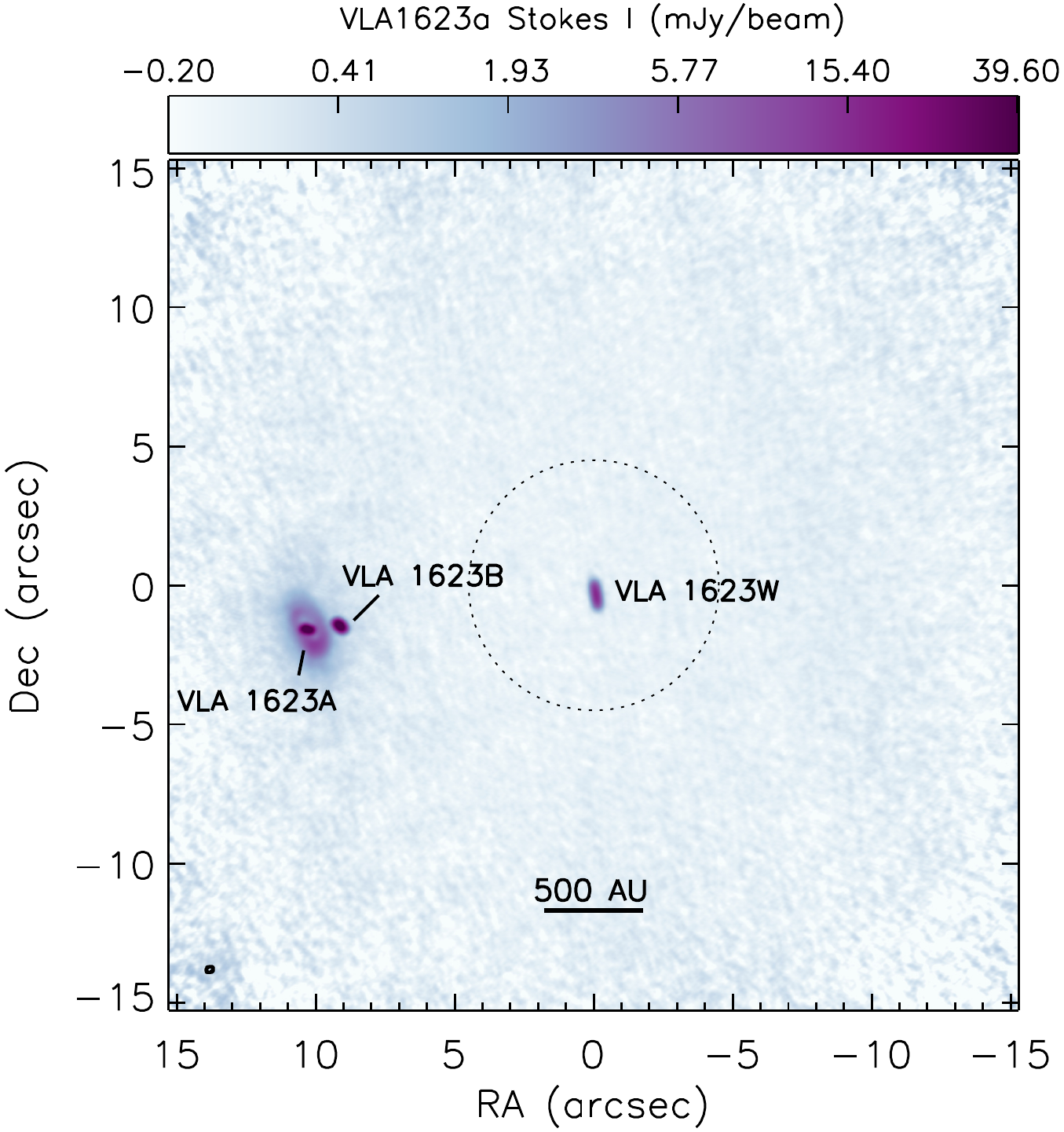}
\qquad
\includegraphics[scale=0.49]{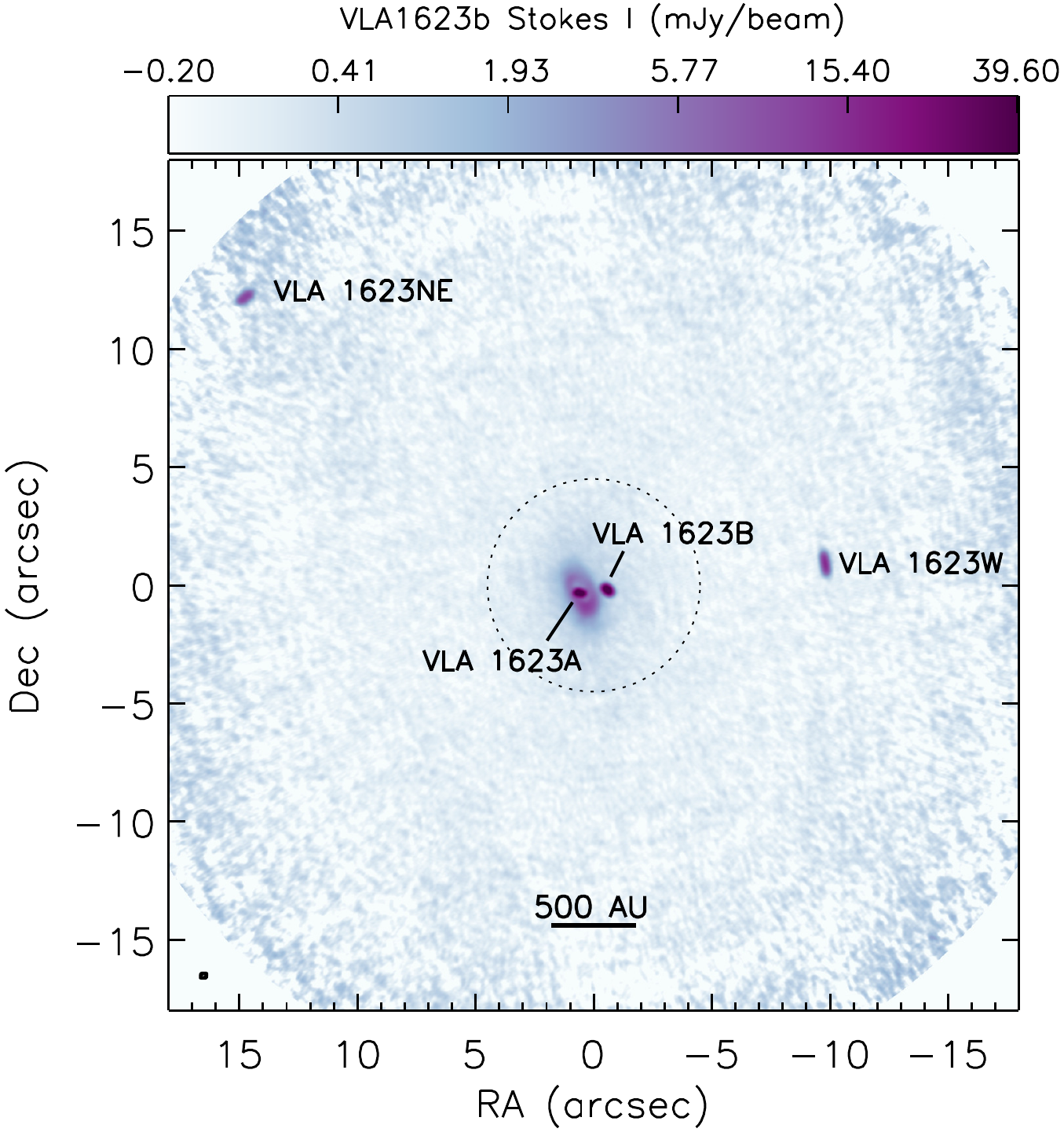}
\qquad
\includegraphics[scale=0.49]{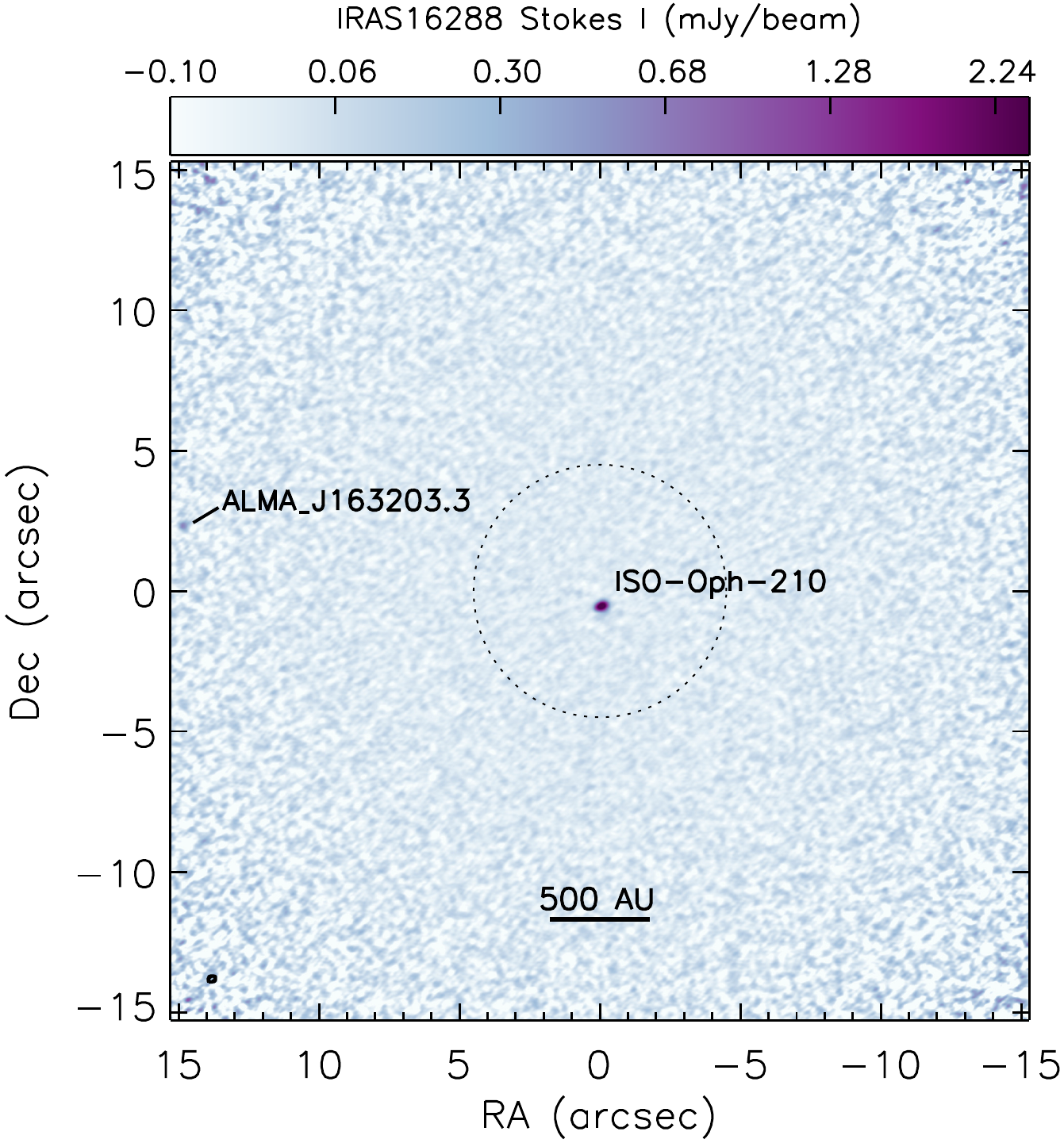}
\caption{Continued -  For fields c2d\_904, c2d\_1003,  c2d\_1008, VLA1623a, VLA1623b, IRAS16288.  For c2d\_1008, we show the inner third of the primary beam from the two fields centered on each protostar.}
\end{figure*}

\subsection{Polarization Overview}\label{pol_overview}

In this section, we give a brief overview of the polarization detection statistics for the entire sample.  We consider a source to have a robust polarization detection if its detection meets the following specific selection criteria:
\begin{enumerate}
\item $I/\sigma_I > 3$ and
\item $\PI/\sPI > 3$.
\item $\sigma_{\theta} < 10\degree$.
\end{enumerate}
These criteria select bright emission above the noise in both Stokes I and polarized intensity.   Of the 41 detected continuum sources, only 14 have polarization measurements that satisfy the above criteria.    

\begin{figure*}
\includegraphics[width=0.88\textwidth]{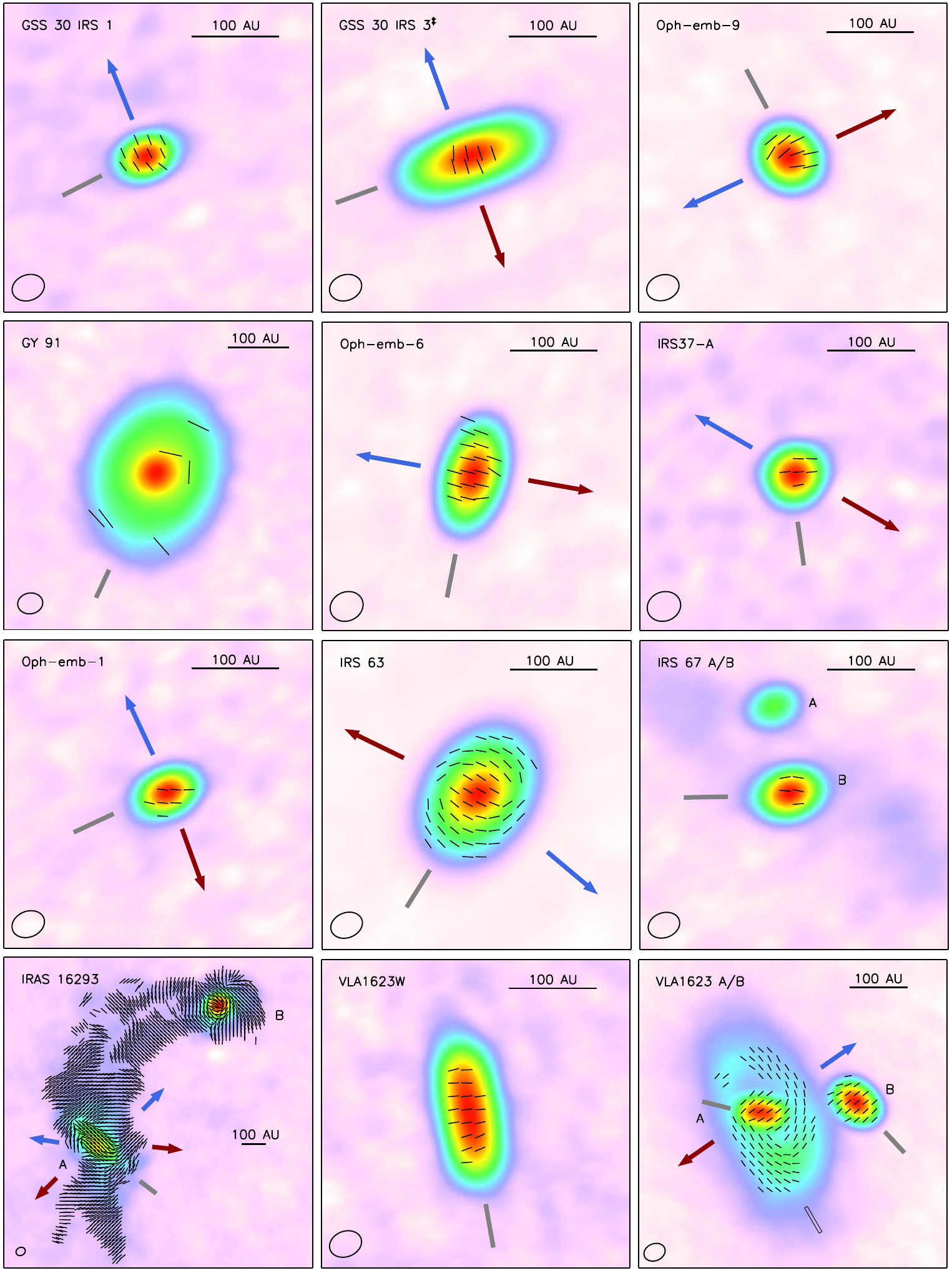}
\caption{The 14 continuum sources with polarization detections.  Background images show the Stokes I maps on a logarithmic color scale (see Appendix \ref{indiv} for the flux scale) and the black line segments show the normalized e-vectors.   Sources with $\ddagger$ are outside of the inner third of the primary beam FWHM. The blue and red arrows indicate the outflow position angle, if known (see Section \ref{morph_desc} for details).  The grey bars show the semi-major axis position angle of the continuum sources detected in polarization, except IRAS 16293B as this source is near face-on and does not have a well constrained continuum position angle.  For VLA 1623 A, we show two grey bars: the solid one shows the position angle of the compact disk from \citet{Harris18} and the open one shows the position angle of the extended disk.   \label{snapshots}}
\vspace{2mm}
\end{figure*}

Figure \ref{snapshots} shows images of the 14 sources with detected polarization.  The background images show the Stokes I continuum maps and the black line segments show normalized polarization e-vectors.  Maps with scaled polarization e-vectors are given in Appendix \ref{indiv}.  The blue and red arrows indicate the direction of the blue-shifted and red-shifted outflow lobes, if known (see Table \ref{disk_scatter} for the associated outflow angles and references) and the grey bars illustrate the source position angle (e.g., see Table \ref{cont_results}).  For most sources, the outflows are perpendicular to the continuum long axis.  This orientation indicates that the dust emission likely originates from a disk or flattened envelope.  

Table \ref{pol_results} gives the debiased polarized intensity and the polarization fraction for each of the 14 well-detected sources as measured at the position of their Stokes I peak.   For GY 91, however, the detected polarized emission is off the source peak and relatively weak ($< 4\sigma$).   We report its results instead at the position of its peak polarized intensity and hereafter consider GY 91 to be marginally detected.  Several sources have low polarization fractions with values $< 0.5$\%\ at their emission peak, but higher fractions at larger radial extents. 

Table \ref{undet_results} lists the 28 compact sources that were undetected in polarization.  This table gives the uncertainty in polarized intensity at each source position and the 3$\sigma$ upper limit to the polarization fraction for the non-detection.  Of the 28 undetected sources, half (14) have 3$\sigma$ upper limits $\lesssim 2\%$ and eight have 3$\sigma$ upper limits of $< 1$\%, indicating that many of the non-detected sources have significantly low polarization fractions.  Sources labeled with a double dagger are outside of the inner third of the primary beam.  We note that the upper limits to the polarization fractions for sources outside of the inner third of the primary beam FWHM may be less reliable (see Appendix \ref{offaxis} for details).

Figure \ref{pol_histograms} compares histograms of log peak Stokes I flux densities for the sources with well-detected polarization and those without polarization detections or with marginal detections.  Unsurprisingly, the sources that are well detected in polarization tend to be the brightest objects, as these sources have higher sensitivity to low polarization fractions.  The seven sources with $I_{peak} > 32$ \mJybeam\ have well-detected polarized intensities, whereas the detection rate drops to 50\%\ (6/12) for the sources with $I_{peak}$ between $10 - 32$ \mJybeam\ and to 9\%\ (1/11) of sources with $I_{peak}$ between $3-10$ \mJybeam.   

Table \ref{pol_region} gives a summary of the polarization statistics by region in Ophiuchus and by source classification, using the updated classifications in Table \ref{cont_results}.  These numbers exclude the 5 objects that were identified as either galaxies or background stars (see Section \ref{gal} and Appendix \ref{indiv}).  Since the younger YSOs in the sample are brighter, they have a higher polarization detection rate than the more evolved objects.  All Class 0 sources are detected in polarization and no Flat or Class II objects detected (excluding the marginal detection for GY 91).  This result is also in agreement with a number of studies that show higher polarization fractions toward younger sources \citep[e.g.,][]{Beckford08, Hull14}.  

\subsection{Deliverables}\label{products}

With this publication, we release the full data data products.  These products include the self-calibrated maps of the Stokes I, Q, U observations and the maps of debiased polarized intensities, polarization position angles, and polarization fraction for each source.  These maps are all primary beam corrected.  We also include their associated error maps and a map of the primary beam.  The Stokes I, Q, U, and polarized intensity errors correspond to the map error at the phase center (see Table \ref{obs_summary}) scaled by the primary beam correction.  The error maps for the polarization position angles and polarization fraction are calculated by propagating the individual errors in the associated maps pixel by pixel.    

The data products are available at Dataverse\footnote{https://doi.org/10.7910/DVN/QYNZRR} for individual fields or for the entire sample.  The maps are provided as FITS files with the form FIELD\_TYPE\_233GHz.fits, where FIELD is the name of the field in Table \ref{obs_summary} and TYPE indicates whether the map is one of the Stokes parameters (StokesI, StokesQ, StokesU), polarized intensity (POLI), polarization position angle (POLA), polarization fraction (POLF), or the field primary beam (pbeam).  Error maps are appended with ``err''.  We also include separate maps for the mosaic of c2d\_1008a and c2d\_1008b used here; these data use the field name of ``c2d\_1008''.

{\setlength{\extrarowheight}{0.8pt}%
\begin{table}[h!]
\caption{Polarization Detection Results}\label{pol_results}
\begin{tabular}{llll}
\hline\hline
Field & Source					& $\PI_{peak}$\tablenotemark{a}	& $\PF_{peak}$\tablenotemark{b} \\
		  &						&(\uJybeam)	&  (\%) \\
\hline
c2d\_811	& GSS 30 IRS 1		& 471 $\pm$ 26	& 3.7 $\pm$ 0.2 \\
		& GSS 30 IRS 3$^{\ddagger}$ & 760 $\pm$ 62	& 1.4 $\pm$ 0.1 \\
c2d\_822	& Oph-emb-9 			& 145 $\pm$ 25	& 0.5 $\pm$ 0.09  \\ 
c2d\_831	& GY 91\tablenotemark{c} & 115 $\pm$ 27	& 12 $\pm$ 3	 \\
c2d\_862	& Oph-emb-6 			& 260 $\pm$ 28	& 1.0 $\pm$ 0.1 \\  
c2d\_885	& IRS 37-A 			& 169 $\pm$ 28	& 1.7 $\pm$ 0.3  \\
c2d\_954	& Oph-emb-1 			& 122 $\pm$ 26	& 1.1 $\pm$ 0.2 \\ 
c2d\_989	& IRS 63 				& 1329 $\pm$ 27	& 1.5 $\pm$ 0.03  \\
c2d\_1003	& IRS 67-B 			& 120  $\pm$ 28	& 0.3 $\pm$ 0.06 \\ 
c2d\_1008 & IRAS 16293B  		& 2460 $\pm$ 26	& 0.5 $\pm$ 0.01 \\ 
		 & IRAS 16293A 		& 895 $\pm$ 27	& 0.5 $\pm$ 0.01 \\
VLA1623a	& VLA 16239W 		& 233 $\pm$ 27	& 1.4 $\pm$ 0.2 \\  
VLA1623b	& VLA 1623B 			& 1285 $\pm$ 27	& 2.0 $\pm$ 0.04 \\ 
		& VLA 1623A 			& 1185 $\pm$ 27	& 2.1 $\pm$ 0.04 \\
\hline
\end{tabular}
\begin{tablenotes}[normal,flushleft]
\item \tablenotemark{a} Debiased polarized intensity at the peak Stokes I position (see Table \ref{cont_results}).  Errors correspond to $\sPI$ for the field scaled by the primary beam correction at the position of the source.  Sources with $\ddagger$ are outside of the inner third of the primary beam.
\item \tablenotemark{b} The corresponding polarization fraction at the same position. Errors corresponds to the statistical error described in Section \ref{pol_overview} and do not include the instrument polarization. 
\item \tablenotemark{c} The values for polarized intensity and polarization fraction for GY 91 are from the position of the peak polarized intensity.
\end{tablenotes}
\end{table}
}

{\setlength{\extrarowheight}{0.8pt}%
\begin{table}[h!]
\caption{Non-detection Upper Limits}\label{undet_results}
\begin{tabular}{llll}
\hline\hline
Field & Source					& $\sigma_{\PI}$\tablenotemark{a}	 & $\PF_{,limit}$\tablenotemark{b}\\
		  &						& (\uJybeam) & (\%) \\
\hline
c2d\_857	& WL 16 				& 26		& 1.8  \\
c2d\_862	& ALMA\_J162705.5 	& 30		& 26 \\
c2d\_867	& WL 17 				& 27.3	& 0.3 \\
c2d\_871	& Elias 29 			& 26.5 	& 0.5 \\
c2d\_885	& IRS 37-B			& 27.5	& 9.3 \\
		& IRS 37-C			& 28		& 10 \\
		& ALMA\_J162717.7		& 29		& 32 \\
		& IRS 39$^{\ddagger}$	& 83.3	& 33  \\ 
c2d\_890	& IRS 42				& 27.5	& 0.7 \\
c2d\_892	& Oph-emb-5\tablenotemark{c} & 26  & $\cdots$ \\
c2d\_894	& Oph-emb-12 			& 26.5	& 1.8 \\
c2d\_899	& IRS 43-A 			& 25.5 	& 0.6 \\
		& IRS 43-B 			& 25.5	& 4.8 \\
		& GY 263$^{\ddagger}$ 	& 30.2	& 1.3 \\
c2d\_901	& IRS 44 				& 25		& 0.7  \\
c2d\_902	& IRS 45	 			& 26.3 	& 3.9 \\
		& VSSG 18 B$^{\ddagger}$ & 43		& 13  \\ 
c2d\_904	& IRS 47 				& 26  	& 1.1 \\
		& ALMA\_J162729.7$^{\ddagger}$ 	& 37 		& 21 \\
c2d\_963	& Oph-emb-18 			& 25.8	& 4.1 \\
c2d\_990	& Oph-emb-4 			& 26.5	& 0.9  \\
c2d\_991	& Oph-emb-25 			& 26.5 	& 0.9 \\
c2d\_996	& Oph-emb-7 			& 26.8  	& 21 \\
c2d\_998	& Oph-emb-15 			& 26 		& 2.4 \\
c2d\_1003	& IRS 67-A 			& 28 		& 1.0 \\
VLA1623b & VLA 1623NE$^{\ddagger}$ 	& 149 	& 3.9 \\ 
IRAS16288 & ISO Oph 210 		& 26 		& 2.0 \\
		   & ALMA\_J163203.3$^{\ddagger}$	& 68.4  	& 33 \\ 
\hline
\end{tabular}
\begin{tablenotes}[normal,flushleft]
\item \tablenotemark{a} The error in the polarized intensity at the position of the source.  Sources with {$\ddagger$} are outside of the inner third of the primary beam.
\item \tablenotemark{b} The 3-$\sigma$ upper limit polarization fraction for a non-detection.
\item \tablenotemark{c} Oph-emb-5 was not detected in Stokes I continuum.
\end{tablenotes}
\end{table}
}

\begin{figure}[h!]
\includegraphics[width=0.475\textwidth]{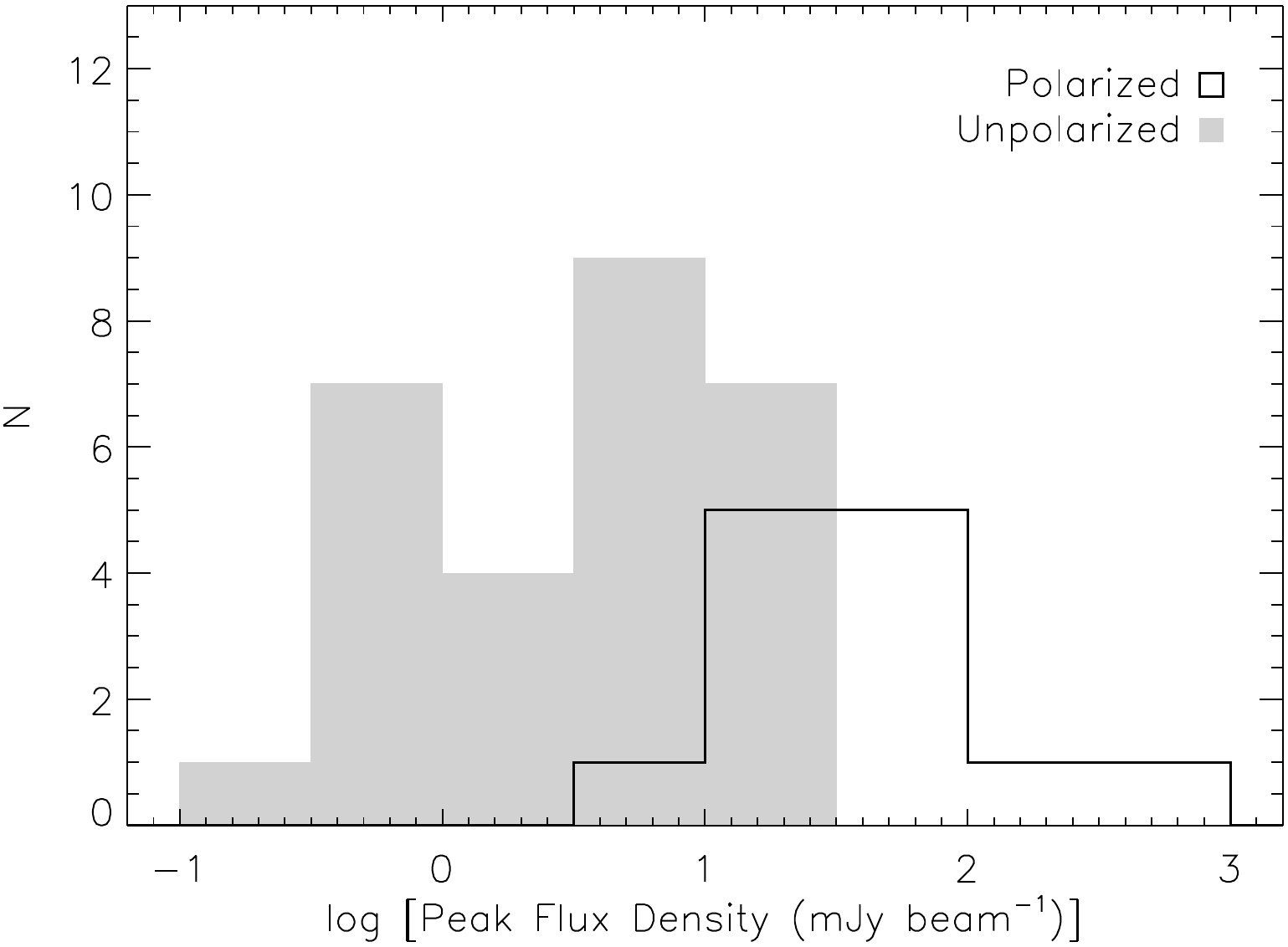}
\caption{Histograms of log peak flux density.  Sources that are well detected in polarized intensity are shown with open histograms and the sources not detected or only marginally detected in polarized intensity by filled histograms. \label{pol_histograms}}
\end{figure}

{\setlength{\extrarowheight}{0.8pt}%
\begin{table}[h!]
\caption{Polarization Detections by Region and Class}\label{pol_region}
\begin{tabular}{lccc}
\hline\hline 
 & Detected\tablenotemark{a}	&  Undetected\tablenotemark{a} & Detection fraction \\
\hline
\multicolumn{4}{c}{Region} \\
\hline
Oph A		&	6	& 	2 	& 0.75 \\
Oph B		& 	1	& 	5 	& 0.17\\
Oph E		& 	1	&	4  	& 0.2 \\
Oph F		& 	0	&	6 	& 0.0 \\
L1688		&	1	&	1	& 0.5 \\
L1689N		& 	2	& 	0 	& 1.0 \\
L1689S		& 	1	& 	5	& 0.17  \\
L1709		&	1	& 	1	& 0.5 \\
\hline
\multicolumn{4}{c}{Class} \\
\hline
Class 0		&	6	& 	0 	& 1.0 \\
Class I		&	7	&  	11	& 0.4 \\
Flat			&	0	&	4	& 0.0 \\
Class II		& 	0	& 	9	& 0.0 \\
\hline
\end{tabular}
\begin{tablenotes}[normal,flushleft]
\item \tablenotemark{a} Excludes objects that are re-classified as galaxies or stars.  The marginal detection for GY 91 is added to the undetected column for this table. 
\end{tablenotes}
\end{table}
}

\section{Polarization Mechanisms} \label{mech_discuss}

Of the 41 continuum sources that we detect, 37 of them are considered to be YSOs between Class 0 and Class II.  Of these 37 YSOs, only 14 ($\sim 38$\%) have polarization detections (35\% if we exclude the marginal detection of GY 91).   In this section, we identify the likely polarization mechanisms for the sources detected in dust polarization.  
We discuss why most of the sources appear undetected in polarization in  Section \ref{non_det}.

\subsection{Morphological Description}\label{morph_desc}

In Section \ref{pol_overview}, we show images of the 14 YSOs that are detected in polarization.  By eye, we see a variety of polarization structures, with some sources showing homogenous polarization position angles, whereas others have circular or complex polarization position angles.  Here, we describe the general polarization e-vector morphology of each source and also the orientation of the e-vectors relative to the Stokes I continuum emission.  

Table \ref{disk_scatter} summarizes the polarization morphological description for each source.  We include a separate entry for the large circumbinary disk around VLA 1623A separately from its smaller compact disk as the two show distinct polarization structures that likely arise from different mechanisms \citepalias[see][]{Sadavoy18b}.  Columns 1 and 2 give the source name and classification from Table \ref{cont_results}.  Column 3 gives the inclination ($i$), estimated from the ratio of the minor to major axis ($\cos{i} = b/a$) from our Gaussian fits (see Table \ref{cont_results}), assuming the dust emission is tracing geometrically thin disks.  Column 4 gives the semi-minor axis position angle ($\phi$).  Column 5 gives the weighted average polarization position angle ($<\theta_P>$)\footnote{We use a weighted average to estimate the typical polarization angle and determine whether or not the polarization angles are uniform.  Most sources have insufficient independent measurements to estimate a mean or standard deviation via a Gaussian distribution.}.  The reported error for $<\theta_P>$ corresponds to the weighted standard deviation.  Column 6 gives the angle difference, $\Delta$, between the weighted average polarization position angle and semi-minor axis, $\Delta = |\phi\ - <\theta_P>|$.  Column 7 describes the overall polarization morphology, where ``U'' is uniform, ``A'' is azimuthal, and ``C'' is complex (see below).  Column 8 indicates whether the polarization is aligned with major or minor axes of the Stokes I continuum source.  Column 9 gives the dust opacity index, $\beta$ (see Appendix \ref{tau} and Section \ref{mech_scat}) and column 10 gives the outflow orientation ($\theta_{out}$) if known.  Since IRAS 16293A and IRAS 16293B are confused with the dense envelope around the stars, we instead fit a Gaussian to their brightest emission defined by an area of $I \gtrsim 100$ m\Jybeam\ to get their general geometries and weighted average polarization angles.

{\setlength{\extrarowheight}{0.8pt}%
\begin{table*}
\caption{Polarization and Disk Orientations}\label{disk_scatter}
\begin{tabular}{lcccccclcl}
\hline\hline
 Source			& Class	& $i$				& $\phi$ 			& $<\theta_P>$  &	 $\Delta$\tablenotemark{a}   &  Morph\tablenotemark{b}  & Align\tablenotemark{b} & $\beta$\tablenotemark{c} & $\theta_{out}$\tablenotemark{d}  \\
				&		& ($\degree$) &($\degree$)		&  ($\degree$) 		& ($\degree$)	& 			    &    					&  	&	($\degree$)			  \\
\hline														
GSS 30 IRS 1		& I	& 74.5 $\pm$ 10 	& 27.0 $\pm$ 6.8	& 28.5 $\pm$ 6.3	&  1.5 $\pm$ 9.0 	& U		& minor		& 0.12 $\pm$ 0.04 & 22 (1*)		\\  
GSS 30 IRS 3 		& I	& 70.8 $\pm$ 0.6 	& 19.6 $\pm$ 0.4 	& 15.7 $\pm$ 4.9	&  3.9 $\pm$ 4.9 	& U		& minor		& 0.19 $\pm$ 0.05 &  20 (1*)		\\  
Oph-emb-9 		& I	& 67.0 $\pm$ 0.9	& -62.2 $\pm$ 0.9	& -64 $\pm$ 15		&  1.8 $\pm$ 15  	& C 		& minor		& 0.18 $\pm$ 0.03 & -65 (2*)	  \\   
GY 91 			& F	& 33.6 $\pm$ 9 	& 65	$\pm$ 29		& 69 $\pm$ 46		&  $\cdots$ 		& A		& $\cdots$ 	& 0.64 $\pm$ 0.44 &	$\cdots$	    \\ 
Oph-emb-6 		& I	& 76.0 $\pm$ 0.5	& 78.6 $\pm$ 0.1 	& 74.6 $\pm$ 4.6	& 4.0 $\pm$ 4.6  	& U		& minor		& 0.27 $\pm$ 0.03 &	 80 (3*)	 \\  
IRS 37-A 			& I	& 69 $\pm$ 8		& -82.5 $\pm$ 3.9 	& -86.9 $\pm$ 4.1	& 4.4 $\pm$ 5.7 	& U		& minor		& $\cdots$ 	     &	60 (4*)	 \\ 
Oph-emb-1 		& 0	& 68.9 $\pm$ 2.5 	& 25.2 $\pm$ 2.0	& 85.5 $\pm$ 3.7	& 60.3 $\pm$ 4.2	& U		& none		& 0.98 $\pm$ 0.06 &	 22 (5) 	  \\  
IRS 63			& I	& 47.2 $\pm$ 1.6 	& 58.4 $\pm$ 2  	& 61 $\pm$ 12		& 2.6 $\pm$ 12.2 	& A		& minor		& 0.35 $\pm$ 0.19 & $50-64$ (6)	 \\ 
IRS 67-B 			& I	& 54.2 $\pm$ 4  	& 0.5 $\pm$ 5.9	& 87.3 $\pm$ 7.9	& 87 $\pm$ 10 		& U		& major		& 0.74 $\pm$ 0.09 & $\cdots$ 	\\  
IRAS 16293A		& 0	& 57.3 $\pm$ 3		& -37 $\pm$ 2.9 	& 63 $\pm$ 21		& 100 $\pm$ 21	& C		& major		& $\cdots$ 	    &  90, -45 (7)   \\ 
IRAS 16293B		& 0	& 20.7 $\pm$ 6		& 35 $\pm$ 42		& 62 $\pm$ 48		& $\cdots$ 		& A		& $\cdots$	& $\cdots$ 	    &   $\cdots$ \\  
VLA 1623W 		& 0	& 81.4 $\pm$ 1.0 	& -79.9 $\pm$ 0.5 	& -81.1 $\pm$ 6.3	& 1.2 $\pm$ 6.3  	& U		& minor		& 0.28 $\pm$ 0.13 &  $\cdots$  \\  
VLA 1623B		& 0	& 70.9 $\pm$ 2.8 	& -47.4 $\pm$ 1.8	& -46.6 $\pm$ 2.3	& 0.8 $\pm$ 2.9 	& U		& minor		& 0.48 $\pm$ 0.08 & -55 (8)	 \\
VLA 1623A (compact)\tablenotemark{e} & 0& 64.1 $\pm$ 3  & -51.8 $\pm$ 3.5	& -53.7 $\pm$ 4.0	& 1.9 $\pm$ 5.3 & U	& minor		& 0.45 $\pm$ 0.29 &  -55 (8)	  \\
VLA 1623A (extended)\tablenotemark{e}& 0 & 54.8	 	   & -60			& 26 $\pm$ 30		& $\cdots$   	 & A	& $\cdots$	& 0.6 $\pm$ 0.7     &  -55 (8) \\  
\hline
\end{tabular}
\begin{tablenotes}[normal,flushleft]
\item \tablenotemark{a} Sources without values have unconstrained values (errors $\gtrsim 30$\degree) for $\phi$ and $<\theta>$.
\item \tablenotemark{b}  Morphological description of the polarization and alignment relative to the major or minor continuum axis.  ``U'' indicates the polarization is uniform, ``A'' indicates the polarization is azimuthal, and ``C'' indicates the morphology is complex.   Multiple entries mean more than one morphology is present.
\item \tablenotemark{c}  Estimated dust opacity index using archival data, if applicable.  See Appendix \ref{tau}.
\item \tablenotemark{d}  Estimated outflow position angle in the literature with reference if applicable (see Figure \ref{snapshots}). For VLA 1623A and VLA 1623B, we give the same outflow orientation for both because only one outflow is seen for both sources.  Numbers in the parentheses indicate the reference for the outflow orientation and stars indicate that the literature reference did state the position angle and as such, we estimated the outflow position angle by eye from integrated intensity maps. References are: (1) \citealt{Friesen18}, (2) \citealt{Kamazaki03}, (3) \citealt{Bussmann07}, (4) \citealt{vanderMarel13}, (5) \citealt{Yen17}, (6) \citealt{Visser02}, (7) \citealt{vanderWiel19}, (8) \citealt{Santangelo15}.
\item \tablenotemark{e}  VLA 1623A is split into two entries, one for the compact disk with uniform polarization and one from the extended disk with azimuthal polarization.  Values for $i$ and $\phi$ are based on a by-eye fit to the continuum data as reported in \citetalias{Sadavoy18b}.  The dust opacity index for the extended emission is estimated from setting \texttt{nterms = 2} in \texttt{clean} to estimate the spectral index, $\alpha$, for $\beta = \alpha - 2$.
\end{tablenotes}
\end{table*}
}

We consider three morphological descriptions for the polarization e-vectors:  ``uniform'', ``azimuthal'', or ``complex''.  While many sources show multiple polarization morphologies, we only report the most dominant one.  The source morphologies are defined as:
\begin{itemize}
\item \textbf{Uniform Polarization:} Uncertainty on the weighted average polarization position angle is $<10\degree$.  
\item \textbf{Azimuthal Polarization:} Polarization e-vectors follow an idealized elliptical pattern \citep[e.g.,][]{Mori19}.  For simplicity, we assume that the idealized elliptical pattern traces the same dimensions as the continuum source (e.g., using the geometry given in Table \ref{cont_results}). 
\item \textbf{Complex Polarization:} The polarization morphology is neither uniform nor azimuthal.
\end{itemize}

From the above definitions, we find that nine YSOs have uniform polarization angles.  These systems are: GSS 30 IRS 1, GSS 30 IRS 3, Oph-emb-6, IRS 37-A, Oph-emb-1, IRS 67-B, VLA 1623W, VLA 1623B, and VLA 1623A (compact).  In most of these systems, the polarization angles are also well aligned with the minor axis, with the exception of IRS 67-B, which is aligned with the major axis, and Oph-emb-1, which is aligned with neither axis.  We consider the polarization aligned with the minor axis if the angle difference, $\Delta$, is consistent with zero within $\sim 1 \sigma$ (or aligned with the major axis if $\Delta$ is consistent with 90\degree\ within $\sim 1 \sigma$).  We do not calculate $\Delta$ for sources with errors $ \gtrsim 30$\degree\ for either their semi-minor axis or weighted average polarization angle as the individual position angles are too unconstrained to measure a meaningful angle difference. 

\begin{figure*}
\begin{tabular}{ll}
\includegraphics[width=0.48\textwidth]{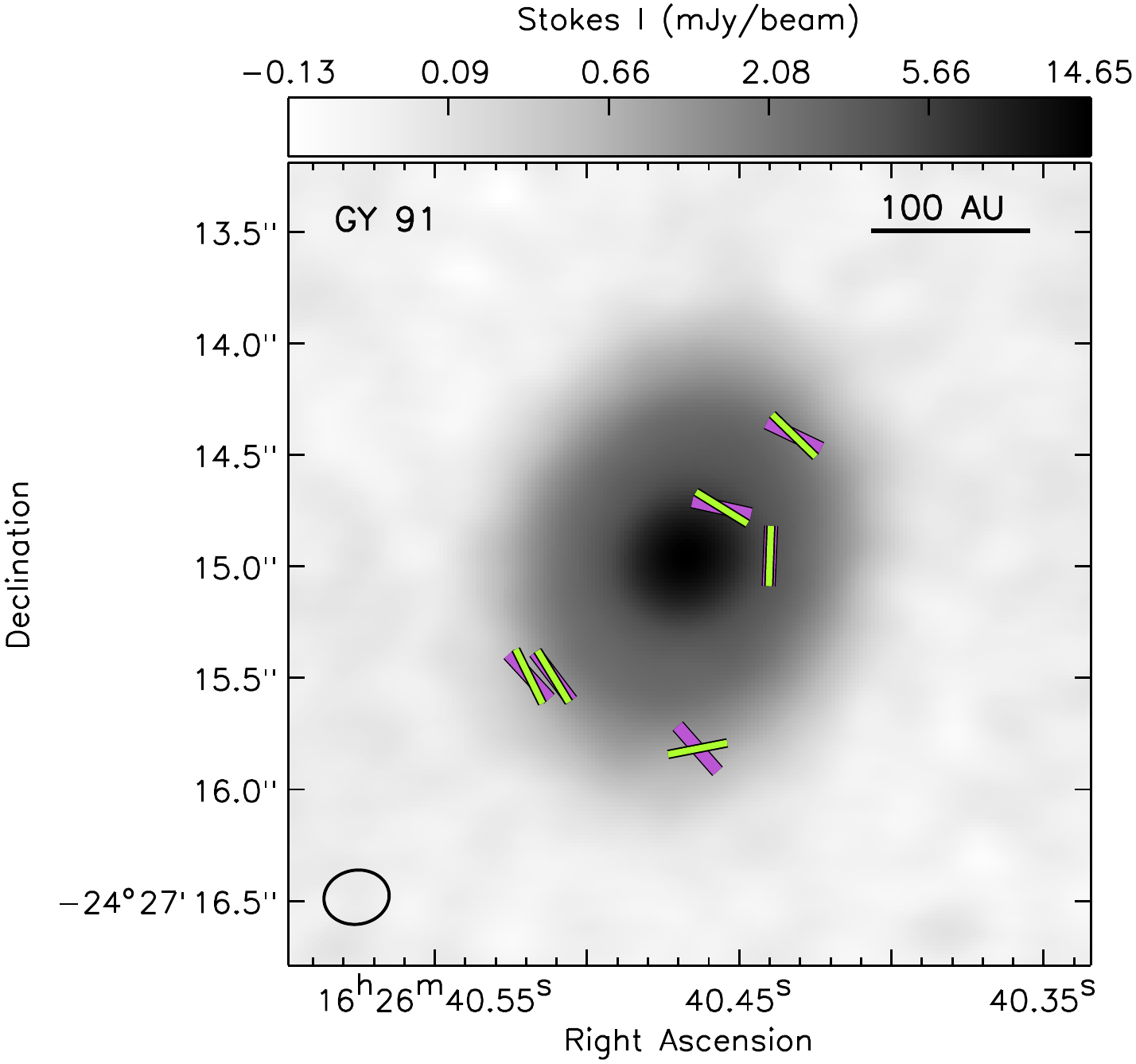} & \includegraphics[width=0.48\textwidth]{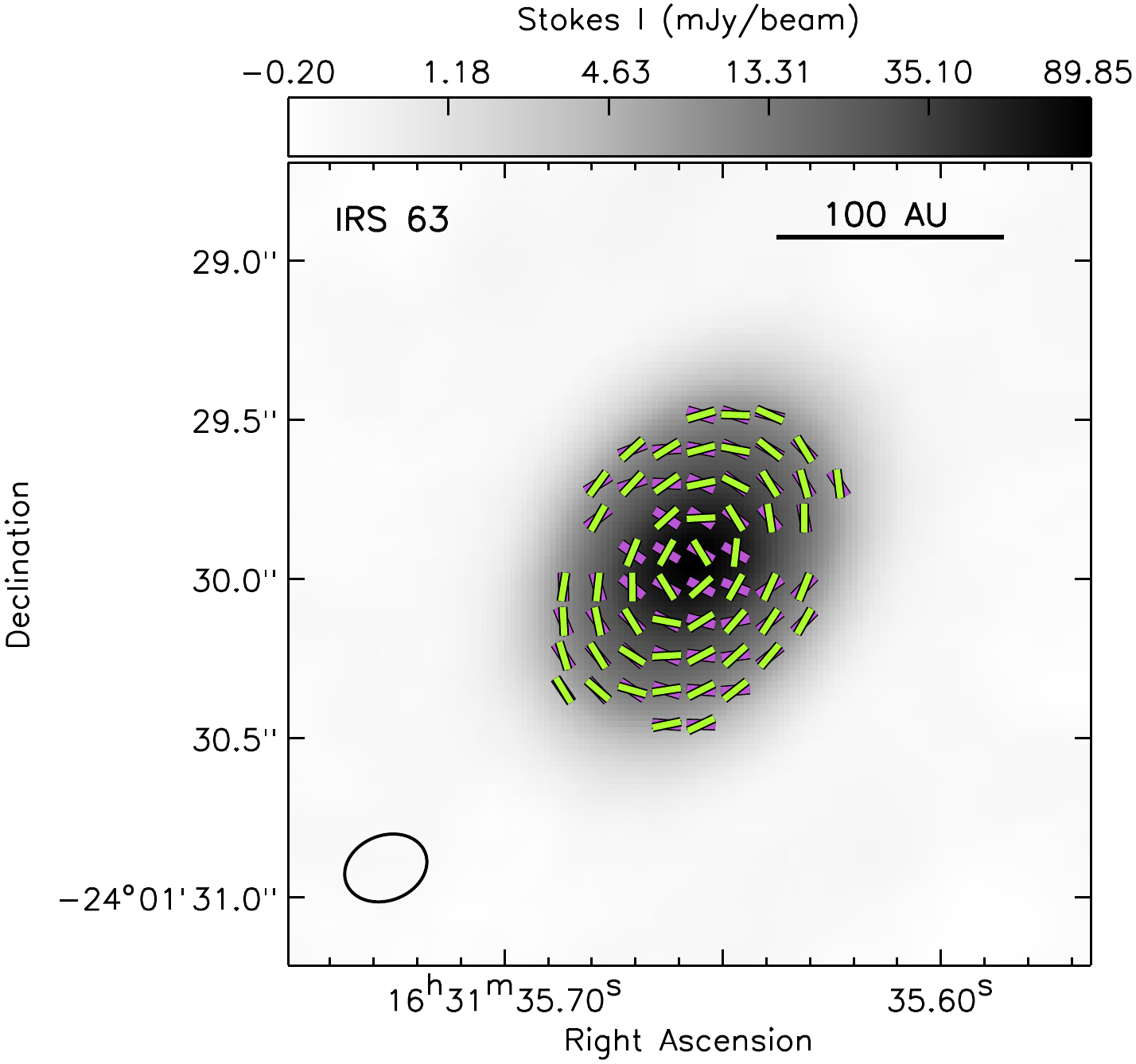} \\
\includegraphics[width=0.48\textwidth]{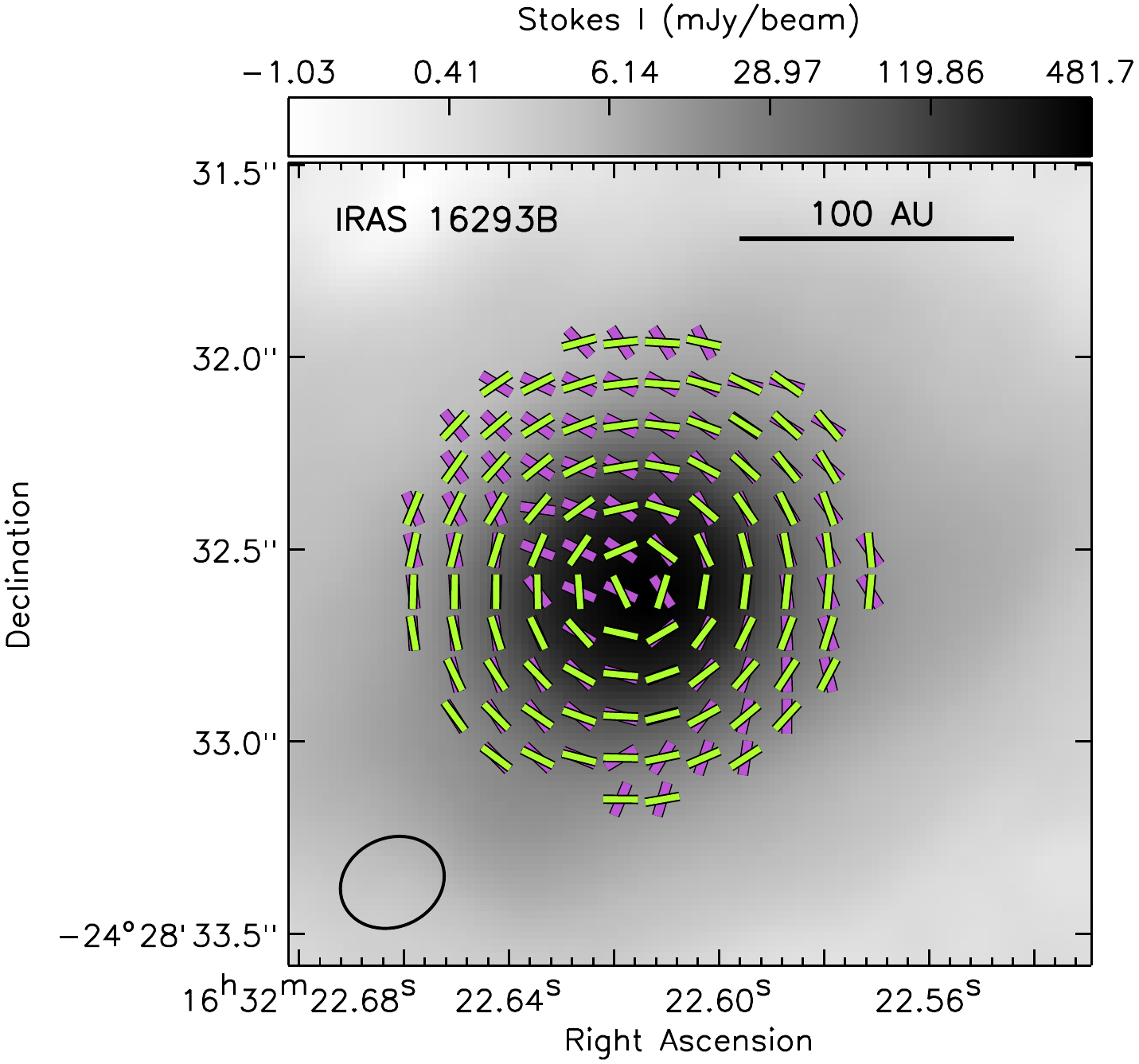} & \includegraphics[width=0.48\textwidth]{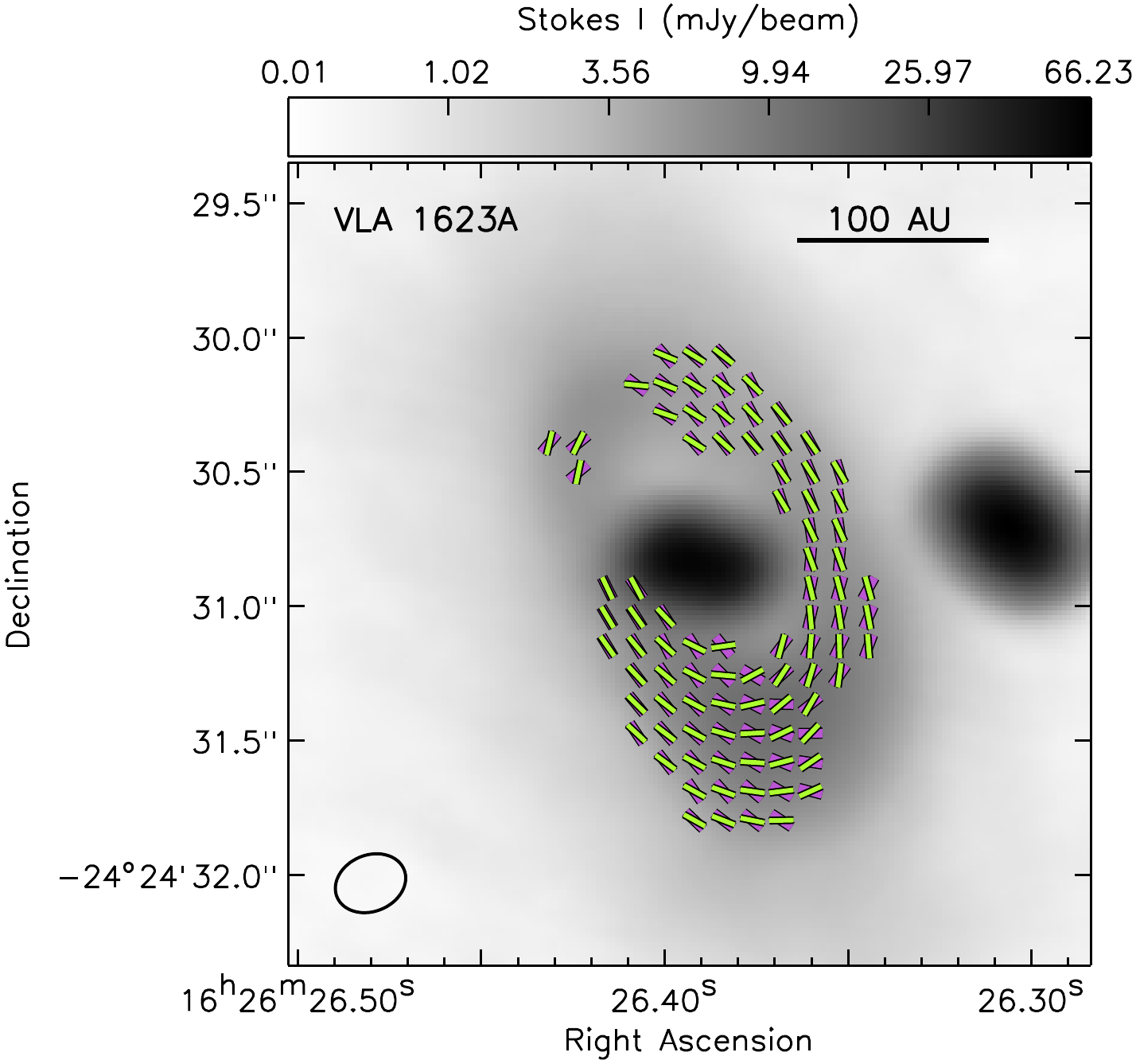}\\
\end{tabular}
\caption{Comparison between the observed polarization e-vectors and an ideal elliptical polarization morphology for the four sources dominated by azimuthal polarization.  The purple line segments show the observed polarization and the green line segments show ideal elliptical polarization.  For simplicity, we show only the polarization vectors toward the inner $\sim 1$\arcmin\ of IRAS 16293B and we mask out the uniform polarization e-vectors toward VLA 1623B and VLA 1623A (compact).  \label{azimuthal}}
\end{figure*}

There are four cases of azimuthal polarization angles.  These sources are GY 91, IRS 63, IRAS 16293B, and VLA 1623A (extended).  We note, however, that none of these sources showed pure circular or elliptical polarization.  Figure \ref{azimuthal} compares the observed polarization (purple line segments) of these four YSOs to their idealized elliptical polarization (green line segments), assuming the idealized elliptical pattern traces the same dimensions as the Stokes I source.  For IRS 63, IRAS 16293B, and VLA 1623A (extended), we see substantial deviations from the elliptical pattern that may be indicative of other polarization mechanisms or changes in optical depth (see Section \ref{mech_scat}).  As such, we cannot quantify the azimuthal structure using the distribution of angular deviations to measure the agreement as in \citet{Mori19}.   Instead, we require that the polarization morphology must be \emph{dominated} by an elliptical component for the system to be considered azimuthal.  We therefore define azimuthal polarization when at least half of the e-vectors have an angular deviation of $< 25\degree$ with the idealized elliptical pattern.  

Finally, there are two YSOs, Oph-emb-9 and IRAS 16293A, that have complex polarization (see Figure \ref{snapshots}).  Both sources appear to have a mix of azimuthal and uniform polarization, but neither morphology dominates.  Nevertheless, its polarization is well aligned with the minor axis.   The polarization of IRAS 16293A is mostly aligned with the major axis, but shows substantial curvature at larger radial extents that result in larger uncertainties on its weighted average polarization position angle (20\degree).  Thus, neither object is well fit with a uniform or elliptical morphology.

\subsection{Physical Interpretation}\label{mech_scat}

In this section, we determine the likely polarization mechanism for the YSOs using the morphological description of the previous section and the physical properties of the source.  Several different mechanisms can cause polarization in disks.  These mechanisms include dust self-scattering processes and grain alignment from either magnetic fields, radiative alignment torques or k-RATs, mechanical torques, or aerodynamic alignment \citep[e.g.,][]{Tazaki17, Hoang18, Kataoka19, Yang19}.  For this analysis, we use morphological arguments to support or reject specific polarization mechanisms.  In particular, we focus on polarization from dust self-scattering and magnetic fields as the theoretical frameworks for these two mechanisms are better established.  

\subsubsection{Polarization from Dust Self-Scattering}

Dust self-scattering is emerging as a common mechanism behind the polarization of young disks \citep[e.g.,][]{Kataoka16hd, Stephens17, Hull18, Harris18, Bacciotti18, Dent19}.  This mechanism is attributed to Rayleigh scattering from large dust grains within the disk \citep{Kataoka15} that produces a polarized signature of a few percent if the disk radiation field is anisotropic, e.g., the scattered emission has a preferred direction based on the flux gradient of the radiation field.  The polarized intensities and polarization morphologies are therefore strongly dependent on the disk properties, such as dust grain size \citep[e.g.,][]{Kataoka15, Kataoka17}, disk geometry \citep[e.g.,][]{Yang16, Kataoka16}, disk substructure \citep[e.g.,][]{Kataoka16hd, Pohl16}, and optical depth \citep{Yang17}.  More recently, models have suggested that the dust grain shape and porosity can also affect the observed polarization structure from dust self-scattering \citep[e.g.,][]{Kirchschlager19}.

To help identify polarization from self-scattering, we also consider the physical properties of the source.  In particular, we use the source inclination, $i$, to infer the geometry and dust opacity index, $\beta$ (see Table \ref{disk_scatter}), to estimate the source optical depth.  A number of theoretical studies have shown that the dust self-scattering signature is highly dependent on the disk geometry and disk optical depth  \citep[e.g.,][]{Yang16, Kataoka16, Kataoka17, Yang17}.  Table \ref{disk_scatter} gives the inclination and dust opacity index for each source.  For the dust opacity index, we first calculate the (sub)millimeter spectral index, $\alpha$, from multi-wavelength data in the literature when available, and then infer $\beta$ under the assumption that $\beta = \alpha - 2$ (see Appendix \ref{tau} for details).  We assume that values of $\beta \approx 0$ are consistent with optically thick dust emission, although we discuss other possibilities in Appendix \ref{tau}.  

We use the following morphological description for polarization from dust self-scattering: inclined ($i > 60\degree$) disks with optically thick dust emission will show uniform polarization angles that are aligned with the disk minor axis, whereas face-on disks ($i \lesssim 20\degree$) will be depolarized toward the center of the disk and show azimuthal polarization at larger radial extents \citep{Kataoka15, Kataoka16, Yang16, Yang17}.  For intermediate inclinations, we assume the scattering polarization morphology can also be a hybrid, with uniform polarization angles toward the center and more azimuthal polarization at large radial extents as shown in \citet{Yang16}.

Table \ref{disk_scatter_sum} summarizes qualitatively the inclination, optical depth, and polarization morphology for the sources.   Qualitatively, nine of the sources are consistent with dust self-scattering.  The majority of cases are associated with optically thick, highly inclined disks that have uniform polarization vectors aligned with their minor axes.  IRS 63 has a moderate inclination (47\degree), and its mix of uniform and azimuthal polarization is consistent with dust scattering in an optically thick, moderately inclined disk.   IRAS 16293B has a low inclination ($\sim 20$\degree), and its azimuthal polarization is consistent with an optically thick, nearly face-on disk, although the polarization structure of this source is highly confused by its surrounding dense envelope.  IRAS 16293B is well known for having high optical depth ($\beta \approx 0$) within its envelope \citep[e.g.,][]{Chandler05} and its near face-on geometry makes disentangling the disk from the envelope complex.  If the polarization signatures arises from self-scattering in a disk, then the self-scattering disk of IRAS 16293B is roughly 1\arcmin\ (140 au) in diameter.  Detailed models of dust self-scattering over multiple wavelengths may be able to fully disentangle the dust polarization from the dense envelope with the disk to enable the first clear measurement of the disk size in IRAS 16293B.

{\setlength{\extrarowheight}{0.8pt}%
\begin{table*}
\caption{Consistency with Dust Self-Scattering}\label{disk_scatter_sum}
\begin{tabular}{lcllll}
\hline\hline
 Source			& Class	& $i$	\tablenotemark{a}	& $\tau$\tablenotemark{b}  & Polarization\tablenotemark{c}	&	Scattering\tablenotemark{d}   \\
\hline														
GSS 30 IRS 1		& I	& high 				& thick			& uniform, minor	& yes			\\  
GSS 30 IRS 3 		& I	& high				& thick		 	& uniform, minor	& yes				\\  
Oph-emb-9 		& I	& high				& thick			& complex, minor	& maybe			  \\   
GY 91 			& F	& moderate			& not thick			& azimuthal, none 	& maybe			    \\ 
Oph-emb-6 		& I	& high				& thick 			& uniform, minor	& yes			 \\  
IRS 37-A 			& I	& high				& $\cdots$ 		& uniform, minor	& yes			 \\ 
Oph-emb-1 		& 0	& high	 			& not thick			& uniform, major 	& no				  \\  
IRS 63			& I	& moderate			& thick	  		& azimuthal, minor	 & yes			 \\ 
IRS 67-B 			& I	& moderate		  	& not thick			& uniform, major	& no			 	\\  
IRAS 16293A		& 0	& moderate			& $\cdots$ 		& complex, major 	& no			   \\ 
IRAS 16293B		& 0	& low				& thick			& azimuthal, none 	& yes		 \\  
VLA 1623W 		& 0	& high	 			& thick 			& uniform, minor 	& yes		  \\  
VLA 1623B		& 0	& high		 		& thick			& uniform, minor 	& yes			 \\
VLA 1623A (compact)\tablenotemark{e} & 0& high	 & thick 			& uniform, minor 	& yes			  \\
VLA 1623A (extended)\tablenotemark{e}& 0 & moderate & not thick		& azimuthal, none 	& no			 \\  
\hline
\end{tabular}
\begin{tablenotes}[normal,flushleft]
\item \tablenotemark{a} Source inclination group, where high is $i > 60$, moderately inclined is $30 < i < 60$, and low inclination is $i < 30$.
\item \tablenotemark{b} Expected optical depth based on the dust opacity index.  We consider a source to be optically thick if $\beta < 0.5$, not thick if $\beta > 0.5$.  For IRAS 16293B, we adopt $\beta \approx 0$ from \citet{Chandler05} for its optical depth.
\item \tablenotemark{c} Polarization morphology and orientation (see Table \ref{disk_scatter}).
\item \tablenotemark{d}  Indicates whether or not the polarized emission is consistent with dust self-scattering.
\item \tablenotemark{e}  VLA 1623A is split into two entries.  See Table \ref{disk_scatter} for details.
\end{tablenotes}
\end{table*}
}

We also identify Oph-emb-9 and GY 91 as partially consistent with dust self-scattering.   Oph-emb-9 has a low value of $\beta$ and high inclination, much like a number of sources consistent with dust self scattering (e.g., GSS 30 IRS 3).  Nevertheless, this source has complex polarization rather than uniform e-vectors (although, we note that its polarization does align with the minor axis on average).  We measure a weighted standard deviation on $<\theta_P>$ of 15\degree, which is more than a factor of 2 higher than most of the disks with uniform polarization position angles.  This more complex morphology could indicate that another polarization mechanism is present in addition to or instead of dust self-scattering.  In the case of GY 91, the disk has moderate inclination and complicated structure.   \citet{SheehanEisner18} found several gaps across the disk \citep[see also,][]{vanderMarel19}.  We detect azimuthal polarization at large radial extents in this disk, which is consistent with models of self-scattering in disks with moderate to low inclinations and gaps \citep[e.g.,][]{Kataoka16hd, Pohl16, Ohashi18}.   We consider GY 91 to be only partially consistent with dust self-scattering, however, because it has a slightly steeper dust opacity index ($\beta \approx 0.64$) and its polarization detections are only marginal.

\subsubsection{Polarization from Magnetic Fields}\label{mech_bfield}

Polarization from dust grains aligned with magnetic fields will have different structures depending on the field morphology.   In general, models of magnetic fields on disk scales can be described as either toroidal (the field curves in the plane of the disk) or a poloidal (the field loops through the plane of the disk).  Since the polarization e-vectors will be perpendicular to the magnetic field direction, a purely toroidal field will give a radial polarization pattern and a purely poloidal field will have the polarization aligned with the disk major axis.  Both of these patterns will vary with disk inclination \citep[e.g.,][]{Tomisaka11, Kataoka12, Reissl14, Bertrang17}, and the field may also have complex structure from hourglass shapes due to the field lines being compressed inward by gravity \citep[e.g.,][]{Mestel66, GalliShu93, Myers18} to multiple field morphologies \citep[e.g.,][]{Lee18,Alves18,Kwon19}.  Since magnetic fields can produce complex polarization patterns, we require that dust polarization attributed to magnetic fields be associated with optically thin dust \citep{Yang17}.     

Excluding the sources identified as having optically thick dust emission (see Table \ref{disk_scatter_sum}), we consider 5 sources\footnote{We also exclude IRS 37-A because it appears to be consistent with dust self-scattering based on its polarization morphology and source inclination.  We were unable to find any complementary high-resolution observations of this source with which to infer an optical depth.} for grain alignment with magnetic fields.  These sources are GY 91, Oph-emb-1, IRS 67-B, IRAS 16293A, and VLA 1623A (extended).   Three of them have $\beta$ values between $\approx 0.6-0.7$, which is still relatively low.  Nevertheless, we suggest that the steeper slope indicates that the dust is not entirely optically thick and that the polarization can be attributed to magnetic fields (see Appendix \ref{tau}).  Oph-emb-1 has the steepest index at $\beta \approx 1$, and for IRAS 16293, we are mainly tracing a dense envelope around the protostars and not disk structures.  Protostellar envelopes are typically less optically thick than disks and we do not expect them to have very large dust grains that are necessary for self-scattering or radiative grain alignment, although some studies suggest that millimeter-sized dust grains could be present in envelopes \citep[e.g.,][]{Miotello14}.  Indeed, the inner 1\arcsec\ region of IRAS 16293B that appears to be optically thick \citep[$\beta \approx 0$, e.g.,][]{Chandler05} is the only section that appears consistent with self scattering \citepalias{Sadavoy18c}.

Assuming the polarization is attributed to magnetic fields, then the polarization position angles must be rotated by 90\degree\ to infer the plane-of-sky magnetic field orientation.  For GY 91, that rotation would imply a radial magnetic field morphology. Such a field morphology could be possible in the case of an hourglass magnetic field that is inclined from the plane-of-the-sky at nearly 90\degree\ \citep{Myers18}.  But the system must have this very special alignment to produce a purely radial field orientation, which seems unlikely.  As such, we do not consider magnetic grain alignment to be likely in the case of GY 91.  For Oph-emb-1, IRS 67-B, IRAS 16293A, and VLA 1623A (extended), we cannot rule out magnetic grain alignment.  We show the inferred field directions for these sources and discuss their implications in more detail in Section \ref{bfield_sources}.

\subsection{Polarization From Other Mechanisms}\label{mech_other}

In this section, we briefly discuss other polarization mechanisms that can affect disk scales.  These mechanisms include aerodynamic grain alignment, mechanical torques, and radiative grain alignment \citep[e.g.,][]{Tazaki17, Yang19, Kataoka19}.   Each of these mechanisms can produce radial or azimuthal polarization patterns depending on gas flow and grain properties and inclination.  These mechanisms can, however, produce uniform polarization if the system is very highly inclined (e.g., near edge-on).   VLA 1623W is our most highly inclined disk detected in polarization at 81\degree, which makes it near edge-on.  Nevertheless, VLA 1623W has relatively high optical depth (see Appendix \ref{tau}), and polarization from grain alignment is suppressed when dust emission is optically thick \citep{Yang17}.  In addition, \citet{Harris18} found identical dust polarization position angles at 872 \um, and we expect the dust emission to be more optically thick at shorter wavelengths.  Based on this analysis, we do not favour any of these alternative polarization mechanisms for our sources with uniform polarization angles.   We instead focus on those sources with azimuthal or complex polarization angles.

While the theoretical framework for these mechanisms is still being developed, we cannot rule out k-RAT alignment, aerodynamic alignment, or mechanical torques for Oph-emb-9, GY 91, IRS 63, IRAS 16293B, and VLA 1623A (extended).   In the case of GY 91, deeper observations are necessary to confirm the polarization structure hinted at in these data.  Ultimately, we lack sufficient sensitivity to properly analyze the polarization for this complicated disk.  For Oph-emb-9, IRS 63, IRAS 16293B, and VLA 1623A (extended), we have a better sampling of their polarization.  These sources also show substantial deviations from idealized elliptical polarization (see Section \ref{morph_desc}), which could imply that multiple mechanisms are contributing the observed polarized structure.  

To determine whether or not these alternative mechanisms are contributed to the observed polarization signatures, we need multi-wavelength dust polarization observations to trace out the polarization structure for different dust grain populations.   Previous multi-wavelength observations of HL Tau have shown a significant change in polarized structure with wavelength that are attributed to different polarization mechanisms.   \citet{Stephens17} found that the polarization morphology of HL Tau transitions from uniform at 870 \um\ to circular at 3 mm \citep[see also,][]{Harrison19}, with the 1.3 mm polarization representing a hybrid of the two patterns.   The uniform polarization at 870 \um\ is well matched by dust self-scattering models \citep{Kataoka15}, whereas the circular polarization at 3 mm is likely due to a different mechanism (see Section \ref{mech_other}).  Initial studies suggested that the circular polarization was from k-RAT alignment \citep{Tazaki17, Kataoka17, Stephens17}, but more recent studies suggest that the signature is inconsistent with that mechanism \citep{Yang19}.  Other mechanisms that may need to be considered are aerodynamic grain alignment \citep{Yang19} or mechanical torques \citep{Kataoka19}.  IRS 63 in particular closely resembles HL Tau at 1.3 mm in both polarization intensity and polarization morphology.  Thus, the polarization we see toward IRS 63 may not be solely due to dust self-scattering.  Oph-emb-9 also shows some similarities to HL Tau and could be another source with multiple mechanisms.


\section{Discussion} \label{discussion}

\subsection{Magnetic Fields on $\lesssim 100$ au Scales}\label{bfield_sources}

The initial goal of this survey was to conduct an unbiased study of dust polarization on $\lesssim 100$ au scales to determine to what extent magnetic fields influence disk formation and fragmentation.  Our study, however, finds that the majority of sources have polarization measurements consistent with dust self-scattering processes rather than magnetic fields.   Out of 37 YSOs detected in continuum emission, only 5 (14\%) of them have polarization morphologies that appear inconsistent with dust self-scattering.  We note, however, that two of these sources also show a mix of dust self-scattering toward the most compact emission, making the detections of magnetic fields down to disk scales more complex.  

Our unbiased survey shows that dust polarization does not appear to be a good tracer of magnetic fields on $\lesssim 100$ au scales on average \citep[see also,][]{Cox18}.  This result is in stark contrast to previous studies of polarization in protostellar cores and envelopes on $> 500$ au scales, which find that nearly all sources are polarized \citep[e.g.,][]{Hull14, Galametz18}.   We note, however, that dust polarization may not be a good tracer of magnetic fields in disks.  For example, larger dust grains from grain growth may be less efficiently aligned with the field \citep[e.g.,][]{Andersson15} or polarization from the magnetic field may be obscured due to dust polarization from competing processes (see Sections \ref{mech_scat} and \ref{mech_other}).  Therefore, we cannot conclude that the undetected disks have no magnetic fields. Their magnetic fields may instead be better detected by other tracers, such as molecular line polarization \citep[e.g.,][Bertrang \& Cortes in prep]{Brauer17}.

\subsubsection{Field Morphologies}\label{diskscales}

\begin{figure*}
\begin{tabular}{ll}
\includegraphics[width=0.48\textwidth]{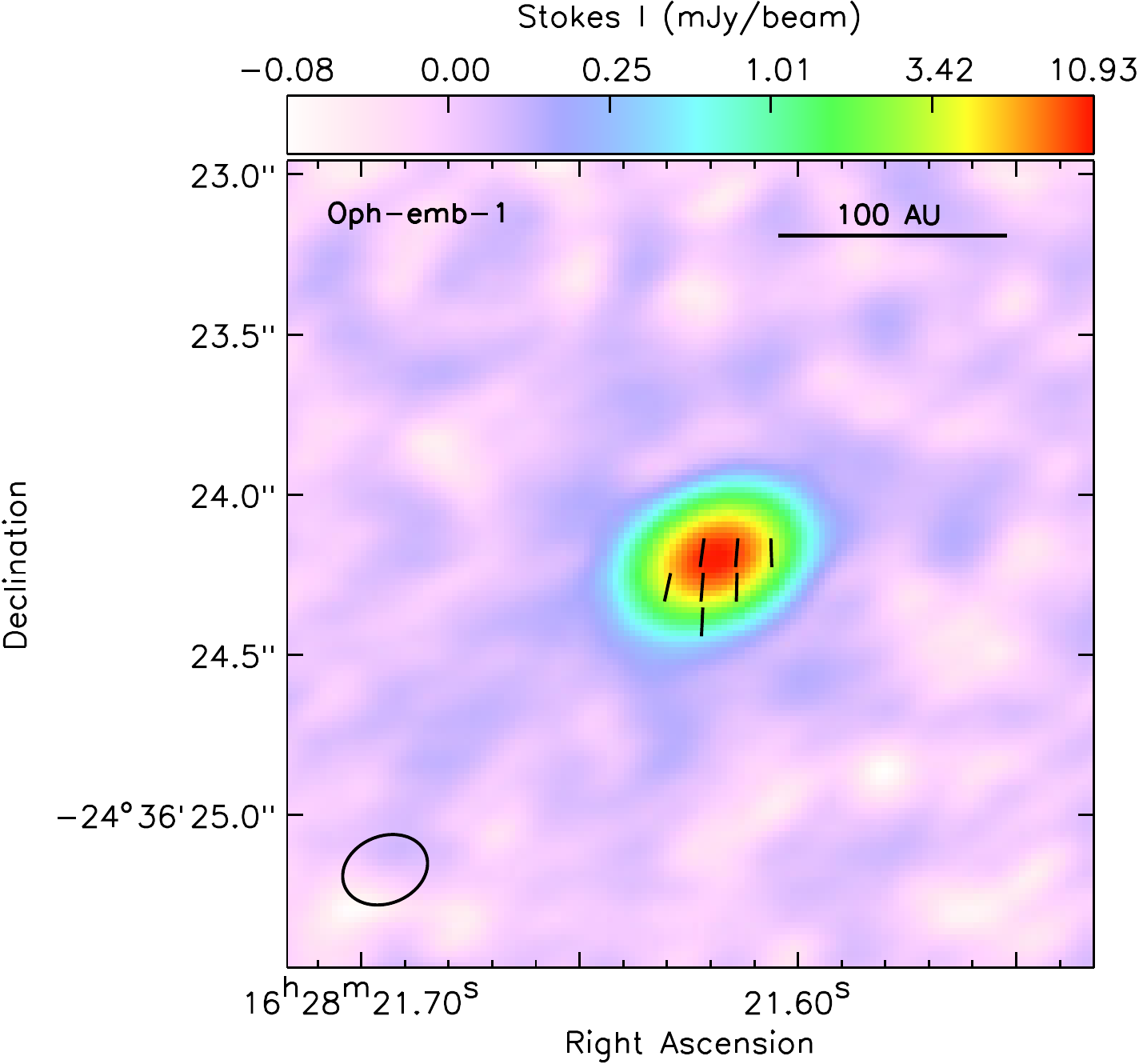} & \includegraphics[width=0.48\textwidth]{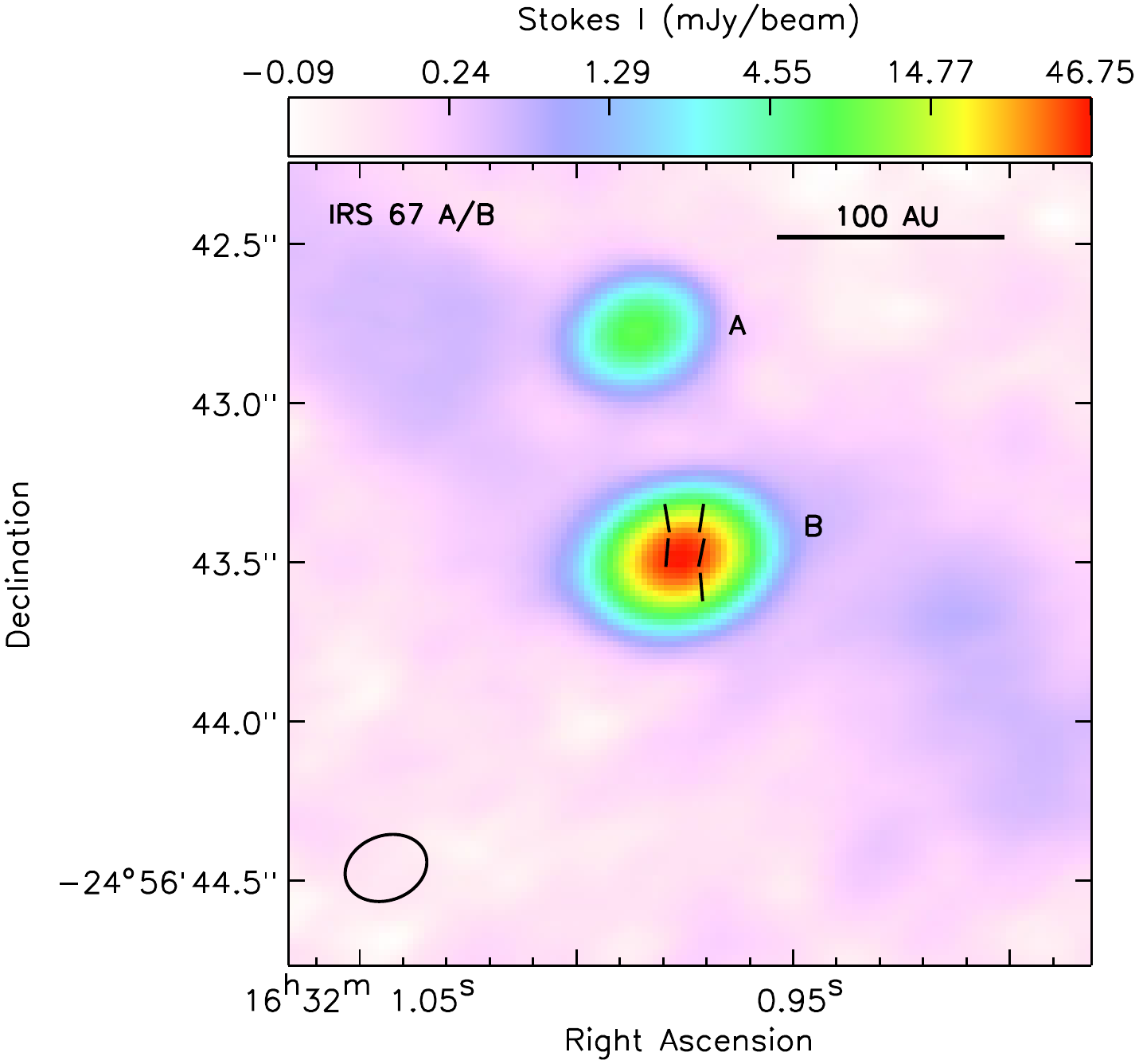} \\
\includegraphics[width=0.5\textwidth]{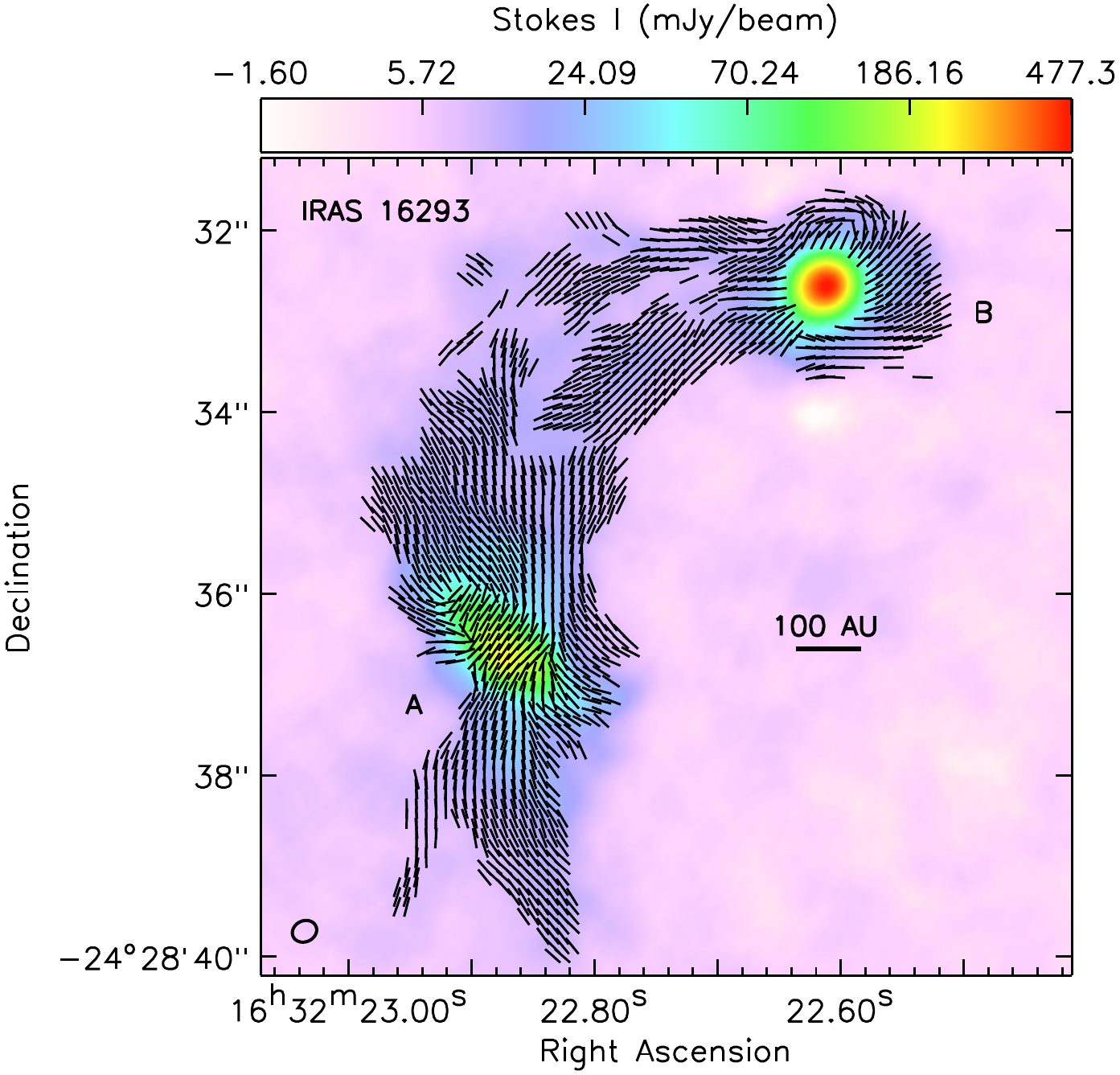} & \includegraphics[width=0.48\textwidth]{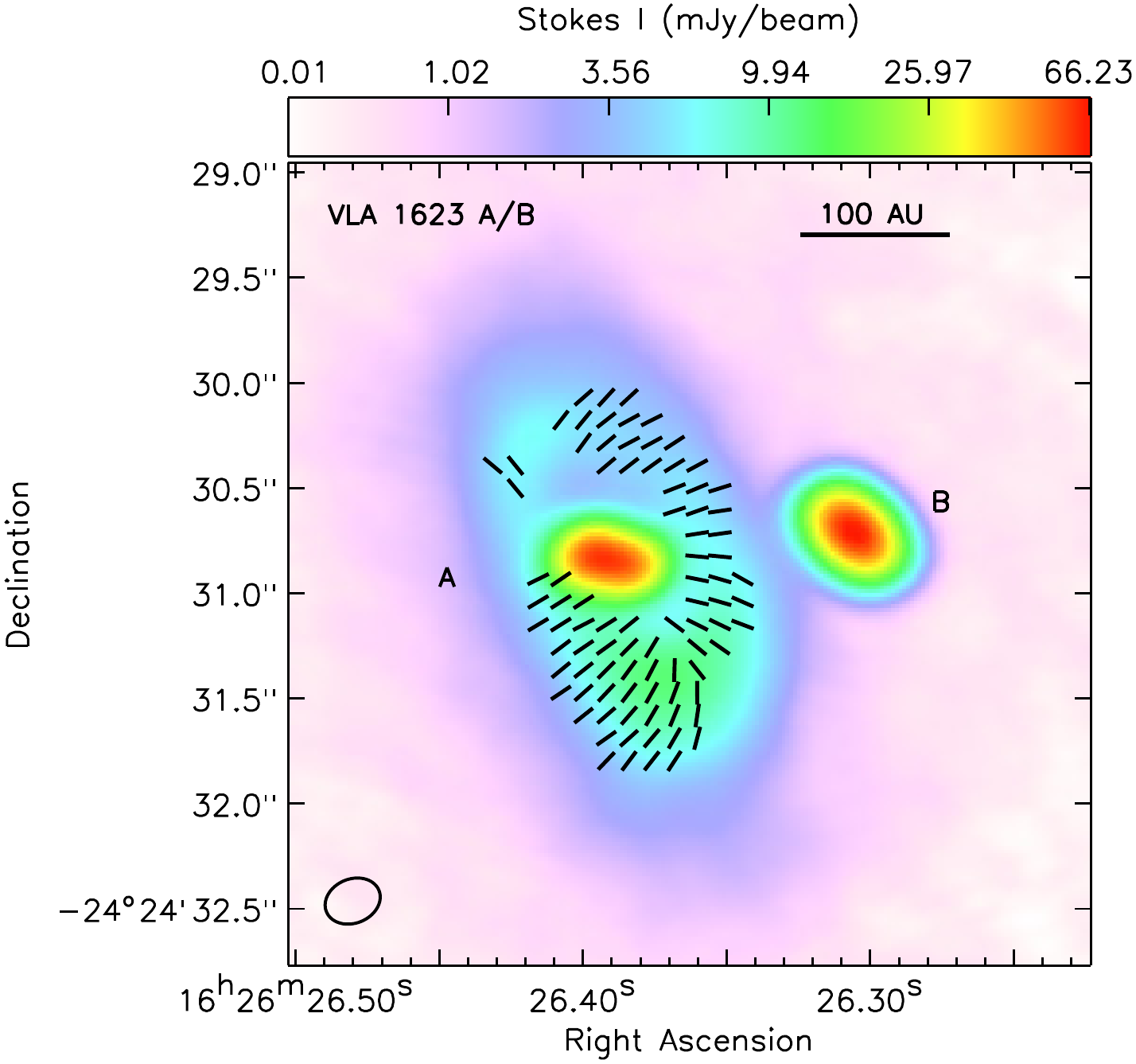}\\
\end{tabular}
\caption{Inferred magnetic field morphology (b-vectors) for the sources that have polarization inconsistent with dust self-scattering.  Line segments represent the same e-vectors as in Section \ref{products} rotated by 90\degree.  For VLA 1623A (compact), VLA 1623B, and IRAS 16293B, we mask out the polarization e-vectors that are associated with self-scattering.  All magnetic field line segments are normalized to show the general field morphology.  \label{bfields}}
\end{figure*}

In this section, we focus on the five sources with polarization morphologies that are inconsistent with expectations from dust self scattering. These sources are IRS 67-B, VLA 1623A (extended), IRAS 16293A, and IRAS 16293B.   (We exclude GY 91 as this source has only a few marginal polarization detections.)   Assuming their dust polarization is attributed to magnetic fields, Figure \ref{bfields} shows their inferred magnetic field orientation obtained from rotating the polarization e-vectors by 90\degree\ (b-vectors).  The line segments in Figure \ref{bfields} are normalized to highlight the broad field morphology.   We also mask out the bright compact disks for VLA 1623A and VLA 1623B, and the bright inner 1$\arcsec$ of IRAS 16293B as these e-vectors appear to be associated with dust self-scattering rather than magnetic fields (see Section \ref{mech_scat}).  

As outlined in Section \ref{mech_bfield}, the polarization around IRAS 16293A and IRAS 16293B mainly trace the inner envelope around the stars and not disks.  The field morphology is also quite different between the two protostars \citepalias[see][for more details]{Sadavoy18c}.  The field around IRAS 16293A shows hints of a pinched morphology, but has a fair degree of disorder that may be attributed to turbulence.  The pinched, hourglass shaped field is more prominent in lower resolution polarization data from the SMA \citep{Rao09,Rao14}, suggesting that gas motions are affecting the field structure only on small scales.  IRAS 16293B also shows hints of a pinched structure at larger radial extents from the protostar, but this morphology may be confused with substantial polarization detected in the dust bridge between the stars.  The dust bridge has relatively uniform polarization where the inferred magnetic field is parallel to the filamentary structure.  In \citetalias{Sadavoy18c}, we suggest that the filamentary structures could be magnetized accretion channels.    

For Oph-emb-1, IRS 67-B and VLA 1623A (extended) the inferred magnetic field is more confined to a compact disk structure.    The inferred fields for IRS 67-B and VLA 1623A (extended) are mainly perpendicular to their disk long axes, whereas the inferred field for Oph-emb-1 is about 60\degree\ offset from the long axis.   Both Oph-emb-1 and IRS 67-B have highly uniform field orientations (see also, Table \ref{disk_scatter}).   VLA 1623A (extended) shows more structure.  Its inferred field morphology is broadly radial with signatures of a pinch. \citetalias{Sadavoy18b} found largely good agreement between the observed field structure and a poloidal, hourglass magnetic field model, although there were some deviations. 

Overall, these protostars have magnetic fields that are primarily poloidal on $<100$ au scales and some of them also have pinched, hourglass-shaped structures.   Such morphologies are generally expected in gravitationally dominated systems where the field is flux-frozen to the gas and dragged inward with the contraction  \citep[e.g.,][]{Mestel66, MestelStrittmatter67, GalliShu93}.  Indeed, each of these sources show evidence of infall or accretion  \citep[e.g.,][]{Mardones97,Pineda12, Evans15, Mottram17, Hsieh19oph1}, which suggests that their larger cores and envelopes are collapsing.  Poloidal fields are important for driving jets and outflows \citep[e.g.,][]{HennebelleCiardi09, Tomisaka11}, but they can also remove angular momentum through magnetic braking and suppress the formation of disks or companion stars \citep[e.g.,][]{Machida05, PriceBate07, HennebelleFromang08, Li11}. 

To first order, we expect magnetic fields in disks to be mainly toroidal, as field lines are wrapped up by rotation  \citep{HennebelleCiardi09, Tomisaka11, Kataoka12}.  This toroidal field is important to stabilize the disk and to promote accretion onto the star \citep[e.g.,][]{BlandfordPayne82, BalbusHawley98, HennebelleTeyssier08}.   Nevertheless, we see no evidence that the magnetic field is dominated by a toroidal component toward these four sources even though all them have signatures of (Keplerian) rotation \citep[e.g.,][]{Murillo13, ArturV18,Calcutt18}.  The inferred magnetic field morphology for a toroidal field would be circular or spiral-shaped \citep[e.g.,][]{Kataoka12, Kataoka17}, whereas the observations show primarily radial or linear field structures.  If there is significant toroidal component within these disks, then our observations suggest that dust polarization may not trace such field morphologies well.

\subsubsection{Comparison with Other Protostars}\label{pfields}

Poloidal and toroidal field morphologies have been seen in a handful of other sources on $\lesssim 100$ au scales.  \citet{Maury18} found extensive polarization in the inner envelope of the Class 0 protostar B335 at $\sim$ 50 au resolution.  The inferred magnetic field shows a pinched structure and is well aligned with the stars' outflow axis.  \citet{Kwon19} detected highly ordered polarization across the inner envelope of the Class 0 protobinary system L1448 IRS 2 at $\sim$ 100 au resolution.  The inferred magnetic field appears to follow an hourglass shape aligned with the outflows in the inner envelope, although the field morphology deviates in the vicinity of the disks.  They suggested that this deviation could be due to an unresolved toroidal component.  \citet{Lee18} also detected complex polarization in the protostellar disk of HH 111. They proposed that the polarization may be tracing a poloidal field at the poles of the disk and a toroidal field in the midplane  \citep[see also HH 211,][]{Lee19}.  Similarly, \citet{Alves18} proposed that their polarization in the circumbinary disk of the Class I protobinary system BHB07-11 is consistent with a mix of poloidal and toroidal field orientations, although there has been some debate \citep[e.g.,][]{Kataoka19}.  Finally, \citet{Ohashi18} find evidence of a toroidal field in the southern half of a protoplanetary disk, HD 142527, where the northern half is consistent with dust self-scattering.  This change in polarization suggests that there may be regional differences in grain populations in the disk.

As discussed in the previous subsection, the magnetic field orientation can have profound consequences for disk formation and stellar multiplicity.  Magnetic fields can suppress formation and fragmentation of the disk via magnetic braking, especially in cases where the disk rotation axis is parallel with the ambient field.  Nevertheless, we find a range of disk and multiplicity results for the systems with mainly vertical fields.  On one hand, IRS 67-B, VLA 1623A, L1448 IRS 2, and BHB07-11 are multiple systems with large disks.  For VLA 1623A (extended) and BHB07-11, their poloidal fields are detected in their circumbinary disks, whereas for L1448 IRS2, the poloidal field is in the extended envelope around the stars.  IRS 67-B also has a large circumbinary disk, but the poloidal field we detect is toward one of the stars and not in the larger disk (see Figure \ref{bfields}).  On the other hand, Oph-emb-1 and B335 are single protostars and their disks are small \citep[e.g.,][]{Imai19}.  \citet{Maury18} proposed that disk formation in B335 may have been suppressed due to magnetic braking from its poloidal magnetic field.   For Oph-emb-1, the disk is larger than B335, but its inferred field orientation is offset from the poloidal axis.  Finally, IRAS 16239A also shows hints of an hourglass poloidal morphology, but its disk properties are unclear due to confusion with the envelope. 

Since these several systems have broadly similar poloidal fields and a range of protostar numbers and disk properties, the degrees of magnetic braking must vary with each system.  For a system like B335, where the disk is small and the star is singular \citep{Evans15, Imai19}, magnetic braking may still be prominent.  Conversely, IRS 67-B, VLA 1623A (extended), L1448 IRS 2, and BHB07-11 have multiple stars and larger disks, and are likely systems where magnetic braking is weakened.  Theoretical studies have shown that magnetic braking can be mitigated if the poloidal magnetic field is misaligned with the rotation axis of the collapsing core \citep[e.g.,][]{HennebelleCiardi09, Joos12}, through magnetic reconnection from turbulence \citep[e.g.,][]{Seifried13, Gray18}, or from non-ideal magnetic hydrodynamic (MHD) processes that decouple the field and gas \citep[e.g.,][]{Tomida15, Masson16, Vaytet18}.  For VLA 1623A (extended), L1448 IRS 2, and BHB07-11, their inferred magnetic fields appear to be aligned with their axes of rotation as traced by outflows \citep[e.g.,][]{Alves17, Sadavoy18b, Kwon19}.  These systems may require either turbulence or non-ideal MHD to circumvent magnetic braking.  A system like B335, however, does not appear to have circumvented its magnetic braking and as a consequence, it's disk is very small.  Oph-emb-1 may be somewhere in the middle.  The apparent misalignment between its field axis and its rotation axis may have reduced the effects of magnetic braking so that it could form a larger disk than what is seen in B335.

\subsection{Comparison with Magnetic Fields on Clump Scales}\label{largescales}

In this section, we compare our high resolution ALMA data to the magnetic field measurements in Ophiuchus on clump scales.  For simplicity, we focus on far-infrared and submillimeter polarization that resolves the Ophiuchus clumps from the James Clerk Maxwell Telescope (JCMT) and the Stratospheric Observatory for Infrared Astronomy (SOFIA) polarimeters.  To first order, we also assume that that far-infrared and submillimeter polarization on clump-scales is generally attributed to grain alignment from a magnetic field.  

The $\rho$ Oph A clump has been observed in dust polarization at 89 \um\ and 154 \um\ with SOFIA/HAWC+ \citep{Santos19} and at 850 \um\ with JCMT/POL-2 \citep{JKwon18}, and the Oph B and C clumps have been observed at 850 \um\ \citep{Soam18, Liu19}.  For Oph A, the 214-850 \um\ data show largely consistent, well-ordered polarization across Oph A such that the inferred magnetic field appears to be mostly uniform.  At 89 \um, there is still broad agreement with the longer wavelength data, but the far-infrared observations includes additional polarization structure east of the main dense clump that is not seen at the longer wavelengths.  This eastern region may be tracing smaller dust grains and cannot be compared to our millimeter observations.  For Oph B and C, the polarization structures and inferred magnetic fields for these two regions are more disordered than what is seen in Oph A and their field strengths are lower than what was obtained for Oph A \citep{Liu19}.  

Eight of our sources lie within Oph A and six of these are well detected in dust polarization (note that GY 91 lies outside of the polarization maps in the aforementioned studies).  By contrast, three of our sources are in Oph B with none of them are detected in polarization, and Oph C is starless.  We therefore focus our large-scale comparison on the six sources in Oph A.   In general, we find that the large-scale polarization and small-scale polarization morphologies are inconsistent.  Some of these differences may be explained by different polarization mechanisms at small scales from magnetic fields on large scales.  Indeed, we find that GSS 30 IRS 1, GSS 30 IRS 4, Oph-emb-6, and the three VLA 1623 circumstellar disks show polarization consistent with dust self-scattering rather than magnetic fields.    Dust self scattering is unlikely to have the same polarization structure as the inferred magnetic field.

For the VLA 1623 region, however, \citet{JKwon18} find an inferred magnetic field direction that is roughly perpendicular to the inferred magnetic field axis from the hourglass model in \citetalias{Sadavoy18b}.  While the dust scattering signatures toward the circumstellar disks can dominate over an hourglass field signature when both components are unresolved \citep[e.g., see the 3\arcsec\ polarization map from][]{Hull14}, these contributions will be localized to VLA 1623A/B  and VLA 1623W, whereas the lower-resolution POL-2 observations show uniform polarization well off of these sources across $\gtrsim 0.05$ pc area of Oph A.  This uniform field structure is in agreement with the slightly higher resolution (7.8-13.6\arcsec) data from SOFIA \citep{Santos19}, but the far-infrared data also show hints of a pinched magnetic field in the vicinity of VLA 1623, although at a different orientation.   If the SOFIA, JCMT, and ALMA observations are each tracing pinched, hourglass-shaped magnetic fields, then the orientation of that field appears to change from the clump-scale to the disk-scale.  This change suggests that either the polarization in the extended disk of VLA 1623A does not trace a magnetic field or the magnetic field toward VLA 1623A may be affected by dynamical processes, such as rotation or outflows, that alters its orientation on small scales.

\subsection{Disk Properties}\label{disk_prop}

We find 37 continuum sources associated with YSOs.  In Appendix \ref{indiv}, we give the source size and mass assuming the compact emission can be fitted with a Gaussian (see also, Table \ref{cont_results}).  We assume these continuum detections are tracing thin disks with smooth density distributions, e.g., most of the continuum sources are compact and ellipsoidal in shape.   We caution, however, that both the Gaussian fits and the mass measurements should be considered approximations since (1) many of these sources are known to have complex disk structure, (2) their dust emission may include flux from the disk and inner envelope, and (3) we adopt the same temperature and opacity for each object.  Nevertheless, applying the same approach to all sources allows us to conduct a broad comparison between them such that the statistical results are still robust.   

\begin{figure*}
\includegraphics[width=0.97\textwidth]{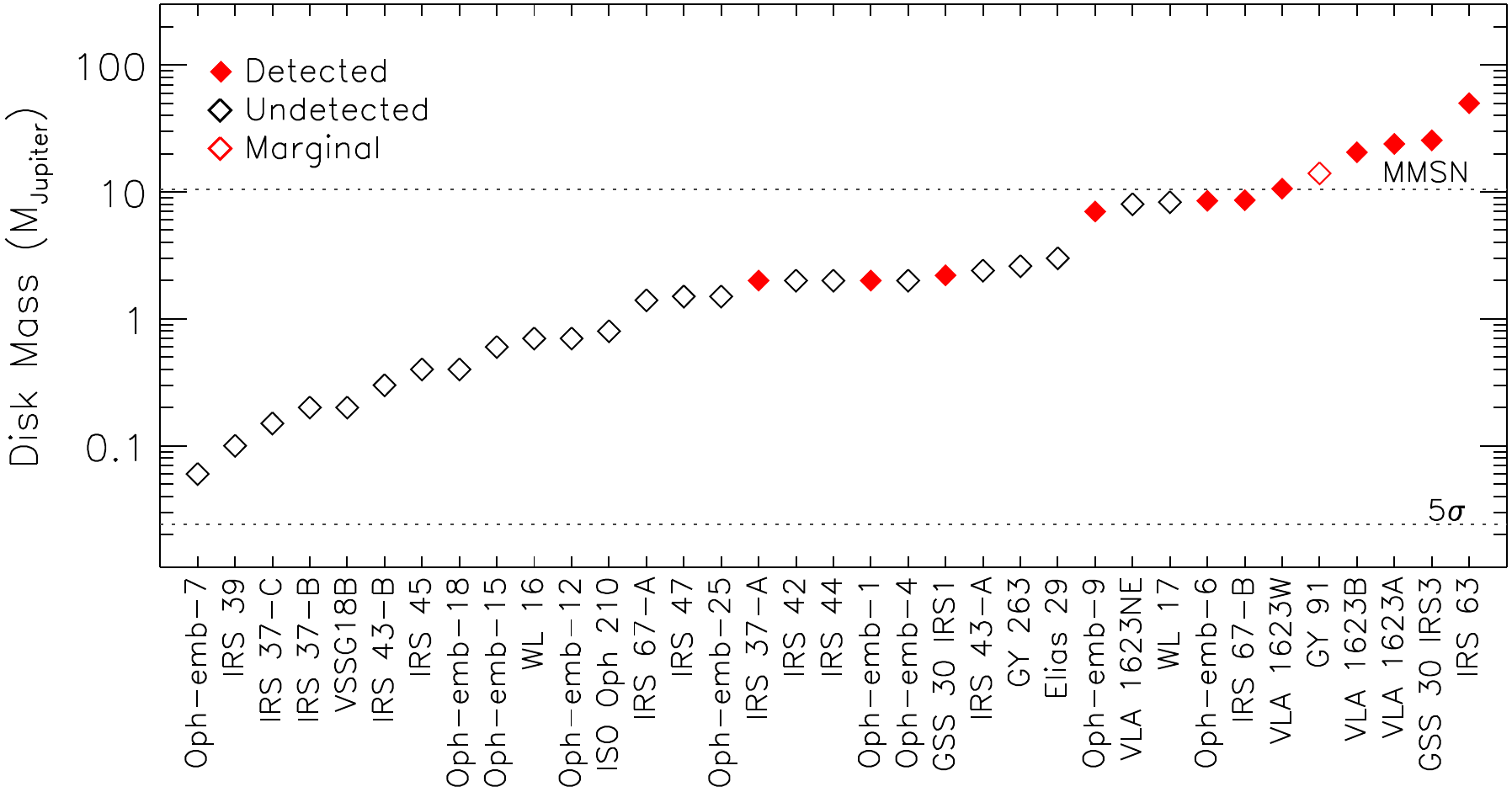}
\caption{Estimated masses for the YSOs in our sample from Gaussian fits to the continuum emission.  To first order, we assume these fits are tracing disks.  Sources detected in polarization are shown in filled red symbols and undetected sources as open black symbols.  The marginal detection, GY 91, is represented by open red symbols and its upper and lower mass estimates in the literature (see text).  IRAS 16239A and IRAS 16293B are not included.   Dotted lines show the 5$\sigma$ mass limit assuming $\sigma = 30$ \uJybeam\ and the minimum mass solar nebula (MMSN) limit.  These are total disk masses, assuming a gas-to-dust mass ratio of 100.  \label{disk_dist}}
\end{figure*}

Figure \ref{disk_dist} shows the distribution of the inferred disk masses in our sample in order of increasing mass.   For simplicity, we exclude IRAS 16293A and IRAS 16293B because their dust continuum is significantly confused with envelope emission and their Gaussian fits are unreliable.  We also exclude the circumbinary disk around VLA 1623A (estimated mass of 0.1 \Msun) and focus instead on the compact circumstellar emission around VLA 1623A and VLA 1623B that can be fitted with a single Gaussian \citepalias{Sadavoy18b}.   We show sources detected in polarization as filled red symbols and the undetected sources as open black symbols.  For GY 91, we use an open red symbol to represent the marginal detection.   

Most of the disks in Ophiuchus have relatively low masses.  Roughly 83\%\ (29/35) have masses $< 10$ \Mjupiter\ (0.01 \Msun), which corresponds to the minimum mass solar nebula \citep[MMSN;][]{Weidenschilling77, Hayashi81}, and roughly 34\% (12/35) have masses $\lesssim 1$ \Mjupiter.   Similar low-mass disks were also seen in T-Tauri stars in Orion \citep{Ansdell17}, Lupus \citep{Ansdell16}, and Chameleon \citep{Long18}.   Our disk masses, however, assume $\gdratio = 100$, whereas  observations of T-Tauri stars generally find $\gdratio < 100$ \citep[e.g.,][]{Ansdell16, Long17, Miotello17}.  We adopt the typical ISM ratio as most of our targets are in the protostellar (Class I) stage, and therefore may still be gas rich. 

 The more massive disks tend to be detected in polarization.  This result matches the detection summary in Section \ref{pol_overview}, where the brightest sources also tended to be detected in polarization.  All sources above 10 \Mjupiter\ are detected in polarization (or marginally detected), although disks with masses down to 2 \Mjupiter\ are also detected.  Of the 20 sources with masses $> 2$ \Mjupiter, 8 are not detected in polarization and 7 of these non-detections have robust 3$\sigma$ upper limits to their polarization fraction of $\lesssim 1$\%, which indicates that these sources have significantly low polarization.   We discuss these non-detections in more detail in Section \ref{non_det}. 

\begin{figure}
\includegraphics[width=0.475\textwidth]{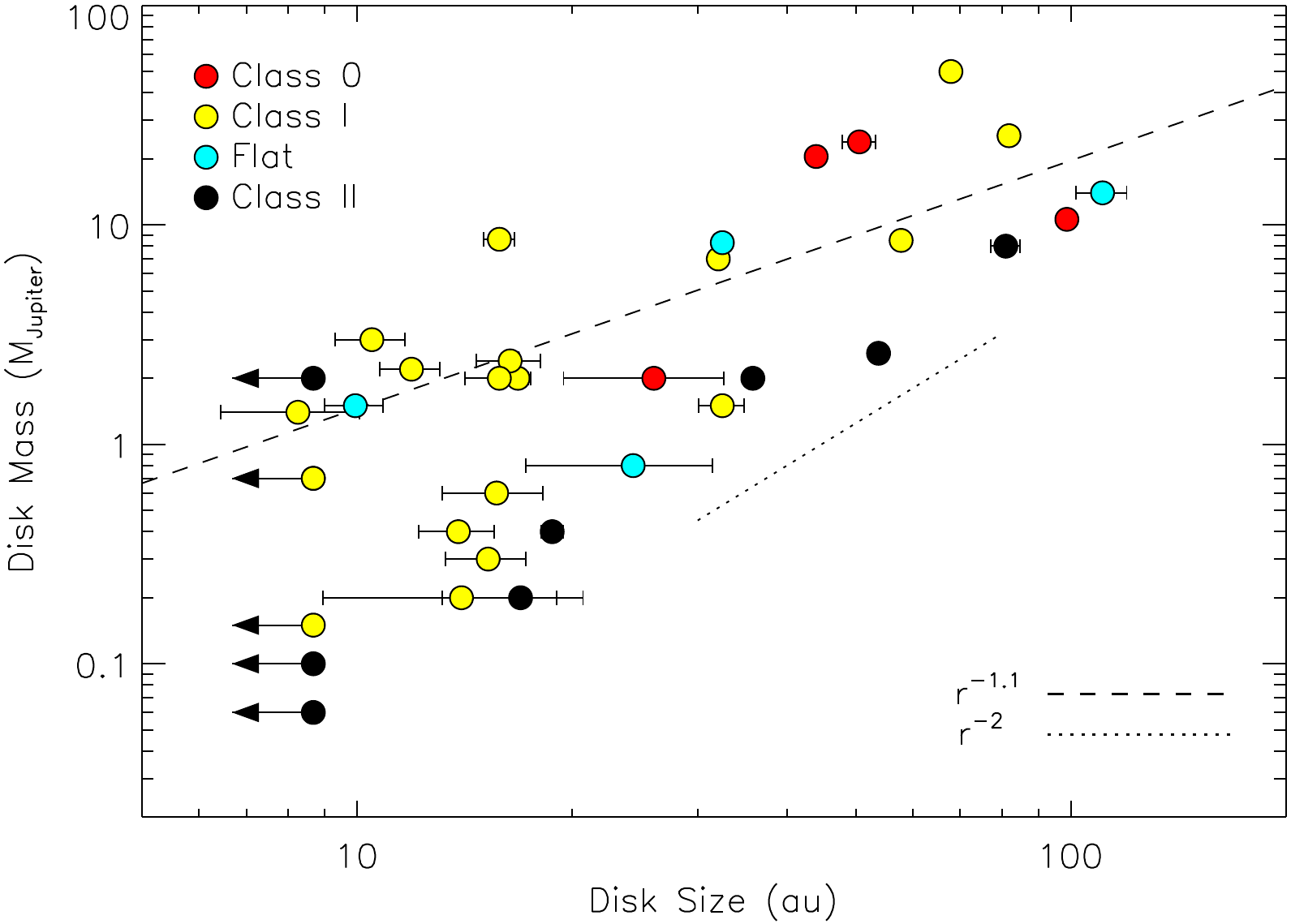}
\caption{Comparison between disk mass and disk size for the YSOs detected in our sample.  Sources are separated by their YSO Classification, with Class 0 disks in red, Class I disks in yellow, Flat disks in cyan, and Class II disks in black.  Unresolved sources are represented by upper limits at 8.7 au (1/4 of the beam).  We do not plot error bars for the mass, because the masses are uncertain within factors of a few.  The dashed line shows a weighted linear least squares fit to the observations.  The dotted line shows an r$^{-2}$ relation for comparison.  \label{mass_size}}
\end{figure}

Figure \ref{mass_size} compares source mass and size (semi-major axis FWHM) that we measure from the Gaussian fits.  Unresolved sources are plotted as upper limits with a fixed size equal to roughly 1/4 of the beam (8.7 au).  The sources are also separated by their classification, with Class 0 in red, Class I in yellow, Flat in cyan, and Class II in black.  For clarity, we do not plot mass error bars as the masses are uncertain within factors of a few\footnote{Mass uncertainties are dominated by error in dust opacity and temperature.  Dust opacity models can vary by factors of $2-4$ at 1.3 mm \citep[e.g.,][]{Ossenkopf94}, and the dust temperature should vary with protostar luminosity \citep[e.g.,][]{Andrews13}, e.g., dust temperatures of 30 K or 50 K would decrease the inferred masses by factors of 1.7 and 3, respectively.}.  Therefore, we only consider broad properties for the continuum detections, assuming they are tracing disks.    

Overall, we find that disk mass and size are correlated.  The dashed line shows the best-fit weighted linear least squares relation to the full sample of resolved disks.  The slope of this relation is 1.13 $\pm$ 0.01.  We note that this linear least squares fit is heavily biased by the more massive disks, as they tend to be brighter and have lower fitting errors.  An unweighted linear least squares fit is steeper with a slope of 1.41 $\pm$ 0.24.    Assuming that $\Sigma \propto M/R^2$, we find that $\Sigma \sim R^{-(0.6-0.9)}$ using both the weighted and unweighted linear least squares fits.   \citet{Andrews09} found a similar surface density of $\Sigma \propto r^{-0.9}$ for nine protoplanetary disks in the Ophiuchus based on SMA observations and radiative transfer models to determine the disk profiles.  This agreement suggests that protostellar disks and protoplanetary disks have similar dust surface densities even though processes such as planet formation, grain growth, and gas depletion may be different.  More detailed disk modeling, however, is necessary to produce more accurate disk masses and test the relationship between disk surface densities as a function of source evolution. 

There is no strong correlation between YSO Class and disk size, although we note that our statistics for the Flat sources and Class II sources are incomplete.  \citet{SeguraCox18} found that Class 0 and Class I disks in Perseus had comparable sizes and masses \citep[see also,][]{Andersen19}, whereas \citet{Maury19} found that Class 0 disks appear smaller than Class I disks.  With only four Class 0 protostars in the Ophiuchus cloud, there are too few sources to support either study.   In general, \citet{Cieza19} found that the majority of Class II disks in Ophiuchus have diameters $< 30$ au, which is comparable to what we see for most objects.   Combining our results with those of Cieza et al., we find that most disks in Ophiuchus have sizes $<30$ au (FWHM) between Class I and Class II, with hints that the Class 0 disks may be larger on average, although we require a larger sample to verify this claim.

\subsection{Non-detected Polarization} \label{non_det}

The majority of our disks are undetected in polarization. Table \ref{undet_results} lists these systems with their 3$\sigma$ upper limits to their undetected polarization.  Roughly half (13/23) of the non-detected disks have $3\sigma$ upper limit polarization fractions $\lesssim 2$\%\ and 7 of these disks have  3$\sigma$ upper limits of $< 1$\%.  Since previous observations of polarization on envelope scales found typical polarization fractions of $\gtrsim 2$\% for YSOs \citep[e.g.,][]{Hull14, Cox18}, many of the disks undetected in polarization have significantly low polarization fractions.   

Indeed, many disks may have very low polarization fractions in general.   \citet{Hughes13} observed three protoplanetary disks in polarization with the SMA and CARMA and found that all were undetected with 3$\sigma$ upper limits of $< 1$\%. Subsequent ALMA observations of one disk, DG Tau, found that these it had uniform polarization consistent with dust self-scattering at low polarization fractions of $\sim 0.4$\%\ \citep[e.g.,][]{Bacciotti18}.  Similarly, \citet{Harrison19} detected very low polarization fractions $< 1$\%\ toward a small sample of protoplanetary disks with ALMA in 3 mm dust polarization.  They also found that two of their six disks remained undetected with 3$\sigma$ upper limits $<1$\%\ with ALMA.  Since the four detected disks show polarization attributed to self-scattering processes or radiative grain alignment, Harrison et al. suggested that the non-detected disks may lack a population of large dust grains which are needed to produce these polarization signatures.   Deeper observations, however, are necessary.

Low polarization fractions may arise if the grains cannot be efficiently aligned (e.g., they lack paramagnetic material necessary to align with the magnetic field) or if the grains are too small to produce a dust self-scattering signature.   \citet{Kataoka15} found that dust grains with sizes of $\approx \lambda/2\pi$ are most efficient at producing polarization from self-scattering processes.  At 1.3 mm, this size corresponds to 200 \um, but a small range in dust grain sizes (e.g., $\sim 50-250$ \um) can produce a polarization signature at 1.3 mm from self scattering.   Simulations of dust grain growth suggest that grains with sizes of $\sim 100$ \um\ can form quickly in the inner radii of disks where the high surface densities are high \citep[e.g.,][]{Brauer08, Birnstiel10}.   We see no evidence of unusually low surface densities in our disks to suggest that they cannot form large dust grains (see Section \ref{disk_prop}).

\begin{figure}[h!]
\includegraphics[width=0.475\textwidth]{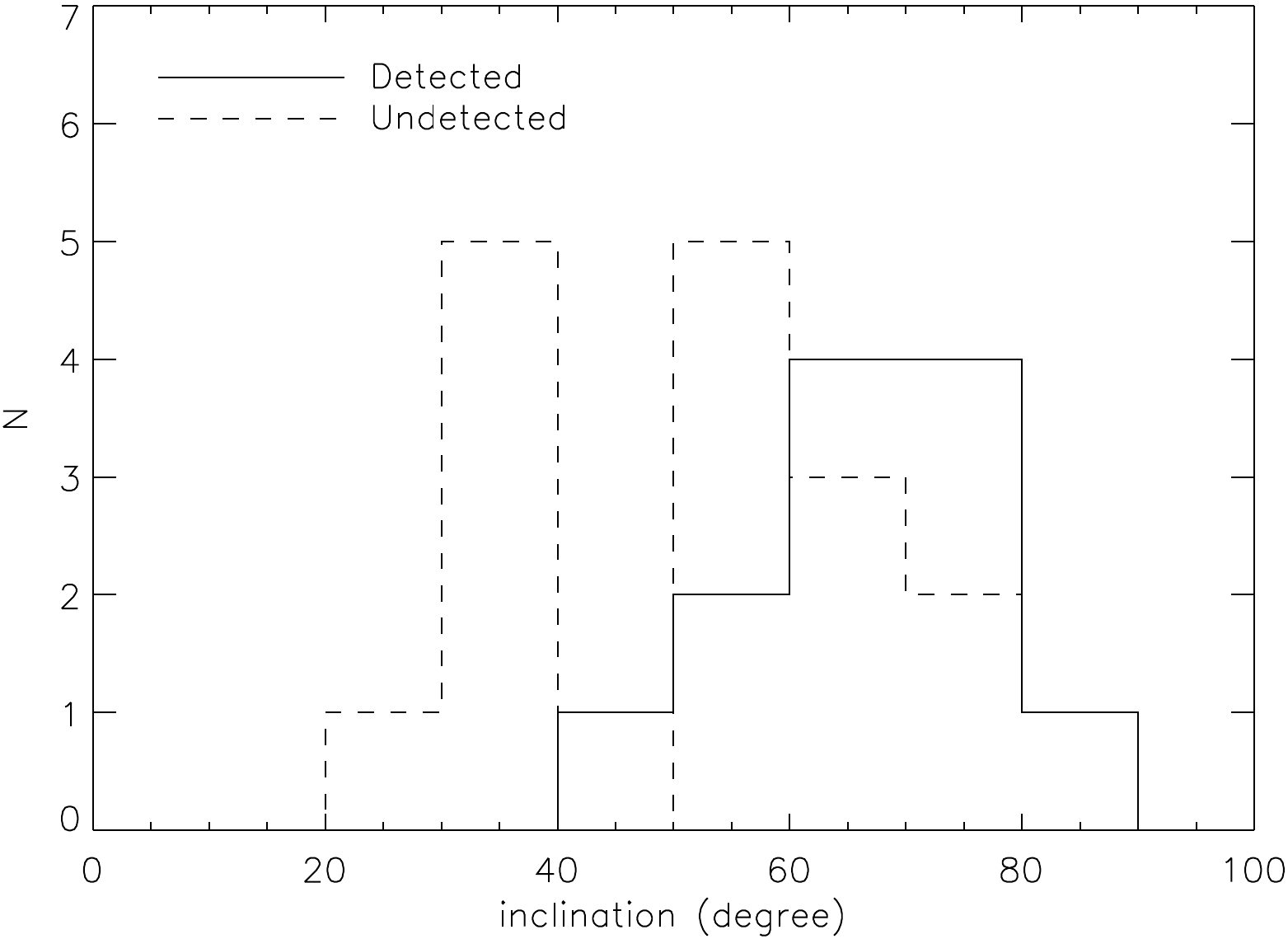}
\caption{Histograms of disk inclination for the resolved disks.  Source detected in polarization are shown by the solid line, and the undetected sources as the dashed line.  These histograms exclude IRAS 16293A, IRAS 16293B, and GY 91 for simplicity.   These data include the compact and extended circumbinary disks of VLA 1623A as two separate objects.   \label{inc_histo}}
\end{figure}

Alternatively, these disks may have unresolved polarization structure that appears depolarized at our resolution.  In Sections \ref{mech_scat} and \ref{mech_bfield}, we outline how disk geometry can affect the observed polarization signatures.  Figure \ref{inc_histo} compares histograms of disk inclination for the resolved sources that are detected in polarization (solid) with those that are undetected (dashed).  For simplicity, we exclude IRAS 16293A and IRAS 16293B, as these sources do not have clear disks with which to measure their inclination, and GY 91, because it is only marginally detected in polarization.  To help improve the statistics, we include two values for VLA 1623A, one for the compact circumstellar material around the unresolved binary and a second entry for the extended circumbinary disk (see Figure \ref{vla1623ab}). We can treat these two components separately because they show different polarization structures attributed to different mechanisms (see Section \ref{mech_discuss}).    

Figure \ref{inc_histo} shows that the disks detected in polarization tend to have inclinations of $> 60$\degree, whereas the undetected disks have more shallow inclinations of $< 60$\degree.   Using an Anderson-Darling test, we find that the two distributions are inconsistent at the 2.5\%\ level.  We note, however, that disk inclinations as measured from $\cos{i} = b/a$ are only robust if the disk is well resolved (e.g., over two beams along the minor axis) and geometrically thin.  Since most of our disks are compact ($< 2$ beams), we may be biased to steeper inclinations.  Nevertheless, higher resolution observations of WL 17, Oph-emb-1, and VLA 1623B \citep{SheehanEisner17, Harris18, Hsieh19oph1} give consistent inclinations within 5\degree\ of our estimate, which suggest that our inclinations are broadly reliable.  

\begin{figure}[h!]
\includegraphics[width=0.475\textwidth]{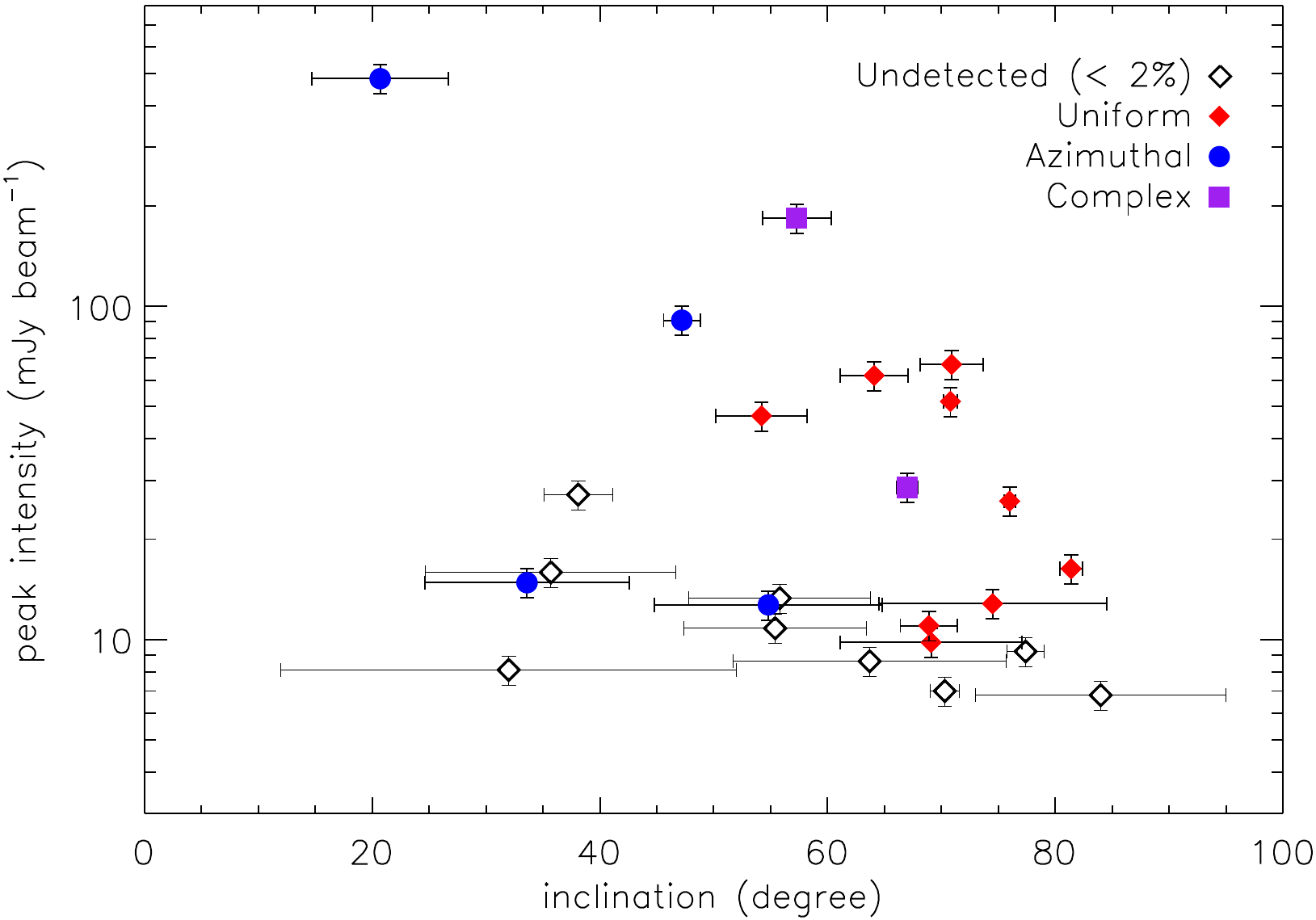} \includegraphics[width=0.475\textwidth]{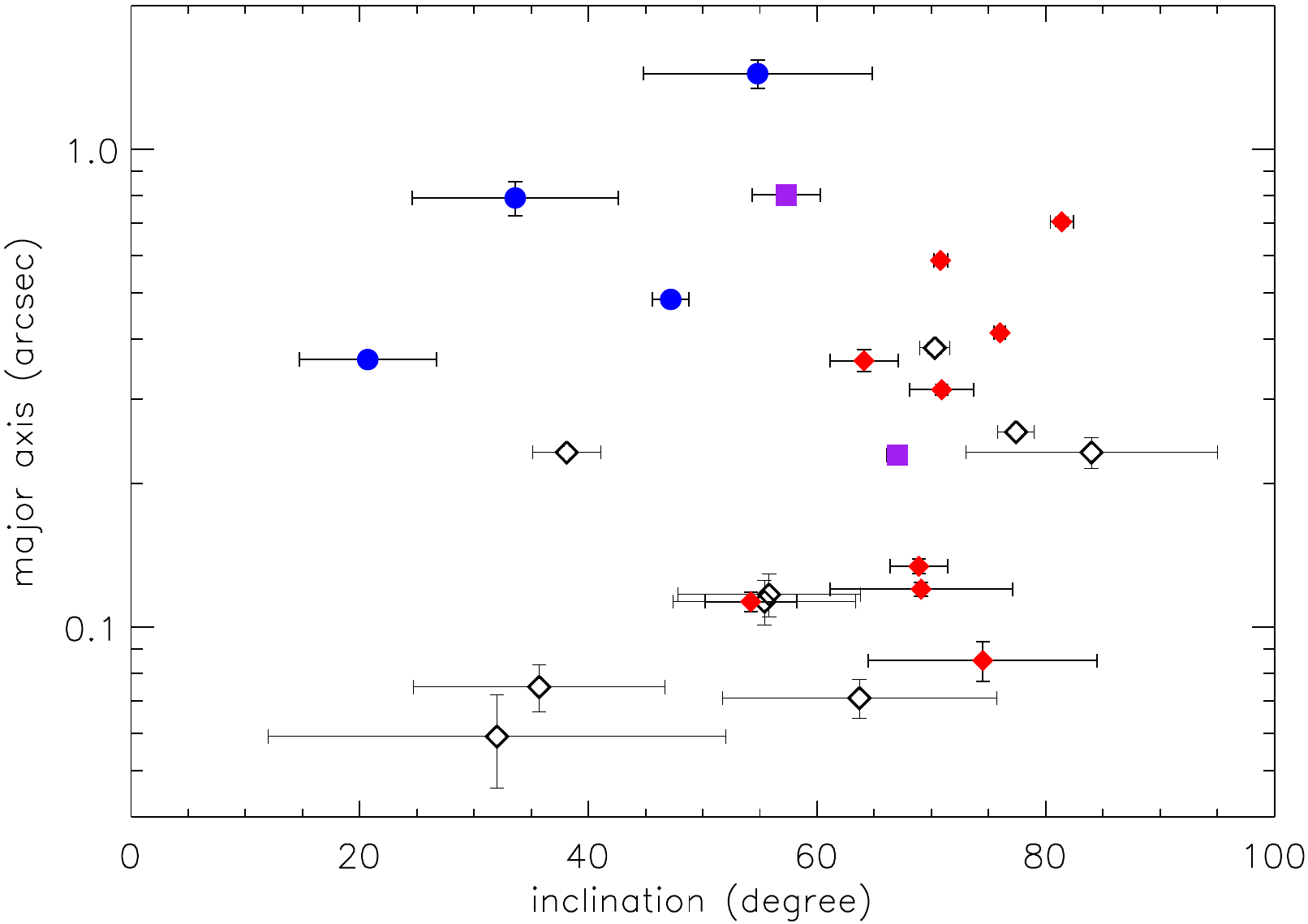}
\caption{Source inclination versus peak Stokes I intensity (top) and deconvolved semi-major axis (bottom).  Red diamonds represent sources with uniform polarization, blue circles represent sources with azimuthal polarization, and purple squares are sources with complex polarization (see Table \ref{disk_scatter}).  Open black diamonds represent sources that are undetected in polarization with 3$\sigma$ upper limits $<2$\%\ (see Table \ref{undet_results}).  For the peak flux, we show 10\%\ error bars representing the flux calibration uncertainty.   \label{inc_size}}
\end{figure}

Figure  \ref{inc_size}  compares disk inclination with peak intensity (top) and deconvolved semi-major axis size (bottom) for the sources with uniform polarization (red diamonds), azimuthal polarization (blue diamonds), and undetected sources (open diamonds).  For simplicity, we only show the undetected sources with 3$\sigma$ upper limits $< 2$\%.  We find that disks with inclinations $> 60$\degree\ tend to have uniform dust polarization morphologies.  Since disks with higher inclinations will be observed through higher columns of dust, we should expect these disks to have higher optical depths on average.  In these cases, we would expect the dust polarization signature to be dominated by self-scattering and the signature to be uniform and along the minor axis due to the high inclination \citep[e.g.,][]{Kataoka16, Yang16, Yang17, Kataoka17}.   

Figure \ref{inc_size} also shows that sources with azimuthal polarization tend to have lower inclinations ($< 50$\degree) and very large sizes.  Sources of comparable inclination but smaller size may therefore appear unpolarized due to unresolved azimuthal structure.  For example, WL 17 is the brightest source in our sample that is undetected in dust polarization.  It has a low dust opacity index of $\beta < 0.3$ (see Appendix \ref{tau}), such that we would expect it to be detected in polarization from self-scattering processes.  Based on its moderate inclination of 38\degree\ and central cavity \citet{SheehanEisner17}, the expected polarization pattern from self-scattering should be azimuthal \citep[e.g.,][]{Pohl16, Ohashi18}.  Indeed, GY 91 and IRS 63 also have moderate inclinations are are dominated by an azimuthal polarization structure.  WL 17, by contrast, is marginally resolved compared to GY 91 and IRS 63, and an azimuthal pattern would be depolarized within the beam \citep[e.g., see the 3 mm observations of HL Tau in][]{Kataoka17}.  We note that smaller and fainter disks than WL 17, e.g., GSS 30 IRS 1 and IRS 37-A, are well detected in polarization likely because they have high inclinations ($\gtrsim 70$\degree) such that their polarization pattern form self-scattering is uniform and can be recovered even if the disk is only marginally resolved.

There are three larger sources ($\gtrsim 0.2$\arcsec) with high inclinations ($> 70$\degree) that are also not detected in polarization with 3$\sigma$ upper limits of $\sim$ 1\%.  These sources are GY 263, IRS 47, and Oph-emb-4.  Based on their inclinations, we would expect these objects to have uniform polarization aligned with their minor axis due to dust self-scattering.  Since other disks in this study and in the literature show very low polarization fractions  of $< 1$\%\ \citep[e.g., like HL Tau or DG Tau;][]{Stephens17, Bacciotti18}, we may need deeper observations to detect the polarization for GY 263, IRS 47, and Oph-emb-4.  Alternatively, the very low polarization fractions may indicate that these disks lack large dust grains or high optical depths necessary to produce a dust self-scattering signature, as seen for similarly inclined but brighter disks like GSS 30 IRS 1 and IRS 37-A.

\subsection{Galaxy Contamination}\label{gal}

Several of our new detections are faint, point-like objects that are also undetected at near-infrared and mid-infrared wavelengths with \emph{Spitzer} and WISE (see Appendix \ref{indiv}).  Since these objects are undetected in the infrared, they are unlikely to be YSOs.  We therefore consider the possibility that they are external galaxies.   Several recent surveys have estimated the background galaxy source counts as a function of brightness using the SMA \citep[e.g.,][]{Hayward13} and ALMA \citep{Carniani15, Hatsukade18}.  These studies typically estimate $> 10^4$ galaxies with intensities $> 0.15$ mJy per square degree at $\sim 1$ mm.  Within the ALMA primary beam, these source counts amount to a non-negligible probability of detecting a galaxy in each field.

The likelihood of detecting a background galaxy is a strong function of its brightness and where it falls  in the primary beam.  To estimate the probability of detecting a background galaxy in our observations, we calculate the likelihood of finding a galaxy above a threshold across the primary beam.  For the galaxy number counts, we use the 1.3 mm results of \citet{Carniani15} based on ALMA observations.   They give a Schechter function of the differential galaxy numbers of,
\begin{equation}
\frac{dN}{dS} = 1800\ \mbox{deg$^{-2}$} \left(\frac{S}{S_0}\right)^{-2.08}\exp\left(-\frac{S}{S_0}\right),
\end{equation}
where $N$ is the number of galaxies with intensity, $S$, and $S_0 = 1.7$ mJy.  For simplicity, we adopt three galaxy intensity thresholds of 5$\sigma(r)$, 10$\sigma(r)$, and 20$\sigma(r)$, where $\sigma(r)$ is the map sensitivity as a function of position in the primary beam.  We adopt $\sigma = 0.03$ mJy at the phase center and allow $\sigma(r)$ to increase at larger radial extents according to the primary beam correction.  Since the threshold varies across the primary beam, we calculate the probability of detecting a galaxy at each threshold in annuli separated by $\sim 0.3$\arcsec, or roughly the synthesized beam width.  

\begin{figure}[h!]
\includegraphics[width=0.475\textwidth]{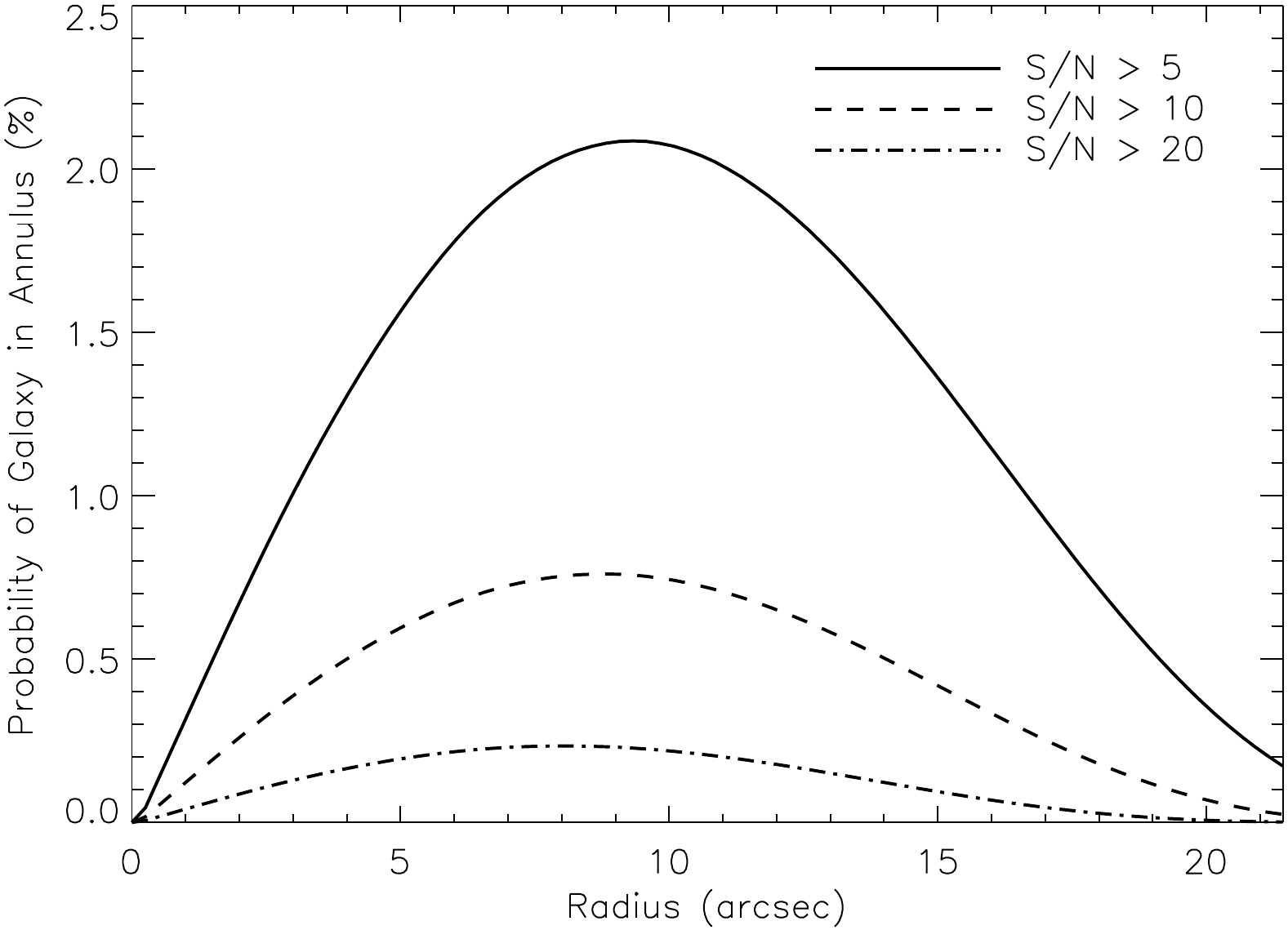}
\caption{Probability of detecting a point-source galaxy at 1.3 mm in different annuli within in the ALMA primary beam, where $R = 0$ corresponds to the phase center.  All annuli have a width roughly equal to the synthesized beam.  The curves show the probability distributions for galaxies detected at $> 5\sigma(r)$ (solid), $> 10\sigma(r)$ (dashed), or $> 20\sigma(r)$ (dot-dashed), assuming $\sigma = 0.03$ mJy at the phase center.   \label{galaxy_diff}}
\end{figure}

Figure \ref{galaxy_diff} shows the probability of detecting a galaxy in these annuli as a function of radial extent in the primary beam.  In all three cases presented, we find the probability differential functions peak near the halfway point between the phase center and the primary beam FWHM.  As outline above, there are two competing factors to the probability of detecting a galaxy across the primary beam.  The search area increases with radius, but the sensitivity to the galaxy decreases with radius.  The distributions peak at slightly different locations for each of the thresholds.  We find a peak probability at 9.3\arcsec\ for the 5$\sigma$ threshold, 8.8\arcsec\ for the 10$\sigma$ threshold, and 8.1\arcsec\ for the 20$\sigma$ threshold. 

\begin{figure}[h!]
\includegraphics[width=0.475\textwidth]{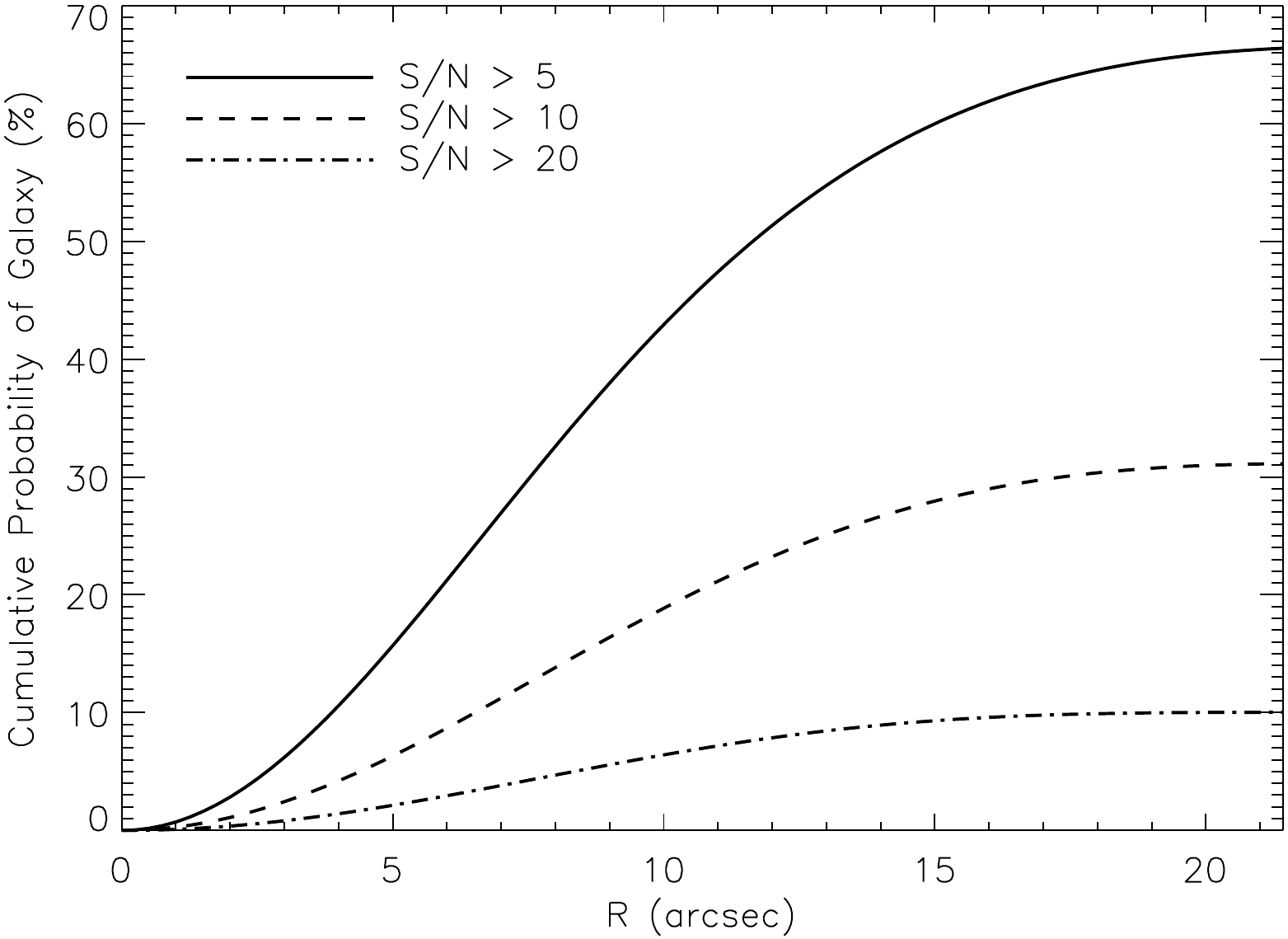}
\caption{Cumulative probability of detecting a point-source galaxy at 1.3 mm as a function of position in the ALMA primary beam for different detection thresholds. The curves show thresholds of $> 5\sigma(r)$ (solid) or $> 10\sigma(r)$ (dashed), and $> 20\sigma(r)$ (dot-dashed), assuming $\sigma = 0.03$ mJy at the phase center.  \label{galaxy_contam}}
\end{figure}

Figure \ref{galaxy_contam} shows the cumulative probability functions for a galaxy detection for each threshold.  These curves are measured from the annuli functions by taking the product of the probability of a non-detection in each annulus, $\Pi (1-P_i)$.  We find that the probability of detecting a galaxy within the primary beam is 66\%\ for the $5\sigma$ threshold, 31\%\ at $10\sigma$, and 10\%\ at $20\sigma$.

We consider galaxy contamination to be most significant for sources with S/N $< 20$ in our observations based on the probability curves in Figure \ref{galaxy_contam}.  Only 6/41 (15\%) of our continuum source have fluxes $< 20\sigma$.  These objects are ALMA\_J162705, ALMA\_J162717.7, IRS 39, ALMA\_J162729.7,  Oph-emb-7, and ALMA\_J162729.7.  Using the source S/N value and their positions within the primary beam, we estimate the probability that these objects are galaxies.  The probabilities are 5\%, 5\%, 35\%, 14\%, $<1$\%, and 42\%, respectively.  Of these faint detections, Oph-emb-7 has a very low probability ($< 1\%$) of being a galaxy and is likely a true YSO.  We also consider IRS 39 to be a YSO, even though it has a high probability ($35$\%) of being a galaxy, because it is also detected at infrared wavelengths \citep[e.g.,][]{Evans09}.  High redshift galaxies detected in the (sub)millimeter have spectral energy distributions (SED)s that peak at far-infrared wavelengths \citep[e.g.,][]{Casey14} and would not be detected at near-infrared or mid-infrared wavelengths. 

Thus, we classify four continuum sources as extragalactic objects.   Based on our analysis, the probability of detecting a galaxy at $\sim$10$\sigma$  in any given field should be $\sim 30$\%.  Excluding those fields with $\sigma \gtrsim 50$ \uJybeam\ at the phase center (e.g., those that are dynamic range limited), we estimate that an extragalactic source should be detected at 10$\sigma$ in roughly 6 fields, which is comparable to the observed number.  We note, however, that our probabilities assume galaxies are distributed randomly in the primary beam, whereas galaxies tend to cluster \citep{Hatsukade18}.  These probabilities should therefore be taken as a first order estimate.

\subsection{Class 0 and Class I Multiplicity}\label{mf}

In this section, we briefly discuss the multiplicity statistics for the Class 0 and Class I systems only.  We exclude the Flat and Class II statistics because those populations statistics incomplete (e.g., we detect Flat and Class II sources only if they were within our field of view or if a Class I sources was previously misclassified).  For simplicity, we consider two sources to be companions if they are within the same primary beam and they are at the same evolutionary stage.  Table \ref{cont_results} lists the adopted evolutionary stages of each YSO based on ancillary data that probe the star SED, outflows, surrounding envelope, disk structure, and chemistry (see Appendix \ref{indiv}).  

We find a total of 6 multiple systems in the Class 0 and Class I populations of the Ophiuchus molecular cloud and 10 single systems.  For the multiple systems, four are binaries (GSS 30, IRS 43, IRS 67, IRAS 16293), 1 is a triple star system (IRS 37), and 1 system has four stars (VLA 1623).  We note that VLA 1623 is identified with four stars using the higher-resolution results from \citet{Harris18} which separate VLA 1623A into two companions with separations of $\sim 14$ au.  For the Class 0 systems, 2/3 (67\%) are in multiple systems, whereas for the Class I systems, 4/13 (31\%) are in multiple systems.  Although we have small numbers, we find a similar decrease in the fraction of multiple systems from the Class 0 to Class I stage that was seen in previous studies \citep[e.g.,][]{Looney00, Chen13, Tobin16}.

\citet{Tobin16} conducted a complete analysis of the Class 0 and Class I multiplicity fraction (MF) and companion star fraction (CSF) for the Perseus molecular cloud using the Karl G. Jansky Very Large Array (VLA).  That study is the largest homogeneous analysis of protostellar multiplicity in a single cloud to date.  By comparison, Ophiuchus has far less Class 0 and Class I objects than Perseus.  \citet{Tobin16} identified $\sim 60$ Class 0 and Class I systems in Perseus (total number depends on how one defines the multiple systems), whereas we find 16 Class 0 and Class I systems in Ophiuchus, and of these, only three are at the Class 0 stage.  Thus, we will evaluate the multiplicity of Ophiuchus as a whole rather than separate the Class 0 and Class I sources.  

Following \citet{Tobin16}, we calculate MF = (B + T + Q)/(S + B + T + Q) and CSF = (B + 2T + 3Q)/(S + B + T + Q), where S is the number of single systems, B is the number of binaries, T is the number of triple systems, and Q is the number of quadruple systems.  Our sample is complete over separation ranges of $\sim 35-1800$ au.  There are higher resolution ($\sim$ 15 au) observations of VLA 1623A \citep{Harris18}, but we lack comparable resolution data for our entire sample.  On larger separations, we are limited by the primary beam FWHM of $\sim 25$\arcsec.  While some sources are detected beyond the primary beam FWHM (e.g., see Figure \ref{multiples}), we may not have complete statistics beyond 1800 au separations.   Thus, we only calculate MF and CSF for separation ranges between 35 - 1800 au. 

For our full sample of Class 0 and Class I systems, we measure MF = 0.29$\pm$0.11 and CSF = 0.41$\pm$0.12 for multiple separations between $35-1800$ au. (Errors in the MF and CSF values are derived from binomial statistics following \citealt{Chen13}.) \citet{Tobin16} reported values of MF = 0.27 and CSF = 0.31 for a separation range of $50-2000$ au and MF = 0.31 and CSF = 0.35 for a separation range of $15-2000$ au. So Ophiuchus and Perseus have broadly consistent multiplicity statistics over similar separation ranges.  The Ophiuchus multiples tend to peak at $\sim$ 100 au and $\sim$ 1200 au.  The first peak matches what has been seen in Perseus \citep{Tobin16, Tobin18} and in Orion (Tobin et al. in preparation), but the latter peak is roughly 3$\times$ smaller than these other clouds.  This smaller second peak may be due to our limited sensitivity to detect companions at separations $>1800$ au or due to the lower number statistics in Ophiuchus compared to Perseus and Orion.  We cannot conclude that the differences in the separation peaks are statistically significant.


\section{Conclusions}\label{summary}

We present high resolution (35 au) ALMA observations at 1.3 mm of full dust polarization of the Class 0 and Class I protostars in the Ophiuchus molecular cloud.  Initial results for VLA 1623 and IRAS 16293-2422 were published in \citet{Sadavoy18b} and \citet{Sadavoy18c}, respectively.  Here, we present the full survey overview, which consisted of 28 pointings toward 26 systems at a sensitivity of $\sim 30$ \uJybeam.   Our main results are:

\begin{enumerate}

\item We identify 41 compact objects in the Stokes I continuum data.  Using ancillary data, we classify 6 compact sources as Class 0, 18 sources as Class I, 4 sources as Flat spectrum objects, 9 sources as Class II objects, and 4 sources as galaxies.  We demonstrate that extragalactic contamination is non-negligible in deep observations of protostellar systems with ALMA, and that we detect galaxies at approximately the expected rate per pointing.  

\item  Roughly one third (14) of our unbiased sample is detected in polarization and roughly half of the undetected sources have $3\sigma$ upper limits to their polarization fractions at $<2\%$, indicating that they have significantly low polarization.    

\item  All 6 of the Class 0 sources are detected in polarization, whereas only 8 (44\%) of the Class I sources are detected in polarization.  The majority of sources detected in polarization have polarization patterns consistent with self-scattering processes in optically thick, inclined disks.  

\item  Nine sources have uniform polarization and four have azimuthal polarization.  The sources with uniform polarization tend to be highly inclined ($>60\degree$), whereas the sources with azimuthal polarization tend to be moderately inclined ($20-50$\degree) and large in size.  We detect uniform polarization toward both compact and extended sources.  

\item Nine sources appear to be consistent with polarization from dust self-scattering and two are partially consistent.  Assuming the sources that are inconsistent with dust self-scattering have polarization from grain alignment with a magnetic field, the inferred magnetic fields are mostly poloidal or pinched in morphology.  We see no evidence of polarization attributed to toroidal magnetic fields.  This result suggests that toroidal components may be difficult to detect toward young protostellar sources.  If the polarization indeed is caused by a poloidal fields, then magnetic braking must be weakened in VLA 1623A and IRS 67-B for these systems to have produced multiples stars and large circumbinary disks.

\item We find no agreement between the polarization morphology on clump scales as seen from single-dish telescopes with the polarization morphology detected on $<100$ au scales from the ALMA data.  Some of the disconnect can be attributed to different polarization mechanisms.  In the case of VLA 1623, the field orientation may be affected by dynamical processes near the protostar.  

\item Most of the disks in our sample have masses below the minimum mass solar nebula and sizes $< 30$ au, in agreement with the Class II disks reported by \citet{Cieza19}.   We find that our sample of disks follow a surface density relation of $\Sigma \sim r^{-(0.9-0.6)}$.  This surface density relation matches what \citet{Andrews09} found for Class II disks in Ophiuchus using radiative transfer models.  This agreement suggests that the protostellar and protoplanetery disks in Ophiuchus have comparable properties.

\item We find similar multiplicity statistics for the Class 0 and Class I protostars in Ophiuchus as previously found for the Perseus molecular cloud by \citet{Tobin16}, although we note that the number counts for Ophiuchus are small.  Ophiuchus has a multiplicity fraction of 0.29 and a companion star fraction of 0.41 for the combined Class 0 and Class I systems between separations of $35-1800$ au.  

\item We also compare ALMA polarization observations that were on axis and off axis using two sets of overlapping fields.  The on-axis and off-axis polarization are well matched for offsets $\lesssim 5$\arcsec from the phase center (within the inner half of the Band 6 primary beam).  For larger offsets of $\sim 10$\arcsec, the off-axis polarization position angles are largely in good agreement with the on-axis polarization position angles, although the polarized intensities and fractions can by off by up to a factor of two.  These results demonstrate that polarization observations at 1.3 mm are very reliable within the inner half ($\lesssim 6\arcsec$) of the primary beam FWHM.

\end{enumerate}

These observations represent the largest, unbiased, homogeneous dust polarization survey of protostellar objects in a single cloud to date.   We include with this data release paper, the full data products from the survey.   For our unbiased survey, we show that most polarization detections of disks are consistent with self-scattering processes rather than magnetic fields.  This result indicates that dust polarization may not be a good tracer of magnetic field structures on disk scales, especially if the disk is highly inclined.  Investigations of magnetic fields down to the scales of disks may be limited to other tracers, such as molecular line polarization.   Dust polarization observations on $< 100$ au scales, however, appear to be an excellent probe of self-scattering processes in disks that have high inclination ($> 60$\degree) or are well-resolved and at moderate inclination.  Such observations will provide invaluable information about the size, structure, and dust grain properties within disks.

\vspace{1cm}
\begin{acknowledgements}
We thank the anonymous referee for comments that helped improved the discussion and analysis.
The authors thank the NAASC and EU-ARC for support with the ALMA observations and data processing.   The authors also thank L. Matra, G. Keating, and A. Kovacs for valuable discussions on galaxy contamination, and S. Andrews for valuable discussions on disk properties.  SIS acknowledges the support for this work provided by NASA through Hubble Fellowship grant HST-HF2-51381.001-A awarded by the Space Telescope Science Institute, which is operated by the Association of Universities for Research in Astronomy, Inc., for NASA, under contract NAS 5-26555.  
W. K. was supported by Basic Science Research Program through the National Research Foundation of Korea (NRF-2016R1C1B2013642). 
This paper makes use of the following ALMA data: ADS/JAO.ALMA\#2015.1.01112.S. ALMA is a partnership of ESO (representing its member states), NSF (USA) and NINS (Japan), together with NRC (Canada), MOST and ASIAA (Taiwan), and KASI (Republic of Korea), in cooperation with the Republic of Chile. The Joint ALMA Observatory is operated by ESO, AUI/NRAO and NAOJ.  The National Radio Astronomy Observatory is a facility of the National Science Foundation operated under cooperative agreement by Associated Universities, Inc.  This research has made use of the SIMBAD database, operated at CDS, Strasbourg, France.

\end{acknowledgements}

\bibliographystyle{apj}

\begin{appendix}

\section{Individual Sources}\label{indiv}

This appendix shows the dust continuum observations for each individual source in our sample.  We describe each object in order of their observation field.  (Note that we use the mosaic c2d\_1008 in place of the individual fields c2d\_1008a and c2d\_1008b.)  For those sources detected in polarization, we show the Stokes I continuum map and a polarized intensity map with polarization e-vectors overlaid.  We plot only those e-vectors that meet the selection criteria (unless stated otherwise) in Section \ref{pol_overview} with Nyquist sampling.    For those sources undetected in polarization, we only show their Stokes I continuum data only.   Unless stated otherwise, the Stokes I maps have a log color scale, whereas the polarized intensity maps have a linear color scale. 

For each continuum detection, we also give a brief description of previous literature observations that were used to help re-classify the detection.  While the sources have bolometric temperatures or infrared spectral indices consistent with Class 0 or Class I protostars \citep[e.g.,][]{Evans09,Enoch09}, subsequent observations have shown that many of the Ophiuchus YSOs are affected by foreground extinction \citep{McClure10}.  To help distinguish genuine embedded protostars from more evolved YSOs that may be affected by foreground extinction, we also include the following criteria:
\begin{enumerate}
\item{Class 0 and Class I protostars have reliable outflows detections and evidence of a surrounding envelope or core,}
\item{Flat spectrum sources have either outflows or a surrounding envelope, but not both, and }
\item{Class II sources lack detectable outflows or a reliable core/envelope.}
\end{enumerate}
We further use the source bolometric temperatures to distinguish between Class 0 and Class I objects \citep{Evans09}.  We note that some sources may not have robust outflow observations, especially from large, shallow surveys with single-dish telescopes due to confusion with core or cloud emission.  These additional criterial are meant to improve the source classifications from the infrared spectral index or bolometric temperature alone using the fact that protostars must be embedded objects and the protostellar stages are when infall and accretion largely take place.

We reclassified 9 of our selected targets as either Flat or Class II objects and identified 4 additional evolved YSOs in the fields.  Our adopted classifications are listed in Table \ref{cont_results} and are also given in the subsections below.  Some of our targets were also observed by \citet{McClure10}.  With our new classifications, we have agreement with \citet{McClure10} for their envelope sources and for several more evolved YSOs (GY 91, WL 17, IRS 39, IRS 42).  Nevertheless, McClure et al. identified several YSOs to be disk sources that we still identify as Class I objects (Elias 29, IRS 37, IRS 45, IRS 47, and IRS 63).  Since these objects have outflows and evidence of envelope emission at (sub)millimeter wavelengths, they may be still accreting material and true protostellar objects.  We keep our Class I designations for these sources but note that some of them may be late stage Class I objects \citep[e.g.,][]{Enoch09} or Flat sources.

We also give the results from Gaussian fits using \texttt{imfit} in CASA in Table \ref{cont_results}.  For each YSO detection, we report the deconvolved sizes from the Gaussian fit and the source masses.  We estimate the mass assuming the thermal dust emission is optically thin \citep{Kauffmann08},
\begin{multline}
M_{disk} = 0.22\ \Mjupiter \left(\frac{S}{1\ \mbox{mJy}}\right) \left(\frac{d}{140\ \mbox{pc}}\right)^2 \\  
\left(\gdratio\frac{0.024\ \cmg}{\kappa_{d}}\right) \left[\exp\left(\frac{11\ \mbox{K}}{T}\right) - 1\right], \label{mass_eq}
\end{multline}
where $S$ is the source flux at 233 GHz, $d$ is the cloud distance, $\kappa_{d}$ is the dust opacity at 233 GHz, $\gdratio$ is the gas-to-dust ratio, and $T$ is the dust temperature.   Unless stated otherwise, we assume a distance of 140 pc, $\kappa_d = 2.4$ \cmg\ at 233 GHz, which is appropriate for dust in protoplanetary disks \citep{Andrews09}, $\gdratio = 100$, and $T = 20$ K.  For the dust temperature, we use the scaling relation $T_{dust} \approx 25(\mbox{L}/\Lsun)^{1/4}$ \citep{Andrews13} with the median bolometric luminosity of 0.52 \Lsun\ for our sample \citep{Enoch09}.

\subsection{Field c2d\_811: GSS 30}

The c2d\_811 field contains two sources, GSS 30 IRS 1 and GSS 30 IRS 3, which are separated by 14.8\arcsec\ ($\sim 2000$ au).  Figure \ref{multiples} shows a wide view of this field with both sources labeled.  These objects are associated with the GSS 30 region of L1688 Oph A, a low-mass core with a bipolar reflection nebula.  The core also contains GSS 30 IRS 2, a T-Tauri star that is likely unrelated to the nebula \citep{Leous91, Weintraub93} but lies within our primary beam, and GSS 30 IRS 4,  an infrared source that lies outside of the main core \citep{Weintraub93} and is outside the primary beam.  GSS 30 IRS 2 is undetected in our ALMA observations and was similarly undetected in previous millimeter observations \citep{Zhang97, Jorgensen09, Friesen18}, although it is bright at 6 cm \citep{Leous91}.  Based on its position in the primary beam, we can put a 3-$\sigma$ upper limit for a point source 1.3 mm continuum flux of $< 0.4$ mJy.

Both GSS 30 IRS 1 and GSS 30 IRS 3 are embedded objects.  GSS 30 IRS 3, however, is considered to be younger, because it is the only object in the core detected at both 1.3 mm and 2.7 mm \citep{AndreMontmerle94, Zhang97}.  Previous observations from the SMA and ALMA show it has a bright elongated disk and a bipolar outflow perpendicular to the long axis of the dust emission \citep{Jorgensen09, Friesen18}.  GSS 30 IRS 1 is considered slightly more evolved than GSS 30 IRS 3 due to less envelope emission \citep{Zhang97, Jorgensen09}.  Nevertheless,  \citet{Andre90} found 1.3 mm continuum around GSS 30 IRS 1 with the IRAM 30m telescope and \citet{vanKempen09} detected 850 \um\ and HCO$^+$ (4-3) emission around both sources with the JCMT.  The single dish observations suggest that GSS 30 IRS 1 is also embedded and may share a common spherical envelope with the younger GSS 30 IRS 3.  Since both sources also appear to drive outflows \citep{White15},  we consider them to be Class I objects for this study.     

\begin{figure*}
\includegraphics[width=0.95\textwidth,trim=0mm 5mm 0mm 9mm,clip=true]{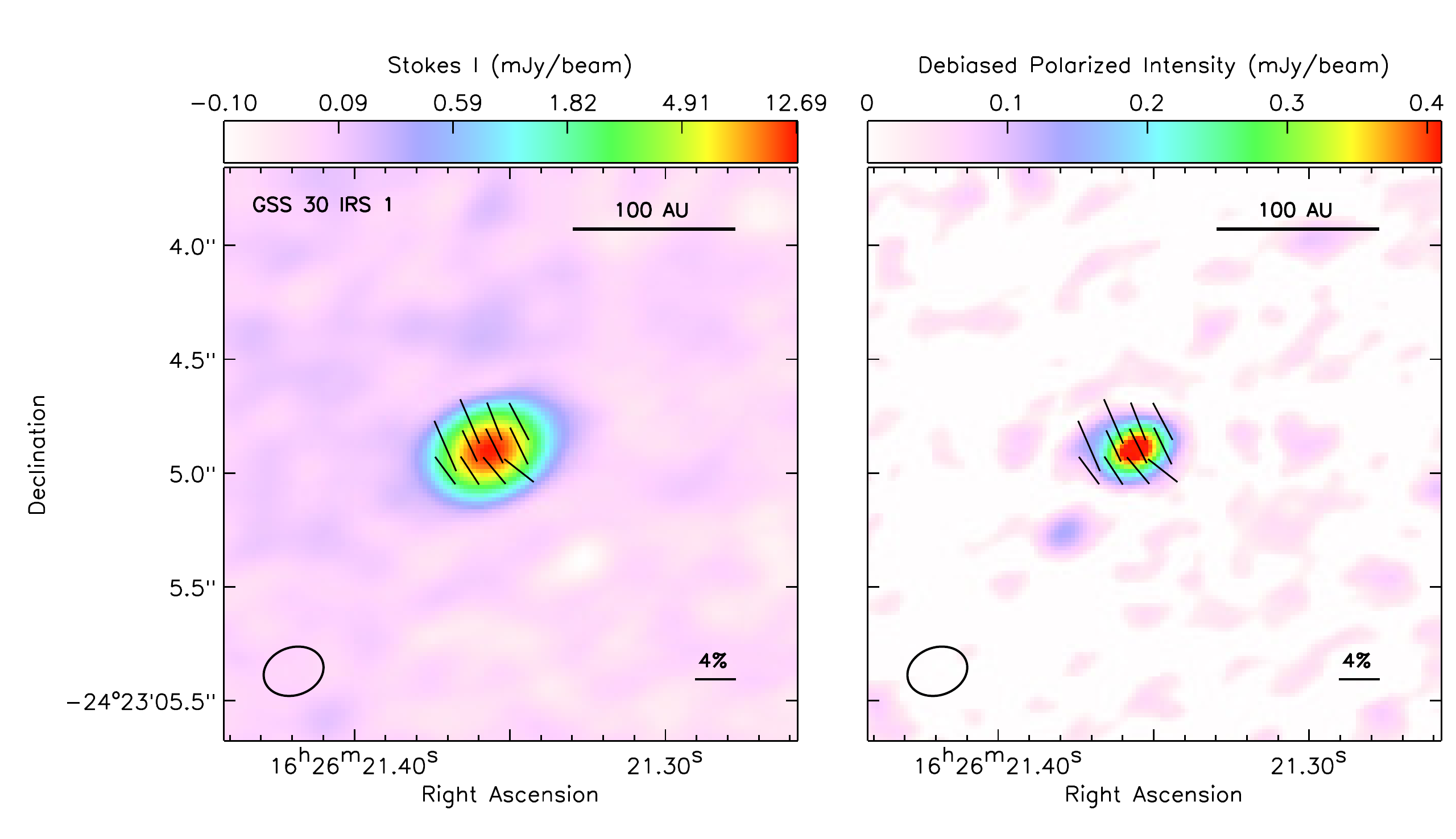}
\caption{Stokes I (left) and debiased polarized intensity (right) of GSS 30 IRS 1 from our ALMA 1.3 mm observations.  The polarization e-vectors are shown as black lines and are scaled by their polarization fraction.  We show e-vectors for those pixels with I$/\sigma_I > 3$, $\PI/\sPI > 3$, and $\sigma_{\theta} < 10\degree$.  The reference polarization fraction is shown in the lower right corner and the beam is shown in the lower left corner.  Unless stated otherwise, the Stokes I maps have a log color scale, whereas the polarized intensity maps have a linear scale.    \label{gss_irs1}}
\end{figure*}

Figure \ref{gss_irs1} shows our ALMA polarization maps for GSS 30 IRS 1.  The polarization angles are mainly uniform at $\approx 29$\degree\ and they are aligned with the minor axis (27\degree).  The disk is very compact with  a (deconvolved) size 12 au $\times$ 3 au (FWHM) and mass of 2.2 \Mjupiter.    We also detect faint diffuse Stokes I continuum that extends from GSS 30 IRS 1 eastward (see Figure \ref{multiples}) that could be associated with an envelope.  This extended dust emission is detected at only $\sim 3\sigma$, but is also faintly detected at 1.1 mm with ALMA in \citet{Friesen18} and it is located in the same direction as the brightest HCO$^+$ (4-3) line emission \citep{vanKempen09}.   

\begin{figure*}
\includegraphics[width=0.95\textwidth,trim=0mm 5mm 0mm 9mm,clip=true]{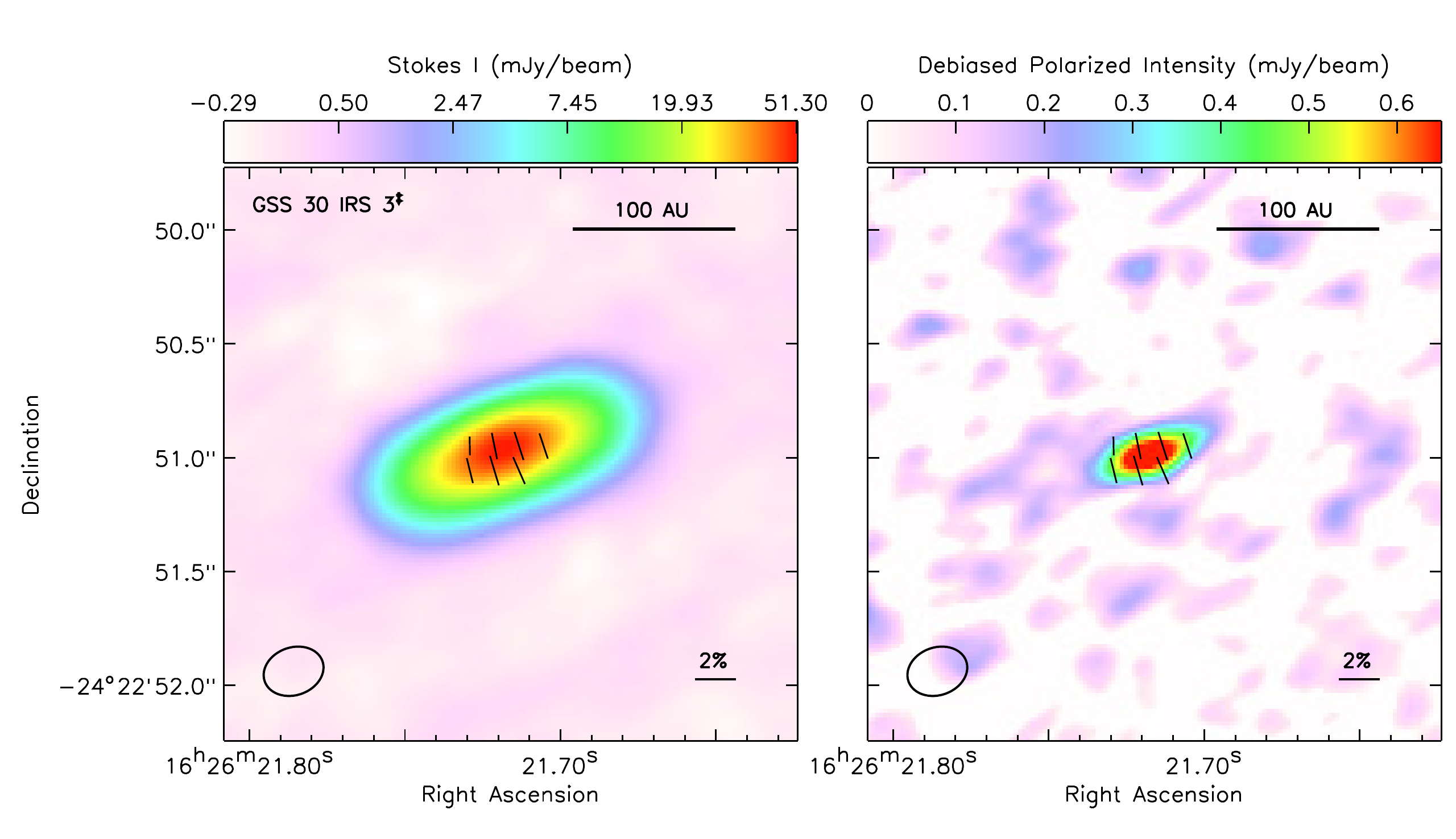}
\caption{Same as Figure \ref{gss_irs1} except for GSS 30 IRS 3.  Since this source is outside of the inner third of the primary beam (see Figure \ref{multiples}), we use more stringent criteria to select robust e-vectors: I$/\sigma_I > 10$ and $\PI/\sPI > 5$.  \label{gss_irs3}}
\end{figure*}

Figure \ref{gss_irs3} shows the ALMA polarization maps for GSS 30 IRS 3, also commonly called LFAM1 \citep{Leous91}.  This source is located well outside the nominal inner third of the primary beam, and in general, such off-axis polarization is considered less robust.    Nevertheless, ALMA commission tests show that off-axis polarization is reliable if the source is strongly polarized \citep{Harris18}.  Indeed, \citet{Harris18} found excellent agreement between off-axis polarization of VLA 1623W in their data and on-axis polarization of VLA 1623W presented here.    In Appendix \ref{offaxis}, we further test the reliability of off-axis polarization using adjacent fields that detect the same sources.  We find excellent agreement between the ``on-axis'' polarization and ``off-axis'' polarization for offsets of $\lesssim 5$\arcsec\ and good agreement for offsets of $\sim 10$\arcsec, indicating that our measurements for GSS 30 IRS 3 can be considered mostly reliable.   To still account for the higher degree of uncertainty for an off-axis source, Figure \ref{gss_irs3} uses more stringent selection criteria for the polarization e-vectors: I$/\sigma_I > 10$ and $\PI/\sPI > 5$. 

As with Figure \ref{gss_irs1}, GSS 30 IRS 3 shows mainly uniform polarization angles that are along the direction of the minor axis.  Here, the polarization position angles are $\approx 16$\degree\ and the minor axis position angle is 20\degree.  This disk is also quite large.  We measure a deconvolved size of 82 au $\times$ 27 au (FWHM) and a mass of 25.5 \Mjupiter.   This mass agrees well with the measurement of 0.026 \Msun\ (27 \Mjupiter) from fitting SMA observations at 1.3 mm with a disk and envelope model in \citet{Jorgensen09}.

\subsection{Field c2d\_822: Oph-emb-9}

This field contains one source, Oph-emb-9.  This source appears to be embedded \citep{Evans09, Enoch09, Dunham15}, and it has red-shifted and blue-shifted emission from CO consistent with a bipolar outflow.  The outflow, however, is partially confused by a redshifted lobe from VLA 1623 \citep{White15}.  Nevertheless, near-infrared nebulosity around Oph-emb-9 shows an opening angle consistent with the blue-shifted emission, indicating that this source is likely driving an outflow with a position angle of roughly $-60$\degree\ to $-70$\degree\ \citep{Kamazaki03}.  We consider Oph-emb-9 to be a Class I object. 

\begin{figure*}
\includegraphics[width=0.95\textwidth,trim=0mm 5mm 0mm 9mm,clip=true]{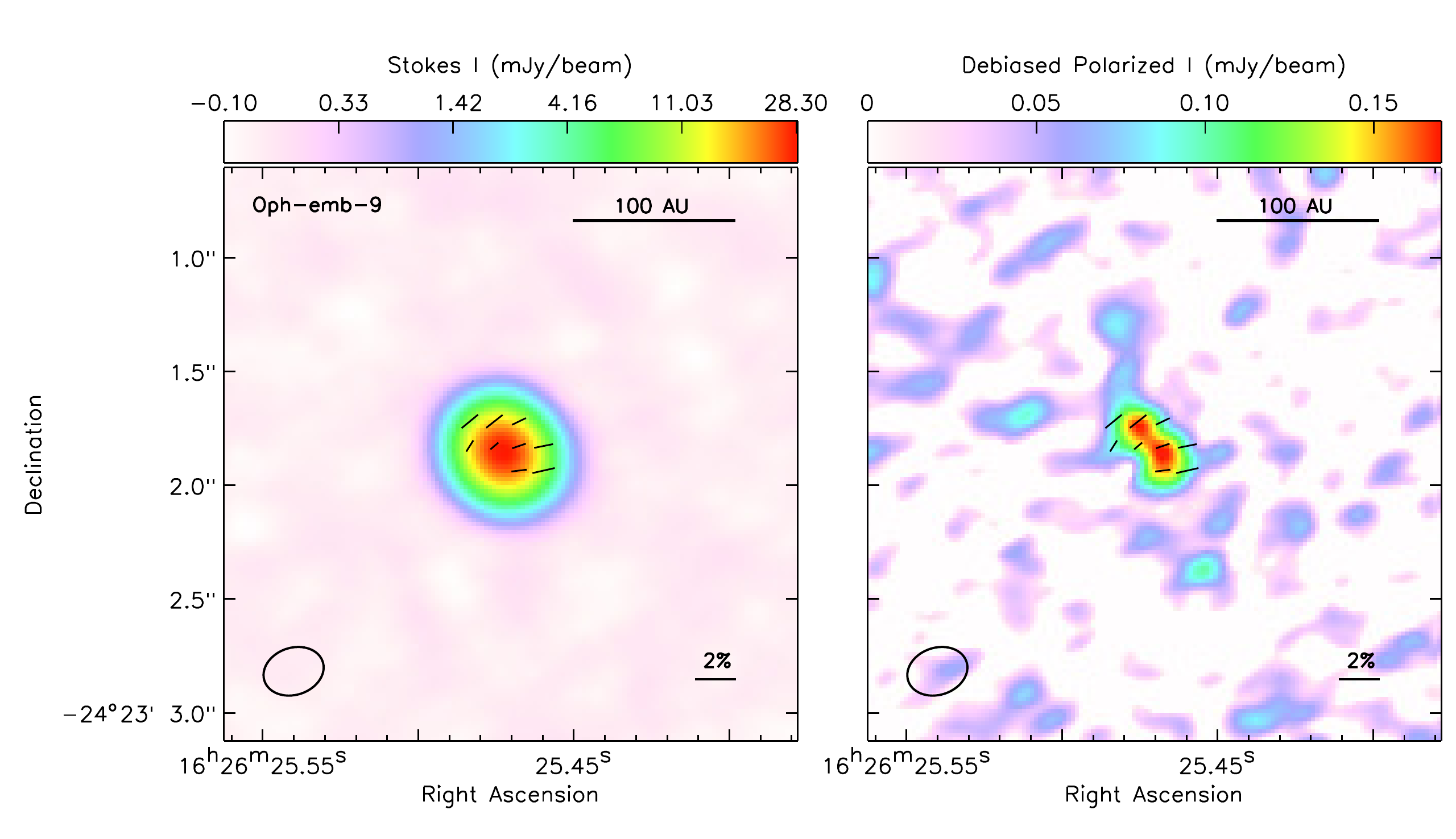}
\caption{Same as Figure \ref{gss_irs1} except for Oph-emb-9.    \label{oph9}}
\end{figure*}

Figure \ref{oph9} shows the polarization results for Oph-emb-9.  The Stokes I image of the source is compact and fairly round, whereas the polarized intensity map is peanut-shaped with minima along the minor axis.  The polarization position angles range from roughly $-40$\degree\ to $-80$\degree\ in an arc-like morphology.    From our Gaussian fit to the continuum emission, we find a disk position angle of 28\degree, which is nearly perpendicular to the position angle of the outflow from \citet{Kamazaki03} and also perpendicular to the general direction of the polarization.   The deconvolved disk size is 32 au $\times$ 12.5 au (FWHM) and the mass is 7 \Mjupiter.    

\begin{figure*}
\includegraphics[width=0.95\textwidth,trim=0mm 5mm 0mm 9mm,clip=true]{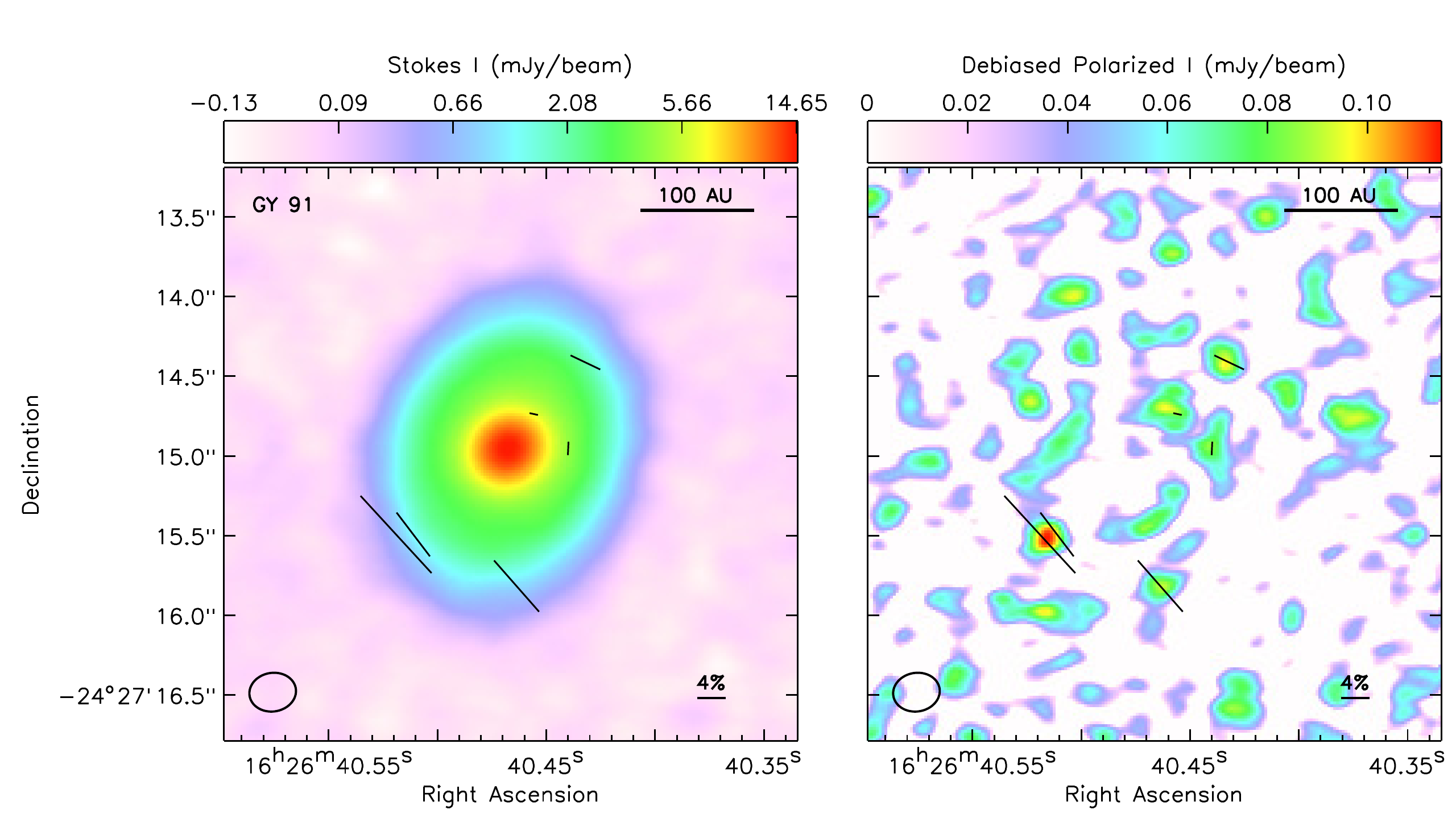}
\caption{Same as Figure \ref{gss_irs1} but for GY 91.  This source has only a marginal detection in polarization.  \label{oph22}}
\end{figure*}

\subsection{Field c2d\_831: GY 91}\label{c2d831}

Field c2d\_831 contains GY 91.   It has been identified as a Class I source \citep{Evans09, Dunham15} and a disk object in \citet{McClure10}.  It also has a spectral type of an M4-star based on infrared spectroscopy \citep{Doppmann05} in spite of being embedded in an envelope \citep{Enoch09}.  Recent high-resolution ALMA observation have shown it to have a very large, slightly inclined disk with multiple rings and gaps.  \citet{SheehanEisner18} observed GY 91 in 3 mm continuum at 0.05\arcsec\ resolution with ALMA.  They found three gaps in the disk and \citet{vanderMarel19} used the same data to identify a fourth gap.  \citet{SheehanEisner18}  measure a slightly larger envelope for GY 91 than HL Tau, suggesting that this source could be younger than HL Tau.   Alternatively, \citet{vanderMarel19} find that GY 91 could be as young as or a bit older than HL Tau, although it was the only source in their sample that was optically extincted.  We classify GY 91 as a Flat source for this study.

GY 91 is the only object with a polarization classified as marginal.  Figure \ref{oph22} shows its polarization results.  Its polarized intensities are weak ($\PI/\sPI < 4$) and off the Stokes I intensity peak.   Based on the non-detection at the continuum peak, we can put a 3$\sigma$ upper limit of $0.6$\%, which suggests it has significantly low polarization fractions at the peak of the continuum.  Nevertheless, the detections are also associated with 5 independent beams and appear to map out a near azimuthal morphology.  

GY 91 is one of the largest disks in our sample.  We find a deconvolved size of 110 au $\times$ 94 au (FWHM), and a mass of 14\ \Mjupiter.  These values, however, are only broadly representative because GY 91 has known substructure \citep[e.g.,][]{SheehanEisner17, vanderMarel19}.   More extensive disk modeling that accounts for the gaps is beyond the scope of the current paper.

\subsection{Field c2d\_857: WL 16}\label{c2d857}

Field c2d\_857 has WL 16.  While this source has been identified as an embedded Class I object \citep{Evans09, Dunham15}, it has also been identified as a more evolved Herbig Ae star \citep{ResslerBarsony03, ConnelleyGreene10}.  This object has a rising red spectral energy distribution that is saturated in some of the \emph{Spitzer} bands \citep{HsiehLai13}, which makes it difficult to classify by its infrared emission.  While previous observations of WL 16 found millimeter emission from Bolocam \citep{Enoch09}, more recent observations have shown that there is no surrounding core detected at 850 \um\ by SCUBA-2 \citep{Pattle15}.  WL 16 also has only faint CO line wings that may be indicative of a weak outflow \citep{White15} and its disk is detected in extended polycyclic aromatic hydrocarbon (PAH) emission that normally traces pre-main sequences stars \citep{Geers07, SeokLi17}.  Thus, WL 16 may be a more evolved source than a Class I object and located behind and heavily extincted by the Oph E cloud.  We reclassify this source as Class II.  

\begin{figure}
\includegraphics[width=0.475\textwidth]{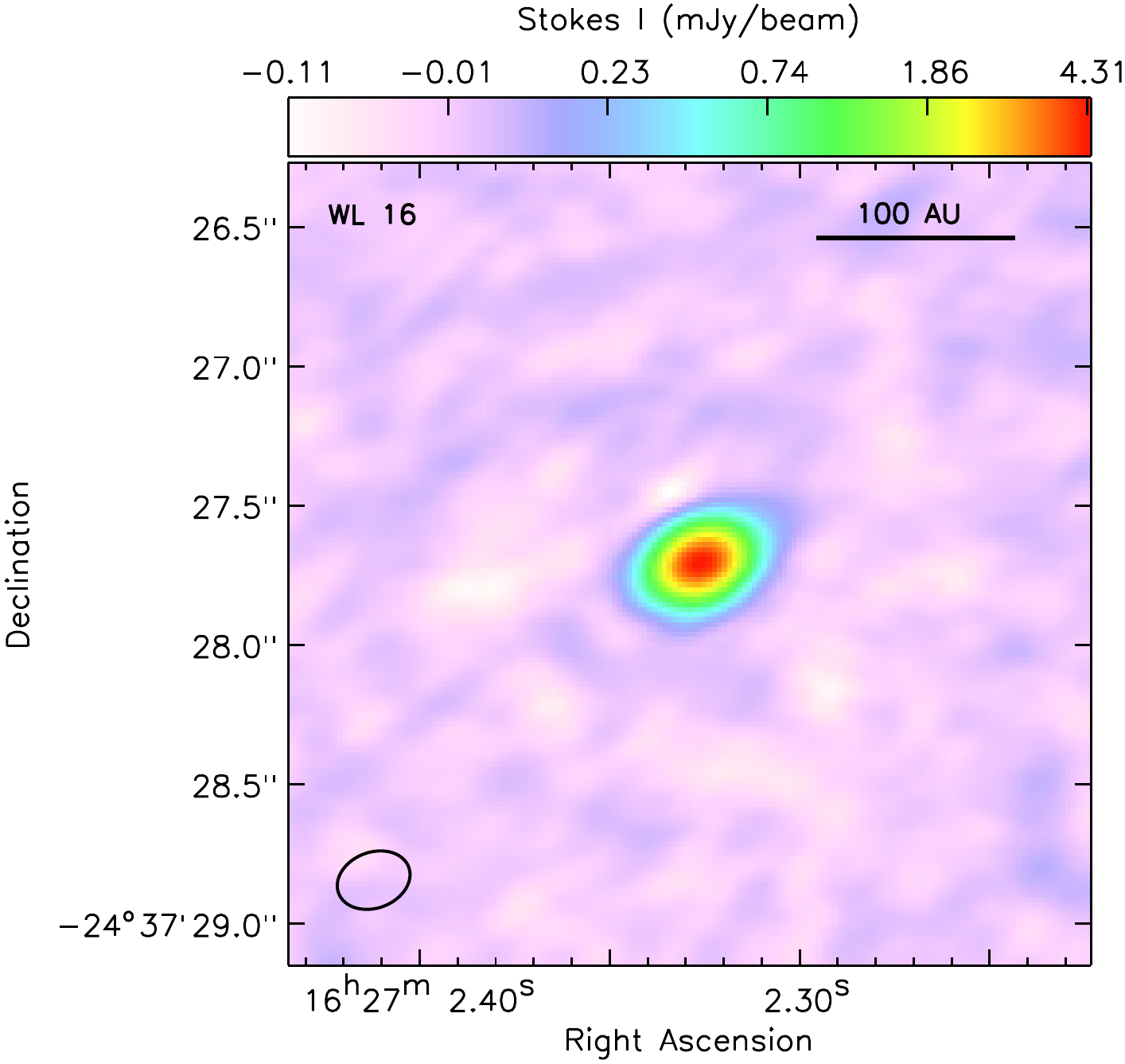}
\caption{Dust continuum map for WL 16.  This source was not detected in polarization.  \label{oph21}}
\end{figure}

Figure \ref{oph21} shows the Stokes I image of WL 16.  This source is undetected in polarization with a 3$\sigma$ upper limit of 1.8\%.  Previously, \citet{Zhang17} found polarization fractions of $1-3$\%\ from spectropolarimetry at 8.7 \um\ and 10.3 \um\ toward WL 16.  The mid-infrared polarization angles were also fairly uniform at $\sim 27-30$\degree\ toward the inner 2\arcsec\ of the mid-infrared disk.   \citet{Zhang17}, however, attribute their mid-infrared polarization to absorption from the foreground cloud rather than emission from the disk.  Their mid-infrared polarization position angles and fractions are consistent with optical and near-infrared polarization of background stars \citep{Sato88, Goodman90, Beckford08} and the expectations that this source is behind the Oph E cloud.  By contrast, our ALMA observations are sensitive to polarization from the disk itself.  

The disk is also unresolved in our ALMA 1.3 mm observations, which suggests that it is very small ($\lesssim 9$ au; or 1/4 the beam) with an estimated mass of 0.7 \Mjupiter.  This disk size and mass is much smaller that what was obtained from near-infrared and mid-infrared observations.  \citet{ResslerBarsony03} measured a disk size of 900 au at mid-infrared wavelengths, which they attributed to very small grains (VSGs) and PAHs.  The 1.3 mm ALMA data will instead be sensitive to larger dust grains, which are more likely than small dust grains to move inward due to radial drift \citep[e.g.,][]{Perez12}.  Thus, we would expect the disk size traced by large dust grains to be more compact than what is traced by small dust grains. WL 16 may  also have a limited population of large dust grains.  \citet{Najita15} report a millimeter-detected disk mass of $\lesssim 0.16$ \Mjupiter\ (scaled by the revised distance to Ophiuchus) using 1.3 mm data in the literature, whereas \citet{ResslerBarsony03} find a VSG disk mass of $\lesssim 10$ \Mjupiter.  This difference indicates that the mass in VSGs may be two orders of magnitude higher than the mass in large dust grains.  We note that our estimated disk mass of 0.7 \Mjupiter\ is higher than \citet{Najita15}, because we use a fixed temperature of 20 K, whereas Najita et al. scale their dust temperatures by the star luminosity.  If we use the same dust temperature as \citet{Najita15}, we get a disk mass of 0.1 \Mjupiter, in agreement with their upper limit.

\subsection{Field c2d\_862: Oph-emb-6}

Field c2d\_862 contains the Class I source, Oph-emb-6.  This source has a well established core \citep{Enoch09, Pattle15} and outflow \citep{White15, Hsieh17}, indicative of a true embedded protostar.  The infrared emission, however, is relatively weak, and Oph-emb-6 has been classified as a candidate Very Low Luminosity Object (VeLLO) in the literature \citep[e.g.,][]{Bussmann07, Dunham08}.  Figure \ref{multiples} shows that we identify two compact objects separately by roughly 4\arcsec\ (560 au).  We label these objects as Oph-emb-6 and ALMA\_J162705.51-243622.27 (ALMA\_J162705.5 for short), where Oph-emb-6 is nearly 80 times brighter than ALMA\_J162705.5 in our 1.3 mm data.  

\begin{figure*}
\includegraphics[width=0.95\textwidth,trim=0mm 5mm 0mm 9mm,clip=true]{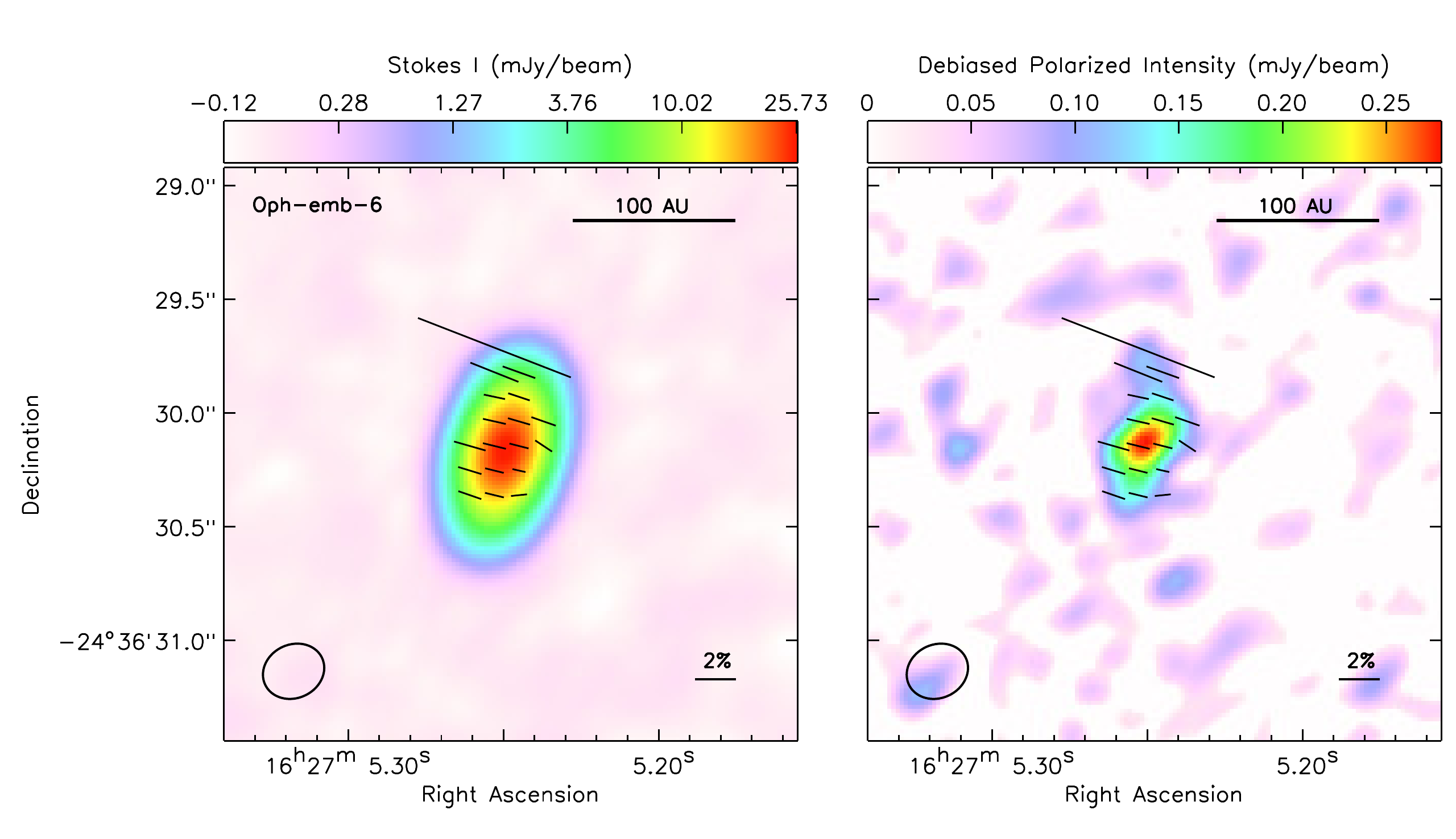}
\caption{Same as Figure \ref{gss_irs1} except for Oph-emb-6.    \label{oph6}}
\end{figure*}

Figure \ref{oph6} shows the dust polarization results for Oph-emb-6.  The Stokes I continuum image is highly elongated and the polarization is uniform with position angles of $\approx 75$\degree\ across the disk with variations of 5\degree\ only at the edges.   These position angles align well with the disk minor axis orientation of 79\degree.  The disk deconvolved size is 58 au $\times$ 14 au (FWHM), and the mass is 8.6 \Mjupiter.  \\

Previous observations of Oph-emb-6 predicted that the protostellar disk would be viewed nearly edge-on based on its outflow cavity.  \citet{Duchene04} found an East-West hourglass shaped cavity toward Oph-emb-6 from near-infrared imaging \citep[see also,][]{Hsieh17}.  Molecular line observations of the outflow are also East-West with a position angle of roughly 80\degree\ North to East \citep{Bussmann07, Nakamura11}.  This orientation places the outflow nearly perpendicular to the long axis of the elongated dust emission in Figure \ref{oph6}, consistent with expectations that the continuum emission is tracing a mostly edge-on disk that is perpendicular to the outflow.

\begin{figure}
\includegraphics[width=0.475\textwidth]{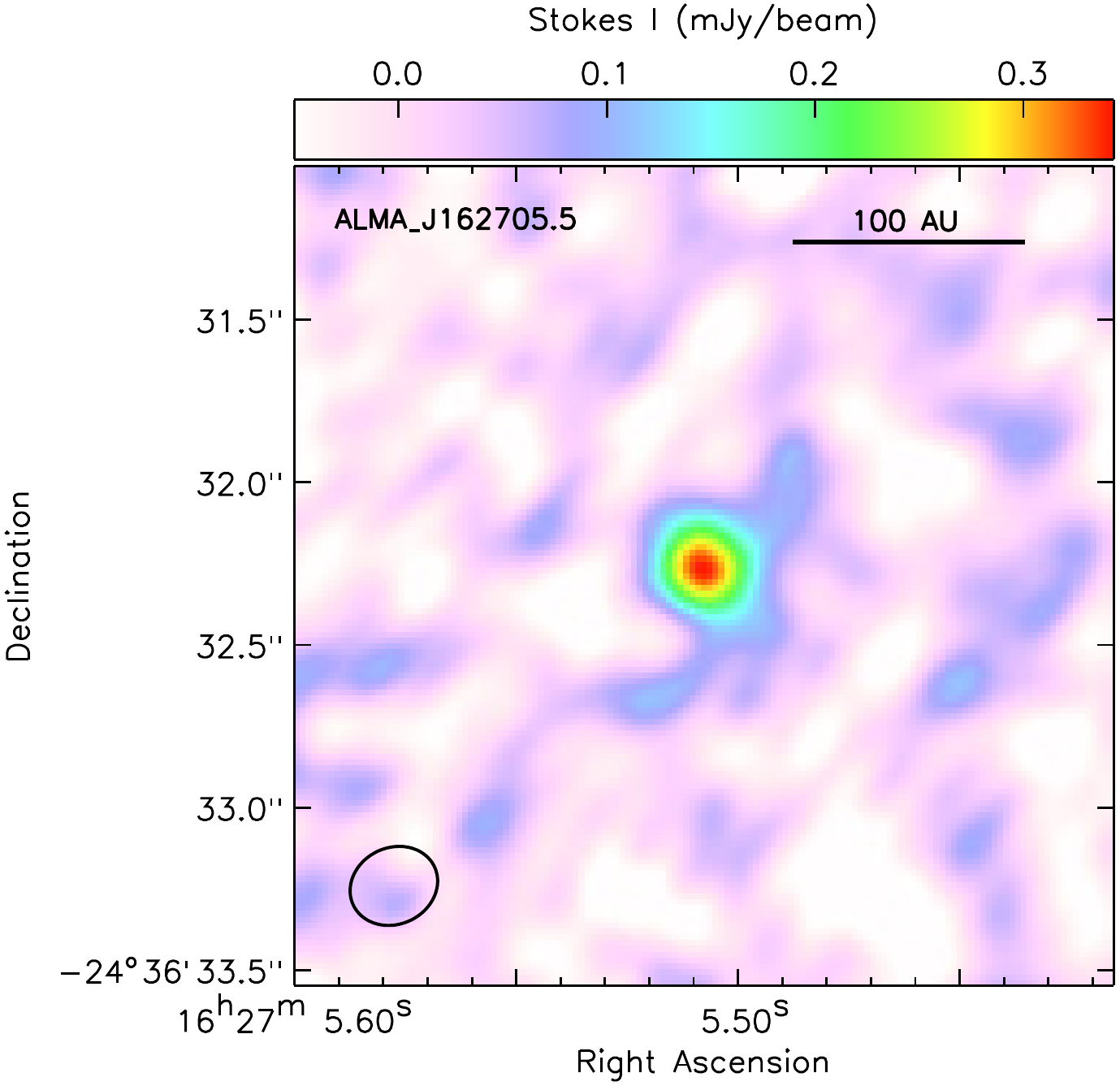}
\caption{Dust continuum map for ALMA\_J162705.5.  This source was not detected in polarization.  Note that the background map uses linear colour scaling. \label{oph6b}}
\end{figure}

Figure \ref{oph6b} shows the Stokes I continuum image of ALMA\_J162705.5.  This source is not detected in polarization, but it is also very faint.  The 1.3 mm continuum peaks at $9\sigma$, resulting in a 3$\sigma$ upper limit of 26\% for polarization.  Thus, we do not have the sensitivity to make any conclusions about the polarization of ALMA\_J162705.5. 

ALMA\_J162705.5 could be a faint binary star companion to Oph-emb-6.  This source has no \emph{Spitzer} counterpart of any classification from the full c2d source catalogue and no object is seen in WISE \citep{Wright10} at this source position.  With a separation of 4\arcsec, ALMA\_J162705.5 may be undetected with \emph{Spitzer} and WISE given its proximity to Oph-emb-6.  Nevertheless, Oph-emb-6 is a relatively low-luminosity protostar and as such, the higher resolution infrared data should have less confusion.  Based on its low flux (peak S/N $= 9$) and galaxy source count statistics (see Section \ref{gal}), we consider ALMA\_J162705.5 to be a background galaxy.

\subsection{Field c2d\_867: WL 17}

Field c2d\_867 contains one object, WL 17.  There are some uncertainties about its classification in the literature.  WL 17 has a rising red spectral energy distribution that peaks at infrared wavelengths \citep{Evans09, Dunham15} and a possible envelope \citep{McClure10} indicative of young, embedded protostars.  Nevertheless, WL 17 has only a weak high velocity CO emission that may trace an outflow, but this emission appears confused by other nearby sources \citep{White15}.  Its disk also appears more evolved.  \citet{SheehanEisner17} observed WL 17 in high resolution (0.05\arcsec) 3 mm continuum with ALMA and found an inner cavity with a radius of 0.1\arcsec\ ($\sim 14$ au).  This cavity is consistent with later-stage transition disks \citep[e.g.,][]{Espaillat07} that are typically associated with more evolved T-Tauri stars.   We therefore consider WL 17 to be more evolved that the canonical Class I stage.  

\begin{figure}[h!]
\includegraphics[width=0.475\textwidth]{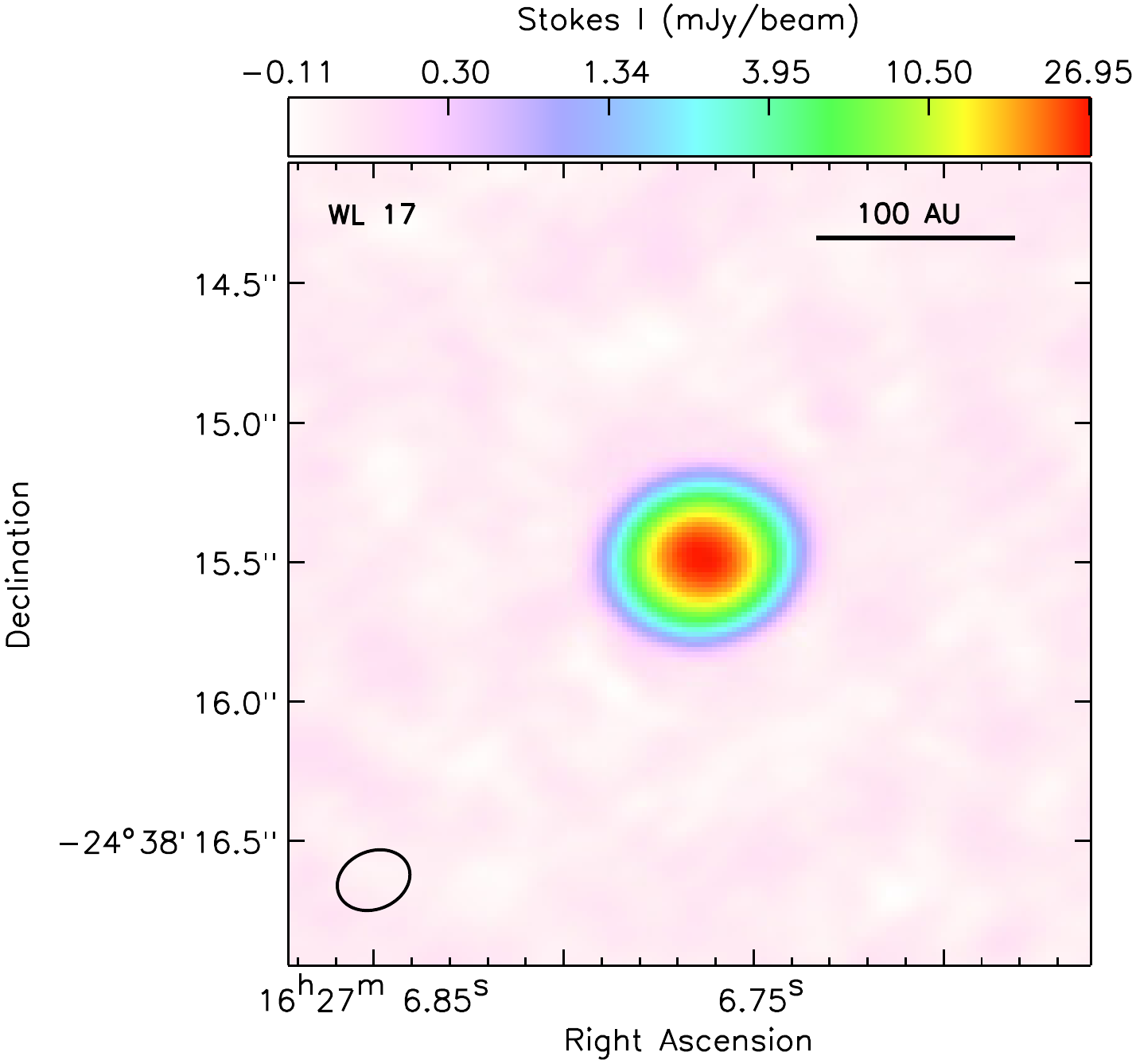}
\caption{Dust continuum map for WL 17.  This source was not detected in polarization.  \label{oph20}}
\end{figure}

Figure \ref{oph20} shows the dust continuum image of WL 17.  This source is not detected in polarization in spite of being one of the brightest objects in the entire sample (peak S/N $\approx 800$).  We measure a 3$\sigma$ upper limit for the polarization fraction of 0.3\%, which is equivalent to a non-detection down to the instrument noise.  WL 17 is also fairly compact.  It has a deconvolved disk size of 32.5 au $\times$ 25.6 au (FWHM) and mass of 8.3 \Mjupiter.  

\subsection{Field c2d\_871: Elias 29}

Field c2d\_871 contains Elias 29, a well-known embedded object with an outflow \citep{White15}.  It also has evidence of a disk from SMA and ATCA observations \citep[e.g.,][]{Lommen08, Jorgensen09, Miotello14}, although these studies did not resolve it.  \citet{Miotello14} combined the 1 mm and 3 mm observations to model the disk and envelope.  They found a disk mass of $\gtrsim$ 10.5 \Mjupiter\ \citep[see also,][]{Jorgensen09} and evidence that dust grains must have reached millimeter sizes within the disk and in its collapsing envelope.  Although \citet{McClure10} identified Elias 29 as a disk source, we consider it a Class I object based on its envelope and outflow detections.

\begin{figure}
\includegraphics[width=0.475\textwidth]{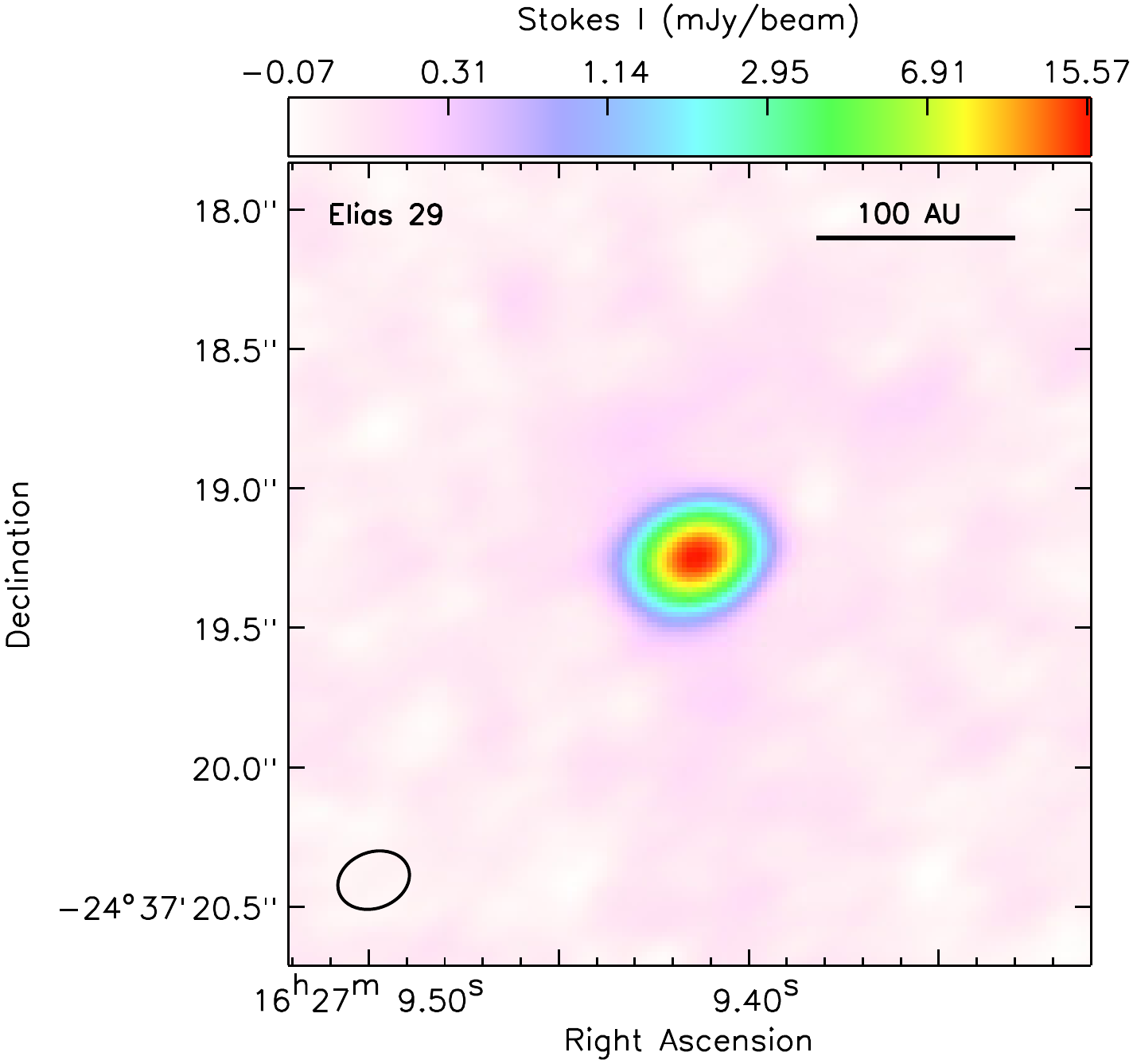}
\caption{Dust continuum map for Elias 29.  This source was not detected in polarization.  \label{oph16}}
\end{figure}

Figure \ref{oph16} shows our dust continuum map for Elias 29.  This source is compact in dust continuum and undetected in polarization with a 3$\sigma$ upper limit of 0.5\%\ for the polarization fraction, which is near the instrument noise limit.   We find a deconvolved size of 10.5 au $\times$ 9 au (FWHM), indicating that this disk is very compact \citep{Miotello14}.  The estimated disk mass is 2.8 \Mjupiter.  

Our estimated mass is a factor of three lower than the mass estimates from \citet{Jorgensen09} and \citet{Miotello14}.  These studies, however, did not resolve the Elias 29 disk and instead fitted a disk and envelope model to the SMA and ATCA visibilities.  We detect substantial extended emission south of the compact disk (see Figure \ref{multiples}) at $> 5 \sigma$.  This extended emission matches the north-south concentration of HCO$^{+}$ (3-2) seen in \citet{Lommen08}, which is mainly attributed to envelope emission.  They find that Elias 29 has a relatively high $M_{env}/M_{disk}$ ratio.   Since this envelope emission is well detected at high resolution with ALMA, the envelope itself may contain substructure that will confuse the disk and envelope models.   Indeed, \citet{Miotello14} find that they can fit the Elias 29 data with either a small optically thick disk of radius 15 au or a large optically thin disk with radius of 50-200 au.  Both disks are much larger than the resolved disk size from these ALMA data, suggesting that previously unresolved envelope structure may have inflated the disk size and mass for this source.

\subsection{Field c2d\_885: IRS 37}

Field c2d\_885 has IRS 37, a well-known YSO that has been identified as Class I \citep{Evans09, Gutermuth09}.  It has a detected outflow \citep{vanderMarel13, White15} and core \citep{Pattle15} even though \citep{McClure10} identified the source as a disk object.  Figure \ref{multiples} shows that the field also contains 4 additional compact objects, including a more evolved YSO, IRS 39, 15\arcsec\ southwest of the phase center.   Several of the aforementioned compact objects were also detected by \citet{Cieza19} in their ALMA 1.3 mm emission.  We use a similar naming scheme as \citet{Cieza19}, and refer to the four sources near the phase center as IRS 37-A, IRS 37-B, IRS 37-C, and ALMA\_J162717.72-242852.84 (hereafter, ALMA\_J162717.7) in order of their brightness.   The faintest object, ALMA\_J162717.7, is a new detection. 

\begin{figure*}
\includegraphics[width=0.95\textwidth,trim=0mm 5mm 0mm 9mm,clip=true]{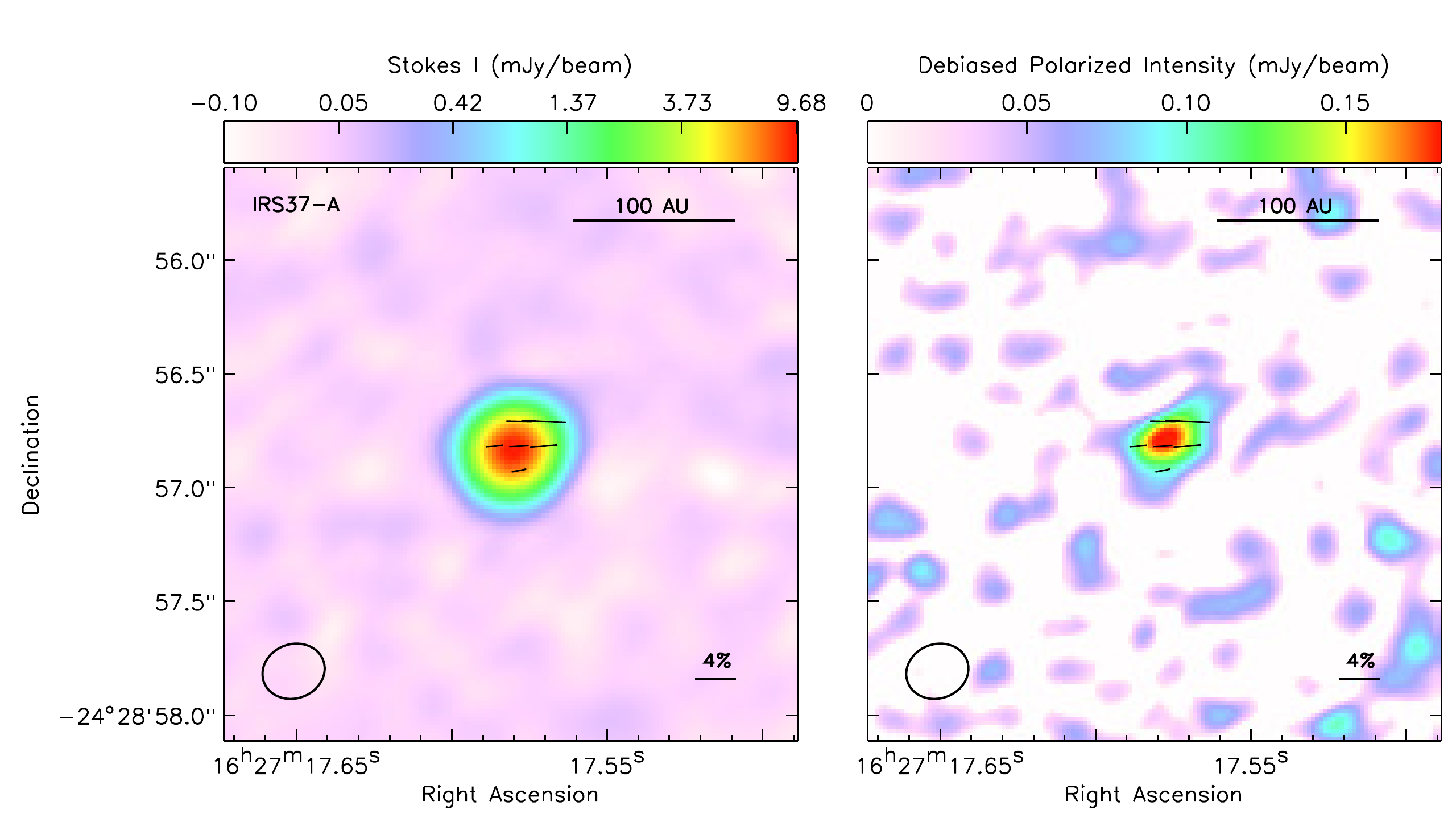}
\caption{Same as Figure \ref{gss_irs1} except for IRS 37-A.    \label{irs37a}}
\end{figure*}

Figure \ref{irs37a} show the polarization results for IRS 37-A, the only source in the field with a robust polarization detection.    The polarization morphology appears fairly uniform with position angles of roughly $-87$\degree.  These polarization angles are roughly parallel to the disk minor axis, which has an orientation of roughly -82\degree.  The disk is marginally resolved, with a deconvolved size of 17 au $\times$ 6 au (FWHM) and mass of 1.8 \Mjupiter.         

We note that the source outflow does not appear to be perpendicular to the long axis of the continuum emission.   \citet{vanderMarel13} and \citet{White15} each used single dish observations to trace CO from outflows throughout L1688.  Since neither study report the outflow position angle, we estimate it by eye using Figure 1 in \citet{vanderMarel13}.  We determine an approximate outflow position angle of 25-30\degree, compared to the disk position angle of 8\degree.   This discrepancy could mean that the continuum source of IRS 37-A is not uniquely tracing a disk, but may include emission from an inner envelope.  Alternatively, the IRS 37 outflow may not be well defined.  Both \citet{vanderMarel13} and \citet{White15} find the outflow confused with neighboring sources and only identify one lobe. 

\begin{figure*}
\centering
\includegraphics[scale=0.55]{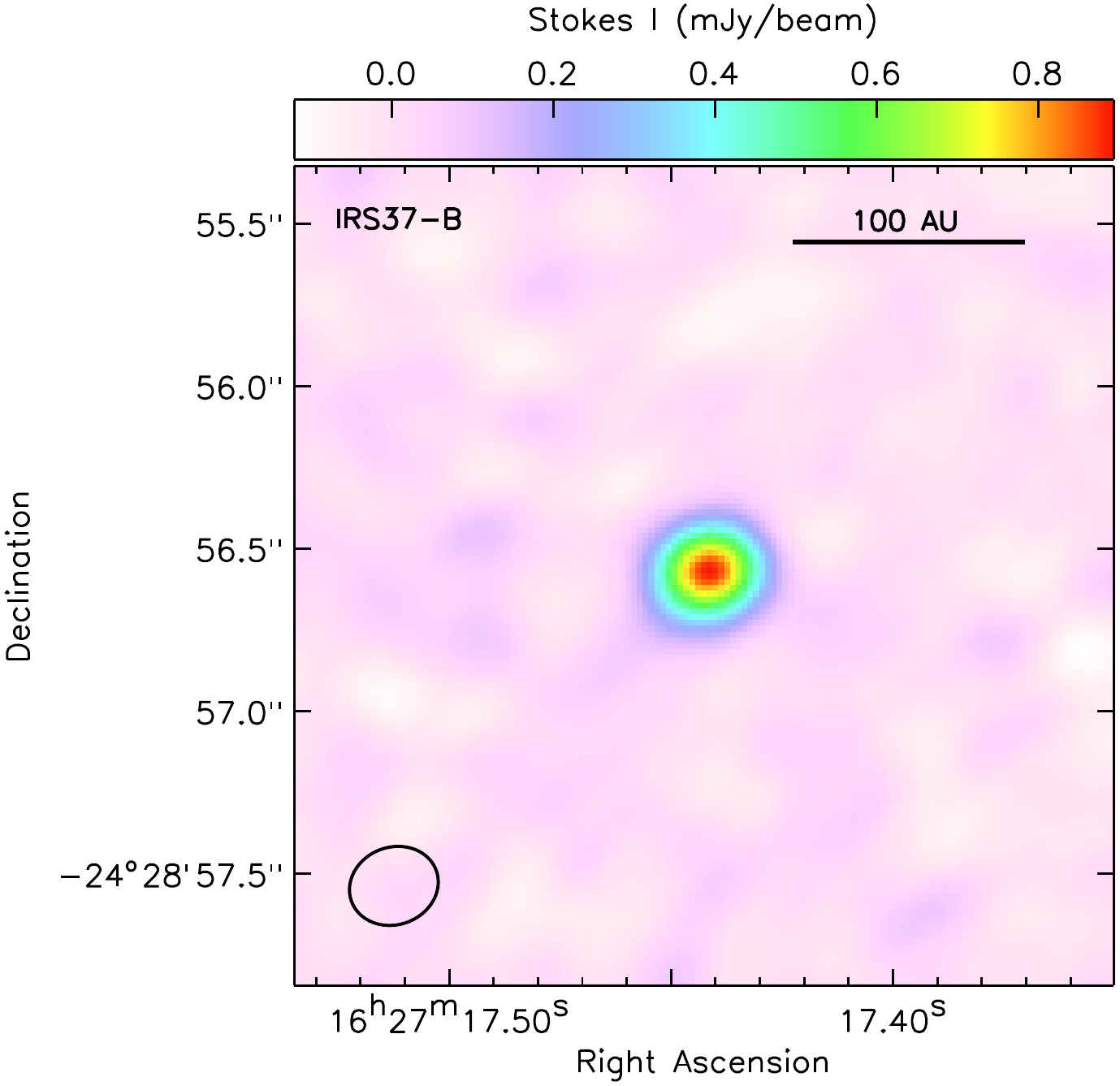}
\qquad
\includegraphics[scale=0.55]{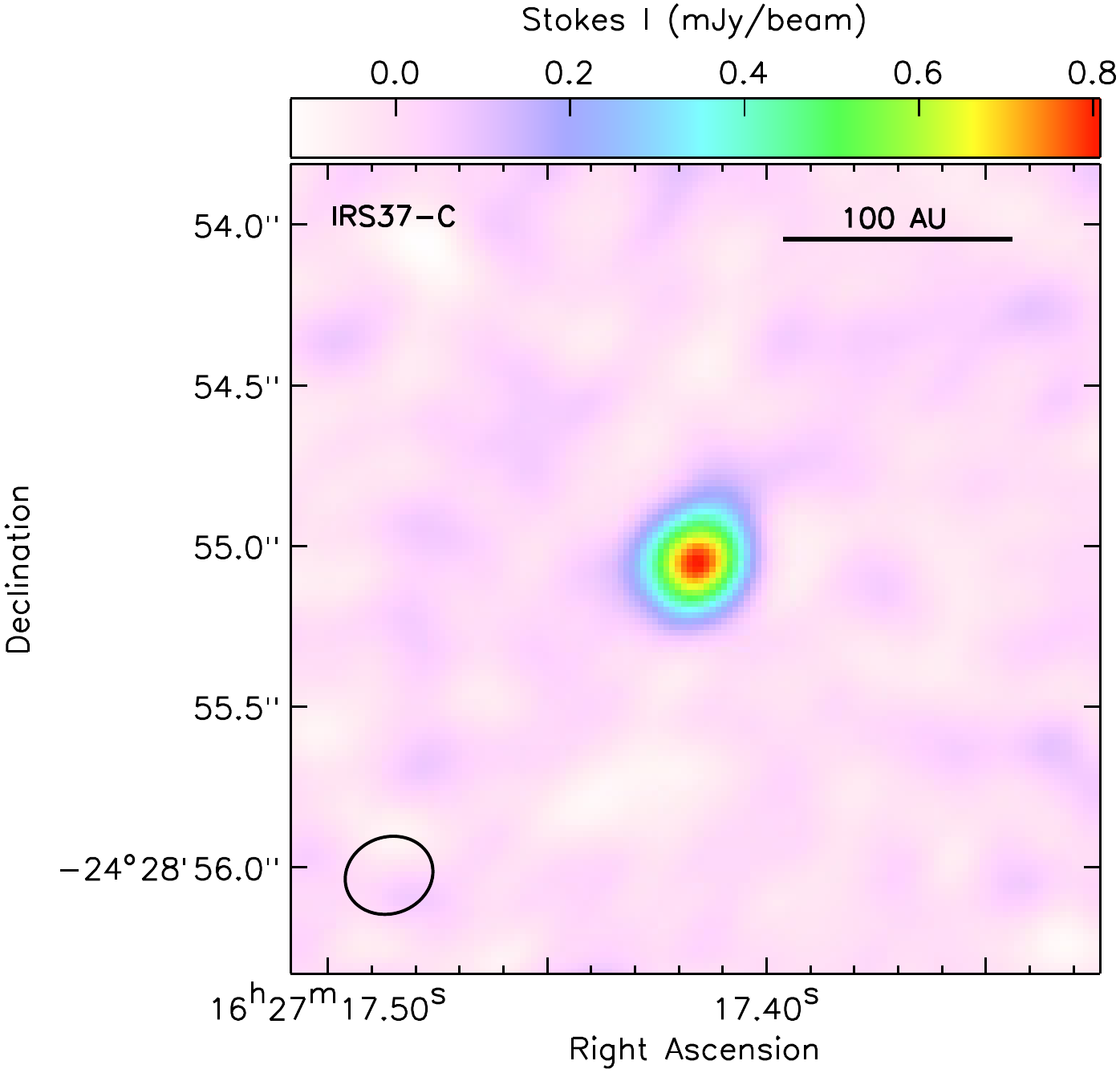}
\qquad
\includegraphics[scale=0.55]{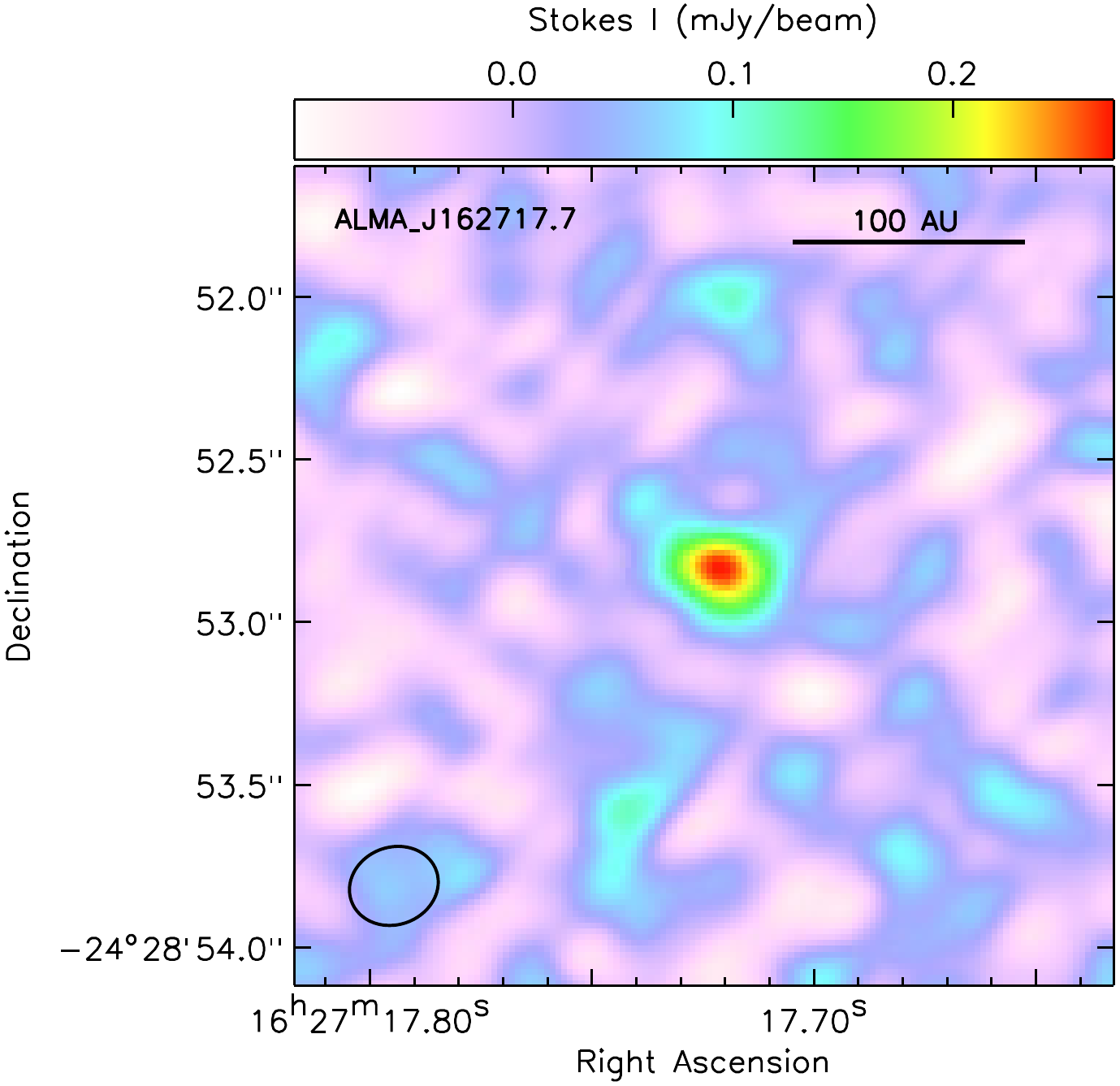}
\qquad
\includegraphics[scale=0.55]{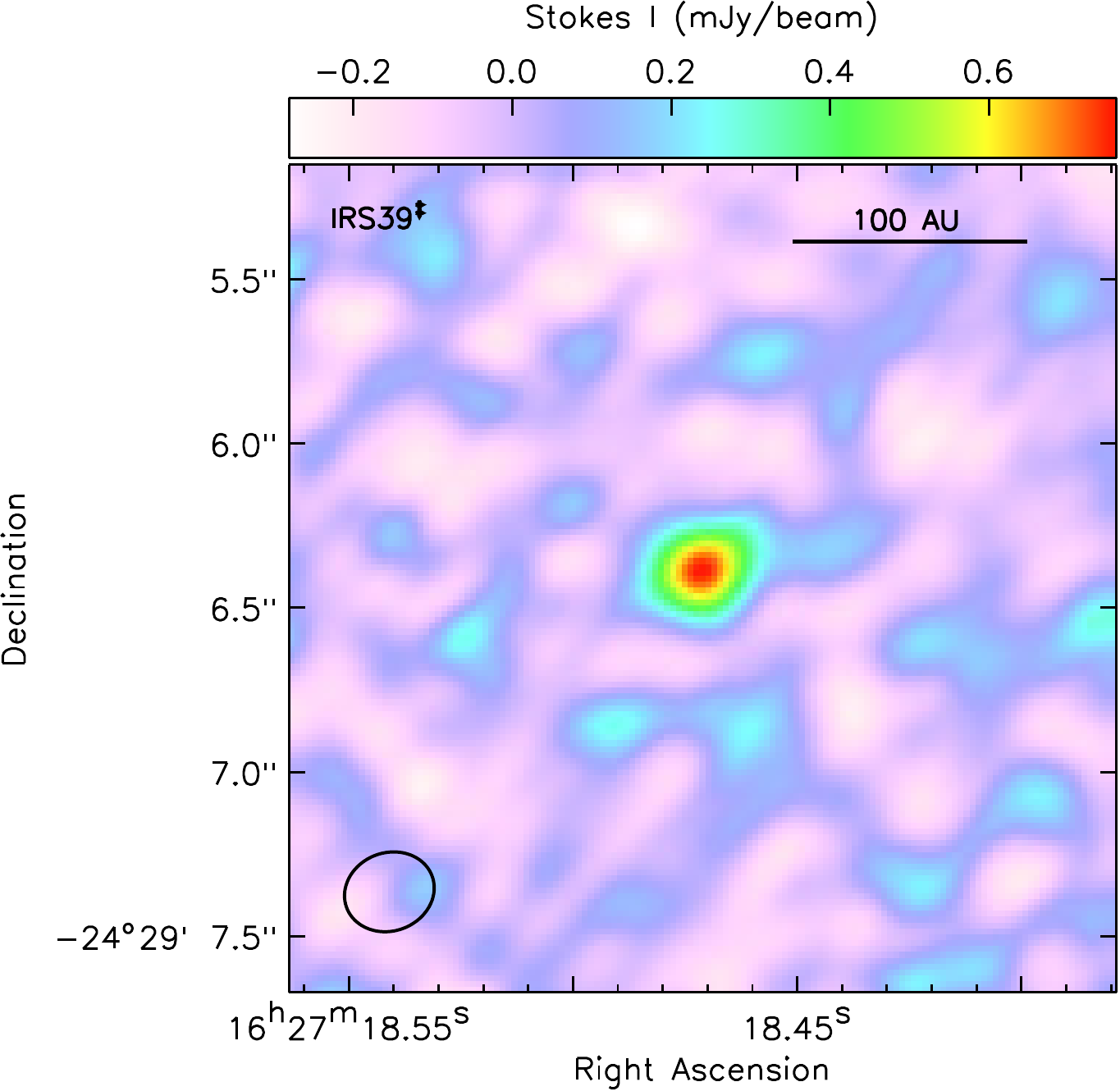}
\caption{Dust continuum map for IRS 37-B (top left), IRS 37-C (top right), ALMA\_J162717.7 (bottom left), and IRS 39 (bottom right), which were all undetected in polarization.  Note that each source uses linear colour scaling.
\label{irs37_nopol}}
\end{figure*}

Figure \ref{irs37_nopol} shows the continuum images for the four sources in the field that are not detected in polarization.  These objects are more than an order of magnitude fainter than IRS 37-A, and the 3$\sigma$ upper limits for their non-detection in polarization range from 9-33\% (see Table \ref{non_det}).    IRS 39 is a more evolved YSO.  It has an infrared spectral index ($\alpha < -1$) and bolometric temperature ($T_{bol} > 1000$ K) consistent with a pre-main sequence star \citep{Evans09, Gutermuth09}.   IRS 37-B, IRS 37-C, and ALMA\_J162717.7 have no \emph{Spitzer} or WISE counterpart, but such emission is likely lost in the PSF wings of the brighter IRS 37-A source.  IRS 37-B and IRS 37-C each have peak 1.3 mm fluxes $> 20\sigma$.  With such high S/N ratios, these two sources have low probabilities of being extragalactic sources (see Section \ref{gal}).  Moreover, IRS 37-A, IRS 37-B, and IRS 37-C roughly align with the long axis of their larger host core \citep{Pattle15}, and this orientation is often seen for wide binary pairs \citep{SadavoyStahler17}. Thus, we consider IRS 37-B and IRS 37-C to be companions to IRS 37-A.  ALMA\_J162717.7, however, is much fainter (peak S/N is 8.5$\sigma$).  Since it has a non-negligible probability of being a background galaxy (see Section \ref{gal}) and it does not align with the host core long axis like the other three sources, we consider it to be an extragalactic object.  

We estimate disk masses of 0.16 \Mjupiter\ for IRS 37-B, 0.15 \Mjupiter\ for IRS 37-C, and 0.12 \Mjupiter for IRS 39.  IRS 37-B is compact, with a deconvolved size of 14 au $\times$ 12.6 au (FWHM), whereas IRS 37-C and IRS 39 are unresolved in our observations.

\subsection{Field c2d\_890: IRS 42}

Field c2d\_890 contains a single source, IRS 42.  IRS 42 is on the edge of the Oph F core outside the main clump.  It has been classified as a Class I object \citep{Evans09, Dunham15} and a Class II object \citep{Gutermuth09} based on its spectral energy distribution.  \citet{vanKempen09} detected only weak HCO$^{+}$ emission and they found no obvious compact core at 850 \um\ \citep[see also,][]{Pattle15}.  IRS 42 also shows only a marginal outflow detection \citep{White15}, suggesting that it is perhaps more evolved.   We consider this object a Class II source in our analysis.

\begin{figure}
\includegraphics[width=0.475\textwidth]{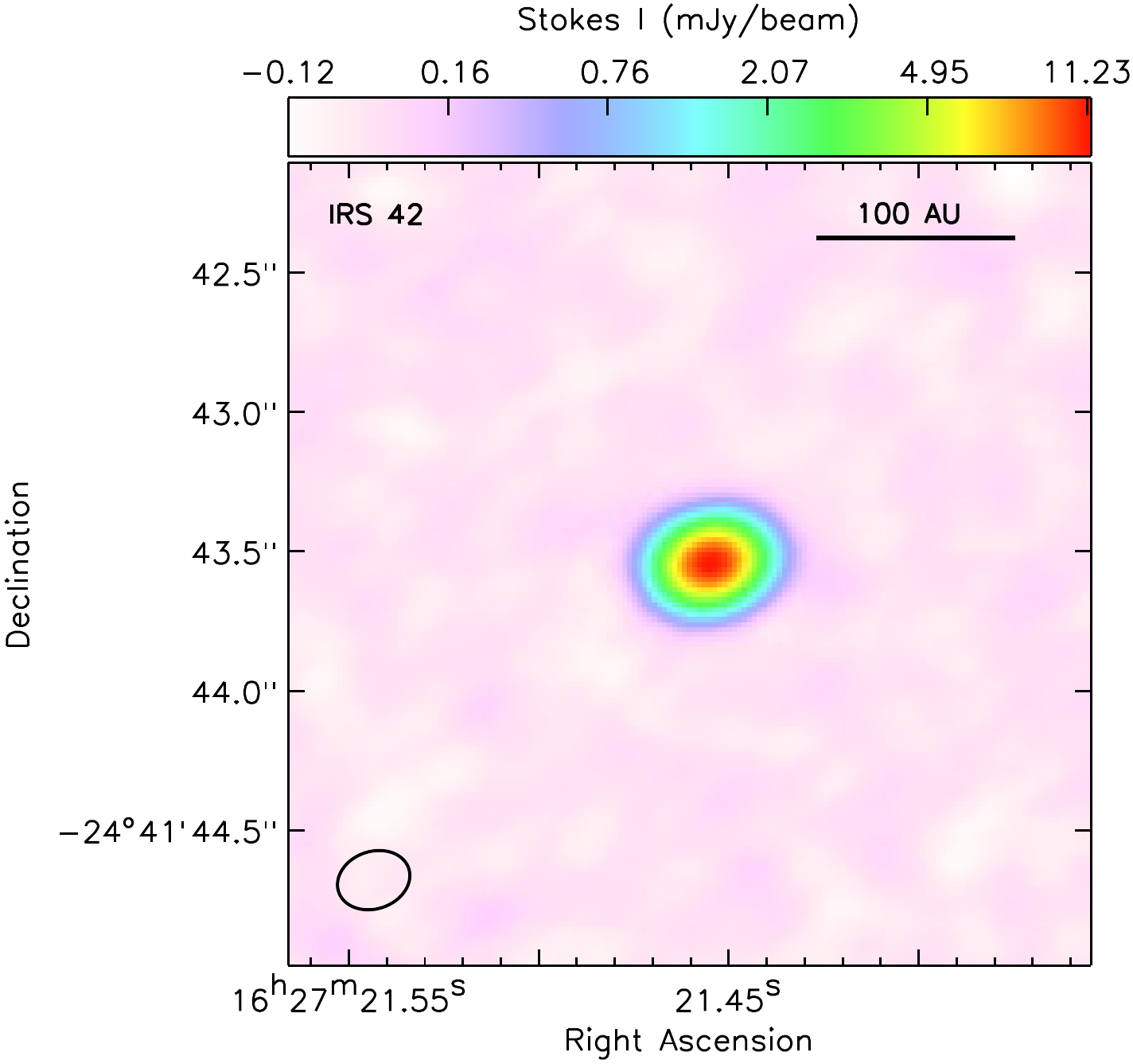}
\caption{Dust continuum map for IRS 42.  This source was not detected in polarization.  \label{irs42}}
\end{figure}

IRS 42 is not detected in polarization.  Figure \ref{irs42} shows its Stokes I continuum image.  Based on the peak intensity, we have a 3$\sigma$ upper limit of 0.7\%, which suggests that this source has very low polarization.  The continuum source is unresolved, however, indicating that the central disk must be compact ($\lesssim 9$ au, 1/4 the beam).  We estimate a disk mass of 2 \Mjupiter.

\subsection{Field c2d\_892}

Field c2d\_892 contains Oph-emb-5, a source identified as a Class I YSO in \citep{Enoch09} based on Bolocam and \emph{Spitzer} detections.  By contrast, \citet{Gutermuth09} classified the \emph{Spitzer} infrared object as a transition disk candidate with an infrared spectral index of $\alpha = -0.93$ and \citet{Sadavoy10} classified the \emph{Spitzer} source as an asymptotic giant branch (AGB) star.   Subsequent observations indicate that Oph-emb-5 is unlikely to be an embedded YSO.  \citet{Pattle15} found no evidence of a dense core at the position of the \emph{Spitzer} source in SCUBA-2 observations, and there is no molecular line emission from dense gas tracers showing an envelope or from CO tracing an outflow \citep{White15, Kamazaki19}.   

\begin{figure}[h!]
\includegraphics[width=0.475\textwidth]{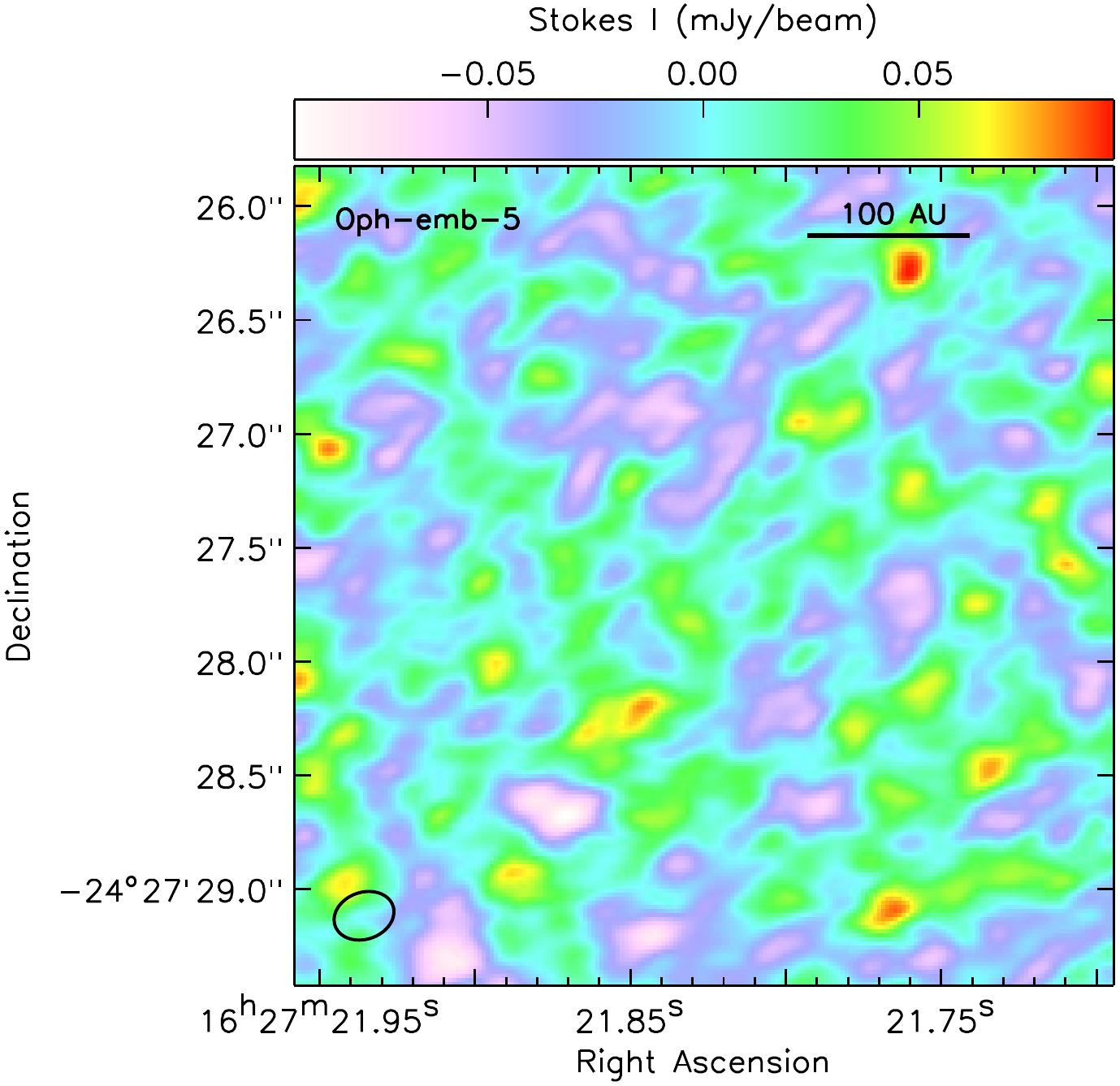}
\caption{Dust continuum map for Oph-emb-5.  This source was not detected in Stokes I continuum.  \label{oph5}}
\end{figure}

We do not detect Oph-emb-5 in Stokes I continuum.  Figure \ref{oph5} shows the noise map at the position of the \emph{Spitzer} infrared source, J162721.82-242727.6.   This is the only object that was completely undetected in our sample.  \citet{Cieza19} and \citet{Kamazaki19} also found no continuum object with 1.3 mm ALMA observations.   In particular, the observations from \citet{Kamazaki19} include both the main ALMA array and the compact array, and show that there is no disk or envelope structure at the position of the Spitzer source.   Our observations have a point source sensitivity of 26 $\mu$Jy, which is nearly a factor of $20$ better than \citet{Kamazaki19} and a factor of 6 better than \citet{Cieza19}.

\citet{Kamazaki19} argued that the non-detection in their continuum data was consistent with low masses in T-Tauri disks.  \citet{Ansdell17} and \citet{Long18} surveyed T-Tauri disks in Orion and Chamaeleon with ALMA and found masses down to $M_d \gtrsim 1.5 \times 10^{-4}$ \Msun\ (assuming a gas-to-dust ratio of 100).   With our sensitivity, we measure a 3$\sigma$ upper limit mass of $10^{-5}$ \Msun\ (0.01 \Mjupiter) with our assumed temperature and opacity.  This upper limit is roughly an order of magnitude lower than the masses in \citet{Ansdell17} and \citet{Long18}.  Even if we adopt a more conservative dust opacity of $\kappa_d = 1.1$ \cmg\ from the relation $\kappa_d = (10\ \cmg)(\nu/1000\mbox{THz})^{\beta}$ and $\beta = 1.5$, which is used in \citet{Kamazaki19}, our 3$\sigma$ mass limit is still considerably lower than the typical dust masses in T-Tauri disks.

With an unclear host core and no detected envelope or disk, Oph-emb-5 is unlikely to be an YSO.  Based on its infrared colours matching an extincted AGB star \citep{Sadavoy10}, we consider this source to be a star that is background to the Ophiuchus cloud.

\subsection{Field c2d\_894: Oph-emb-12}

Field c2d\_894 contains a single source, Oph-emb-12.  This source is an embedded YSO based on its infrared spectral energy distribution shape \citep{Evans09, McClure10} and its proximity to a dense core \citep{Evans09, Pattle15}, although the infrared source is offset from the core continuum peak \citep{vanKempen09}.  It also has a compact bipolar nebula at near-infrared wavelengths that is indicative of a nearly edge-on disk \citep{Brandner00}.  It has a marginal outflow detection \citep{White15}, such that we consider Oph-emb-12 to be a Class I object.    Using near- and mid-infrared spectroscopy, \citet{Pontoppidan05} conducted radiative transfer models for the disk.  They estimated a disk mass of 1.6 \Mjupiter, radius of 90 au, and inclination of 69\degree. 

\begin{figure}[h!]
\includegraphics[width=0.475\textwidth]{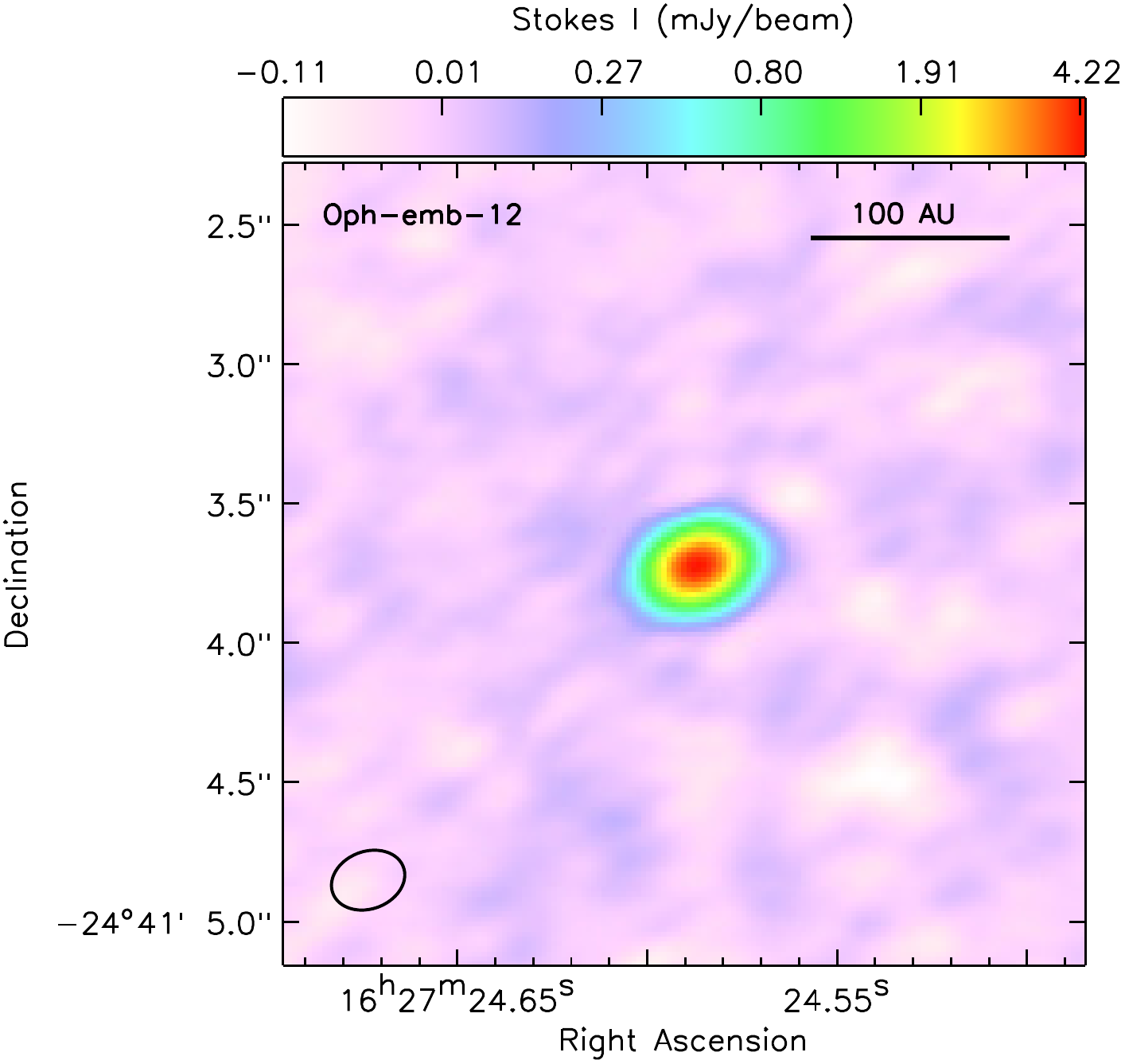}
\caption{Dust continuum map for Oph-emb-12.  This source was not detected in polarization.  \label{oph12}}
\end{figure}

Figure \ref{oph12} shows the continuum image of Oph-emb-12.  It is undetected in polarization with a $3\sigma$ upper limit of 1.8\%.  The disk in very compact and unresolved.  We estimate a disk mass of 0.7 \Mjupiter\ with a disk size of $\lesssim 9$ au (1/4 the beam).  \citet{Cieza19} similarly did not resolve Oph-emb-12 in their ALMA data at slightly higher resolution.     

\subsection{Field c2d\_899: IRS 43}

Field c2d\_899 contains IRS 43, a well studied protobinary system.  \citet{Girart00} first identified two sources (VLA1 and VLA2) separated by $0.6$\arcsec\ using the VLA.  We refer to these sources as IRS 43-A and IRS 43-B, respectively.   The IRS 43 system are embedded protostars.  \citet{Evans09} measured an infrared spectral index and bolometric temperature consistent with deeply embedded Class I protostars, and the stars have a clear outflow and envelope \citep[e.g.,][]{McClure10, White15, Pattle15}.  \citet{Girart04} proposed that the system was in transition between Class 0 and Class I based on non-detections at near-infrared wavelengths, but subsequent deep near-infrared observations detect this source \citep[e.g.,][]{Parks14}.    For this study, we consider the IRS 43 system to be at the Class I stage.  

\begin{figure}[h!]
\includegraphics[width=0.475\textwidth]{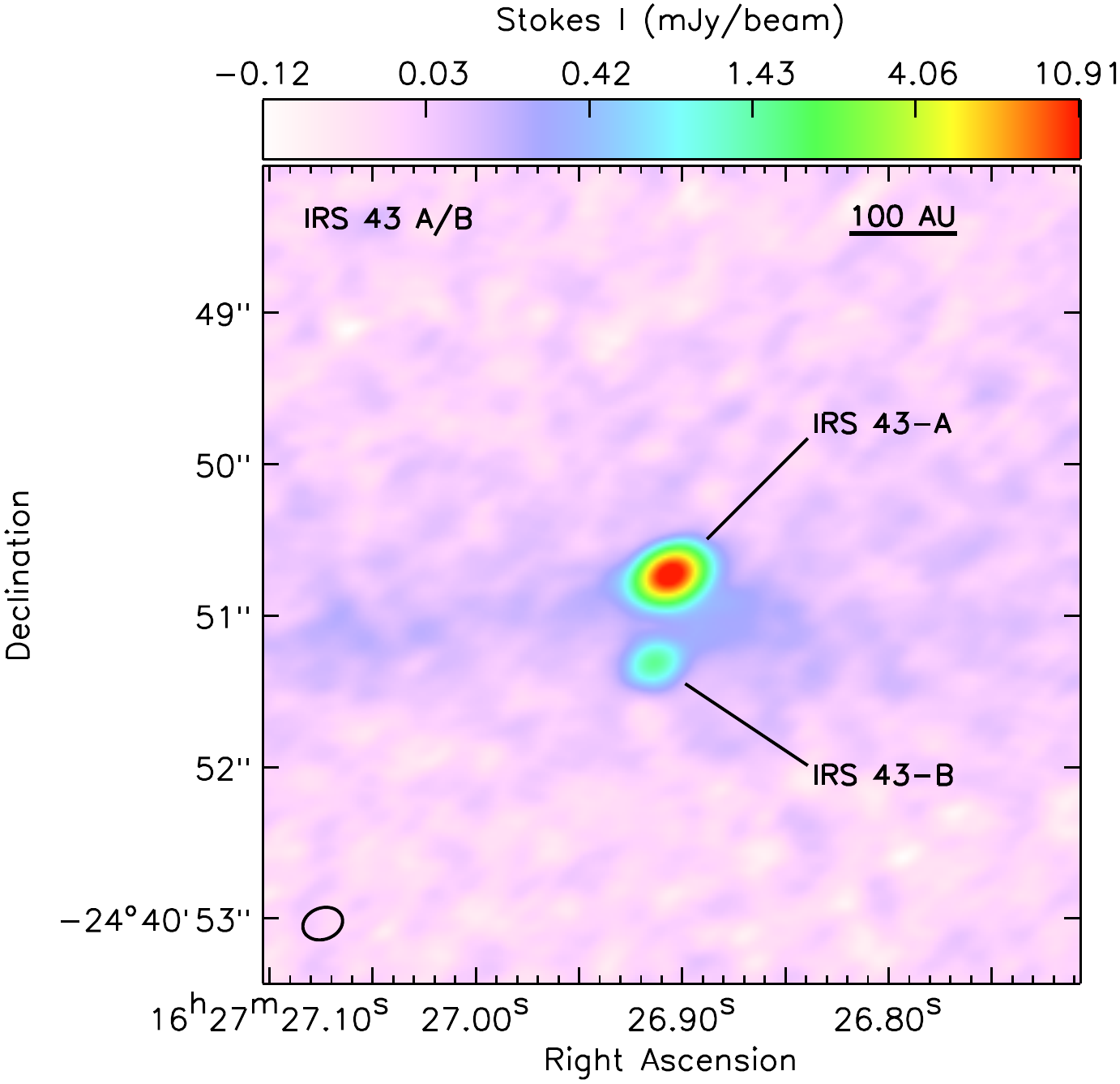}
\caption{Dust continuum map for IRS 43.  This source was not detected in polarization.  \label{irs43}}
\end{figure}

Figure \ref{irs43} shows the continuum image of IRS 43.  We do not detect polarization toward either object in the protobinary.  The 3$\sigma$ upper limits for the non-detections are 0.6\%\ for IRS 43-A and 4.8\%\ for IRS 43-B.  Thus, we can only conclude that IRS 43-A has significantly low polarization.  Both sources are compact in our observations.  We find deconvolved sizes of 16 au $\times$ 9 au (FWHM) for IRS 43-A and 15 au $\times$ 13 au (FWHM) for IRS 43-B.  The inferred masses are 2.4 \Mjupiter\ and 0.3 \Mjupiter, respectively.   These masses agree with the measurements from \citet{Brinch16} for both sources from 1.1 mm ALMA observations at slightly higher resolution.   

The circumbinary emission around the stars has been seen previously in molecular line emission \citep{BrinchJorgensen13} and in dust \citep{Brinch16}.  \citet{BrinchJorgensen13} found near-Keplerian motions through the circumbinary disk, and they modeled the molecular line emission to find a mass of $\sim 0.004$ \Msun\ (4.2 \Mjupiter) and an inclination of 70\degree\ for the circumbinary material.  \citet{Brinch16} later resolved the compact disks around IRS 43-A and IRS 43-B and estimated their orbital parameters.  They found that the stellar orbits (inclined at 30\degree) are misaligned with the axis of the circumbinary material and proposed that the system orientation either formed by turbulent fragmentation or ejection of a third component.

\begin{figure}
\includegraphics[width=0.475\textwidth]{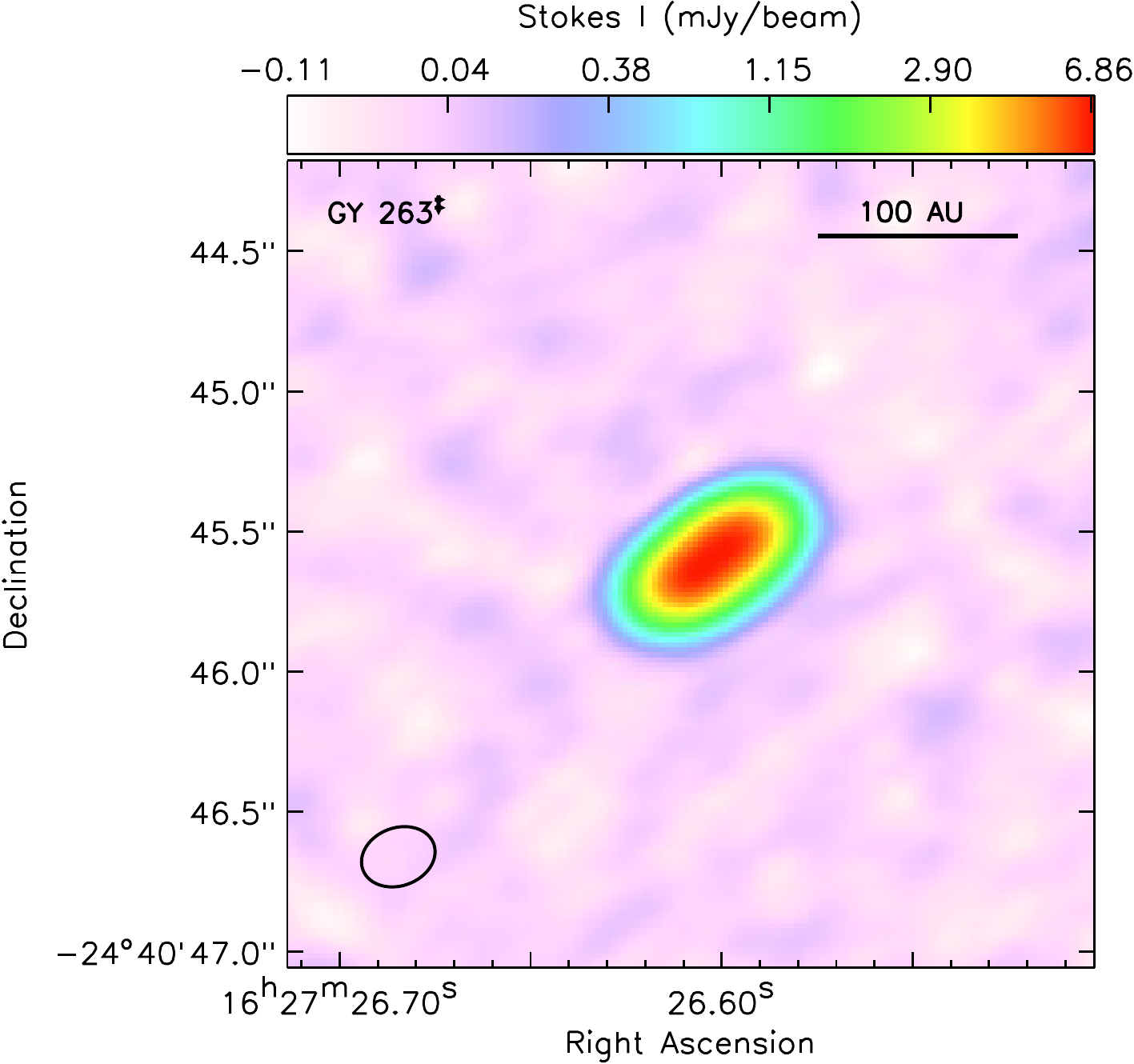}
\caption{Dust continuum map for GY 263.  This source was not detected in polarization.  \label{gy263}}
\end{figure}

The field contains a third source, GY 263, which is located roughly 6.5\arcsec\ northwest from IRS 43 (see Figure \ref{multiples}).  Figure \ref{gy263} shows the continuum emission for this object.  It was not detected in polarization at a 3$\sigma$ upper limit of 1.3\%.   GY 263 was also detected by \citet{Brinch16} in their ALMA data, but it has no \emph{Spitzer} counterpart in the full c2d source catalogue.  The 2MASS detection of GY 263 appears to be an extension of the brighter IRS 43 \citep[see also,][]{Wilking08, Beckford08}, indicating that any \emph{Spitzer} emission may have been confused with the much brighter IRS 43 system.  For example, \citet{Barsony05} observed GY 263 at 10.8 \um\ and 12.5 \um\ with Keck at $\sim$0.25\arcsec\ resolution, and detected the source weakly at 10.8 \um\ only.  They subsequently classified GY 263 as a  Class II object.   We also consider GY 263 to be a Class II object and measure a deconvolved disk size of 54 au $\times$ 18 au (FWHM) and mass of 2.6 \Mjupiter.

\subsection{Field c2d\_901: IRS 44}

Field c2d\_901 contains IRS 44.  IRS 44 is a well-established Class I object based on its infrared spectral energy distribution \citep[e.g.,][]{Evans09, Gutermuth09}, embeddedness in a dense core \citep[e.g.,][]{vanKempen09, McClure10, Pattle15}, and outflows \citep[e.g.,][]{White15}.   Nevertheless, IRS 44 does not appear to have any prior observations at high resolution with an interferometer in the literature and no prior estimates of a disk mass or size.

\begin{figure}[h!]
\includegraphics[width=0.475\textwidth]{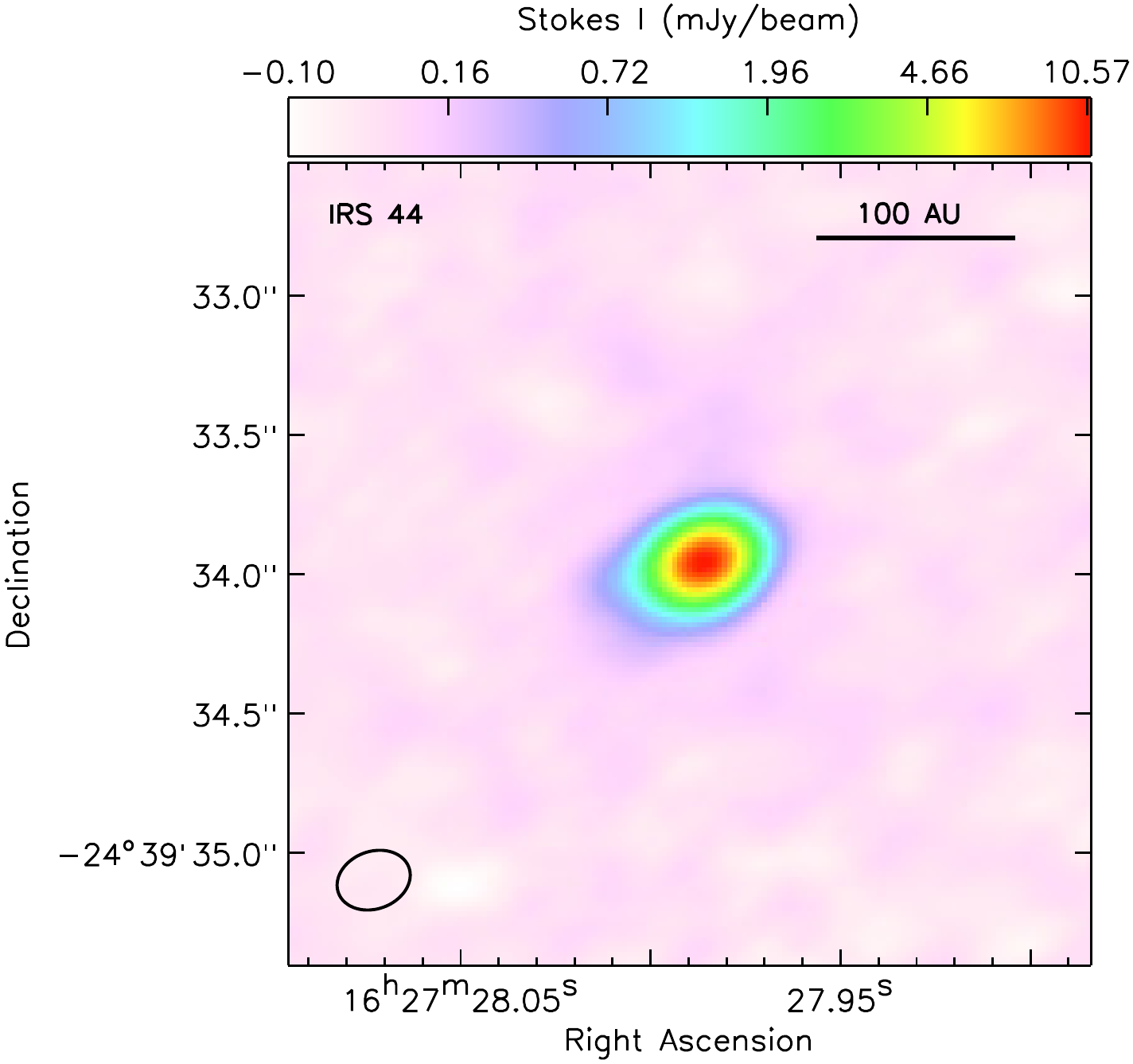}
\caption{Dust continuum map for IRS 44.  This source was not detected in polarization.  \label{irs44}}
\end{figure}

Figure \ref{irs44} shows our continuum image of IRS 44.  The central continuum source is marginally resolved by the beam, but is not detected in polarization with a 3$\sigma$ upper limit of 0.7\%.  This upper limit indicates that IRS 44 is significantly unpolarized.  The continuum source is compact, with a deconvolved size of 16 au $\times$ 9 au (FWHM) and mass of 1.9 \Mjupiter.  

\subsection{Field c2d\_902: IRS 45}

Field c2d\_902 is centered on IRS 45.   This source has varied classifications in the past.  Using \emph{Spitzer} observations, \citet{Evans09} and \citet{Gutermuth09} both found relatively flat infrared spectral indices for IRS 45 that indicate a Flat spectral source and \citet{McClure10} identified it as a disk object from infrared spectroscopy.  Nevertheless, IRS 45 is located on the edge of a dense core \citep[e.g.,][]{vanKempen09, Pattle15} suggestive of an embedded source.  Moreover, \citet{Kamazaki19} observed this source at 1.3 mm with the ALMA main array and the ACA.  They detected a compact continuum source toward IRS 45 with red- and blue-shifted CO outflow lobes.  Thus, we consider IRS 45 to be a Class I object in this study.

\begin{figure}[h!]
\includegraphics[width=0.475\textwidth]{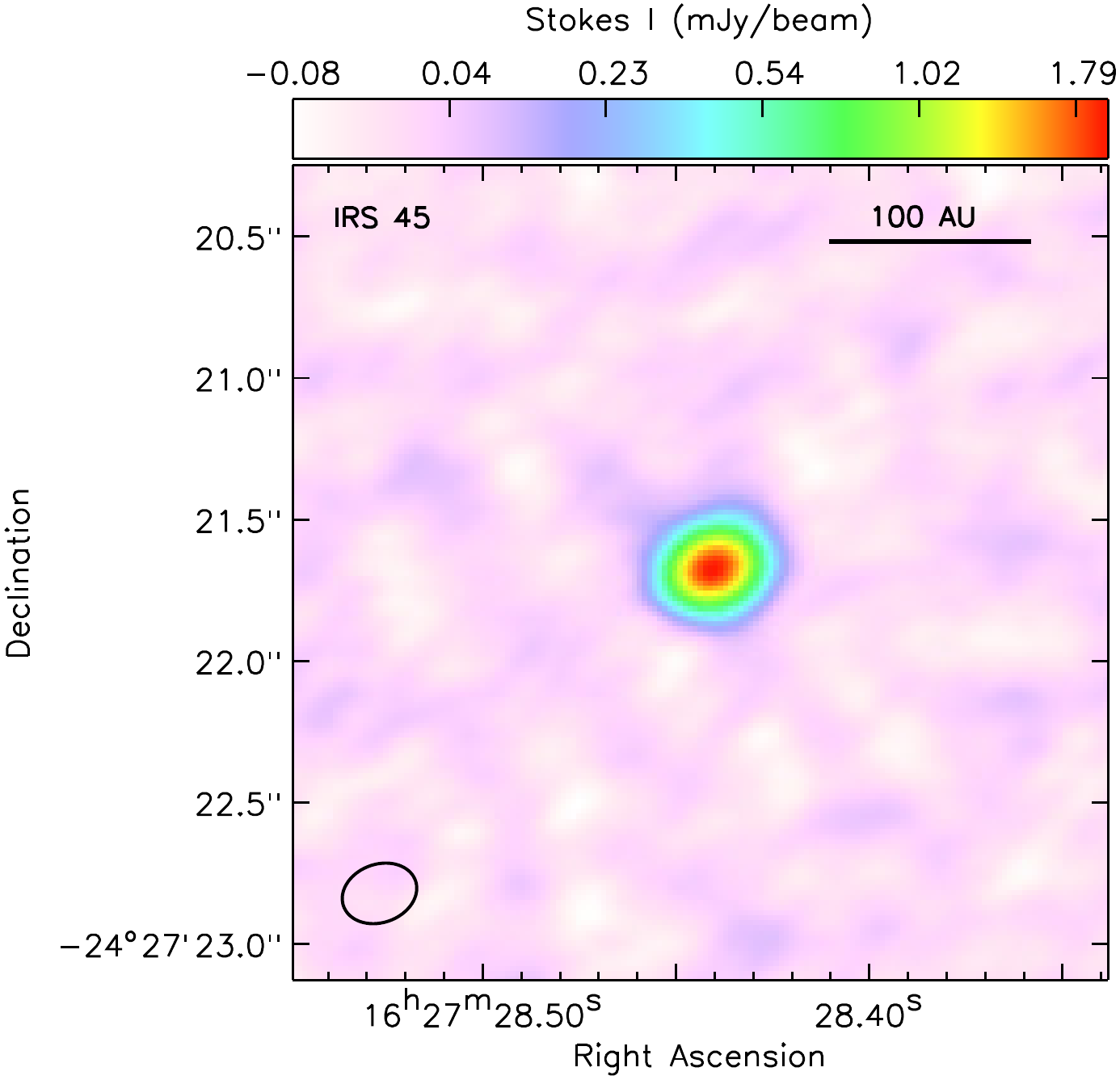}
\caption{Dust continuum map for IRS 45.  This source was not detected in polarization.  \label{irs45}}
\end{figure}

Figure \ref{irs45} shows the continuum image of IRS 45.  We do not detect this object in dust polarization, with a 3$\sigma$ upper limit of 3.9\%.   The continuum is also very compact and only marginally resolved with a deconvolved size of 14 au $\times$ 8 au (FWHM).  The source mass is 0.4 \Mjupiter, making IRS 45 one of the lower-mass disk candidates in our sample.   

This field also contains a second object, VSSG 18 B, which is 11\arcsec\ northeast of IRS 45 (see Figures \ref{multiples}).  This source was also undetected in polarization, with a 3$\sigma$ upper limit of 13\%.  This source is listed in the full c2d catalogue as a ``red'' spectrum object, but was not originally classified as a YSO in \citet{Evans09}.  More recently, VSSG 18 B was identified as a Flat spectral source in the \emph{Sptizer} YSO variable catalogue \citep[SSTYSV J162729.21-242716.9,][]{Gunther14}.   Since VSSG 18 B does not have a clear outflow or 70 \um\ emission \citep{Kamazaki19}, it is more likely an evolved YSO.  Thus, we consider this object to be an Class II object.  This source has a deconvolved size of 17 au $\times$ 14 au (FWHM) and a mass of 0.2 \Mjupiter.     

\begin{figure}[h!]
\includegraphics[width=0.475\textwidth]{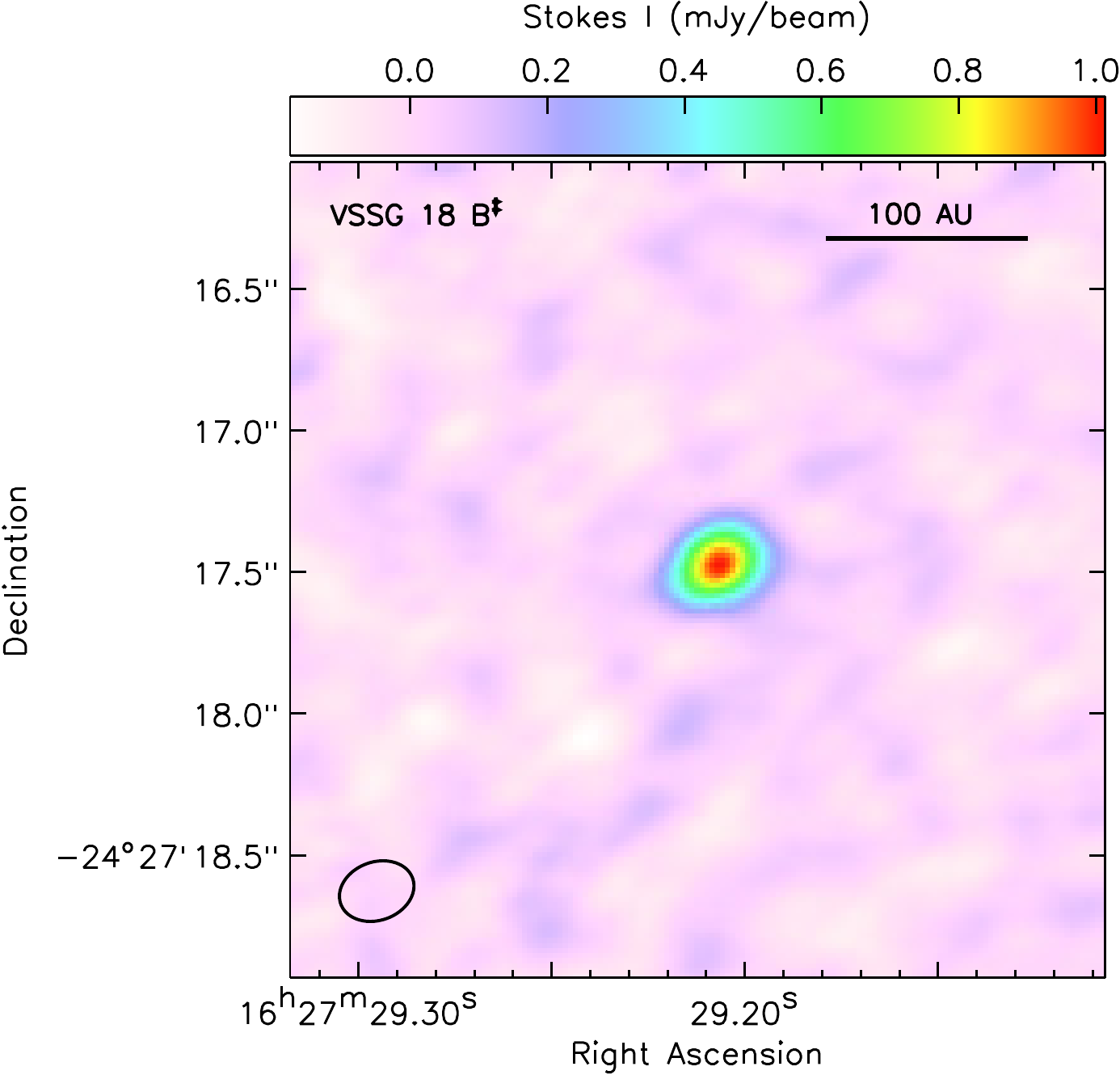}
\caption{Dust continuum map for VSSG 18 B.  This source was not detected in polarization.  \label{vssg}}
\end{figure}

\subsection{Field c2d\_904: IRS 47}

Field c2d\_904 is centered on IRS 47.  This object has an infrared spectral energy distribution consistent with a Flat spectral source or Class II object \citep{Evans09, Gutermuth09} and \citet{McClure10} identified its infrared SED as being dominated by a disk.  Nevertheless, IRS 47 appears to be driving a large outflow \citep{White15, Kamazaki19} and it is well detected in mid- and far-infrared wavelengths, with hints of a surrounding core structure from single-dish data \citep{Enoch09, Kamazaki19}.  Thus, we consider this source to be a Class I object. 

\begin{figure}[h!]
\includegraphics[width=0.475\textwidth]{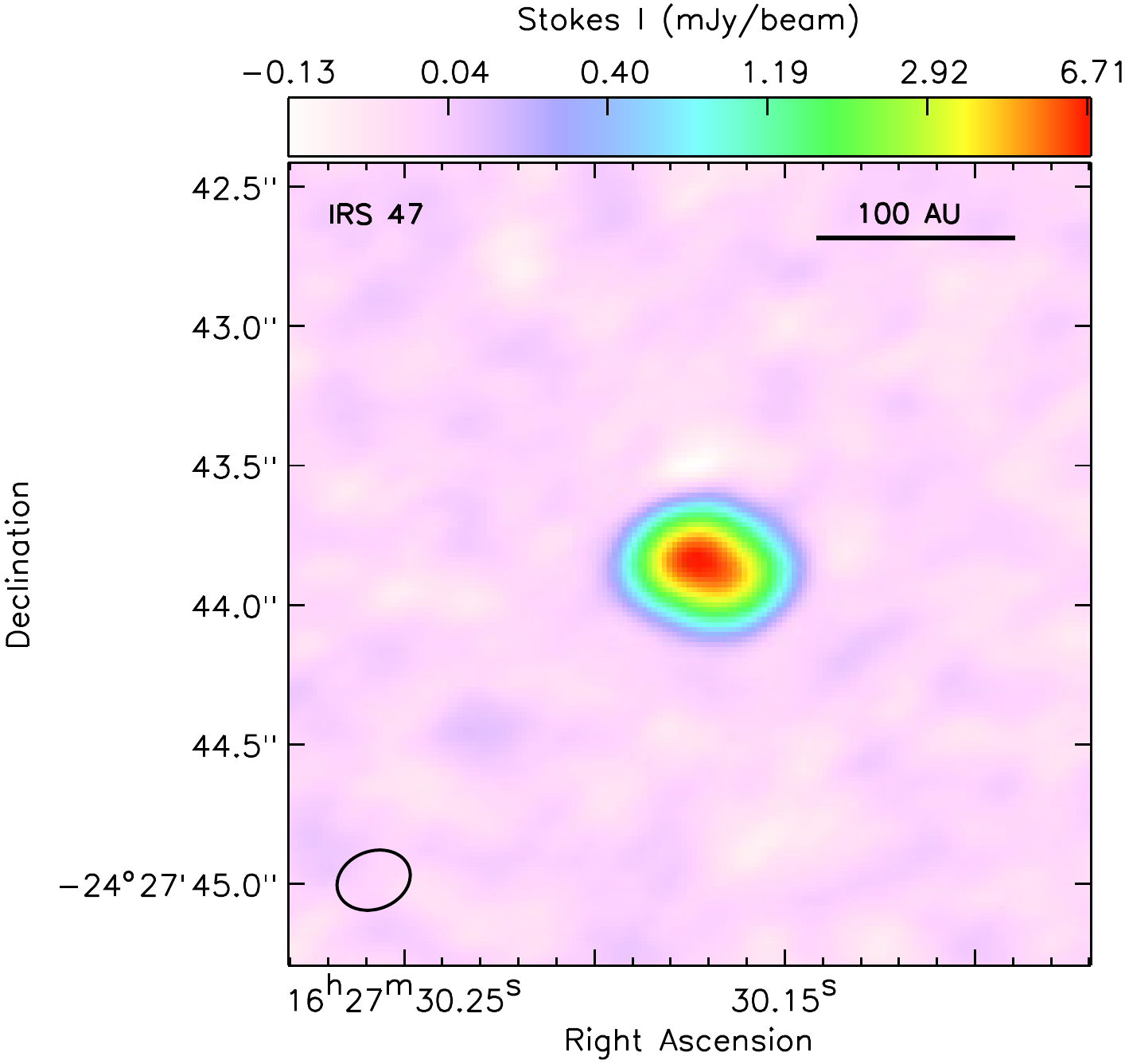}
\caption{Dust continuum map for IRS 47.  This source was not detected in polarization.  \label{irs47}}
\end{figure}

Figure \ref{irs47} shows the continuum image of IRS 47.  We do not detect it in polarization with a 3$\sigma$ upper limit of 1.1\%.  The source also appears to be highly inclined.  The deconvolved size is roughly 32.5 au $\times$ $\lesssim 1.8$ au (FWHM), where the minor axis dimension depends heavily on the elliptical region we use to fit the source with \texttt{imfit}.  This minor axis limit gives a disk inclination of $\gtrsim 86$\degree.  The estimated disk mass is 1.6 \Mjupiter.

\begin{figure}[h!]
\includegraphics[width=0.475\textwidth]{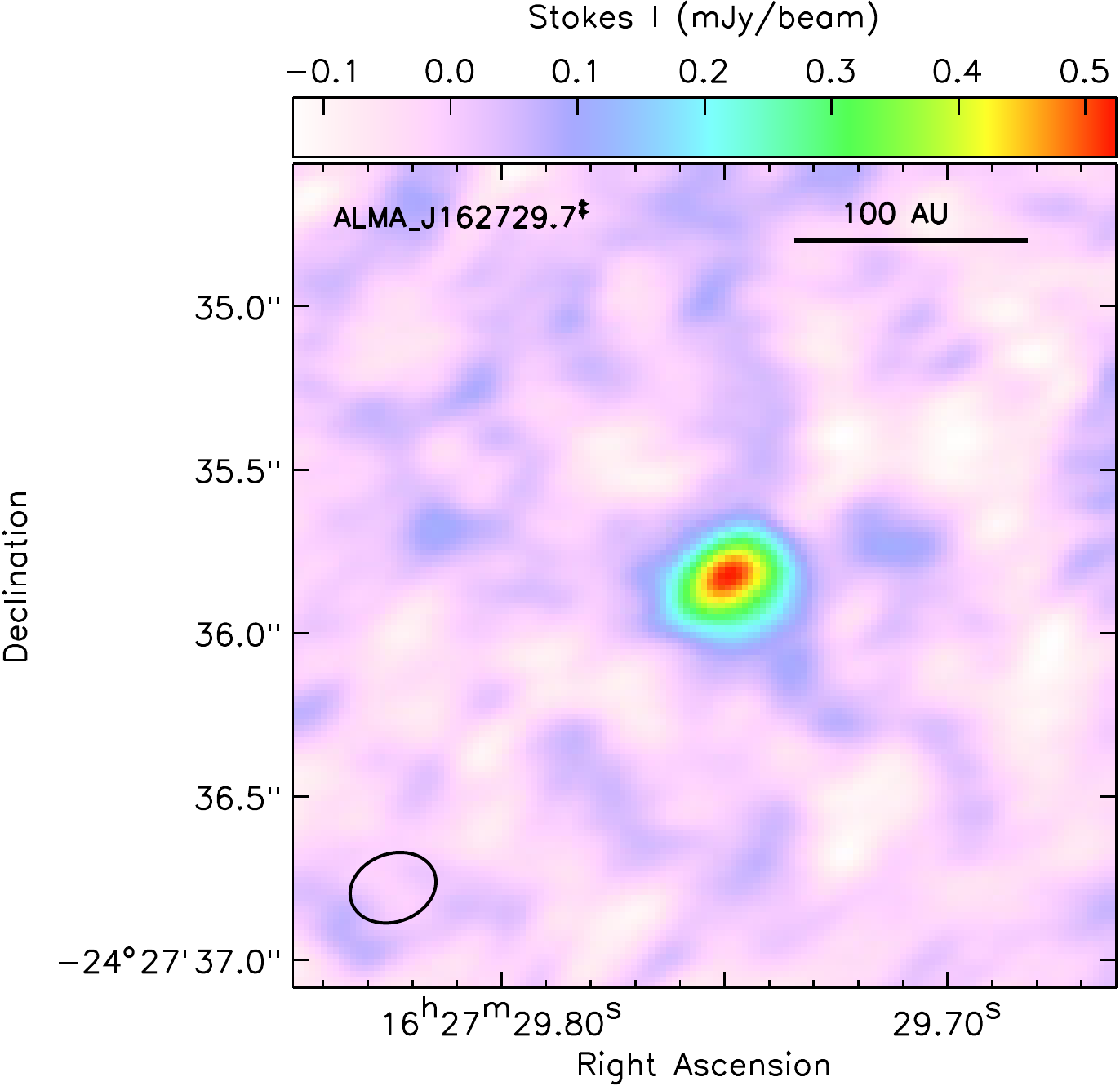}
\caption{Dust continuum map for ALMA\_J162729.7.  This source was not detected in polarization.  \label{mms}}
\end{figure}

We also detect a second compact source 10\arcsec\ northwest of IRS 47.   Figure \ref{mms} shows the continuum emission for this source, ALMA\_J162729.75-242735.83 (hereafter, ALMA\_J162729.7).  It is not detected in polarization with a 3$\sigma$ upper limit of 21\%.   ALMA\_J162729.7 was not previously detected in the literature and there is no counterpart in the 2MASS, \emph{Spitzer}, or WISE catalogues.  Moreover, this source was undetected by \citet{Kamazaki19} in both dust and gas with their ALMA data.  With a peak flux of 12.5$\sigma$, ALMA\_J162729.7 has a non-negligible probability of being a galaxy (see Section \ref{gal}).  Thus, we consider it to be an extragalactic object.     

Figure \ref{multiples} shows that there is substantial extended emission between IRS 47 and ALMA\_J162729.7.   \citet{Kirk17} weakly detected similar extended emission at 3 mm in $\sim 2$\arcsec\ resolution ALMA data, and \citet{Kamazaki19} strongly detected this structure in 1.3 mm continuum, $^{13}$CO (2-1), and C$^{18}$O (2-1) in their ALMA observations that combine the 12 m main array and the compact ACA.  The arc structure curves further around IRS 47 in the two line tracers, forming a near bubble and the arc also coincides with 70 \um\ emission from \emph{Herschel} \citep{Kamazaki19}.  The arc is highly filtered out in our observations, but its shape matches previous data.

\subsection{Field c2d\_954: Oph-emb-1}

Field c2d\_954 contains the deeply embedded Class 0 protostar, Oph-emb-1.  This object has a low bolometric temperature \citep{Evans09} and a bipolar, well collimated outflow \citep{Stanke06}.  \emph{Spitzer} observations further show a prominent jet and an extensive outflow cavity in scattered light and shocked H$_2$ emission that is viewed nearly face-on \citep{Barsony10, Hsieh17}.   Oph-emb-1 also has a dense, symmetric, and mostly spheroidal envelope that shows a small gradient normal to the outflow and also traces the outflow cavity \citep{Tobin10, Tobin11}.  Previous high resolution observations of Oph-emb-1 \citep{Chen13, Yen17, Hsieh19oph1} revealed a compact, elongated structure perpendicular to the outflow that is indicative of a disk.  \citet{Hsieh19oph1} further show that this structure has red- and blue-shifted C$^{18}$O (2-1) emission indicative of both infall and rotation.  They suggest that Oph-emb-1 is likely to form a brown dwarf.   

\begin{figure*}
\includegraphics[width=0.95\textwidth,trim=0mm 5mm 0mm 9mm,clip=true]{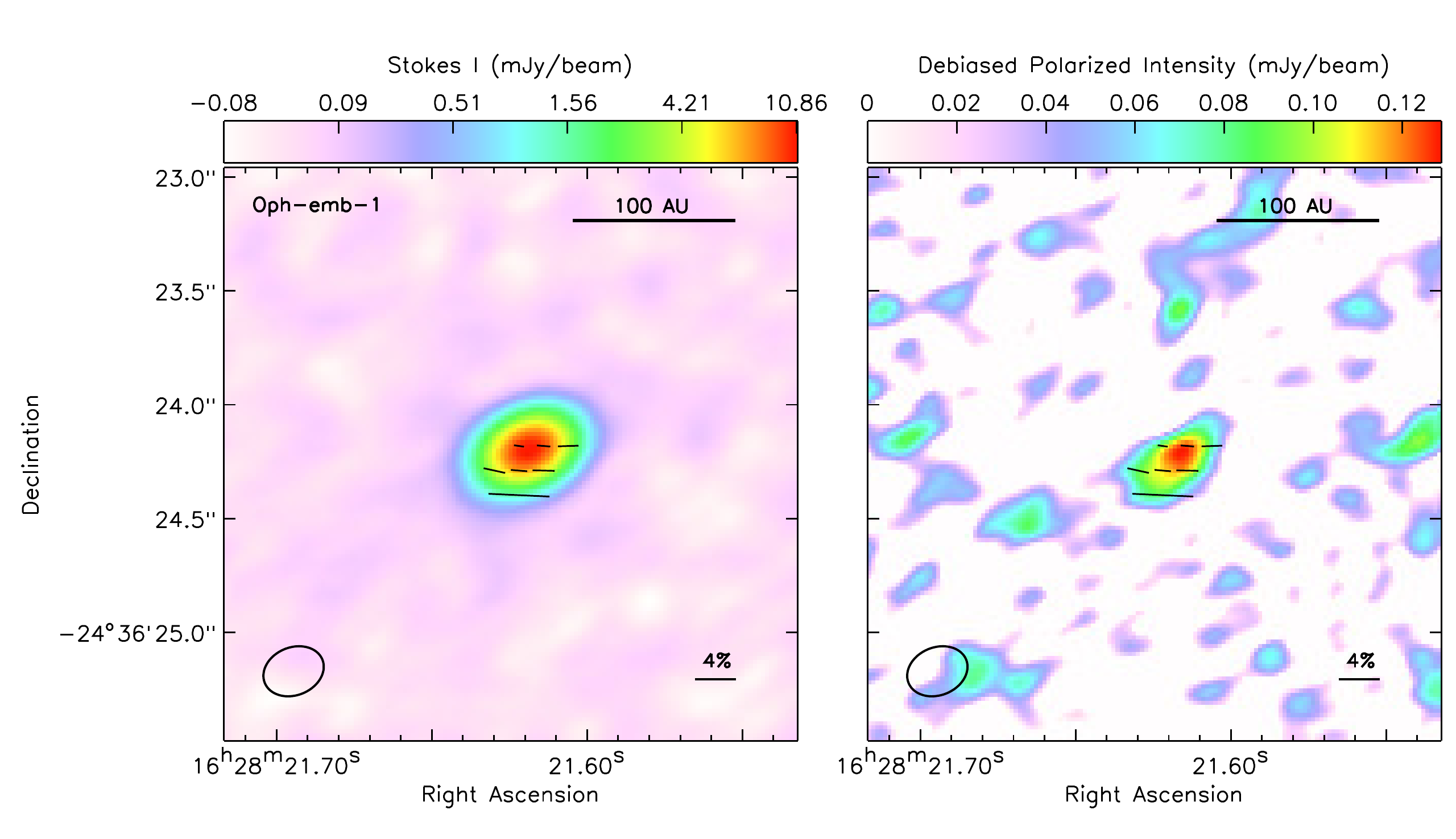}
\caption{Same as Figure \ref{gss_irs1} except for Oph-emb-1.   \label{oph1}}
\end{figure*}

Figure \ref{oph1} shows the polarization results for Oph-emb-1.  The source is very compact with  uniform polarization e-vectors.  The polarization position angles are $\approx$ 85\degree\ over the entire compact structure.  By comparison, the disk position angle is -65\degree, which means that the polarization angles are misaligned with both the major axis and minor axis of the disk.    Oph-emb-1 is compact with a deconvolved size of 18.8 au $\times$ 6.7 au (FWHM).  This disk size is a factor of two smaller than the disk size from lower resolution ALMA data in \citet[][]{Yen17}, which suggests that their two-Gaussian fit to a compact and extended component may have overestimated the size of the compact component.  The inferred disk position angle of -65\degree\ is nearly perpendicular to the outflow position angle of 20-25\degree\ \citep{Yen17}.  We measure a disk mass of 2 \Mjupiter. 

 \citet{Hsieh17} proposed that Oph-emb-1 may be a binary system based on its outflow showing an S-shape morphology in scattered light and H$_2$ emission.  Binary systems induce tidal interactions that can change the momentum vector of the outflowing source and cause the outflow to precess.  To test for a possible companion, we re-imaged the c2d\_954 field with \texttt{robust=-2} and no UV taper.  The resulting map has a resolution of $0.2\arcsec\ \times\ 0.11\arcsec$ (28 au $\times$ 15 au), but we find no evidence of substructure toward Oph-emb-1 in this higher resolution map.   Therefore, Oph-emb-1 appears to be a single protostar, although we cannot rule out a companion object at $< 20$ au scales.

\subsection{Field c2d\_963: Oph-emb-18}

Field c2d\_963 is centered on Oph-emb-18.  This object located on the outskirts of L1688 and has not been well studied in the literature.  It is a low luminosity source with a spectral energy distribution indicative of a Class I protostar \citep[e.g.,][]{Evans09, HsiehLai13}.  \citet{Antoniucci14}, however, identified it as a candidate for eruptive EXor accretion based on observations from \emph{Spitzer} and WISE.  EXor events are marked by significant IR excess and redder emission when the source is fading and are typically associated with T-Tauri stars \citep{Herbig08}.  \citet{vanKempen09} also classified this object as a later-stage YSO based on the lack of envelope emission in dust and gas.   We consider it a Class II object in this study.

\begin{figure}[h!]
\includegraphics[width=0.475\textwidth]{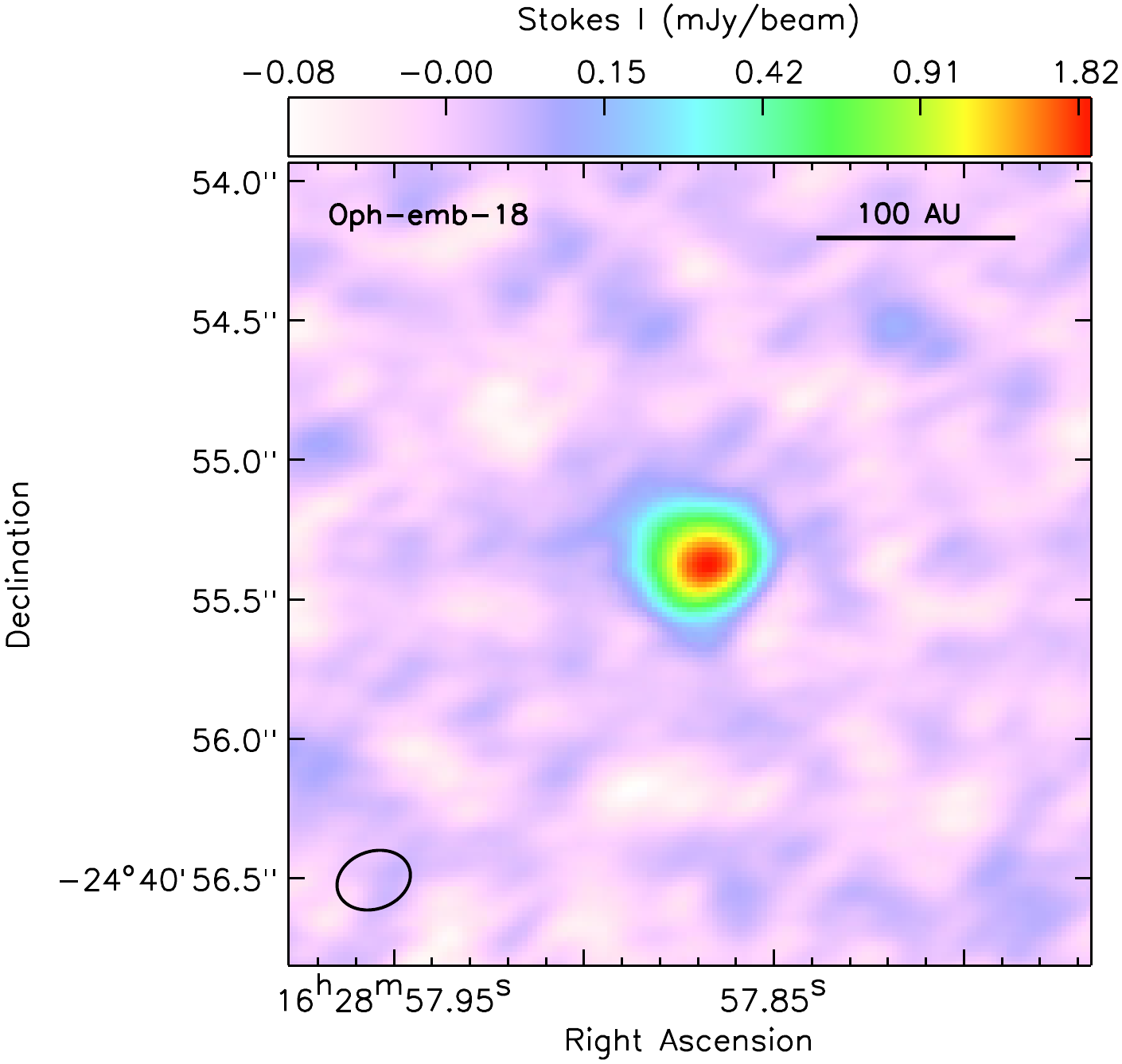}
\caption{Dust continuum map for Oph-emb-18.  This source was not detected in polarization.  \label{oph18}}
\end{figure}

Figure \ref{oph18} shows the continuum image of Oph-emb-18.  This is a relatively fainter source that is not detected in polarization.  We estimate a 3$\sigma$ upper limit for the dust polarization of 4.1\%.  Despite its faint emission, Oph-emb-18 is marginally resolved in our observations.  We find a deconvolved size of 20 au $\times$ 11.6 au with a mass of 0.4 \Mjupiter.  

\subsection{Field c2d\_989: IRS 63}

Field c2d\_989 contains a very bright, well known Class I source, IRS 63.  This source has a spectral energy distribution consistent with a Class I object \citep{Evans09}, but its disk emission is significant \citep{McClure10}.   Nevertheless, IRS 63 has a bipolar outflow \citep{Visser02,Dunham14} and a surrounding dense envelope \citep{vanKempen09}, which are expected for an embedded protostar.  We consider it a Class I object here.  IRS 63 has been well studied in the past, in particular for its large disk \citep{Andrews07, Lommen08, Miotello14}.  Previous studies using SMA observations estimated a disk size of 165 au and mass of 100 \Mjupiter\ from radiative transfer models \citep[e.g.,][]{BrinchJorgensen13}.  These values indicate that IRS 63  has a very high disk mass for a Class I protostar \citep[e.g.,][]{Jorgensen09, Aso15}.  

\begin{figure*}
\includegraphics[width=0.95\textwidth,trim=0mm 5mm 0mm 9mm,clip=true]{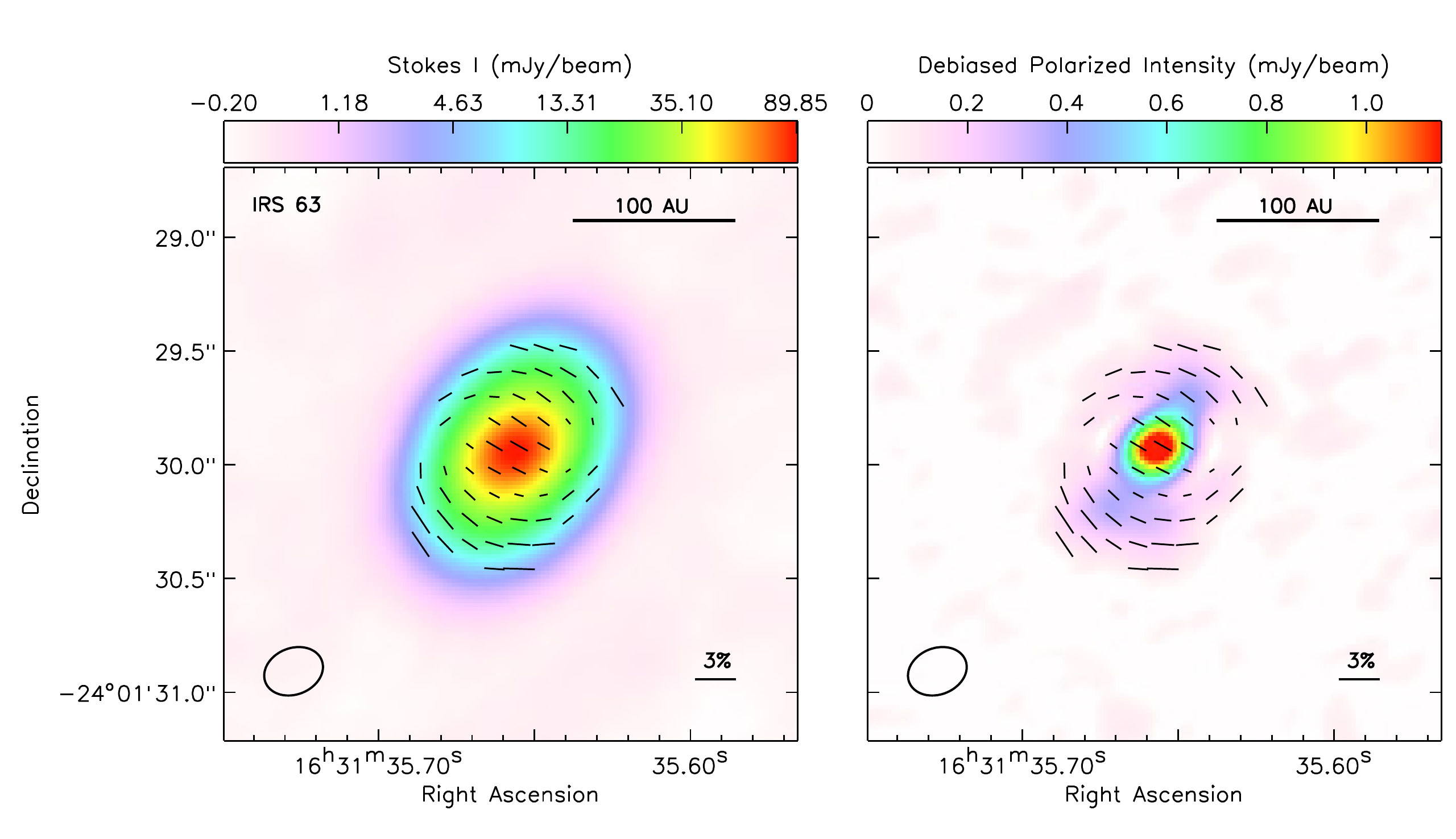}
\caption{Same as Figure \ref{gss_irs1} except for IRS 63.    \label{irs63}}
\end{figure*}

Figure \ref{irs63} shows the polarization results of IRS 63.   We detect substantial polarization across the entire disk and a clear change in the polarization structure as a function of radius.  Toward the disk center, the polarization position angles are uniform at $\approx 60$\degree\ that are nearly parallel to the minor axis of the disk (58\degree).  At larger radial extents, however, the polarization transitions to an azimuthal morphology.  This morphology is also captured in the polarized intensity map which looks peanut-shaped.  

IRS 63 is one of the best resolved sources in our sample.  We find a deconvolved size of 68 au $\times$ 46 au (FWHM) and a mass of 50 \Mjupiter.  More recently, \citet{Cox17} observed IRS 63 at $\sim 0\farcs2$ resolution with ALMA in 0.87 mm dust continuum.  They find a smaller disk size of 73 au $\times$ 50 au (FWHM) and a smaller mass of 47 \Mjupiter\ from a Gaussian fit to the continuum data.  These values should only be considered a first order estimate, however.  First, IRS 63 appears to have a relatively high-mass, dense disk.  The dust emission toward such a disk may be optically thick (see Appendix \ref{tau}).  Indeed, we find a higher disk mass at 1.3 mm than Cox et al. at 0.87 mm.  Second, higher resolution observations indicate that this disk has at least one bright ring (Segura-Cox et al. in preparation).  If confirmed, then IRS 63 may not be well fit with a simple Gaussian profile and will require more complex disk models to determine its size, mass, and geometry.

\subsection{Field c2d\_990: Oph-emb-4}\label{c2d990}

Field c2d\_990 contains a single object, Oph-emb-4, located roughly 3\arcmin\ south of IRS 63 in L1709.    The classification of this source is uncertain.  Its \emph{Spitzer} infrared spectral index has rising blue emission, suggesting that its a Class II object  \citep{vanderMarel16}, but it has the bolometric temperature is indicative of a Class I object \citep[e.g.,][]{Evans09}.  Moreover, this source is faint.  \citet{Dunham08} identified Oph-emb-4 as a candidate low luminosity embedded object with no known associated high density material. \citet{Riaz18} classified it as a proto-brown dwarf based on a low ($< 0.1$ \Msun) dust mass from SCUBA-2 observations and \citet{vanderMarel16} identify it as a low-mass ($< 5$ M$_{Jupiter}$) transition disk with a small cavity.  We consider this source to be a Class II object.

\begin{figure}[h!]
\includegraphics[width=0.475\textwidth]{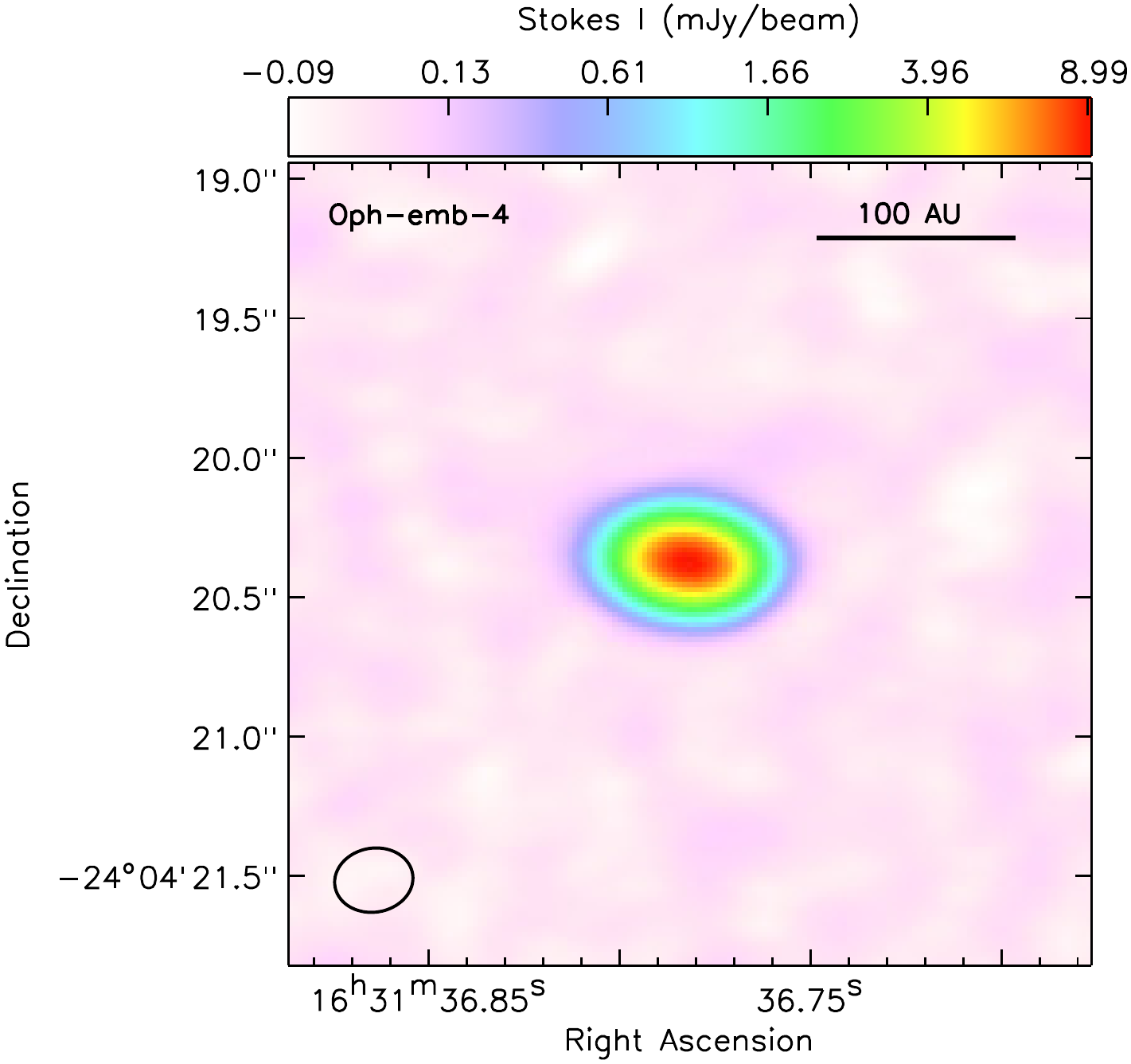}
\caption{Dust continuum map for Oph-emb-4.  This source was not detected in polarization.  \label{oph4}}
\end{figure}

Figure \ref{oph4} shows the continuum image of Oph-emb-4.  The continuum emission is fairly bright and compact, but it is not detected in polarization.  We measure a 3$\sigma$ upper limit of 0.9\%, indicating that the continuum source appears significantly unpolarized compared to typical disks.   The source also appears highly inclined.  Its deconvolved size is 36 au $\times$ 8 au (FWHM), resulting in an estimated inclination of 77\degree.   We also find a mass is 2 \Mjupiter, which is in agreement with the transition disk limit of $< 5$ M$_{Jupiter}$ from \citet{vanderMarel16}.

\subsection{Field c2d\_991: Oph-emb-25}

Field c2d\_991 is centered on Oph-emb-25.    It has an infrared spectral index consistent with a Flat spectral source and a bolometric temperature consistent with a Class I object \citep{Evans09}.   Oph-emb-25 is another proto-brown dwarf candidate \citep{Riaz18, Whelan18}, although \citet{Dunham08} do not list this object as a candidate low luminosity source.  \citet{Whelan18} observed a jet and outflow toward this source with near-infrared spectroscopy.  They find a relatively narrow (40\degree\ opening angle) and high outflow mass, which suggests that the outflow is young.  Nevertheless, \citet{Pattle15} do not detect a core around the infrared source. We therefore consider Oph-emb-25 to be a Flat object.  

\begin{figure}[h!]
\includegraphics[width=0.475\textwidth]{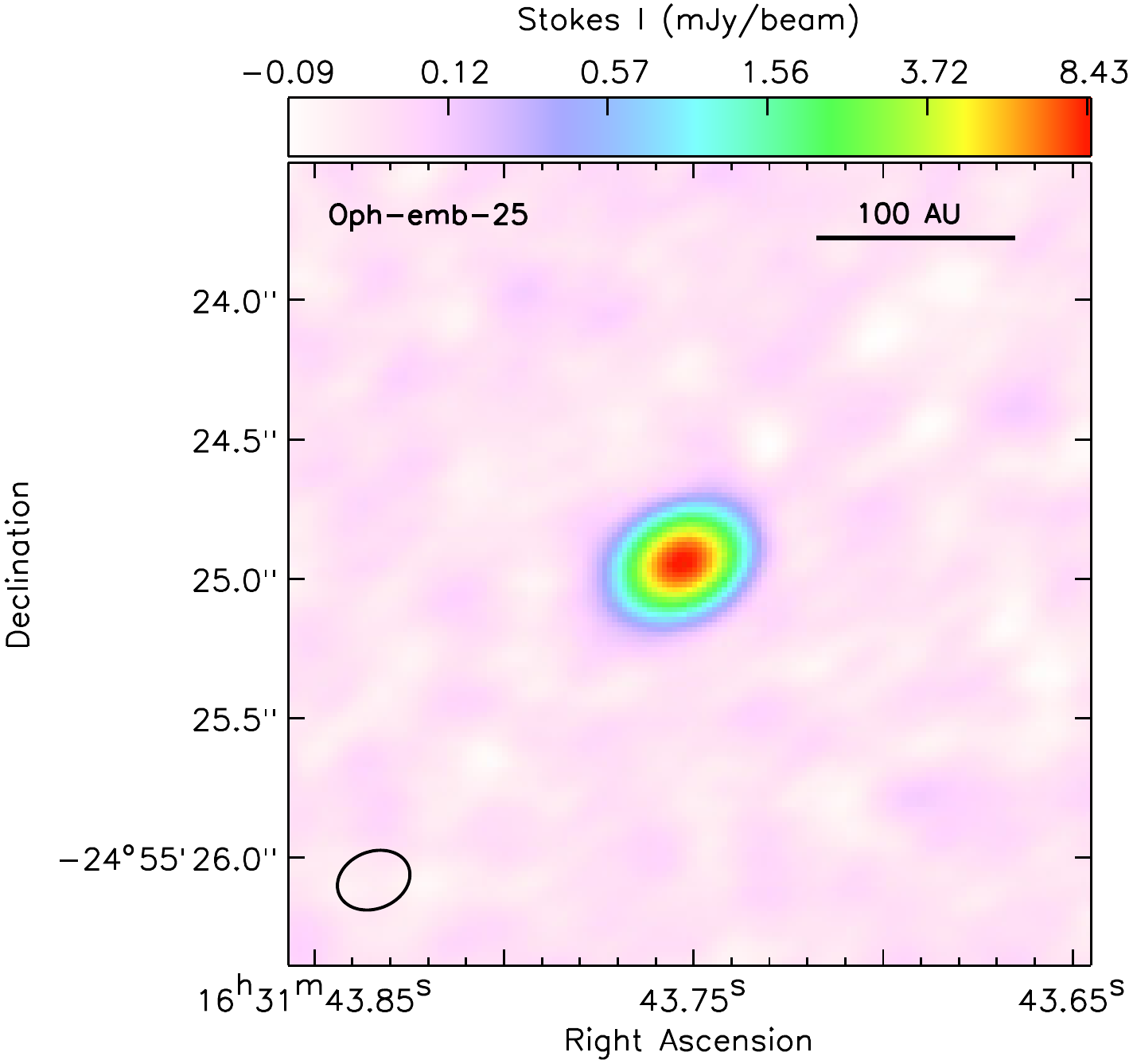}
\caption{Dust continuum map for Oph-emb-25.  This source was not detected in polarization.  \label{oph25}}
\end{figure}

Figure \ref{oph25} shows the continuum image of Oph-emb-25, which we do not detect in polarization.  The 3$\sigma$ upper limit for the non-detection is 0.9\%.  We find a deconvolved size of 10 au $\times$ 4.5 au (FWHM) with a position angle of -38\degree.  Since the outflow position angle is 50\degree, the major axis is roughly perpendicular to the direction of the outflow, indicating that the compact object is likely tracing a disk.  We estimate a disk mass of roughly 1.5 \Mjupiter.   

\subsection{Field c2d\_996: Oph-emb-7}

Field c2d\_996 is centered on Oph-emb-7.  \citet{Bontemps01} first identified this source with mid-infrared observations and classified it as a Class II object.  \emph{Spitzer} observations, however, yield an infrared spectral index and bolometric temperature that is normally associated with a Class I object \citep[e.g.,][]{Evans09, Dunham15}.  Nevertheless,  \citet{Jorgensen08} identified this source as a candidate edge on disk and single-dish (sub)millimeter observations found no clear surrounding core for this object.   There is no corresponding SCUBA or SCUBA-2 core \citep{Jorgensen08, Pattle15} at the position of this source, and the nearest Bolocam source, Bolo 33, is roughly half an arcminute south of the infrared detection \citep[][]{Young06, Enoch09}.  Thus, we consider Oph-emb-7 to be a Class II source.

\begin{figure}[h!]
\includegraphics[width=0.475\textwidth]{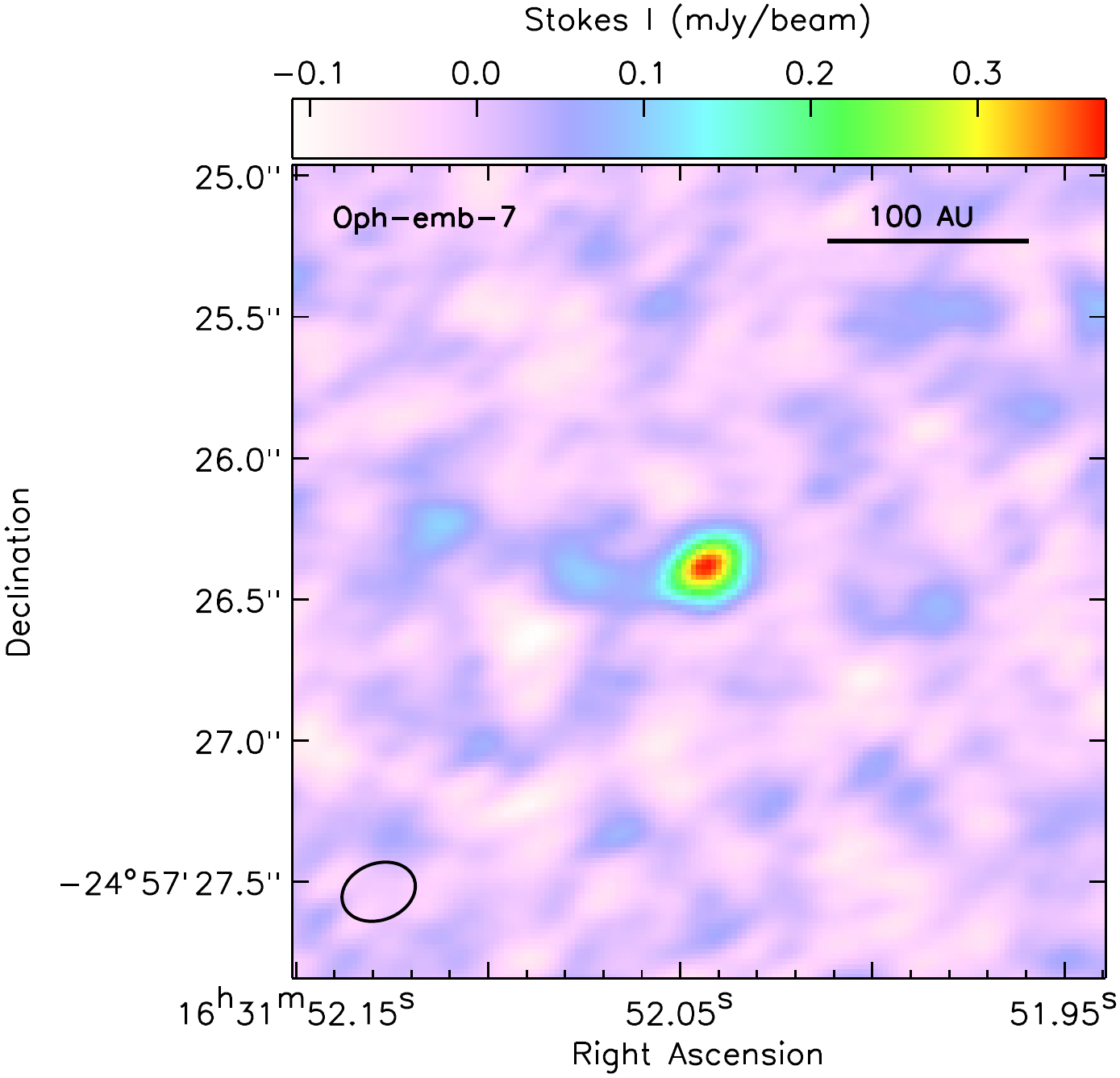}
\caption{Dust continuum map for Oph-emb-7.  This source was not detected in polarization.  \label{oph7}}
\end{figure}

Figure \ref{oph7} shows the continuum image of Oph-emb-7.  It is the faintest of the main targets in our sample, with a peak flux of only 12$\sigma_{peak}$.   We do not detect this source in polarization with a 3$\sigma$ upper limit of 21\%.  The continuum emission is unresolved in our data, indicating that the disk is compact ($\lesssim 9$ au; 1/4 the beam).  We also find a very low disk mass of 0.05 \Mjupiter.  If we instead calculate the mass in dust only, we find a \emph{dust mass} of roughly 0.2 M$_{\oplus}$.  Thus, the Oph-emb-7 disk mass appears to be a factor of 9 lower than the lowest-mass Class II disk in  \citet{Ansdell17} for Orion and a factor of 2 lower than the lowest-mass disk in \citep{Long18} for Chameleon.

\subsection{Field c2d\_998: Oph-emb-15}

Field c2d\_998 contains Oph-emb-15.  This source has an uncertain classification.  It has been identified as a Class II object \citet{Bontemps01} and a Class I object \citep[e.g.,][]{Evans09, Dunham15} based on its spectral energy distribution.  We could find no previous outflow observation in the literature for this source, however.  Oph-emb-15 appears to have a surrounding core  \citep{Jorgensen08, Pattle15}, although the infrared source is located toward the edge of a core (roughly 17\arcsec\ north of the core peak).  This offset is still considered within the core boundary \citep{Sadavoy10} and optical observations of Oph-emb-15 show extended nebulosity indicative of a surrounding envelope \citep{Duchene04}.   Thus, we classify Oph-emb-15 as a Class I source.

\begin{figure}[h!]
\includegraphics[width=0.475\textwidth]{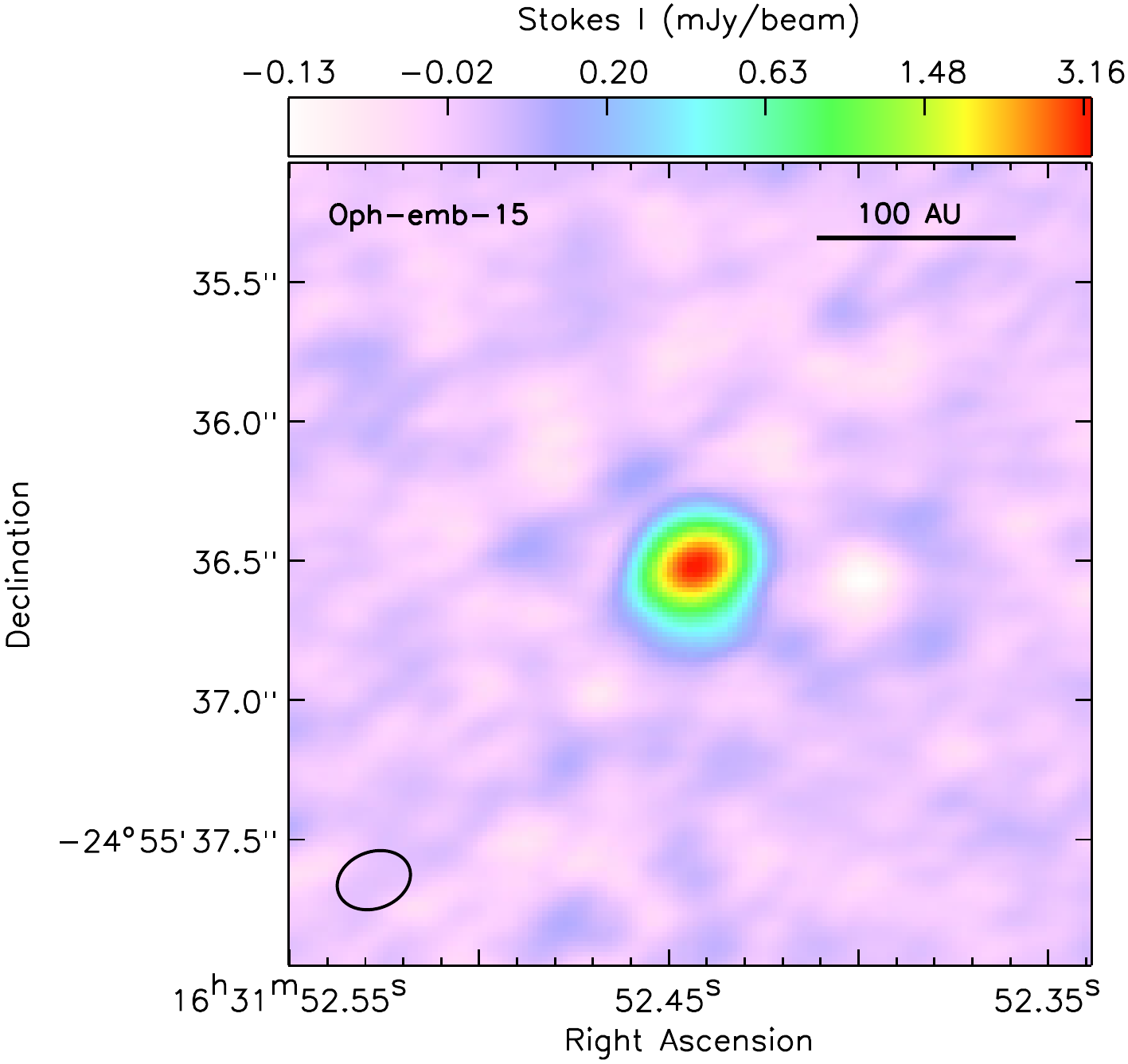}
\caption{Dust continuum map for Oph-emb-15.  This source was not detected in polarization.  \label{oph15}}
\end{figure}

Figure \ref{oph15} shows the continuum image of Oph-emb-15.  It is not detected in polarization with a 3$\sigma$ upper limit of 2.4\%\ for the polarization fraction.  The dust emission is compact, however, with a deconvolved size of 16 au $\times$ 6 au and a disk mass of 0.6 \Mjupiter.

\subsection{Field c2d\_1003: IRS 67}

Field c2d\_1003 is centered on the well known Class I system, IRS 67.  This object has an established dense core \citep{Young06, Pattle15} and its infrared emission is consistent with a Class I object \citep{Evans09, McClure10}.  It also has signatures of both infall and outflows \citep{Bontemps96, Mottram17} and strong nebulosity \citep{Duchene04}, which signify its youth.  Due to its embeddedness, IRS 67 was first identified as a binary system separated by $\sim 0.6$\arcsec\ through deep infrared observations \citep{McClure10}.   \citet{McClure10} called the components A and B, where the brighter A source was a disk candidate and the fainter B source was a younger envelope candidate.  More recently, \citet{ArturV18} observed IRS 67 with ALMA in Band 7 at $\sim$ 0.4\arcsec\ resolution.  They spatially resolved the A and B components and also detected a circumbinary disk around them in dust and gas emission.  The circumbinary structure also shows a clear velocity gradient that is consistent with Keplerian rotation and infall.

\begin{figure*}
\includegraphics[width=0.95\textwidth,trim=0mm 5mm 0mm 9mm,clip=true]{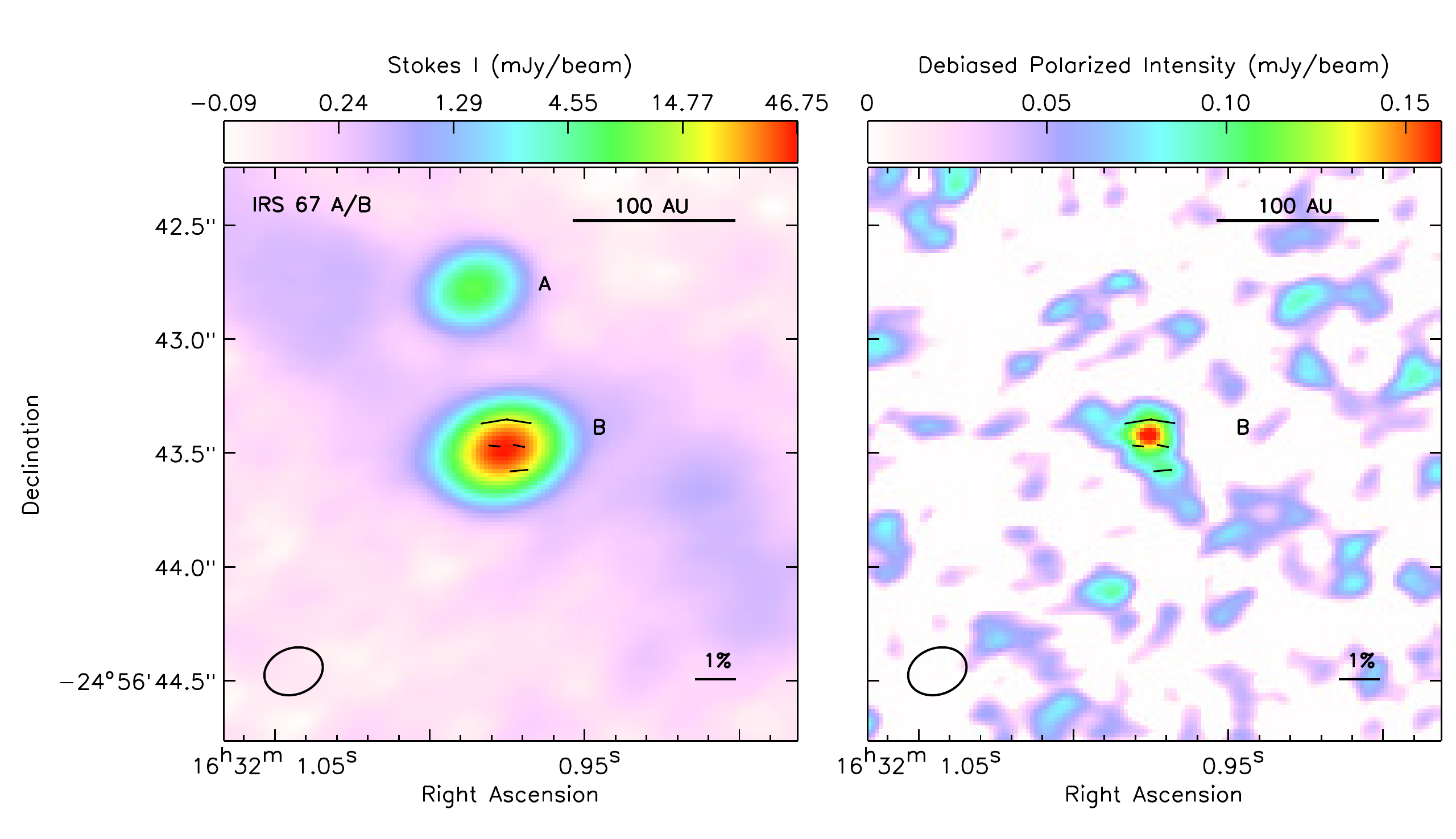}
\caption{Same as Figure \ref{gss_irs1} except for IRS 67.  Source IRS 67-A located south of IRS 67-B is not detected in polarized intensity.   \label{irs67}}
\end{figure*}

Figure \ref{irs67} shows the polarization results for IRS 67.  IRS 67-B is much brighter than IRS 67-A at millimeter wavelengths \citep[see also,][]{ArturV18}, and subsequently,  we detect uniform dust polarization only toward IRS 67-B.  The polarization is fairly uniform, with typical position angles of 87\degree\ that are nearly aligned with the long axis of the disk at 89\degree.  IRS 67-B also has relatively low polarization fractions of $\lesssim 1$\%.  IRS 67-A is not detect in dust polarization with a 3$\sigma$ upper limit of 1\%.  If IRS 67-A has similar polarization fractions as IRS 67-B, it would be below our sensitivity.   

Both sources are compact.  We find a deconvolved disk size of 8 au $\times$ 7 au (FWHM) and mass of 1.4 \Mjupiter\ for IRS 67-A and 15.8 au $\times$ 9.3 au (FWHM) and 8.6 \Mjupiter\ for IRS 67-B.   We also detect the circumbinary disk around the stars in Stokes I continuum, although the extended dust emission is highly filtered out in our observations.

\subsection{Field c2d\_1008: IRAS 16293}

Field c2d\_1008 contains the well-studied Class 0 protostellar system, IRAS 16293-2422 (hereafter, IRAS 16293).  This is a deeply embedded protostellar system in L1689N containing two main complexes, IRAS 16239A to the south and IRAS 16293B to the north, separated by roughly 5\arcsec\ \citep[e.g.,][]{Chen13, Jorgensen16}.  Both sources have been the target of extensive study for their hot core chemistry \citep[e.g.,][]{Schoier02, Jorgensen11, Pineda12}, including a dedicated ALMA survey to sample this molecular line emission of this source over 40 GHz in Band 7 \citep[e.g.,][and references therein]{Jorgensen16, Manigand19}.  The ALMA observations have also revealed an extensive dust Bridge between the stars \citep[e.g.,][]{Pineda12, Jacobsen18, vanderWiel19}.  

\begin{figure*}
\includegraphics[width=0.95\textwidth,trim=0mm 5mm 0mm 9mm,clip=true]{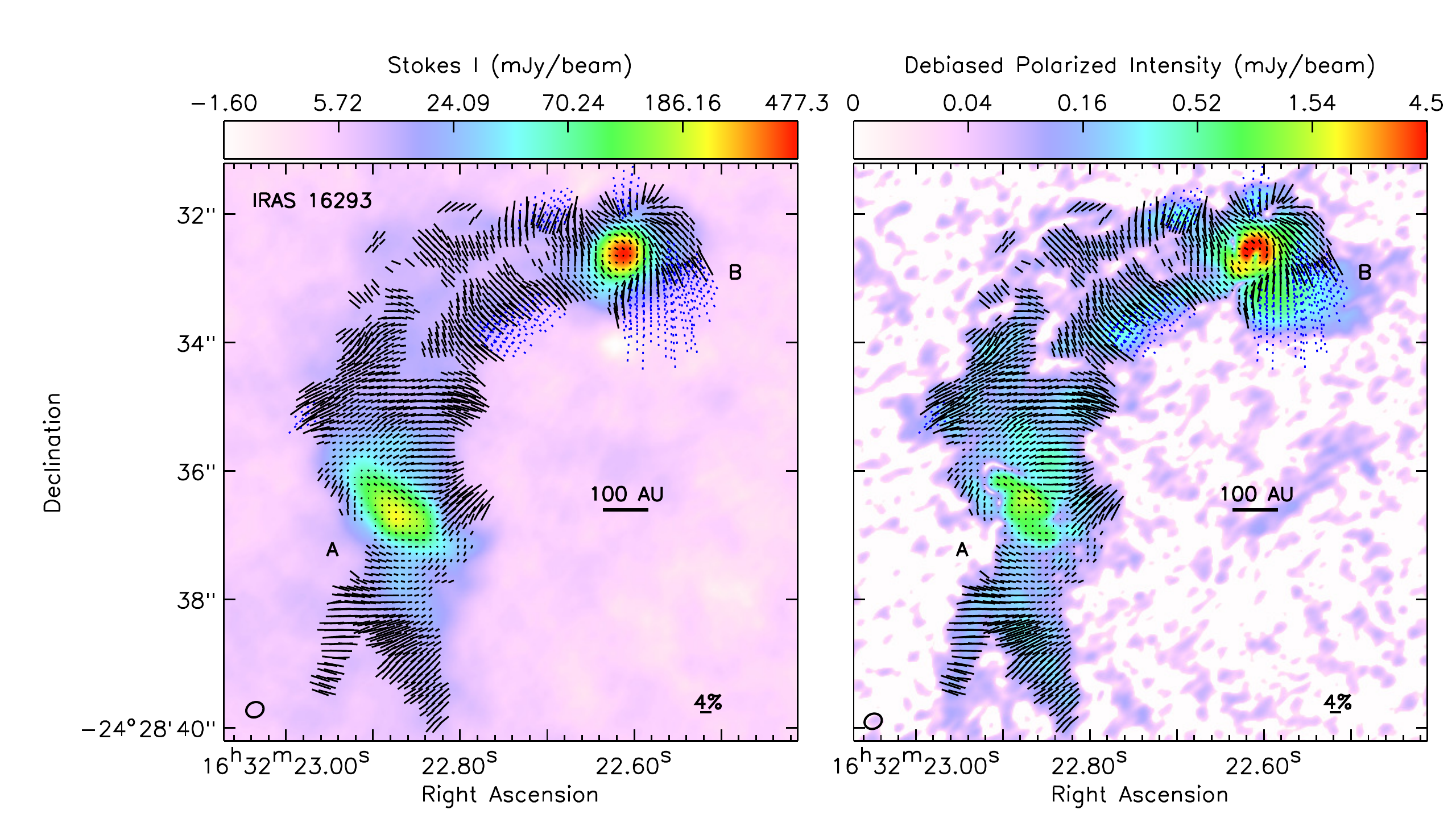}
\caption{Same as Figure \ref{gss_irs1} except for IRAS 16293A and IRAS 16293B.  We also added the criterion that $\PF/\sPF > 3$ to remove spurious e-vectors at the edges and off source. The dotted blue line segments indicate polarization fractions $>10$\%.  IRAS 16293A is located to the south and IRAS 16293B is located to the north (see Figure \ref{multiples}). \label{iras16293}}
\end{figure*}

Figure \ref{iras16293} shows the polarization results for IRAS 16293.  These data were previously discussed in \citetalias{Sadavoy18c}.    Briefly, this map shows the most extensive and complex polarization structure of the entire sample.  We see distinct and resolved polarization morphologies for IRAS 16293A and IRAS 16293B and polarization in the dust Bridge between them, as well as the dust streamers from each of the stars.  We also resolve a depolarization zone between the northern streamer from IRAS 16239B and the Bridge between the stars \citepalias{Sadavoy18c}.  This depolarization region also aligns well with the bluer dust emission from the three-colour image of IRAS 16293 from \citet{Jorgensen16}.   

We note that Figure \ref{iras16293} shows more polarization e-vectors than \citetalias{Sadavoy18c} due to the updated polarization debiasing method (see Section \ref{debias}) that reliably corrects polarization e-vectors with $3 < \PI/\sPI < 4$.  Most of the new e-vectors are in the lower emission regions and do not alter the conclusions in \citetalias{Sadavoy18c}.  We also see evidence of very high polarization fractions $> 10$\%\ toward the periphery of IRAS 16293B and the Bridge.  

The dust emission and dust polarization seen toward IRAS 16293A and IRAS 16293B are likely tracing material from the inner envelope around and between the two stars rather than disks.  As such, we use a Gaussian fit to only the brightest emission ($ > 100$ \mJybeam) to estimate the general morphology of each source.  Since this emission primarily traces envelopes rather than disks, we do not attempt to estimate their disk masses.

\subsection{Field VLA1623a: VLA 1623 West}\label{vla1623a}

Field VLA1623a is centered on VLA 1623W (see Figure \ref{multiples}), one of the protostellar objects embedded within the VLA 1623.4-2418 core, the canonical Class 0 object \citep{Andre93}.  \citet{BontempsAndre97} first identified this component in VLA radio emission at 3.6 cm and 6 cm observations and initially called it clump B.  The source name changed to VLA 1623 West (or VLA 1623W) in \citet{Chen13} following the naming contention of  VLA 1623A and VLA 1623B for the eastern sources \citep{Looney00}.  VLA 1623A/B are also in the field, but will be discussed in Section \ref{vla1623b}.

The classification of VLA 1623W has been under debate \citep{Maury12, Murillo18}.   The spectral energy distribution of the VLA 1623 core indicates that VLA 1623A/B protostars are deeply embedded \citep[e.g.,][]{Evans09, Gutermuth09} with a prominent bipolar outflow \citep[e.g.,][]{White15}, but the spectral energy distribution of VLA 1623W peaks at shorter wavelengths.  VLA 1623W also has a lower envelope mass compared to VLA 1623A/B \citep{MurilloLai13, Murillo18}, and it has no obvious outflow \citep{Santangelo15, Nisini15}.  Nevertheless, we consider this source to be a Class 0 object in this study.  First, VLA 1623W appears to be co-moving with VLA 1623A/B \citep{Harris18} and associated with the same dense core \citep{Pattle15}.  Second, VLA 1623W still has a substantial envelope, although it may be more tenuous than the envelope around VLA 1623A/B.  \citet{Murillo18} note that their ALMA data recover scales out to $\sim 400$ au, whereas the envelope can extend to 1000 au scales.   For example, \citet{Kirk17} measured a more comparable envelope mass of 0.1 \Msun\ for VLA 1623W with 3 mm ALMA data that recover emission out to 3000 au.  

\begin{figure*}
\includegraphics[width=0.95\textwidth,trim=0mm 5mm 0mm 9mm,clip=true]{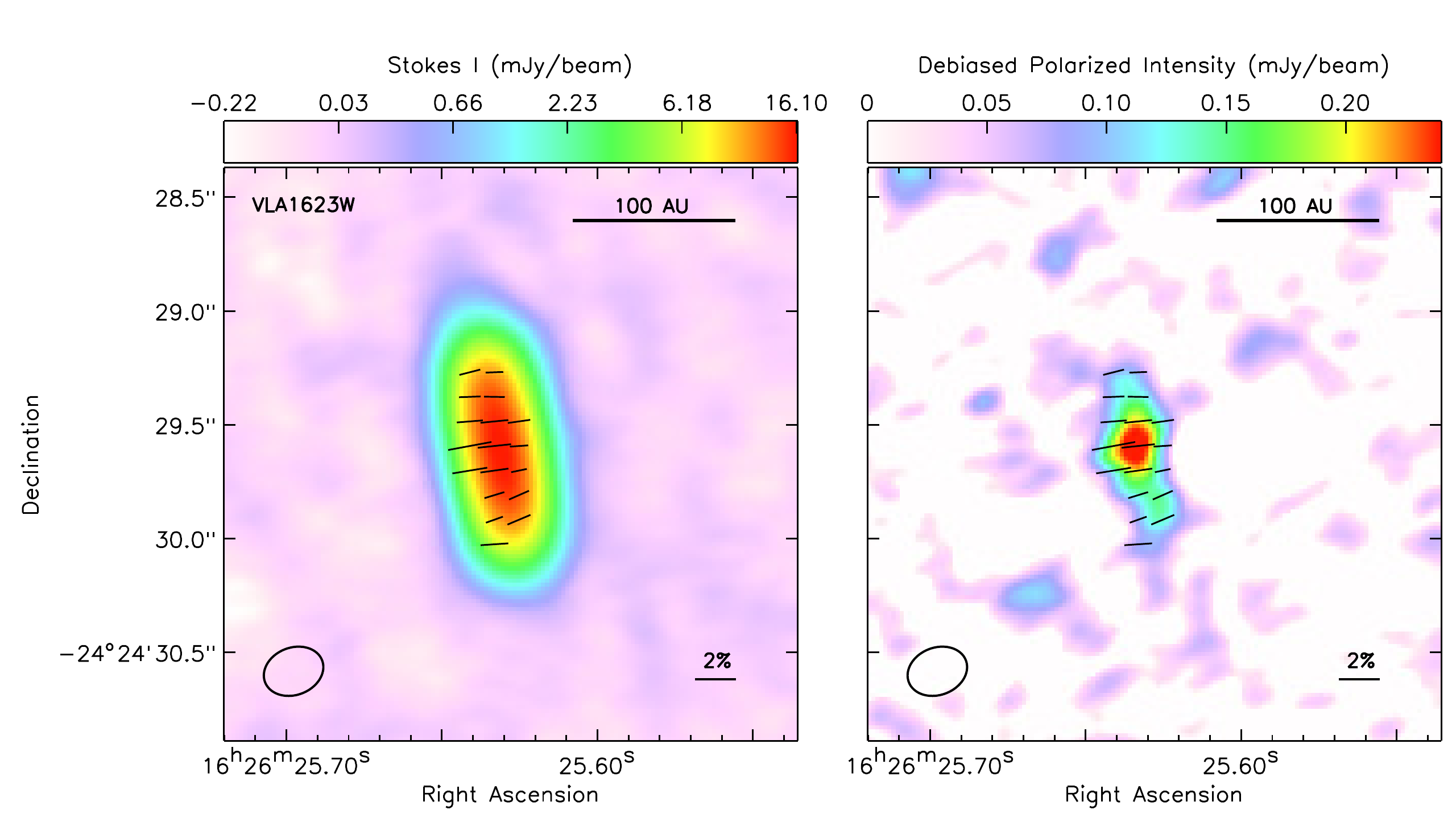}
\caption{Same as Figure \ref{gss_irs1} except for VLA 1623W.    \label{vla1623w}}
\end{figure*}

Figure \ref{vla1623w} shows the polarization results for VLA 1623W.   We see uniform polarization structure, with position angles of roughly $-81$\degree\ across the disk that are aligned with the disk minor axis of -80\degree.  The disk also appears to be highly elongated.  We measure a disk size of 99 au $\times$ 14.7 au (FWHM) and mass of 10.6 \Mjupiter.

The 1.3 mm polarization in Figure \ref{vla1623w} is in good agreement with 872 \um\ polarization from \citet{Harris18}.  The average 872 \um\ polarization position angle was -80\degree\ across VLA 1623W.  This agreement is significant to not only help identify the polarization mechanism (see Section \ref{mech_discuss}), but to also trust the 872 \um\ polarization observations of VLA 1623W, which was not located at the phase center of the primary beam.  \citet{Harris18} centered their map between VLA 1623W and VLA 1623A/B, such that each system was roughly 5\arcsec\ from the phase center.  This positioning placed VLA 1623W outside the inner third of the primary beam FWHM ($R \approx 3$\arcsec\ at 345 GHz).    Although off-axis polarization is considered less reliable, \citet{Harris18} argued that the agreement between their off-axis polarization data at 872 \um\ and the on-axis 1.3 mm polarization data presented here indicate that their measurements are robust.  In Appendix \ref{offaxis}, we test the polarization measurements for adjacent fields and indeed find that the dust polarization is largely consistent for separations $\lesssim 10$\arcsec.

\subsection{Field VLA1623b: VLA 1623 East}\label{vla1623b}

Field VLA1623b is centered on VLA 1623A/B.  Both sources are well studied in the literature and are separated by $\approx$ 1\arcsec\ \citep{Chen13}.  We consider these sources to be Class 0 objects \citep{Andre93} as both sources are deeply embedded in a dense core \citep[e.g.,][]{Pattle15}, they have cold spectral energy distributions \citep[e.g.,][]{Evans09, Gutermuth09, Murillo18}, and at least one of them is driving a powerful, bipolar outflow \citep[e.g.,][]{MurilloLai13, White15}.  VLA 1623A further has a large ($R \approx 180$ au), massive disk that shows evidence of Keplerian rotation \citep{Murillo13, Hsieh19}.  \citet{Harris18} used very high resolution data to show that the VLA 1623A source is itself a tight binary system (VLA 1623Aa and VLA 1623Ab) separated by $\sim 14$ au such that the large disk around VLA 1623A is a circumbinary disk.  For simplicity, we use VLA 1623A to refer to the unresolved circumstellar material from VLA 1623Aa and VLA 1623Ab.  The field also contains VLA 1623W, which we discuss in Section \ref{vla1623a}.

\begin{figure*}
\includegraphics[width=0.95\textwidth,trim=0mm 5mm 0mm 9mm,clip=true]{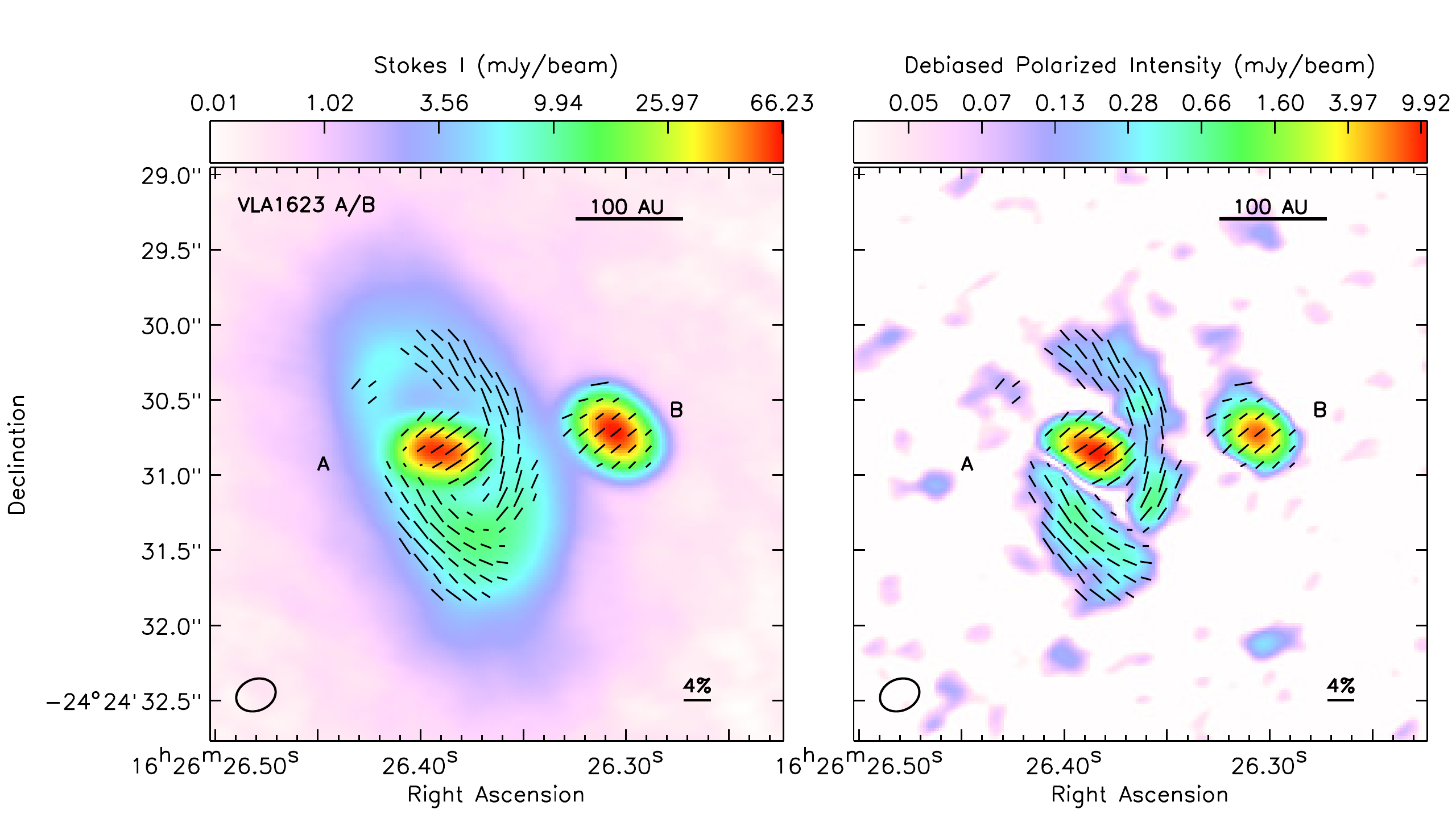}
\caption{Same as Figure \ref{gss_irs1} except for VLA 1623A and VLA 1623B.  We used a stricter selection criteria of $I/\sigma_I > 50$ to exclude the noise-like features within the extended emission of the larger VLA 1623-A circumbinary disk but are unlikely to be real because they are smaller than the beam.    \label{vla1623ab}}
\end{figure*}

Figure \ref{vla1623ab} shows extensive polarization across VLA 1623A/B.  These observations were first discussed in \citetalias{Sadavoy18b} and are consistent with that study.  In brief, the polarization structure of VLA 1623A and VLA 1623B show two distinct morphologies.  The dust polarization toward the compact, circumstellar material around VLA 1623A and VLA 1623B are both uniform with angles of $\approx -50$\degree\ that are roughly parallel to the minor axes of the dust emission.  In the extended dust emission around VLA 1623A (e.g., toward the larger Keplerian circumbinary disk), the dust polarization is azimuthal.  These dual polarization morphologies are also seen at 872 \um\ by \citet{Harris18}.    The deconvolved sizes are 50.5 au $\times$ 22 au for the circumstellar material of VLA 1623A (e.g., excluding the extended dust emission) and 44 au $\times$ 14.4 au for the circumstellar disk of VLA 1623B \citepalias[see also,][]{Sadavoy18b}.  Their estimated mass are 22 \Mjupiter\ and 20.6 \Mjupiter, respectively.

Figure \ref{vla1623b} also shows several additional polarization e-vectors than \citetalias{Sadavoy18b}.  Since we use a more robust debiasing method (see Section \ref{debias}), we are able to include these polarization e-vectors that have lower S/N ($3 < \PI/\sPI < 4$).   In particular, there are three e-vectors in the upper-left quadrant that were previously below the selection criteria in \citetalias{Sadavoy18b}.   These new e-vectors are consistent with the overall azimuthal polarization structure seen in the circumbinary disk.  

\begin{figure}
\includegraphics[width=0.475\textwidth]{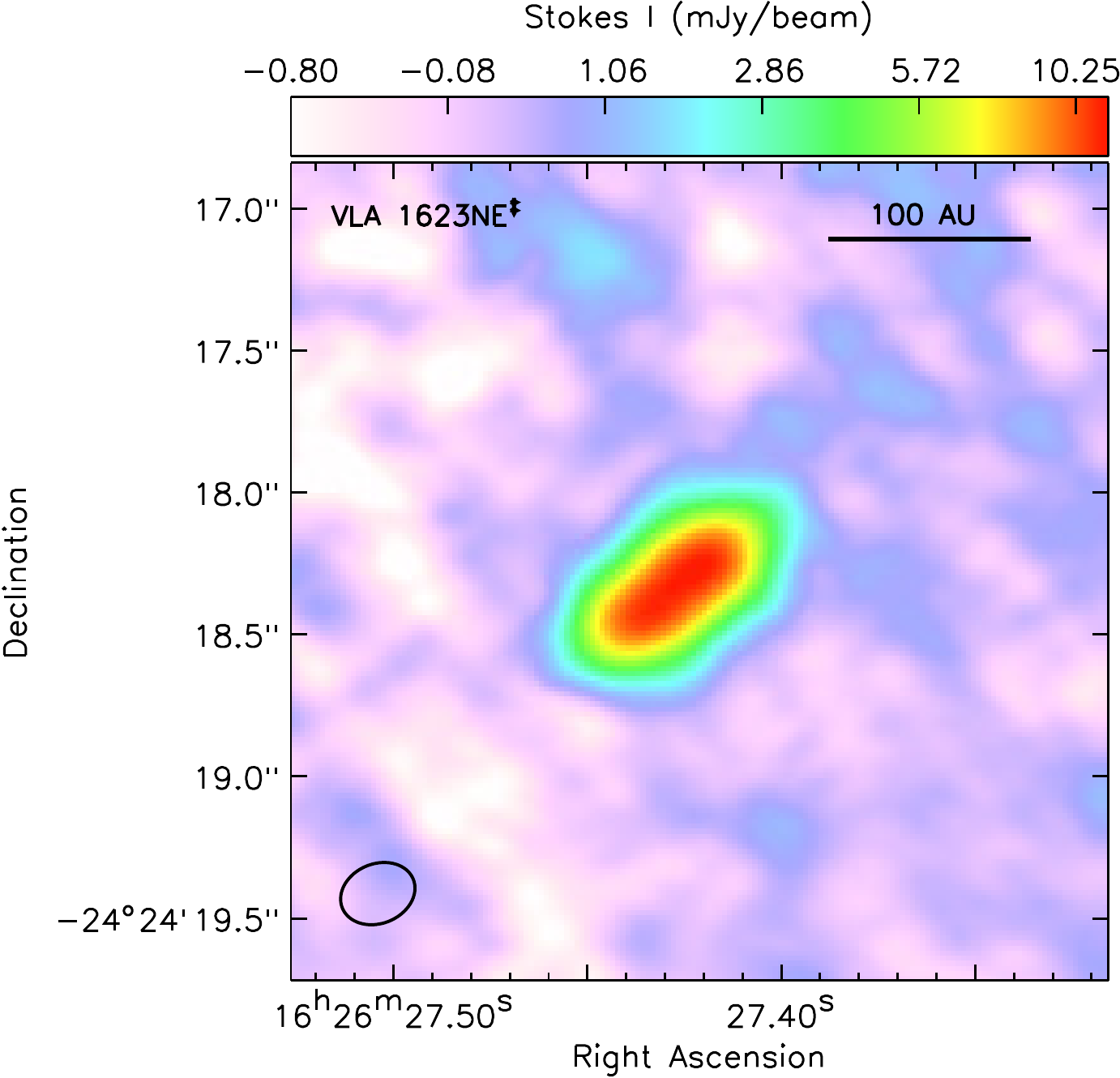}
\caption{Dust continuum map for VLA 1623NE.  This source was not detected in polarization.  \label{vla1623ne}}
\end{figure}

This field also contains a fourth object, VLA 1623NE, 19\arcsec\ northeast of VLA 1623A/B (see Figure  \ref{multiples}).  Figure \ref{vla1623ne} shows the continuum emission for this object.  It is not detected in polarization with a 3$\sigma$ upper limit of 3.9\%.  The continuum source is well resolved, with a deconvolved size of 81 au $\times$ 37 au and a mass of 8 \Mjupiter.  \citet{Kirk17} found a much higher mass of 0.071 \Msun\ (74 \Mjupiter) in lower resolution 3 mm observations, whereas \citet{Kawabe18} found $\sim 25$ \Mjupiter\ using ALMA and VLA observations between 1 mm and 6 cm.  As these observations have lower resolution, the dust masses may be affected by envelope emission.  

VLA 1623NE has unclear classification.  It is also called VLA 1623N1 \citep{ChenHirano18} and Source X \citep{Kawabe18}.  It has been previously seen at (sub)millimeter wavelengths \citep[e.g.,][]{Kirk17, ChenHirano18} and in X-rays with Chandra \citep[][]{Imanishi03, Gagne04}, but it is not seen in dense gas tracers \citep{ChenHirano18} or in the infrared \citep{Evans09, Gutermuth09}.  There is no corresponding object in the entire c2d catalogue at this position, although there is extended mid-infrared emission at its position that could be obscuring a fainter point source. \citet{Kawabe18} detected high velocity blue-shifted and red-shifted CO (2-1) emission near VLA 1623NE, but the emission overlap northeast of the source and there is no corresponding lobe southwest.  If this emission is from an outflow, \citet{Kawabe18} suggest the outflow axis must be along the plane of the sky and that VLA 1623NE is either a proto-brown dwarf or a very young low-mass protostar.  Since VLA 1623NE coincides with X-ray emission but lacks a clear outflow and dense gas, we consider it to be a Class II YSO.   Based on our estimated disk mass of 8 \Mjupiter, we suggest that VLA 1623NE is a Class II source that will form a low-mass star rather than a proto-brown dwarf.

\subsection{Field IRAS16288: ISO Oph 210}

Field IRAS 16288 contains the infrared source ISO Oph 210.  This source has not been well studied in the literature.  It was first identified in mid-infrared emission with ISO by \citet{Bontemps01} and subsequently detected in near-infrared emission with 2MASS \citep{Cutri03}.  This source was also detected in \emph{Spitzer} observations of Ophiuchus, but was classified as a star (``star+dust(IR1)'') in \citet{Evans09}.  \citet{HsiehLai13} revisited the \emph{Spitzer} observations and found that this source is consistent with an embedded YSO and  \citet{Duchene04} consider this object a Flat spectrum source based on its infrared spectral index.  But ISO Oph 210 has no corresponding dense core \citep[e.g.,][]{Pattle15} and unknown outflow properties.  Given its lack of complimentary information, we classify this object as a Flat spectrum source \citep{Duchene04}. 

\begin{figure}[h!]
\includegraphics[width=0.475\textwidth]{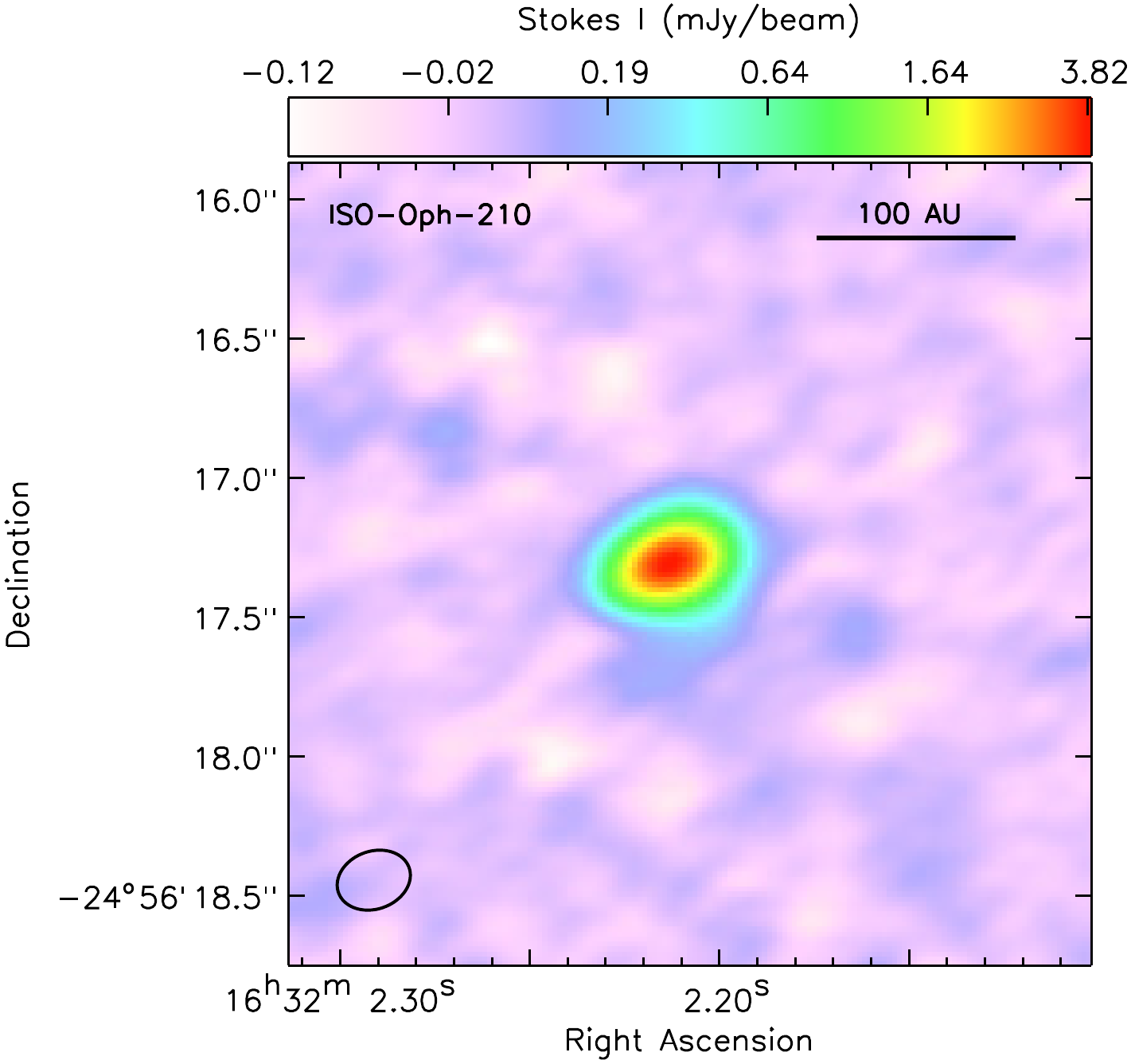}
\caption{Dust continuum map for ISO Oph 210.  This source was not detected in polarization.  \label{isoA}}
\end{figure}

Figure \ref{isoA} shows the continuum source for ISO Oph 210.  It is undetected in polarization with a 3$\sigma$ upper limit of 2\%, which is marginally significant given the typical polarization fraction for young stars.  The continuum source is also compact.  We find a deconvolved source size of 26 au $\times$ 14.4 au, with a mass of 0.8 \Mjupiter.  

\begin{figure}[h!]
\includegraphics[width=0.475\textwidth]{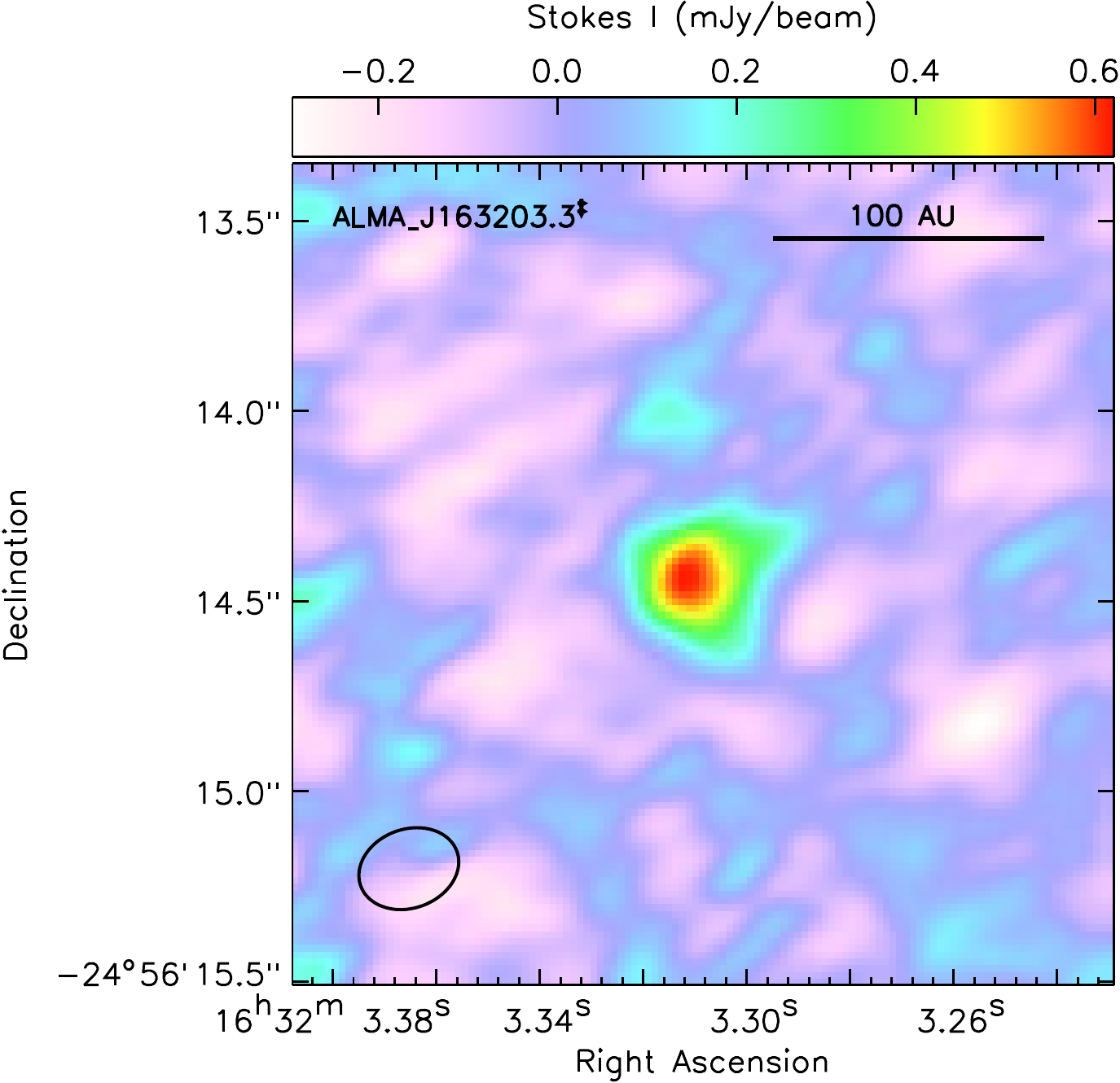}
\caption{Dust continuum map for ALMA\_J163203.3.  This source was not detected in polarization.  \label{isoB}}
\end{figure}

\citet{Duchene04} also detect a second near-infrared object 7.8\arcsec\ south of ISO Oph 210.  We do not see any object at this position.  Instead, we find a new detection roughly 15\arcsec\ east of the YSO.  Figure \ref{isoB} shows the continuum image of this source, which we call ALMA\_J163203.31-245614.43 (hereafter, ALMA\_J163203.3).  This object is much fainter than ISO Oph 210, and it is undetected in polarization with a 3$\sigma$ upper limit of 33\%.  We could find no corresponding emission within 5\arcsec\ of this source position at any other wavelength in a literature search with SIMBAD \citep{Wenger00} and Vizier \citep{Ochsenbein00}.  Based on its faint detection (peak Stokes I  S/N $\approx 7$) and its lack of complementary data, we classify ALMA\_J163203.3 as a background galaxy (see Section \ref{gal}).

\section{Off Axis Polarization}\label{offaxis}

In this appendix, we test the on-axis and off-axis polarization observations for IRAS 16293 and VLA 1623.  The ALMA Observatory limits reliable polarization measurements to the inner third of the primary beam FWHM, where polarization measurements at larger radial extents are considered less reliable.  At 1.3 mm, this area corresponds to the inner 4\arcsec\ of the primary beam.  Nevertheless, we have overlapping fields that are centered on different sources separated by roughly 5\arcsec\ (IRAS 16293) and 10\arcsec\ (VLA 1623), which puts the phase centers of the two fields outside of the nominal inner third of the primary beam FWHM.  We use the independent polarization measurements for both of these cases to determine whether or not we can use the off-axis polarization measurements or mosaic the overlapping fields.

\begin{figure}[h!]
\includegraphics[width=0.49\textwidth]{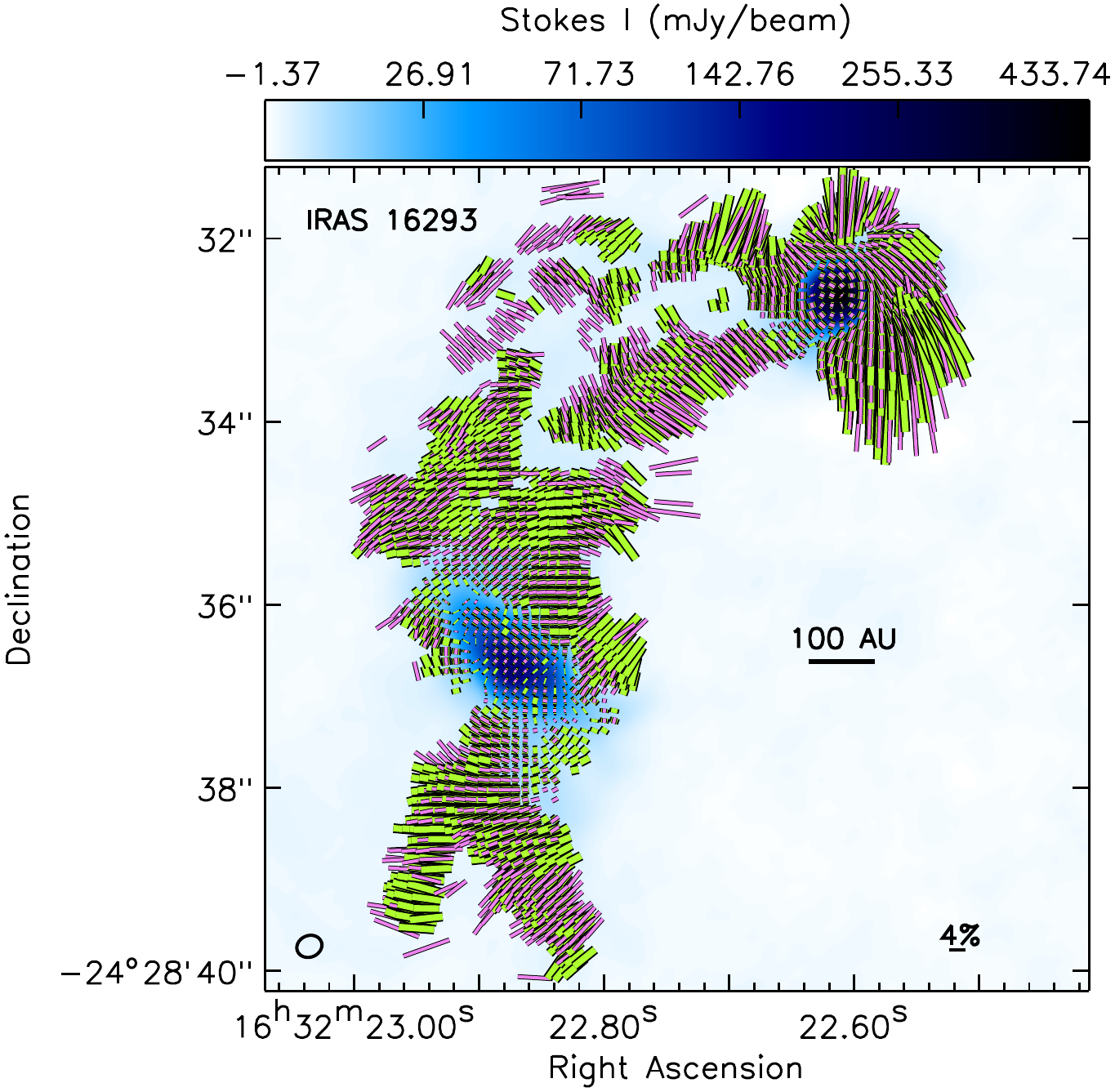}
\caption{Polarization for IRAS 16293A and IRAS 16293B from the individual fields.  Field c2d\_1008b (centered on IRAS 16293A) is shown by green line segments and Field c2d\_1008b (centered on IRAS 16293B) is shown in magenta line segments.  We also added the criterion that $\PF/\sPF > 3$ to remove spurious e-vectors at the edges and off source.  IRAS 16293A is located to the south and IRAS 16293B is located to the north (see Figure \ref{multiples}). \label{iras16293_axis}}
\end{figure}

Figure \ref{iras16293_axis} compares the polarization observations for Field c2d\_1008a (green line segments, centered on IRAS 16293A) and Field c2d\_1008b (purple line segments, centered on IRAS 16293B).  We use the same criteria listed in Section \ref{pol_overview}, with the additional criterion that $\PF/\sPF > 3$ to avoid spurious e-vectors at the edges of the map (see also, Figure \ref{iras16293}).  The polarization structure from the two maps have excellent agreement, even though the maps are on axis for one source and slightly off axis (e.g., positioned outside the inner third of the primary beam FWHM) for the other source. 

\begin{figure*}
\includegraphics[scale=0.9]{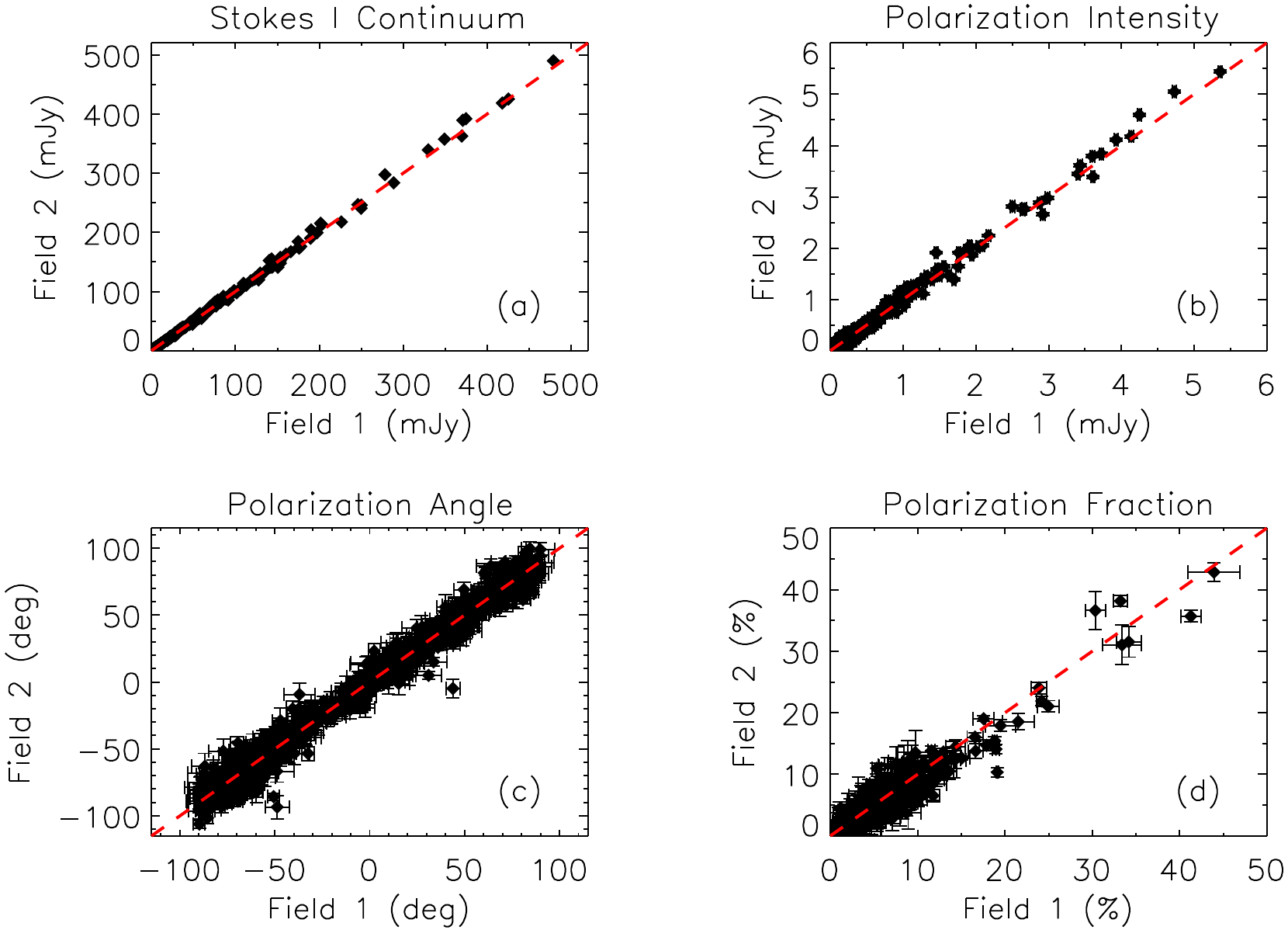}
\caption{Comparison between Fields c2d\_1008a (Field 1) and c2d\_1008b (Field 2) in (a) Stokes I continuum, (b) debiased polarized intensity, (c) polarization position angle, and (d) polarization fraction.  The data points only correspond to e-vectors that are measured in both fields in Figure \ref{iras16293_axis}.  Dashed lines show a one-to-one relation.  \label{iras16293_axis_results}}
\end{figure*}

To quantify the agreement, we compare both fields directly.  Figure \ref{iras16293_axis_results} shows the continuum, polarized intensity, polarization position angle, and polarization fraction from both fields, with c2d\_1008a as ``Field 1'' and c2d\_1008b as ``Field 2''.   We only show the data points for the 959 e-vectors given in Figure \ref{iras16293_axis} which have robust detections in both fields.   The dashed lines in each panel show a perfect one-to-one relation.  We find excellent agreement between the two fields, even though the two fields contain both compact and extended emission and both fields contain emission that extends beyond the inner third of the primary beam FWHM (see Figure \ref{multiples}).  

We also have two overlapping fields of VLA 1623, with one field centered on VLA 1623A and VLA 1623B, and the other field centered on VLA 1623W.  The VLA 1623A/B and VLA 1623W field centers are separated by a larger distance ($\sim 10$\arcsec) than the two IRAS 16293 fields (5\arcsec) and extend over an area beyond the inner half of the primary beam (see Figure \ref{multiples}).  Thus, the off axis uncertainties may be more significant for this region.

\begin{figure*}[h!]
\centering
\includegraphics[scale=0.55]{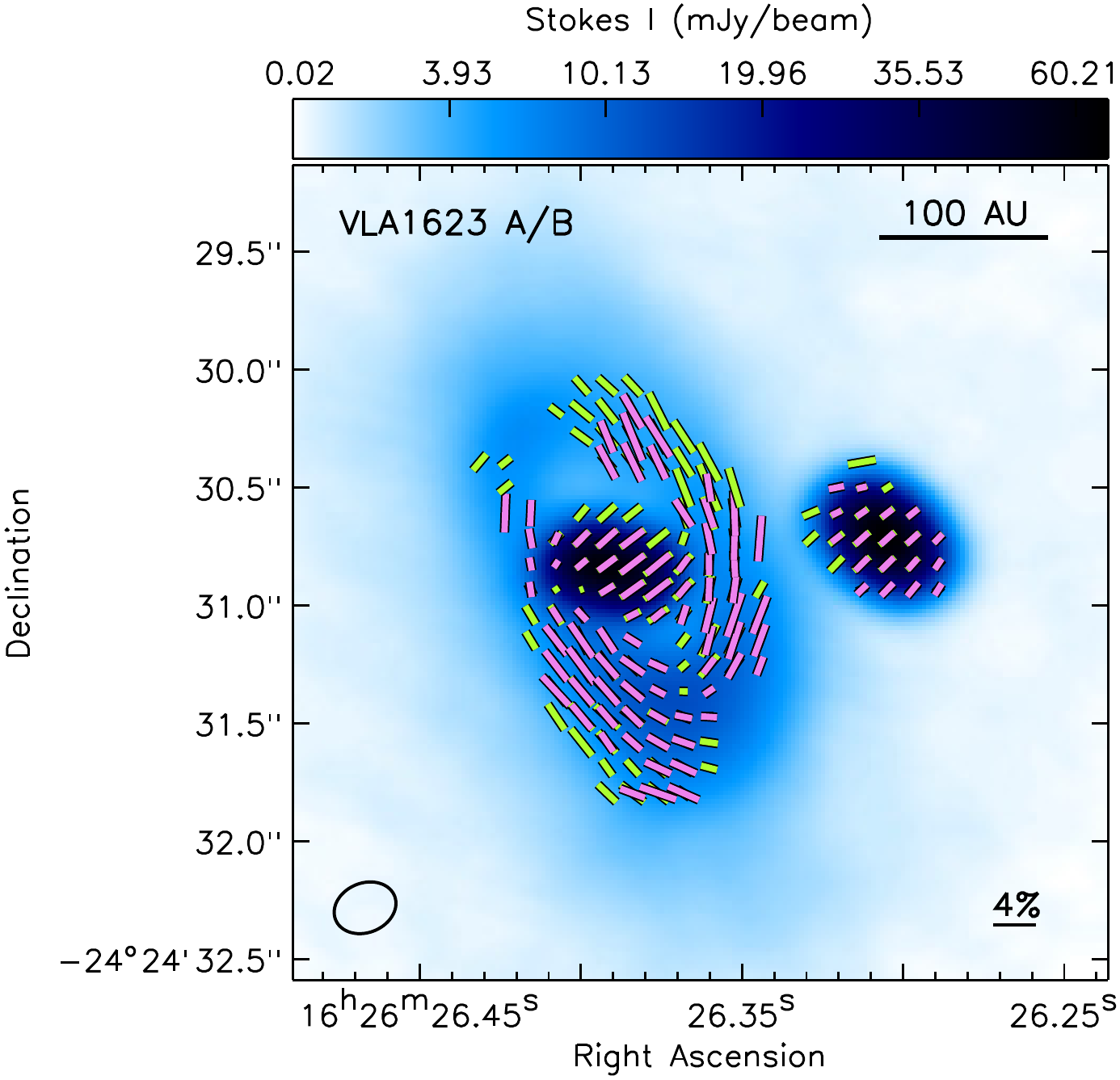}
\qquad
\includegraphics[scale=0.55]{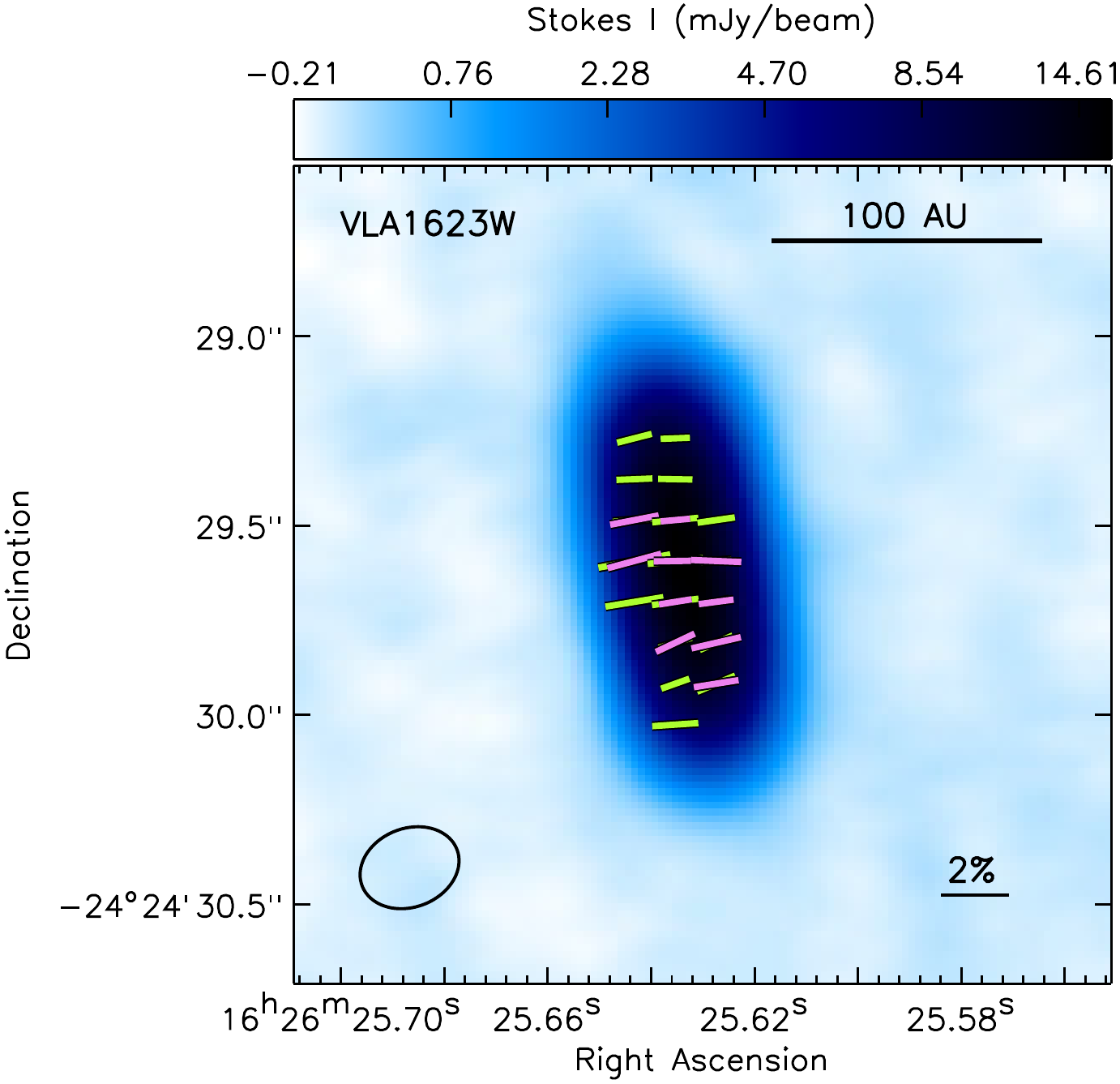}
\caption{Dust polarization maps for VLA 1623A/B (left) and VLA 1623W (right).   In both panels, the on-axis e-vectors are shown in green and the off-axis e-vectors are shown in purple.  We show only those e-vectors that match the same criteria as in Figures \ref{vla1623ab} and \ref{vla1623w}, respectively.
\label{vla1623_axis}}
\end{figure*}

Figure \ref{vla1623_axis} compares the polarization observations from the two fields containing VLA 1623.  In both cases, the green line segments show the on-axis polarization results and the purple line segments show the off-axis results using the same criteria listed in Section \ref{pol_overview}, with the additional criterion that $I/\sigma_I > 50$ to avoid spurious e-vectors at the edges of the disk in VLA 1623A (see also, Figure \ref{vla1623ab}).  As with IRAS 16293, we find good agreement between the on-axis and off-axis polarization structure, suggesting that the off-axis data are reliable.

\begin{figure*}
\includegraphics[scale=0.9]{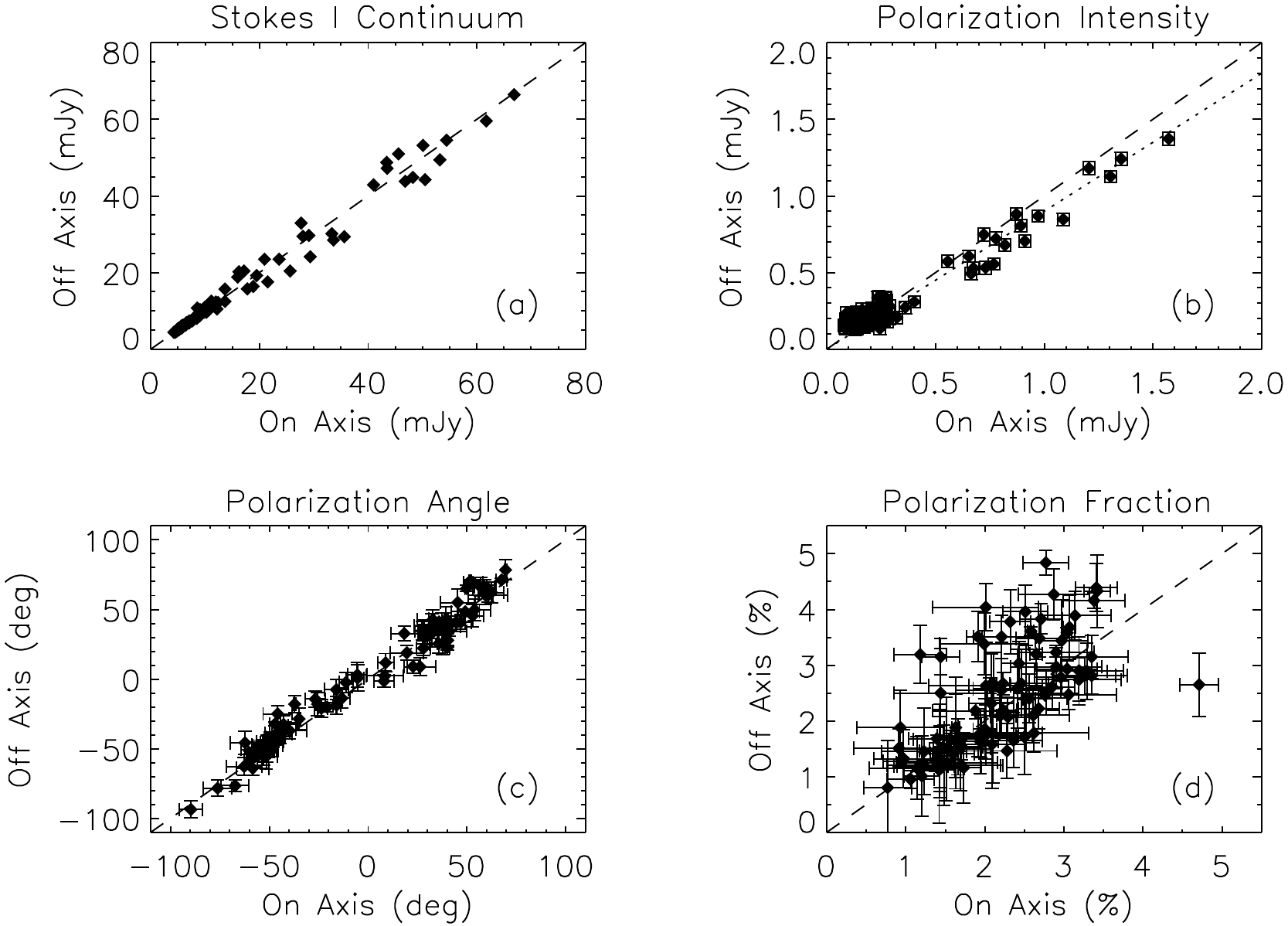}
\caption{Comparison between VLA 1623A/B from the on-axis field and off-axis field in (a) Stokes I continuum, (b) debiased polarized intensity, (c) polarization position angle, and (d) polarization fraction.  The data points only correspond to e-vectors that are measured in both fields in Figure \ref{vla1623_axis}.  Dashed lines show a one-to-one relation.  The dotted line in (b) shows a linear least squares fit to the observations, with a slope of $0.90 \pm 0.01$. \label{vla1623_axis_results}}
\end{figure*}

Figure \ref{vla1623_axis_results} compares the polarization measurements for VLA 1623A/B in the on-axis (Field VLA1623b) data and the off-axis (Field VLA1623a) data.   This figure contains data points for only the 93 e-vectors given in Figure \ref{vla1623_axis} which have robust detections in both fields.   In general, we find broad agreement between the on-axis and off-axis measurements, with near one-to-one relations for the Stokes I continuum and polarization position angle.  The polarization intensity, however, appears to be slightly underestimated in the off-axis map over the on-axis map at the high intensity end and overestimated at the low intensity end.  The dotted line in Figure \ref{vla1623_axis_results}b shows a best-fit linear least squares relation with a slope of $0.90 \pm 0.01$.  As a consequence, the polarization fractions can vary by roughly a factor of two between the on-axis and off-axis measurements.    

\begin{figure*}
\includegraphics[scale=0.9]{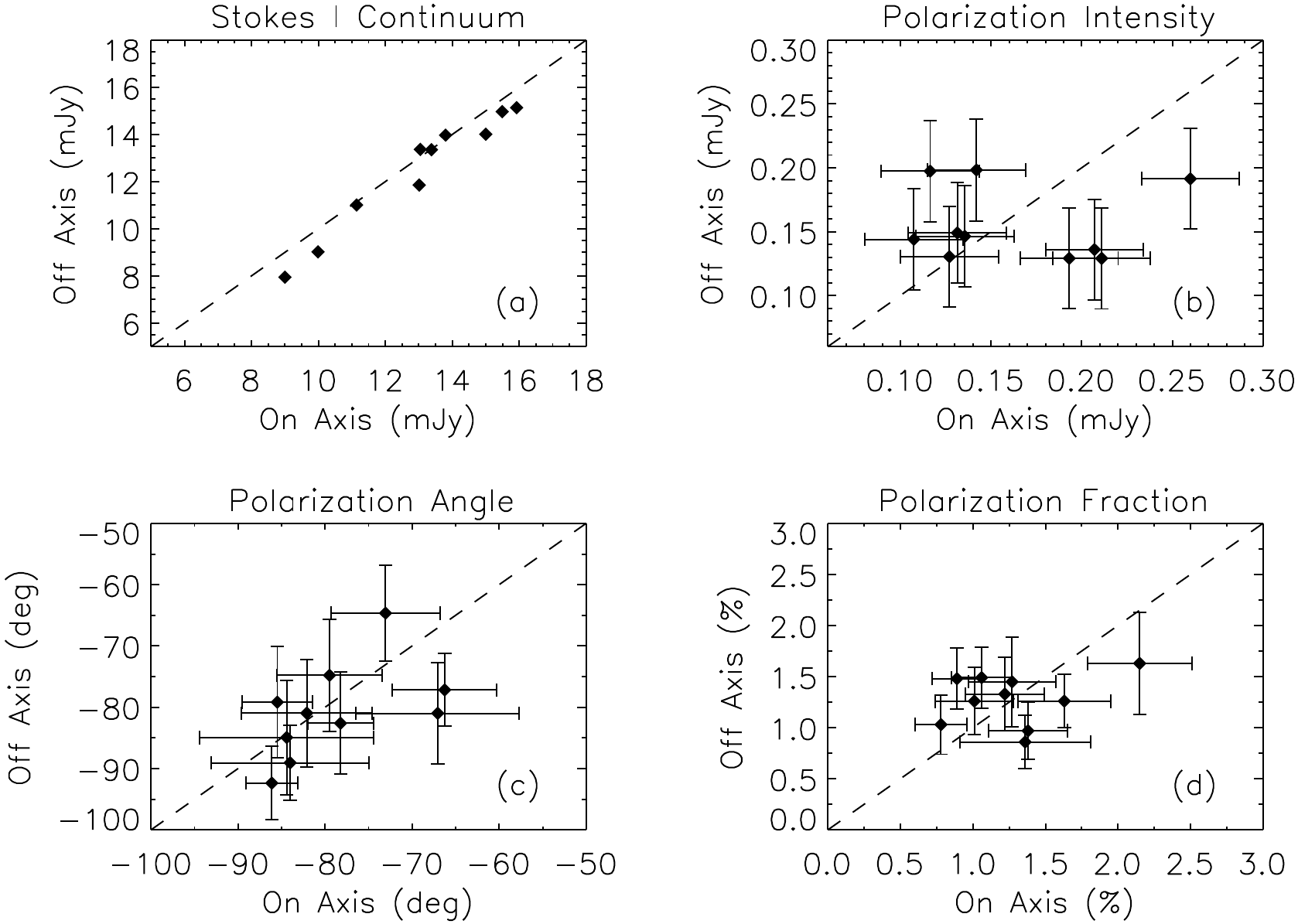}
\caption{Same as Figure \ref{vla1623_axis_results} but for VLA 1623W. \label{vla1623w_axis_results}}
\end{figure*}

Figure \ref{vla1623w_axis_results} shows the corresponding results for VLA 1623W.  VLA 1623W is considerably fainter and smaller than VLA 1623A/B resulting in only 10 matching e-vectors between the on-axis and off-axis maps.  With such small numbers, it is difficult to draw any conclusions about the reliability of the off-axis observations.  We use the on-axis and off-axis results for VLA 1623A/B instead and include this figure for completeness.

The IRAS 16293 fields are separated by 5\arcsec, whereas the VLA 1623 fields are separated by 10\arcsec.  Thus, we can expect the off-axis VLA 1623A/B sources to be more susceptible to uncertainties in the polarization calibration than the off-axis IRAS 16293 observations.  The strong agreement in Figure \ref{iras16293_axis_results} suggests that our polarization measurements should be considered robust within roughly the inner half of the primary beam FWHM (e.g., a radius of 6\arcsec).  While the polarization measurements are less reliable with larger separations, we note that a 10\arcsec\ offset from the phase center still produces good polarization position angles.  These results indicate that we can reliably mosaic the two IRAS 16293 fields, but we do not attempt to mosaic the two VLA 1623 fields.

\section{Disk Optical Depth}\label{tau}

In this appendix, we calculate the spectral index, $\alpha$, for the disks with multiple continuum wavelengths in the literature as a proxy for the disk optical depth.  We use optical depth to help determine why some disks are undetected in polarization and to also investigate the polarization mechanism for those disks that are detected.  For simplicity, we infer optical depths from the dust opacity index, $\beta$, assuming $\beta \rightarrow 0$ as the dust emission becomes optically thick (e.g., the dust emits like a perfect black body).  We note that $\beta$ is only a proxy for optical depth.  It can also vary from changes in dust grain size, shape, composition, and structure \citep[e.g.,][]{Ossenkopf94, Ormel11}.  In addition, a temperature gradient in the disk can broaden the spectral energy distribution and flatten $\beta$  \citep[e.g.,][]{Shetty09}.  Thus, a true characterization of dust opacity for these disks will require detailed modeling of their intensity profiles and geometry over multiple wavelengths, which is beyond the scope of the current paper. 

We measure $\beta$ by combining our 233 GHz fluxes with complementary fluxes in the literature.  We only consider emission $> 10$ GHz from our 233 GHz observations to ensure there is a large enough lever arm to constrain the spectral index and $\beta$.  We also require comparable resolution $\lesssim 0.5$\arcsec\ to our ALMA observations so that the data cover spatial scales similar to our 233 GHz data.  This last criterion is necessary as most of our disks are embedded in dense envelopes.  Lower-resolution observations may therefore be biased toward the envelope emission and have elevated fluxes.   

Table \ref{disk_beta} lists the disks where we found complementary, high-resolution observations in the literature.  Column 2 lists our 233 GHz flux density from Table \ref{cont_results}.  Columns 3 and 4 give the frequency and total flux density from the literature (reference in the fifth column).  The fluxes are similarly obtained from Gaussian fits.  If the reference did not quote an error for their source fluxes, we assume 10\%.  Columns 6 and 7 give our estimated values of $\alpha$ and $\beta$ from combining our 233 GHz fluxes with the literature values and Column 7 gives our assessment of the optical depth.  We measure $\beta$ using the flux density spectral index, $S_{\nu} \sim \nu^{\alpha}$, where $\alpha = \beta + 2$ \citep[e.g.,][]{BeckwithSargent91} for dust emission along the Rayleigh-Jeans tail of the SED. 

{\setlength{\extrarowheight}{0.8pt}%
\begin{table*}
\caption{Dust Opacity Estimations}\label{disk_beta}
\begin{tabular}{lccclccl}
\hline\hline
Source			& $S_{1.3}$	& $\nu$\tablenotemark{a} 		& $S_{\nu}$\tablenotemark{a}  &	 Ref\tablenotemark{a} & $\alpha$ & $\beta$	&  Comments\\
				&(mJy)		&  (GHz) 		& (mJy)	     &  		&    &  \\
\hline														 
GSS 30 IRS 1		& 13.4 $\pm$ 0.13	 & 349.6	& 31.1 $\pm$ 0.2	& 1	& 2.12 $\pm$ 0.04 & 0.12 $\pm$ 0.04 &  Appears optically thick \\  
GSS 30 IRS 3 		& 158.5 $\pm$ 1.7	 & 349.6	& 379.3 $\pm$ 3.6 	& 1	& 2.19 $\pm$ 0.05 & 0.19 $\pm$ 0.05 & Appears optically thick \\  
Oph-emb-9		& 45.0 $\pm$ 0.2	& 343.8	& 105.0 $\pm$ 0.5	& 2	& 2.18 $\pm$ 0.03 & 0.18 $\pm$ 0.03 & Appears optically thick \\
GY 91			& 88.8 $\pm$ 6.5	 & 348.6	& 258 $\pm$ 26	& 3	& 2.64 $\pm$ 0.44 & 0.64 $\pm$ 0.44 & Not optically thick \\
Oph-emb-6		& 53.1 $\pm$ 0.3	& 348.6	& 128.3 $\pm$ 0.6	& 2	& 2.27 $\pm$ 0.03 & 0.27 $\pm$ 0.03 & Appears optically thick \\
WL 17			& 51.26 $\pm$ 0.83	 & 349.6	& 130 $\pm$ 0.5	& 1	& 2.29 $\pm$ 0.05 & 0.29 $\pm$ 0.05 & Appears optically thick  \\
Elias 29			& 17.2 $\pm$ 0.2	& 348.6	& 41.2 $\pm$ 0.6	& 2	& 2.25 $\pm$ 0.06 & 0.25 $\pm$ 0.06 & Appears optically thick \\
IRS 43-A			& 15.08 $\pm$ 0.33	& 252	& 18.5 $\pm$ 1.9	& 4 	& 2.61 $\pm$ 1.5 & 0.61 $\pm$ 1.5 &  Not optically thick \\
$\cdots$			& 15.08 $\pm$ 0.33 & 348.6	& 41.6 $\pm$ 1.9	& 2	& 2.61 $\pm$ 0.17 & 0.61 $\pm$ 0.17 & Not optically thick \\
IRS 43-B			& 1.92 $\pm$ 0.07	& 348.6	& 7.5 $\pm$ 1.0 	& 2	& 3.5 $\pm$ 0.4 & 1.5 $\pm$ 0.4 	& Not optically thick \\
IRS 44			& 12.0 $\pm$ 0.3	& 348.6	& 38.6 $\pm$ 1.2	& 2	& 3.00 $\pm$ 0.15	& 1.00 $\pm$ 0.15	& Not optically thick \\
Oph-emb-1		& 12.5 $\pm$ 0.13 	 & 217.24	&  10.1 $\pm$ 0.4	& 5	& 3.04 $\pm$ 0.72 	& 1.04 $\pm$ 0.72 	& Not optically thick \\
$\cdots$			& 12.5 $\pm$ 0.13	& 344.6	&  36.7 $\pm$ 1.9	& 5	& 2.75 $\pm$ 0.16   & 0.75 $\pm$ 0.16   & Not optically thick \\
$\cdots$			& 12.5 $\pm$ 0.13	& 348.6	&  39.8 $\pm$ 0.7	& 2	& 2.98 $\pm$ 0.06	& 0.98 $\pm$ 0.06	& Not optically thick \\
IRS 63			& 312.2 $\pm$ 8.8	& 343.48	& 776 $\pm$ 35	& 6	& 2.35 $\pm$ 0.19 	& 0.35 $\pm$ 0.19 	& Appears optically thick \\
Oph-emb-15		& 3.7 $\pm$ 0.1	& 343.8	& 11.3 $\pm$ 0.3	& 2	& 2.87 $\pm$ 0.13	& 0.87 $\pm$ 0.13	& Not optically thick \\
IRS 67-A			& 8.59 $\pm$ 0.2  	 & 343.8	& 35.7 $\pm$ 7		& 2	& 3.66 $\pm$ 0.64 	& 1.66 $\pm$ 0.64 	& Not optically thick \\
IRS 67-B 			& 53.4 $\pm$ 0.4  	 & 343.8	& 155.5 $\pm$ 4	& 2	& 2.74 $\pm$ 0.09 	& 0.74 $\pm$ 0.09 	& Not optically thick \\  
VLA 1623W 		& 65.5 $\pm$ 1.2 	 & 343.8	& 159 $\pm$ 5 		& 7	& 2.28 $\pm$ 0.13 	& 0.28 $\pm$ 0.13 	&Appears optically thick \\  
VLA 1623B		& 123.7 $\pm$ 5 	& 343.8	&  324 $\pm$ 5		& 7	& 2.48 $\pm$ 0.08 	& 0.48 $\pm$ 0.08 	&Appears optically thick \\
VLA 1623A		& 141.8 $\pm$ 5.9	 & 343.8	&  368 $\pm$ 26	& 7	& 2.45 $\pm$ 0.29 	& 0.45 $\pm$ 0.29 	& Appears optically thick \\ 
\hline
\end{tabular}
\begin{tablenotes}[normal,flushleft]
\item \tablenotemark{a} Frequency and flux for select disks in the literature.  We assume 10\%\ errors on flux if the errors are not reported.  We exclude IRS 43-B because it was too faint to robustly measure its source flux relative to its larger circumbinary disk.  References correspond to (1) P. Sheehan private communication,  (2) \citealt{ArturV19}, (3) \citealt{vanderMarel19}, (4) \citealt{Brinch16}, (5) \citealt{Hsieh19oph1}, (6) \citealt{Cox17}, (7) \citealt{Harris18}
\end{tablenotes}
\end{table*}
}

 We note, however, that if the dust emission is cold (e.g., $\lesssim 10$ K), even the 1.3 mm emission may not be on the Rayleigh-Jeans tail, such that spectral index will appear flatter.  We require an estimate of the dust temperature to properly correct the SED, however.   Since we do not have accurate dust temperatures, we use the spectral index slope alone to give a first order estimate of $\beta$.  The spectral index may also appear steeper or flatter (or negative) from  dust scattering \citep{Liu19scat, Zhu19}, which can increase or decrease $\alpha$ depending on the albedo of the dust grains, the grain size distribution, and whether or not the dust emission is optically thick.  Since several disks show polarization signatures consistent with dust self-scattering, we expect the dust scattering to also affect the SED slope.  Nevertheless, correcting for dust scattering in the SED would require extensive multi-wavelength observations and radiative transfer modeling, which is beyond the scope of this paper and beyond the current datasets available for most sources.  We therefore assume the effects of dust scattering on $\alpha$ (and $\beta$) are negligible in our analysis.

Eighteen disks have complementary high-resolution data in the literature for at least one alternative wavelength.  Oph-emb-1 and IRS 43-A have several complementary datasets in the literature, and we report the values of $\alpha$ and $\beta$ for each of them separately.  Fifthteen of the sampled disks (83\%) have $\beta < 1$ ($\alpha < 3$) and ten (56\%) have $\beta < 0.5$ ($\alpha < 2.5$).  IRS 43-B, IRS 44, Oph-emb-1, and IRS 67-A are the only two disks with $\beta \gtrsim 1$.  Oph-emb-1, however, has a slightly shallower value of $\beta \sim 0.75$ using  344.6 GHz data from \citet{Hsieh19oph1}.  For simplicity, we report the median slope of 0.98 $\pm$ 0.06 for Oph-emb-1 in Section \ref{morph_desc}.  Finally, we note that IRAS 16293B has a known spectral index of $\alpha \approx 2$ down to radio frequencies \citep[e.g.,][]{Chandler05}, which indicate that $\beta \approx 0$ for this object.  We do not include this measurement in Table \ref{disk_beta} as we do not have a reliable measurement of the 1.3 mm flux density for its disk.

The disks with uniform polarization aligned with their minor axes have $\beta < 0.5$ ($\alpha < 2.5$).  If $\beta < 0.5$ indicates these disks are optically thick, then the polarization morphology should be dominated by self-scattering processes over grain alignment.  Polarization from grain alignment with a magnetic field is suppressed relative to dust self-scattering if the emission is optically thick \citep{Yang17}.  When dust grains are aligned in a specific direction (e.g., with a magnetic field), their thermal emission is expected to be preferentially aligned with the long axis and will appear polarized \citep[e.g.,][]{Hildebrand00, ChoLazarian07}.  If the emission originates from an isothermal slab (a rough approximation of a disk midplane), we can expect to detect emission both parallel to and perpendicular to the dust grain long axis as the optical depth increases.   These two contributions will then become equal in the optically thick ($\tau \gg 1$) case and their resulting emission will appear unpolarized \citep{Yang17}.   As a result, one will not detect any polarization from grain alignment in the optically thick limit, but polarization from self-scattering can be detected. 

To first order, we consider disks with $\beta < 0.5$ to be optically thick sources, whereas the sources with $\beta > 0.5$ do not appear to be optically thick.  Sources with $\beta > 0.5$ also do not show clear signatures of dust self-scattering, even if they are highly inclined.  In particular, Oph-emb-1 and IRS 67-B are inconsistent with dust self-scattering for inclined, optically thick disks (see Section \ref{mech_discuss}).   We note that there are several optically thick disks ($\beta < 0.5$) that is not detected in polarization (e.g., WL 17).  We discuss these source in more detail in Section \ref{non_det}.

\end{appendix}


\bibliography{references}

\end{document}